\magnification \magstep1
\raggedbottom
\hsize=15truecm
\vsize=23truecm
\openup 2\jot
\voffset6truemm
\def\II{{\rm 1\!\hskip-1pt I}}
\def\cstok#1{\leavevmode\thinspace\hbox{\vrule\vtop{\vbox
{\hrule\kern1pt
\hbox{\vphantom{\tt/}\thinspace{\tt#1}\thinspace}}
\kern1pt\hrule}\vrule}\thinspace}
\centerline {\bf COMPLEX GEOMETRY OF NATURE AND}
\centerline {\bf GENERAL RELATIVITY}
\vskip 1cm
\leftline {Giampiero Esposito}
\vskip 1cm
\noindent 
{\it INFN, Sezione di Napoli,
Mostra d'Oltremare Padiglione 20, 80125 Napoli, Italy}
\vskip 0.3cm
\noindent
{\it Dipartimento di Scienze Fisiche, Universit\`a degli Studi di
Napoli Federico II, Complesso Universitario di Monte S. Angelo,
Via Cintia, Edificio G, 80126 Napoli, Italy}
\vskip 1cm
\noindent
{\bf Abstract.} An attempt is made of giving a self-contained
introduction to holomorphic ideas in general relativity, following
work over the last thirty years by several authors. The main topics
are complex manifolds, spinor and twistor methods, heaven spaces. 
\vskip 100cm
\centerline {\it CHAPTER ONE}
\vskip 1cm
\centerline {\bf INTRODUCTION TO COMPLEX SPACE-TIME}
\vskip 1cm
\noindent
The physical
and mathematical motivations for studying complex space-times
or real Riemannian four-manifolds in gravitational physics
are first described.
They originate from algebraic geometry, Euclidean quantum
field theory, the path-integral approach to quantum
gravity, and the theory of conformal gravity.
The theory of complex manifolds is then briefly outlined.
Here, one deals with paracompact Hausdorff spaces where
local coordinates transform by complex-analytic
transformations. Examples are given such as complex
projective space $P_{m}$, non-singular sub-manifolds
of $P_{m}$, and orientable surfaces. The
plan of the whole paper is eventually presented, with
emphasis on two-component spinor calculus, Penrose transform
and Penrose formalism for spin-${3\over 2}$ potentials.
\vskip 100cm
\centerline {\bf 1.1 From Lorentzian space-time to complex space-time}
\vskip 1cm
\noindent
Although Lorentzian geometry is the mathematical framework
of classical general relativity and can be seen as a good
model of the world we live in (Hawking and Ellis 1973,
Esposito 1992, Esposito 1994), the theoretical-physics
community has developed instead many models based on a complex
space-time picture. We postpone until section 3.3 the
discussion of real, complexified or complex manifolds, and
we here limit ourselves to say that the main motivations for studying
these ideas are as follows.

(1) When one tries to make sense of quantum field theory
in flat space-time, one finds it very convenient to
study the Wick-rotated version of Green functions,
since this leads to well defined mathematical calculations
and elliptic boundary-value problems. At the end, quantities
of physical interest are evaluated by analytic continuation
back to {\it real} time in Minkowski space-time.

(2) The singularity at $r=0$ of the Lorentzian Schwarzschild 
solution disappears on the real Riemannian section of the
corresponding complexified space-time, since $r=0$ no longer
belongs to this manifold (Esposito 1994). Hence there are
real Riemannian four-manifolds which are singularity-free,
and it remains to be seen whether they are the most 
fundamental in modern theoretical physics.

(3) Gravitational instantons shed some light on possible
boundary conditions relevant for path-integral quantum
gravity and quantum cosmology 
(Gibbons and Hawking 1993, Esposito 1994).

(4) Unprimed and primed spin-spaces 
are not (anti-)isomorphic
if Lorentzian space-time is replaced by a complex or real
Riemannian manifold. Thus, for example, the Maxwell field
strength is represented by two independent symmetric spinor
fields, and the Weyl curvature is also represented by two
independent symmetric spinor fields (see (2.1.35) and (2.1.36)). 
Since such spinor
fields are no longer related by complex conjugation
(i.e. the (anti-)isomorphism between the two spin-spaces), one of
them may vanish without the other one having to vanish
as well. This property gives rise to the so-called self-dual
or anti-self-dual gauge fields, as well as to self-dual or
anti-self-dual space-times (section 4.2).

(5) The geometric study of this special class of space-time
models has made substantial progress by using twistor-theory
techniques. The underlying idea (Penrose 1967, Penrose 1968,
Penrose and MacCallum 1973, Penrose 1975, Penrose 1977,
Penrose 1980, Penrose and Ward 1980, Ward 1980a--b, Penrose 1981,
Ward 1981a--b, Huggett 1985, Huggett and Tod 1985, Woodhouse 1985,
Penrose 1986, Penrose 1987, Yasskin 1987, Manin 1988,
Bailey and Baston 1990, Mason and Hughston 1990, 
Ward and Wells 1990, Mason and Woodhouse 1996)
is that conformally invariant concepts
such as null lines and null surfaces are the basic
building blocks of the world we live in, whereas space-time
points should only appear as a derived concept. By using
complex-manifold theory, twistor theory provides an
appropriate mathematical description of this key idea.

A possible mathematical motivation for twistors can be
described as follows (papers 99 and 100 in Atiyah (1988)). In
two real dimensions, many interesting problems are best tackled
by using complex-variable methods. In four real dimensions,
however, the introduction of two complex coordinates is not,
by itself, sufficient, since no preferred choice exists. In other
words, if we define the complex variables
$$
z_{1} \equiv x_{1}+ix_{2},
\eqno (1.1.1)
$$
$$
z_{2} \equiv x_{3}+ix_{4},
\eqno (1.1.2)
$$
we rely too much on this particular coordinate system, and a
permutation of the four real coordinates $x_{1},x_{2},x_{3},x_{4}$
would lead to new complex variables not well related to the
first choice. One is thus led to introduce three complex variables
$\Bigr(u,z_{1}^{u},z_{2}^{u}\Bigr)$: the first variable $u$ tells
us which complex structure to use, and the next two are the
complex coordinates themselves. In geometric language, we start
with the complex projective three-space $P_{3}(C)$ (see section 1.2)
with complex homogeneous coordinates $(x,y,u,v)$, and we remove the
complex projective line given by $u=v=0$. Any line in
$\Bigr(P_{3}(C)-P_{1}(C)\Bigr)$ is thus given by a pair of
equations
$$
x=au+bv,
\eqno (1.1.3)
$$
$$
y=cu+dv.
\eqno (1.1.4)
$$
In particular, we are interested in those lines for which
$c=-{\overline b},d={\overline a}$. The determinant $\Delta$ of
(1.1.3) and (1.1.4) is thus given by
$$
\Delta=a{\overline a}+b{\overline b}={|a|}^{2}
+{|b|}^{2},
\eqno (1.1.5)
$$
which implies that the line given above never intersects
the line $x=y=0$, with the obvious exception of the case when 
they coincide. Moreover, no two lines intersect, and they fill
out the whole of $\Bigr(P_{3}(C)-P_{1}(C)\Bigr)$. This leads to
the fibration $\Bigr(P_{3}(C)-P_{1}(C)\Bigr)\longrightarrow
R^{4}$ by assigning to each point of $\Bigr(P_{3}(C)-P_{1}(C)\Bigr)$
the four coordinates $\Bigr({\rm Re}(a),{\rm Im}(a),
{\rm Re}(b),{\rm Im}(b)\Bigr)$.
Restriction of this fibration to a plane of the form
$$
\alpha u + \beta v =0,
\eqno (1.1.6)
$$
yields an isomorphism $C^{2} \cong R^{4}$, which depends on the
ratio $(\alpha,\beta) \in P_{1}(C)$. This is why the picture
embodies the idea of introducing complex coordinates.

Such a fibration depends on the conformal structure of $R^{4}$.
Hence, it can be extended to the one-point compactification
$S^{4}$ of $R^{4}$, so that we get a fibration $P_{3}(C)
\longrightarrow S^{4}$ where the line $u=v=0$, previously 
excluded, sits over the point at $\infty$ of
$S^{4}=R^{4} \cup \Bigr \{ \infty \Bigr \}$. This fibration is
naturally obtained if we use the quaternions $H$ to identify
$C^{4}$ with $H^{2}$ and the four-sphere $S^{4}$ with $P_{1}(H)$,
the quaternion projective line. We should now recall that the
quaternions $H$ are obtained from the vector space $R$ of real
numbers by adjoining three symbols $i,j,k$ such that
$$
i^{2}=j^{2}=k^{2}=-1,
\eqno (1.1.7)
$$
$$
ij=-ji=k, \; 
jk=-kj=i, \; 
ki=-ik=j .
\eqno (1.1.8)
$$
Thus, a general {\it quaternion} $\in H$ is defined by
$$
x \equiv x_{1}+x_{2}i+x_{3}j+x_{4}k,
\eqno (1.1.9)
$$
where $\Bigr(x_{1},x_{2},x_{3},x_{4}\Bigr) \in R^{4}$,
whereas the conjugate quaternion $\overline x$ is given by
$$
{\overline x} \equiv x_{1}-x_{2}i-x_{3}j-x_{4}k .
\eqno (1.1.10)
$$
Note that conjugation obeys the identities
$$
{\overline {(xy)}}={\overline y} \; {\overline x},
\eqno (1.1.11)
$$
$$
x {\overline x}={\overline x} x =\sum_{\mu=1}^{4}
x_{\mu}^{2} \equiv {|x|}^{2}.
\eqno (1.1.12)
$$
If a quaternion does not vanish, it has a unique inverse
given by
$$
x^{-1} \equiv {{\overline x} \over {|x|}^{2}}.
\eqno (1.1.13)
$$
Interestingly, if we identify $i$ with $\sqrt{-1}$,
we may view the complex numbers $C$ as
contained in $H$ taking $x_{3}=x_{4}=0$. Moreover,
every quaternion $x$ as in (1.1.9) has a unique
decomposition
$$
x=z_{1}+z_{2}j,
\eqno (1.1.14)
$$
where $z_{1} \equiv x_{1}+x_{2}i$, $z_{2} \equiv x_{3}+x_{4}i$,
by virtue of (1.1.8). This property enables one to identify
$H$ with $C^{2}$, and finally $H^{2}$ with $C^{4}$, as we
said following (1.1.6).

The map $\sigma: P_{3}(C) \longrightarrow P_{3}(C)$
defined by
$$
\sigma(x,y,u,v)=(-{\overline y},{\overline x},-{\overline v},
{\overline u}),
\eqno (1.1.15)
$$
preserves the fibration because $c=-{\overline b},
d={\overline a}$, and induces the antipodal map on each
fibre. We can now lift problems from $S^{4}$ or $R^{4}$
to $P_{3}(C)$ and try to use complex methods.
\vskip 1cm
\centerline {\bf 1.2 Complex manifolds}
\vskip 1cm
\noindent
Following Chern (1979), we now describe some basic ideas and
properties of complex-manifold theory. The reader should
thus find it easier (or, at least, less difficult) to
understand the holomorphic ideas used in the rest of the
paper.

We know that a manifold is a space which is locally similar
to Euclidean space in that it can be covered by coordinate
patches. More precisely (Hawking and Ellis 1973), we say
that a {\it real} $C^{r}$ $n$-dimensional manifold 
$\cal M$ is a set $\cal M$ together with a $C^{r}$ atlas
$\Bigr \{U_{\alpha},\phi_{\alpha} \Bigr \}$, i.e. a collection
of charts $\Bigr(U_{\alpha},\phi_{\alpha}\Bigr)$, where the
$U_{\alpha}$ are subsets of $\cal M$ and the $\phi_{\alpha}$
are one-to-one maps of the corresponding $U_{\alpha}$ into open
sets in $R^{n}$ such that

(i) $\cal M$ is covered by the $U_{\alpha}$, i.e.
${\cal M}=\bigcup_{\alpha} U_{\alpha}$

(ii) if $U_{\alpha} \cap U_{\beta}$ is non-empty, the map
$$
\phi_{\alpha} \circ \phi_{\beta}^{-1}:
\phi_{\beta} \Bigr(U_{\alpha} \cap U_{\beta}\Bigr)
\rightarrow \phi_{\alpha}\Bigr(U_{\alpha}\cap U_{\beta}
\Bigr)
$$
is a $C^{r}$ map of an open subset of $R^{n}$ into an open
subset of $R^{n}$. In general relativity, it is of considerable
importance to require that the Hausdorff separation axiom
should hold. This states that if $p,q$ are any two distinct
points in $\cal M$, there exist disjoint open sets $U,V$ in
$\cal M$ such that $p \in U$, $q \in V$. The space-time
manifold $(M,g)$ is therefore taken to be a connected,
four-dimensional, Hausdorff $C^{\infty}$ manifold $M$ with a
Lorentz metric $g$ on $M$, i.e. the assignment of a 
symmetric, non-degenerate bilinear form 
$g_{\mid p}:T_{p}M \times T_{p}M \rightarrow R$ with diagonal
form $(-,+,+,+)$ to each tangent space. Moreover, a time
orientation is given by a globally defined, timelike vector
field $X:M\rightarrow TM$. This enables one to say that a
timelike or null tangent vector $v \in T_{p}M$ is
future-directed if $g(X(p),v)<0$, or past-directed if
$g(X(p),v)>0$ (Esposito 1992, Esposito 1994).

By a complex manifold we mean a paracompact Hausdorff space
covered by neighbourhoods each homeomorphic to an open set
in $C^{m}$, such that where two neighbourhoods overlap, the
local coordinates transform by a complex-analytic 
transformation. Thus, if $z^{1},...,z^{m}$ are local coordinates
in one such neighbourhood, and if $w^{1},...,w^{m}$ are local
coordinates in another neighbourhood, where they are both 
defined one has $w^{i}=w^{i}\Bigr(z^{1},...,z^{m}\Bigr)$, where
each $w^{i}$ is a holomorphic function of the $z$'s, and the
determinant $\partial \Bigr(w^{1},...,w^{m}\Bigr)/
\partial \Bigr(z^{1},...,z^{m}\Bigr)$ does not vanish. Various
examples can be given as follows (Chern 1979).
\vskip 0.3cm
\noindent
{\bf E1.} The space $C^{m}$ whose points are the 
$m$-tuples of complex numbers $\Bigr(z^{1},...,z^{m}\Bigr)$.
In particular, $C^{1}$ is the so-called Gaussian plane.
\vskip 0.3cm
\noindent
{\bf E2.} Complex projective space $P_{m}$, also denoted by
$P_{m}(C)$ or $CP^{m}$. Denoting by $\{ 0 \}$ the origin
$(0,...,0)$, this is the quotient space obtained by identifying
the points $\Bigr(z^{0},z^{1},...,z^{m}\Bigr)$ in 
$C^{m+1}- \{ 0 \}$ which differ from each other by a factor. The
covering of $P_{m}$ is given by $m+1$ open sets $U_{i}$ defined
respectively by $z^{i} \not =0$, $0 \leq i \leq m$. In $U_{i}$
we have the local coordinates $\zeta_{i}^{k} \equiv
z^{k}/z^{i}$, $0 \leq k \leq m$, $k \not = i$.
In $U_{i} \cap U_{j}$, transition of local coordinates is
given by $\zeta_{j}^{h} \equiv \zeta_{i}^{h}/ \zeta_{i}^{j}$,
$0 \leq h \leq m$, $h \not =j$, which are holomorphic
functions. A particular case is the Riemann sphere $P_{1}$.
\vskip 0.3cm
\noindent
{\bf E3.} Non-singular sub-manifolds of $P_{m}$, in particular,
the non-singular hyperquadric
$$
{\Bigr(z^{0}\Bigr)}^{2}+...+{\Bigr(z^{m}\Bigr)}^{2}=0.
\eqno (1.2.1)
$$
A theorem of Chow states that every compact sub-manifold
embedded in $P_{m}$ is the locus defined by a finite 
number of homogeneous polynomial equations. Compact 
sub-manifolds of $C^{m}$ are not very important, since
a connected compact sub-manifold of $C^{m}$ is a point.
\vskip 0.3cm
\noindent
{\bf E4.} Let $\Gamma$ be the discontinuous group generated
by $2m$ translations of $C^{m}$, which are linearly 
independent over the reals. The quotient space 
$C^{m}/ \Gamma$ is then called the complex torus. Moreover,
let $\Delta$ be the discontinuous group generated by
$z^{k} \rightarrow 2z^{k}$, $1 \leq k \leq m$. The quotient
manifold ${\Bigr(C^{m}- \{0 \} \Bigr)}/ \Delta$ is the
so-called Hopf manifold, and is homeomorphic to
$S^{1} \times S^{2m-1}$. Last but not least, we consider the
group $M_{3}$ of all matrices
$$
E_{3}= \pmatrix {1&z_{1}&z_{2}\cr
0&1&z_{3}\cr 0&0&1\cr},
\eqno (1.2.2)
$$
and let $D$ be the discrete group consisting of those
matrices for which $z_{1},z_{2},z_{3}$ are Gaussian
integers. This means that $z_{k}=m_{k}+in_{k}$,
$1 \leq k \leq 3$, where $m_{k},n_{k}$ are rational
integers. An Iwasawa manifold is then defined as the
quotient space $M_{3}/D$.
\vskip 0.3cm
\noindent
{\bf E5.} Orientable surfaces are particular complex
manifolds. The surfaces are taken to be $C^{\infty}$,
and we define on them a positive-definite Riemannian
metric. The Korn--Lichtenstein theorem ensures that local
parameters $x,y$ exist such that the metric locally takes
the form
$$
g=\lambda^{2} \Bigr(dx \otimes dx + dy \otimes dy \Bigr),
\; \lambda>0,
\eqno (1.2.3)
$$
or
$$
g=\lambda^{2} dz \otimes d{\overline z},
\; z \equiv x+iy.
\eqno (1.2.4)
$$
If $w$ is another local coordinate, we have
$$
g=\lambda^{2} dz \otimes d{\overline z}
=\mu^{2} dw \otimes d{\overline w},
\eqno (1.2.5)
$$
since $g$ is globally defined. Hence $dw$ is a multiple of
$dz$ or $d{\overline z}$. In particular, if the complex
coordinates $z$ and $w$ define the same orientation, then
$dw$ is proportional to $dz$. Thus, $w$ is a holomorphic
function of $z$, and the surface becomes a complex manifold.
Riemann surfaces are, by definition, one-dimensional complex
manifolds.

Let us denote by $V$ an $m$-dimensional real vector space.
We say that $V$ has a {\it complex structure} 
if there exists a linear
endomorphism $J:V \rightarrow V$ such that $J^{2}=-\II$, where
$\II$ is the identity endomorphism. An eigenvalue of $J$ is a 
complex number $\lambda$ such that the equation $Jx=\lambda x$
has a non-vanishing solution $x \in V$. Applying $J$ to both
sides of this equation, one finds $-x=\lambda^{2}x$. Hence
$\lambda=\pm i$. Since the complex eigenvalues occur in
conjugate pairs, $V$ is of even dimension $n=2m$. Let us now
denote by $V^{*}$ the dual space of $V$, i.e. the space of all
real-valued linear functions over $V$. The pairing of $V$ and
$V^{*}$ is $\langle x,y^{*}\rangle$, 
$x \in V$, $y^{*} \in V^{*}$, so that
this function is $R$-linear in each of the arguments. Following
Chern 1979, we also consider $V^{*} \otimes C$, i.e. the space
of all complex-valued $R$-linear functions over $V$. By
construction, $V^{*} \otimes C$ is an $n$-complex-dimensional
complex vector space. Elements $f \in V^{*} \otimes C$ are
of type $(1,0)$ if $f(Jx)=if(x)$, and of type $(0,1)$
if $f(Jx)=-if(x)$, $x \in V$.

If $V$ has a complex structure $J$, an {\it Hermitian structure}
in $V$ is a complex-valued function $H$ acting on $x,y \in V$
such that
$$
H \Bigr(\lambda_{1}x_{1}+\lambda_{2}x_{2},y\Bigr)
=\lambda_{1} H(x_{1},y)+\lambda_{2}
H(x_{2},y)
\; \; x_{1},x_{2},y \in V
\; \; 
\lambda_{1},\lambda_{2} \in R ,
\eqno (1.2.6)
$$
$$
{\overline {H(x,y)}}=H(y,x),
\eqno (1.2.7)
$$
$$
H(Jx,y)=iH(x,y) \Longleftrightarrow
H(x,Jy)=-iH(x,y).
\eqno (1.2.8)
$$
By using the split of $H(x,y)$ into its real and imaginary
parts
$$
H(x,y)=F(x,y)+iG(x,y),
\eqno (1.2.9)
$$
conditions (1.2.7) and (1.2.8) may be re-expressed as
$$
F(x,y)=F(y,x), \;
G(x,y)=-G(y,x),
\eqno (1.2.10)
$$
$$
F(x,y)=G(Jx,y), \;
G(x,y)=-F(Jx,y).
\eqno (1.2.11)
$$

If $\cal M$ is a $C^{\infty}$ manifold of dimension $n$,
and if $T_{x}$ and $T_{x}^{*}$ are tangent and cotangent
spaces respectively at $x \in {\cal M}$, an 
{\it almost complex structure} on $\cal M$ is a
$C^{\infty}$ field of endomorphisms $J_{x}:T_{x}
\rightarrow T_{x}$ such that $J_{x}^{2}=-\II_{x}$, 
where $\II_{x}$ is the identity endomorphism in $T_{x}$.
A manifold with an almost complex structure is called
{\it almost complex}. If a manifold is almost complex,
it is even-dimensional and orientable. However, this
is only a necessary condition. Examples can be found
(e.g. the four-sphere $S^{4}$) of even-dimensional,
orientable manifolds which cannot be given an almost 
complex structure.
\vskip 1cm
\centerline {\bf 1.3 An outline of this work}
\vskip 1cm
\noindent
Since this paper is devoted to the geometry of complex
space-time in spinor form, chapter two presents the
basic ideas, methods and results of two-component 
spinor calculus. Such a calculus is described in terms
of spin-space formalism, i.e. a complex vector space
endowed with a symplectic form and some fundamental 
isomorphisms. These mathematical
properties enable one to raise and lower indices, define
the conjugation of spinor fields in Lorentzian or
Riemannian four-geometries, translate tensor fields into
spinor fields (or the other way around). The standard
two-spinor form of the Riemann curvature tensor is then obtained
by relying on the (more) familiar tensor properties of the
curvature. The introductory analysis ends with the Petrov
classification of space-times, expressed in terms of the
Weyl spinor of conformal gravity.

Since the whole of twistor theory may be viewed as a holomorphic 
description of space-time geometry in a conformally invariant
framework, chapter three studies the key results of conformal
gravity, i.e. $C$-spaces, Einstein spaces and complex Einstein
spaces. Hence a necessary and sufficient condition for a space-time
to be conformal to a complex Einstein space is obtained,
following Kozameh {\it et al}. (1985). Such a condition involves the
Bach and Eastwood--Dighton spinors, and their occurrence is
derived in detail. The difference between Lorentzian space-times,
Riemannian four-spaces, complexified space-times and complex
space-times is also analyzed.

Chapter four is a pedagogical introduction to twistor spaces,
from the point of view of mathematical physics and relativity
theory. This is obtained by defining twistors as $\alpha$-planes
in complexified compactified Minkowski space-time, and as 
$\alpha$-surfaces in curved space-time. In the former case, one
deals with totally null two-surfaces, in that the complexified
Minkowski metric vanishes on any pair of null tangent vectors
to the surface. Hence such null tangent vectors have the
form $\lambda^{A}\pi^{A'}$, where $\lambda^{A}$ is varying
and $\pi^{A'}$ is covariantly constant. This definition can be
generalized to complex or real Riemannian four-manifolds, provided
that the Weyl curvature is anti-self-dual. An alternative definition of
twistors in Minkowski space-time is instead based on the vector
space of solutions of a differential equation, which involves
the symmetrized covariant derivative of an unprimed spinor field.
Interestingly, a deep correspondence exists between flat space-time
and twistor space. Hence complex space-time points correspond to
spheres in the so-called projective twistor space, and this
concept is carefully formulated. Sheaf cohomology is then presented
as the mathematical tool necessary to describe a conformally
invariant isomorphism between the complex vector space of
holomorphic solutions of the wave equation on the forward tube
of flat space-time, and the complex vector space of 
complex-analytic functions of three variables. These are arbitrary,
in that they are not subject to any differential equation.
Eventually, Ward's one-to-one correspondence between complex
space-times with non-vanishing cosmological constant, and
{\it sufficiently small} deformations of flat projective 
twistor space, is presented.

An example of explicit construction of anti-self-dual space-time
is given in chapter five, following Ward (1978). This generalization
of Penrose's non-linear graviton (Penrose 1976a-b) combines
two-spinor techniques and twistor theory in a way very instructive
for beginning graduate students. However, it appears necessary
to go beyond anti-self-dual space-times, since they are only a
particular class of (complex) space-times, and they do not
enable one to recover the full physical content of (complex)
general relativity. This implies going beyond the original
twistor theory, since the three-complex-dimensional space of
$\alpha$-surfaces only exists in anti-self-dual space-times.
After a brief review of alternative ideas, attention is focused
on the recent attempt by Roger Penrose to define twistors as
{\it charges} for massless spin-${3\over 2}$ fields. Such an
approach has been considered since a vanishing Ricci tensor 
provides the consistency condition for the existence and
propagation of massless helicity-${3\over 2}$ fields in
curved space-time. Moreover, in Minkowski space-time the space 
of charges for such fields is naturally identified with the
corresponding twistor space. The resulting geometric scheme
in the presence of curvature is as follows. First, define a
twistor for Ricci-flat space-time. Second, characterize the 
resulting twistor space. Third, reconstruct the original Ricci-flat
space-time from such a twistor space. 
One of the main technical difficulties
of the program proposed by Penrose is to obtain a {\it global}
description of the space of potentials for massless
spin-${3\over 2}$ fields. The corresponding {\it local} theory is
instead used, for other purposes, in our chapter eight (see below).
 
The two-spinor description of complex space-times with torsion
is given in chapter six. These space-times are studied since
torsion is a naturally occurring geometric property of
relativistic theories of gravitation, the gauge theory of the
Poincar\'e group leads to its presence and the occurrence of
cosmological singularities can be less generic than in 
general relativity (Esposito 1994 and references therein).
It turns out that, before studying the complex theory, many
differences already arise, since the Riemann tensor has 36
independent real components at each point (Penrose 1983),
rather than 20 as in general relativity. This happens since the
connection is no longer symmetric. Hence the Ricci tensor acquires
an anti-symmetric part, and the reality conditions for the
trace-free part of Ricci and for the scalar curvature no longer
hold. Hence, on taking a complex space-time with non-vanishing
torsion, all components of the Riemann curvature are given by
{\it independent} spinor fields and scalar fields, not related 
by any conjugation. Torsion is, itself, described by two
independent spinor fields. The corresponding integrability
condition for $\alpha$-surfaces is shown to involve the
self-dual Weyl spinor, the torsion spinor with three primed
indices and one unprimed index (in a non-linear way), and
covariant derivatives of such a torsion spinor. The key
identities of two-spinor calculus within this framework, including 
in particular the spinor Ricci identities, are derived in a
self-consistent way for pedagogical reasons.

Chapters seven and eight of our paper are devoted to the
application of two-spinor techniques to problems motivated by
supersymmetry and quantum cosmology. For this purpose, chapter
seven studies spin-${1\over 2}$ fields in real Riemannian
four-geometries. After deriving the Dirac and Weyl equations
in two-component spinor form in Riemannian backgrounds, we focus
on boundary conditions for {\it massless} fermionic fields motivated
by local supersymmetry. These involve the normal to the boundary
and a pair of independent spinor fields $\psi^{A}$ and
${\widetilde \psi}^{A'}$. In the case of flat Euclidean four-space
bounded by a three-sphere, they eventually imply that the classical modes
of the massless spin-${1\over 2}$ field multiplying harmonics 
having positive eigenvalues for the intrinsic three-dimensional
Dirac operator on $S^{3}$ should vanish on $S^{3}$. Remarkably,
this coincides with the property of the classical boundary-value
problem when global boundary conditions are imposed on the
three-sphere in the massless case. The boundary term in the action
functional is also derived. Our analysis makes it necessary to
use part of the analysis in section 5.8 of Esposito (1994), to prove
that the Dirac operator subject to supersymmetric boundary conditions
on the three-sphere admits self-adjoint extensions. The proof relies
on the Euclidean conjugation and on a result originally
proved by von Neumann for complex scalar fields.
Chapter seven ends with a mathematical introduction to the
global theory of the total Dirac operator in Riemannian four-geometries,
described as a first-order elliptic operator mapping smooth
sections (i.e. the spinor fields) 
of a complex vector bundle into smooth sections of the 
same bundle. Its action on the sections is obtained by composition
of Clifford multiplication with covariant differentiation, and 
provides an intrinsic formulation of the spinor covariant derivative
frequently used in our paper.

The {\it local} theory of potentials for massless 
spin-${3\over 2}$ fields is applied to the classical boundary-value
problems relevant for quantum cosmology in chapter eight
(cf. chapter five). For this purpose, we first study local
boundary conditions involving field strengths and the normal
to the boundary, originally considered  in anti-de Sitter
space-time, and recently applied in one-loop quantum cosmology.
Following Esposito (1994) and Esposito and Pollifrone (1994),
we derive the conditions under which spin-lowering and
spin-raising operators preserve these local boundary conditions
on a three-sphere for fields of spin $0,{1\over 2},1,{3\over 2}$
and $2$. Second, the two-component spinor analysis of the four
Dirac potentials of the totally symmetric and independent field
strengths for spin ${3\over 2}$ is applied to the case of a three-sphere
boundary. It is found that such boundary conditions can only be
imposed in a flat Euclidean background, for which the gauge 
freedom in the choice of the massless 
potentials remains. Third, we study
the alternative, Rarita--Schwinger form of the spin-${3\over 2}$
potentials. They are no longer symmetric in the pair of unprimed or
primed spinor indices, and their gauge freedom involves a spinor
field which is no longer a solution of the Weyl equation. 
Gauge transformations on the potentials are shown to be compatible
with the field equations provided that the background is
Ricci-flat, in agreement with well known results in the literature.
However, the preservation of boundary conditions under such 
gauge transformations is found to restrict the gauge freedom. The
construction by Penrose of a second set of potentials which
supplement the Rarita--Schwinger potentials is then applied.
The equations for these potentials, jointly with the boundary
conditions, imply that the background four-manifold is further
restricted to be totally flat. In the last part of chapter eight,
massive spin-${3\over 2}$ potentials in conformally flat Einstein
four-manifolds are studied. The analysis of supergauge transformations
of potentials for spin ${3\over 2}$ shows that the gauge freedom
for massive spin-${3\over 2}$ potentials is generated by 
solutions of the supertwistor equations. Interestingly, the
supercovariant form of a partial connection on a non-linear
bundle is obtained, and the basic equation obeyed by the second
set of potentials in the massive case is shown to be the
integrability condition on super $\beta$-surfaces of a differential
operator on a vector bundle of rank three.

The mathematical foundations 
of twistor theory are re-analyzed in chapter nine.
After a review of various definitions of twistors in curved
space-time, we present the Penrose transform and the
ambitwistor correspondence in terms of the double-fibration
picture. The Radon transform in complex geometry is also
defined, and the Ward construction of massless fields as
bundles is given. The latter concept has motivated the recent
work by Penrose on a second set of
potentials which supplement the Rarita--Schwinger
potentials in curved space-time. Recent progress on quantum
field theories in the presence of boundaries is then
described, since the boundary conditions of chapters seven
and eight are relevant for the analysis of mixed boundary
conditions in quantum field theory and quantum gravity.
Last, chapter ten reviews old and new ideas in complex general
relativity: heaven spaces and heavenly equations, complex relativity
and real solutions, multimomenta in complex general relativity.
\vskip 100cm
\centerline {\it CHAPTER TWO}
\vskip 1cm
\centerline {\bf TWO-COMPONENT SPINOR CALCULUS}
\vskip 1cm
\noindent
Spinor calculus is presented by relying on
spin-space formalism. 
Given the existence of unprimed and primed spin-space,
one has the isomorphism between such vector spaces and
their duals, realized by a symplectic form. Moreover, for
Lorentzian metrics, complex conjugation is the 
(anti-)isomorphism between unprimed and primed spin-space.
Finally, for any space-time point, its tangent space is
isomorphic to the tensor product of unprimed and primed 
spin-spaces via the Infeld--van der Waerden symbols.
Hence the correspondence between tensor fields and
spinor fields. Euclidean conjugation in Riemannian
geometries is also discussed in detail.
The Maxwell field strength is written in this language,
and many useful identities are given. The curvature spinors
of general relativity are then constructed explicitly, and
the Petrov classification of space-times 
is obtained in terms of the Weyl
spinor for conformal gravity.
\vskip 100cm
\centerline {\bf 2.1 Two-component spinor calculus}
\vskip 1cm
\noindent
Two-component spinor calculus is a powerful tool for
studying classical field theories in four-dimensional
space-time models. Within this framework,
the basic object is spin-space,
a two-dimensional complex vector space $S$ with a
symplectic form $\varepsilon$, i.e. an antisymmetric
complex bilinear form. Unprimed spinor indices 
$A,B,...$ take the values $0,1$ whereas primed spinor
indices $A',B',...$ take the values $0',1'$ since there
are actually two such spaces: unprimed spin-space
$(S,\varepsilon)$ and primed spin-space $(S',\varepsilon')$.
The whole two-spinor calculus in {\it Lorentzian}
four-manifolds relies on three fundamental properties
(Veblen 1933, Ruse 1937, Penrose 1960,
Penrose and Rindler 1984, 
Esposito 1992, Esposito 1994):

(i) The isomorphism between $\Bigr(S,\varepsilon_{AB}\Bigr)$ and
its dual $\Bigr(S^{*},\varepsilon^{AB}\Bigr)$. This is provided
by the symplectic form $\varepsilon$, which raises and
lowers indices according to the rules
$$
\varepsilon^{AB} \; \varphi_{B}=\varphi^{A} \; \in \; S,
\eqno (2.1.1)
$$
$$
\varphi^{B} \; \varepsilon_{BA}=\varphi_{A} \; \in \; S^{*}.
\eqno (2.1.2)
$$
Thus, since
$$
\varepsilon_{AB}=\varepsilon^{AB}=\pmatrix {0&1\cr -1&0 \cr},
\eqno (2.1.3)
$$
one finds in components $\varphi^{0}=\varphi_{1},
\varphi^{1}=-\varphi_{0}$. 

Similarly, one has the
isomorphism $\Bigr(S',\varepsilon_{A'B'}\Bigr)
\cong \Bigr((S')^{*},\varepsilon^{A'B'}\Bigr)$, which implies
$$
\varepsilon^{A'B'} \; \varphi_{B'}=\varphi^{A'} \; \in \; S',
\eqno (2.1.4)
$$
$$
\varphi^{B'} \; \varepsilon_{B'A'}=\varphi_{A'} \; \in 
\; (S')^{*},
\eqno (2.1.5)
$$
where
$$
\varepsilon_{A'B'}=\varepsilon^{A'B'}=\pmatrix
{0'&1'\cr -1'&0'\cr}.
\eqno (2.1.6)
$$

(ii) The (anti-)isomorphism between $\Bigr(S,\varepsilon_{AB}\Bigr)$
and $\Bigr(S',\varepsilon_{A'B'}\Bigr)$, called complex conjugation,
and denoted by an overbar. According to a standard convention,
one has
$$
{\overline {\psi^{A}}} \equiv {\overline \psi}^{A'}
\; \in \; S',
\eqno (2.1.7)
$$
$$
{\overline {\psi^{A'}}} \equiv {\overline \psi}^{A}
\; \in \; S.
\eqno (2.1.8)
$$
Thus, complex conjugation maps elements of a spin-space to
elements of the {\it complementary} spin-space. Hence 
some authors say it is an anti-isomorphism.
In components, if $w^{A}$ is thought as
$w^{A}=\pmatrix {\alpha \cr \beta \cr}$, the action of (2.1.7)
leads to
$$
{\overline {w^{A}}} \equiv {\overline w}^{A'}
\equiv \pmatrix {{\overline \alpha} \cr {\overline \beta}\cr},
\eqno (2.1.9)
$$
whereas, if $z^{A'}=\pmatrix {\gamma \cr \delta \cr}$, then
(2.1.8) leads to
$$
{\overline {z^{A'}}} \equiv {\overline z}^{A}
=\pmatrix {{\overline \gamma}\cr {\overline \delta}\cr} .
\eqno (2.1.10)
$$
With our notation, $\overline \alpha$ denotes complex
conjugation of the function $\alpha$, and so on. Note that
the symplectic structure is preserved by complex conjugation,
since ${\overline \varepsilon}_{A'B'}=\varepsilon_{A'B'}$.

(iii) The isomorphism between the tangent space $T$ at a
point of space-time and the tensor product of the 
unprimed spin-space $\Bigr(S,\varepsilon_{AB}\Bigr)$ and the
primed spin-space $\Bigr(S',\varepsilon_{A'B'}\Bigr)$:
$$
T \cong \Bigr(S,\varepsilon_{AB}\Bigr) \otimes
\Bigr(S',\varepsilon_{A'B'}\Bigr).
\eqno (2.1.11)
$$
The Infeld--van der
Waerden symbols $\sigma_{\; \; AA'}^{a}$ and
$\sigma_{a}^{\; \; AA'}$ express this isomorphism, and the
correspondence between a vector $v^{a}$ and a spinor 
$v^{AA'}$ is given by
$$
v^{AA'} \equiv v^{a} \; \sigma_{a}^{\; \; AA'},
\eqno (2.1.12)
$$
$$
v^{a} \equiv v^{AA'} \; \sigma_{\; \; AA'}^{a}.
\eqno (2.1.13)
$$
These mixed spinor-tensor symbols obey the identities
$$
{\overline \sigma}_{a}^{\; \; AA'}=\sigma_{a}^{\; \; AA'},
\eqno (2.1.14)
$$
$$
\sigma_{a}^{\; \; AA'} \; \sigma_{\; \; AA'}^{b}
=\delta_{a}^{\; \; b} ,
\eqno (2.1.15)
$$
$$
\sigma_{a}^{\; \; AA'} \; \sigma_{\; \; BB'}^{a}
=\varepsilon_{B}^{\; \; A} \; \varepsilon_{B'}^{\; \; \; A'} ,
\eqno (2.1.16)
$$
$$
\sigma_{[a}^{\; \; AA'} \; \sigma_{b]A}^{\; \; \; \; \; B'}
=-{i\over 2} \; \varepsilon_{abcd} \; \sigma^{cAA'}
\; \sigma_{\; \; A}^{d \; \; B'}.
\eqno (2.1.17)
$$
Similarly, a one-form $\omega_{a}$ has a spinor equivalent
$$
\omega_{AA'} \equiv \omega_{a} \; \sigma_{\; \; AA'}^{a} ,
\eqno (2.1.18)
$$
whereas the spinor equivalent of the metric is
$$
\eta_{ab} \; \sigma_{\; \; AA'}^{a}
\; \sigma_{\; \; BB'}^{b} \equiv
\varepsilon_{AB} \; \varepsilon_{A'B'} .
\eqno (2.1.19)
$$
In particular, in Minkowski space-time, the above equations
enable one to write down a coordinate system
in $2 \times 2$ matrix form
$$
x^{AA'}={1\over \sqrt{2}}
\pmatrix {{x^{0}+x^{3}}&{x^{1}-ix^{2}}\cr
{x^{1}+ix^{2}}&{x^{0}-x^{3}}\cr}.
\eqno (2.1.20)
$$
More precisely, in a (curved) space-time, one should write the
following equation to obtain the spinor equivalent of a vector:
$$
u^{AA'}=u^{a} \; e_{a}^{\; {\hat c}} \; 
\sigma_{{\hat c}}^{\; \; \; AA'} ,
$$
where $e_{a}^{\; {\hat c}}$ is a standard notation for the tetrad,
and $e_{a}^{\; {\hat c}}\sigma_{{\hat c}}^{\; \; \; AA'} \equiv
e_{a}^{\; AA'}$ is called the {\it soldering form}. This is, by
construction, a spinor-valued one-form, which encodes the relevant
information about the metric $g$, because $g_{ab}=e_{a}^{\; {\hat c}}
e_{b}^{\; {\hat d}} \eta_{{\hat c}{\hat d}}$, $\eta$ being the
Minkowskian metric of the so-called ``internal space".

In the Lorentzian-signature case, the Maxwell 
two-form $F \equiv F_{ab}dx^{a} \wedge dx^{b}$ can be
written spinorially (Ward and Wells 1990) as
$$
F_{AA'BB'}={1\over 2}\Bigr(F_{AA'BB'}-F_{BB'AA'}\Bigr)
=\varphi_{AB} \; \varepsilon_{A'B'}
+\varphi_{A'B'} \; \varepsilon_{AB} ,
\eqno (2.1.21)
$$
where
$$
\varphi_{AB} \equiv {1\over 2} 
F_{AC'B}^{\; \; \; \; \; \; \; \; \; C'}
=\varphi_{(AB)} ,
\eqno (2.1.22)
$$
$$
\varphi_{A'B'} \equiv {1\over 2}
F_{CB' \; \; A'}^{\; \; \; \; \; \; C}
=\varphi_{(A'B')} .
\eqno (2.1.23)
$$
These formulae are obtained by applying the identity
$$
T_{AB}-T_{BA}=\varepsilon_{AB} \; T_{C}^{\; \; C}
\eqno (2.1.24)
$$
to express ${1\over 2}\Bigr(F_{AA'BB'}-F_{AB'BA'}\Bigr)$
and ${1\over 2}\Bigr(F_{AB'BA'}-F_{BB'AA'}\Bigr)$.
Note also that round brackets $(AB)$ denote (as usual)
symmetrization over the spinor indices $A$ and $B$, and that
the antisymmetric part of $\varphi_{AB}$ vanishes by virtue
of the antisymmetry of $F_{ab}$, since (Ward and Wells 1990)
$\varphi_{[AB]}={1\over 4}\varepsilon_{AB} \;
F_{CC'}^{\; \; \; \; \; CC'}={1\over 2}\varepsilon_{AB}
\; \eta^{cd} \; F_{cd}=0$. Last but not least, in the 
Lorentzian case
$$
{\overline {\varphi_{AB}}} \equiv {\overline \varphi}_{A'B'}
=\varphi_{A'B'}.
\eqno (2.1.25)
$$
The symmetric spinor fields $\varphi_{AB}$ and
$\varphi_{A'B'}$ are the anti-self-dual and self-dual parts
of the curvature two-form, respectively.

Similarly, the Weyl curvature $C_{\; \; bcd}^{a}$, i.e. the
part of the Riemann curvature tensor invariant under conformal
rescalings of the metric, may be expressed spinorially,
omitting soldering forms 
for simplicity of notation, as
$$
C_{abcd}=\psi_{ABCD} \; \varepsilon_{A'B'} \;
\varepsilon_{C'D'}
+{\overline \psi}_{A'B'C'D'} \;
\varepsilon_{AB} \; \varepsilon_{CD}.
\eqno (2.1.26)
$$

In canonical gravity (Ashtekar 1988, Esposito 1994) 
two-component spinors
lead to a considerable simplification of calculations. Denoting
by $n^{\mu}$ the future-pointing unit timelike normal to a
spacelike three-surface, its spinor version obeys the relations
$$
n_{AA'} \; e_{\; \; \; \; \; i}^{AA'}=0,
\eqno (2.1.27)
$$
$$
n_{AA'} \; n^{AA'}=1,
\eqno (2.1.28)
$$
where $e_{\; \; \; \; \; \mu}^{AA'} \equiv e_{\; \; \mu}^{a}
\; \sigma_{a}^{\; \; AA'}$ is the two-spinor version of the tetrad,
i.e. the soldering form introduced before. 
Denoting by $h$ the induced metric on the three-surface, other
useful relations are (Esposito 1994)
$$
h_{ij}=-e_{AA'i} \; e_{\; \; \; \; \; j}^{AA'} ,
\eqno (2.1.29)
$$
$$
e_{\; \; \; \; \; 0}^{AA'}=N \; n^{AA'}
+N^{i} \; e_{\; \; \; \; \; i}^{AA'} ,
\eqno (2.1.30)
$$
$$
n_{AA'} \; n^{BA'}={1\over 2} \varepsilon_{A}^{\; \; B} ,
\eqno (2.1.31)
$$
$$
n_{AA'} \; n^{AB'}={1\over 2} \varepsilon_{A'}^{\; \; \; B'} ,
\eqno (2.1.32)
$$
$$
n_{[EB'} \; n_{A]A'}={1\over 4}\varepsilon_{EA} \;
\varepsilon_{B'A'} ,
\eqno (2.1.33)
$$
$$
e_{AA'j} \; e_{\; \; \; \; \; k}^{AB'}
=-{1\over 2}h_{jk} \; \varepsilon_{A'}^{\; \; \; B'}
-i \varepsilon_{jkl}\sqrt{{\rm det} \; h} \;
n_{AA'} \; e^{AB'l} .
\eqno (2.1.34)
$$
In Eq. (2.1.30), $N$ and $N^{i}$ are the lapse and shift
functions respectively (Esposito 1994).

To obtain the space-time curvature, we first need to define
the spinor covariant derivative $\nabla_{AA'}$. 
If $\theta,\phi,\psi$ are spinor fields, $\nabla_{AA'}$
is a map such that (Penrose and Rindler 1984, Stewart 1991)
\vskip 0.3cm
\noindent
(1) $\nabla_{AA'}(\theta+\phi)=\nabla_{AA'}\theta
+\nabla_{AA'}\phi$ (i.e. linearity).
\vskip 0.3cm
\noindent
(2) $\nabla_{AA'}(\theta \psi)=\Bigr(\nabla_{AA'}\theta\Bigr)\psi
+\theta \Bigr(\nabla_{AA'}\psi\Bigr)$ (i.e. Leibniz rule).
\vskip 0.3cm
\noindent
(3) $\psi=\nabla_{AA'}\theta$ implies
${\overline \psi}=\nabla_{AA'}{\overline \theta}$
(i.e. reality condition).
\vskip 0.3cm
\noindent
(4) $\nabla_{AA'}\varepsilon_{BC}=\nabla_{AA'}\varepsilon^{BC}=0$,
i.e. the symplectic form may be used to raise or lower indices
within spinor expressions acted upon by $\nabla_{AA'}$, in
addition to the usual metricity condition
$\nabla g=0$, which involves instead the product of two
$\varepsilon$-symbols (see also section 6.3).
\vskip 0.3cm
\noindent
(5) $\nabla_{AA'}$ commutes with any index substitution
not involving $A,A'$.
\vskip 0.3cm
\noindent
(6) For any function $f$, one finds
$\Bigr(\nabla_{a}\nabla_{b}-\nabla_{b}\nabla_{a}\Bigr)f
=2S_{ab}^{\; \; \; c} \; \nabla_{c}f$, where 
$S_{ab}^{\; \; \; c}$ is the torsion tensor.
\vskip 0.3cm
\noindent
(7) For any derivation $D$ acting on spinor fields, a spinor
field $\xi^{AA'}$ exists such that $D \psi=\xi^{AA'}
\; \nabla_{AA'} \psi, \forall \psi$.
\vskip 0.3cm
\noindent
As proved in Penrose and Rindler (1984), such a spinor covariant
derivative exists and is unique.

If Lorentzian space-time is replaced by a complex or
real Riemannian four-manifold, an important modification
should be made, since the (anti-)isomorphism between
unprimed and primed spin-space no longer exists. This
means that primed spinors can no longer be regarded as
complex conjugates of unprimed spinors, or viceversa,
as in (2.1.7) and (2.1.8). In particular, Eqs. (2.1.21)
and (2.1.26) should be re-written as
$$
F_{AA'BB'}=\varphi_{AB} \; \varepsilon_{A'B'}
+{\widetilde \varphi}_{A'B'} \; \varepsilon_{AB} ,
\eqno (2.1.35)
$$
$$
C_{abcd}=\psi_{ABCD} \; \varepsilon_{A'B'}
\; \varepsilon_{C'D'}
+{\widetilde \psi}_{A'B'C'D'} \; 
\varepsilon_{AB} \; \varepsilon_{CD} .
\eqno (2.1.36)
$$
With our notation, $\varphi_{AB},{\widetilde \varphi}_{A'B'}$,
as well as $\psi_{ABCD},{\widetilde \psi}_{A'B'C'D'}$
are {\it completely independent} symmetric spinor fields,
not related by any conjugation.

Indeed, a conjugation can still be defined in the real
Riemannian case, but it no longer 
relates $\Bigr(S,\varepsilon_{AB}\Bigr)$
to $\Bigr(S',\varepsilon_{A'B'}\Bigr)$. It is instead an
anti-involutory operation which maps elements of a spin-space
(either unprimed or primed) to elements of the {\it same}
spin-space. By anti-involutory we mean that, when applied twice
to a spinor with an odd number of indices, 
it yields the same spinor with the opposite
sign, i.e. its square is minus the identity, whereas the square
of complex conjugation as defined in (2.1.9) and (2.1.10) equals
the identity. Following Woodhouse (1985) and Esposito (1994),
Euclidean conjugation, denoted by a {\it dagger}, is defined by
$$
{\Bigr(w^{A}\Bigr)}^{\dagger} \equiv 
\pmatrix {{\overline \beta}\cr -{\overline \alpha}\cr} ,
\eqno (2.1.37)
$$
$$
{\Bigr(z^{A'}\Bigr)}^{\dagger} \equiv
\pmatrix {-{\overline \delta}\cr {\overline \gamma}\cr} .
\eqno (2.1.38)
$$
This means that, in flat Euclidean four-space, a unit
$2 \times 2$ matrix $\delta_{BA'}$ exists such that
$$
{\Bigr(w^{A}\Bigr)}^{\dagger} \equiv
\varepsilon^{AB} \; \delta_{BA'} \;
{\overline w}^{A'} .
\eqno (2.1.39)
$$
We are here using the freedom to regard $w^{A}$ either as an
$SL(2,C)$ spinor for which complex conjugation can be defined,
or as an $SU(2)$ spinor for which Euclidean conjugation is
instead available. The soldering forms for $SU(2)$ spinors
only involve spinor indices of the same spin-space, i.e.
${\widetilde e}_{i}^{\; \; AB}$ and 
${\widetilde e}_{i}^{\; \; A'B'}$
(Ashtekar 1991). More precisely, denoting by $E_{a}^{i}$
a real {\it triad}, where $i=1,2,3$, and by 
$\tau_{\; \; A}^{a \; \; \; B}$ the three Pauli matrices,
the $SU(2)$ soldering forms are defined by 
$$
{\widetilde e}_{\; \; A}^{j \; \; \; B} \equiv
-{i \over \sqrt{2}} \; E_{a}^{j} \; \tau_{\; \; A}^{a \; \; \; B}.
\eqno (2.1.40)
$$
Note that our conventions differ from the ones in Ashtekar (1991),
i.e. we use ${\widetilde e}$ 
instead of $\sigma$, and $a,b$ for Pauli-matrix
indices, $i,j$ for tangent-space indices on a three-manifold $\Sigma$,
to agree with our previous notation. The soldering form in (2.1.40)
provides an isomorphism between the three-real-dimensional tangent
space at each point of $\Sigma$, and the three-real-dimensional
vector space of $2 \times 2$ trace-free Hermitian matrices.
The Riemannian three-metric on $\Sigma$ is then given by
$$
h^{ij}=-{\widetilde e}_{\; \; A}^{i \; \; \; B} \; 
{\widetilde e}_{\; \; B}^{j \; \; \; A}.
\eqno (2.1.41)
$$
\vskip 1cm
\centerline {\bf 2.2 Curvature in general relativity}
\vskip 1cm
\noindent
In this section, following Penrose and Rindler (1984), we
want to derive the spinorial form of the Riemann curvature
tensor in a Lorentzian space-time with vanishing torsion, 
starting from the well-known symmetries of Riemann. In
agreement with the abstract-index translation of tensors
into spinors, soldering forms will be omitted in the
resulting equations (cf. Ashtekar (1991)).

Since $R_{abcd}=-R_{bacd}$ we may write
$$ \eqalignno{
R_{abcd}&=R_{AA'BB'CC'DD'}\cr
&={1\over 2} 
R_{AF'B \; \; \; cd}^{\; \; \; \; \; \; \; \; \; F'}
\; \varepsilon_{A'B'} 
+{1\over 2} 
R_{FA' \; \; B'cd}^{\; \; \; \; \; \; F}
\; \varepsilon_{AB} .
&(2.2.1)\cr}
$$
Moreover, on defining
$$
X_{ABCD} \equiv {1\over 4} 
R_{AF'B \; \; \; CL'D}^{\; \; 
\; \; \; \; \; \; \; F' \; \; \; \; \; \; \; \; \; \; L'} ,
\eqno (2.2.2)
$$
$$
\Phi_{ABC'D'} \equiv {1\over 4}
R_{AF'B \; \; \; LC' \; \; D'}^{\; \; 
\; \; \; \; \; \; \; F' \; \; \; \; \; \; L} ,
\eqno (2.2.3)
$$
the anti-symmetry in $cd$ leads to
$$ \eqalignno{
R_{abcd}&=X_{ABCD} \; \varepsilon_{A'B'} \; \varepsilon_{C'D'}
+\Phi_{ABC'D'} \; \varepsilon_{A'B'} \; \varepsilon_{CD} \cr
&+{\overline \Phi}_{A'B'CD} \; \varepsilon_{AB} \; \varepsilon_{C'D'}
+{\overline X}_{A'B'C'D'} \; \varepsilon_{AB} \;
\varepsilon_{CD} .
&(2.2.4)\cr}
$$
According to a standard terminology, the spinors (2.2.2) and (2.2.3)
are called the {\it curvature spinors}. In the light of the
(anti-)symmetries of $R_{abcd}$, they have the following
properties:
$$
X_{ABCD}=X_{(AB)(CD)} ,
\eqno (2.2.5)
$$
$$
\Phi_{ABC'D'}=\Phi_{(AB)(C'D')} ,
\eqno (2.2.6)
$$
$$
X_{ABCD}=X_{CDAB} ,
\eqno (2.2.7)
$$
$$
{\overline \Phi}_{ABC'D'}=\Phi_{ABC'D'} .
\eqno (2.2.8)
$$
Remarkably, Eqs. (2.2.6) and (2.2.8) imply that
$\Phi_{AA'BB'}$ corresponds to a trace-free and real tensor:
$$
\Phi_{a}^{\; \; a}=0 , \; 
\Phi_{AA'BB'}=\Phi_{ab}={\overline \Phi}_{ab} .
\eqno (2.2.9)
$$
Moreover, from Eqs. (2.2.5) and (2.2.7) one obtains
$$
X_{A(BC)}^{\; \; \; \; \; \; \; \; \; A}=0 .
\eqno (2.2.10)
$$
Three duals of $R_{abcd}$ exist which are very useful and are
defined as follows:
$$
R_{\; \; abcd}^{*} \equiv {1\over 2}
\varepsilon_{cd}^{\; \; \; \; pq} \; R_{abpq}
=i \; R_{AA'BB'CD'DC'} ,
\eqno (2.2.11)
$$
$$
{ }^{*}R_{abcd} \equiv {1\over 2} 
\varepsilon_{ab}^{\; \; \; \; pq} \; R_{pqcd}
=i \; R_{AB'BA'CC'DD'} ,
\eqno (2.2.12)
$$
$$
{ }^{*}R_{\; \; abcd}^{*} \equiv {1\over 4}
\varepsilon_{ab}^{\; \; \; \; pq} \;
\varepsilon_{cd}^{\; \; \; \; rs} \;
R_{pqrs} = -R_{AB'BA'CD'DC'} .
\eqno (2.2.13)
$$
For example, in terms of the dual (2.2.11), the familiar equation
$R_{a[bcd]}=0$ reads
$$
R_{\; \; ab}^{* \; \; \; \; bc}=0 .
\eqno (2.2.14)
$$
Thus, to derive the spinor form of the cyclic identity, one
can apply (2.2.14) to the equation
$$ \eqalignno{
R_{\; \; abcd}^{*}&=-i \; X_{ABCD} \; \varepsilon_{A'B'}
\; \varepsilon_{C'D'}
+i \; \Phi_{ABC'D'} \; \varepsilon_{A'B'} \; \varepsilon_{CD} \cr
&-i \; {\overline \Phi}_{A'B'CD} \; \varepsilon_{AB} \;
\varepsilon_{C'D'}
+i \; {\overline X}_{A'B'C'D'} \; \varepsilon_{AB}
\; \varepsilon_{CD} .
&(2.2.15)\cr}
$$
By virtue of (2.2.6) and (2.2.8) one thus finds
$$
X_{AB \; \; \; C}^{\; \; \; \; \; B} \; \varepsilon_{A'C'}
={\overline X}_{A'B' \; \; \; C'}^{\; \; \; \; \; \; \; B'}
\; \varepsilon_{AC} ,
\eqno (2.2.16)
$$
which implies, on defining
$$
\Lambda \equiv {1\over 6} X_{AB}^{\; \; \; \; \; AB} ,
\eqno (2.2.17)
$$
the reality condition
$$
\Lambda={\overline \Lambda} .
\eqno (2.2.18)
$$

Equation (2.2.1) enables one to express the Ricci tensor
$R_{ab} \equiv R_{acb}^{\; \; \; \; \; c}$ in spinor form as
$$
R_{ab}=6\Lambda \; \varepsilon_{AB} \; \varepsilon_{A'B'}
-2\Phi_{ABA'B'} .
\eqno (2.2.19)
$$
Thus, the resulting scalar curvature, trace-free part of Ricci
and Einstein tensor are
$$
R=24 \Lambda ,
\eqno (2.2.20)
$$
$$
R_{ab}-{1\over 4} R \; g_{ab}=-2 \Phi_{ab}
=-2 \Phi_{ABA'B'} ,
\eqno (2.2.21)
$$
$$
G_{ab}=R_{ab}-{1\over 2} R \; g_{ab}
=-6 \Lambda \; \varepsilon_{AB} \; \varepsilon_{A'B'}
-2 \Phi_{ABA'B'} ,
\eqno (2.2.22)
$$
respectively.

We have still to obtain a more suitable form of the Riemann
curvature. For this purpose, following again Penrose and
Rindler (1984), we point out that the curvature spinor
$X_{ABCD}$ can be written as
$$ \eqalignno{
X_{ABCD}&={1\over 3} \Bigr(X_{ABCD}+X_{ACDB}+X_{ADBC}\Bigr)
+{1\over 3} \Bigr(X_{ABCD}-X_{ACBD}\Bigr)\cr
&+{1\over 3} \Bigr(X_{ABCD}-X_{ADCB}\Bigr)\cr
&=X_{(ABCD)}+{1\over 3} \varepsilon_{BC} \;
X_{AF \; \; D}^{\; \; \; \; \; F}
+{1\over 3} \varepsilon_{BD} \;
X_{AFC}^{\; \; \; \; \; \; \; \; F} .
&(2.2.23)\cr}
$$
Since $X_{AFC}^{\; \; \; \; \; \; \; F}=3 \Lambda \;
\varepsilon_{AF}$, Eq. (2.2.23) leads to
$$
X_{ABCD}=\psi_{ABCD}+\Lambda \Bigr(\varepsilon_{AC} \;
\varepsilon_{BD}+\varepsilon_{AD} \; \varepsilon_{BC} \Bigr) ,
\eqno (2.2.24)
$$
where $\psi_{ABCD}$ is the Weyl spinor.

Since $\Lambda={\overline \Lambda}$ from (2.2.18), the
insertion of (2.2.24) into (2.2.4), jointly with the
identity
$$
\varepsilon_{A'B'} \; \varepsilon_{C'D'}
+\varepsilon_{A'D'} \; \varepsilon_{B'C'}
-\varepsilon_{A'C'} \; \varepsilon_{B'D'}=0 ,
\eqno (2.2.25)
$$
yields the desired decomposition of the Riemann curvature as
$$ \eqalignno{
R_{abcd}&=\psi_{ABCD} \; \varepsilon_{A'B'} \; \varepsilon_{C'D'}
+{\overline \psi}_{A'B'C'D'} \; \varepsilon_{AB} \;
\varepsilon_{CD} \cr
&+\Phi_{ABC'D'} \; \varepsilon_{A'B'} \; \varepsilon_{CD}
+{\overline \Phi}_{A'B'CD} \; \varepsilon_{AB} \;
\varepsilon_{C'D'} \cr
&+2 \Lambda \Bigr(\varepsilon_{AC} \; \varepsilon_{BD}
\; \varepsilon_{A'C'} \; \varepsilon_{B'D'}
- \varepsilon_{AD} \; \varepsilon_{BC} \; \varepsilon_{A'D'}
\; \varepsilon_{B'C'} \Bigr) .
&(2.2.26)\cr}
$$
With this standard notation, the conformally invariant part of
the curvature takes the form $C_{abcd}={ }^{(-)}C_{abcd}
+{ }^{(+)}C_{abcd}$, where
$$
{ }^{(-)}C_{abcd} \equiv \psi_{ABCD} \; \varepsilon_{A'B'} \;
\varepsilon_{C'D'} ,
\eqno (2.2.27)
$$
$$
{ }^{(+)}C_{abcd} \equiv {\overline \psi}_{A'B'C'D'}
\; \varepsilon_{AB} \; \varepsilon_{CD} ,
\eqno (2.2.28)
$$
are the anti-self-dual and self-dual Weyl tensors,
respectively.
\vskip 1cm
\centerline {\bf 2.3 Petrov classification}
\vskip 1cm
\noindent
Since the Weyl spinor is totally symmetric, we may use a
well known result of two-spinor calculus, according to
which, if $\Omega_{AB...L}$ is totally symmetric, then there
exist univalent spinors $\alpha_{A},\beta_{B},...,\gamma_{L}$
such that (Stewart 1991)
$$
\Omega_{AB...L}=\alpha_{(A} \; \beta_{B} ... \gamma_{L)} ,
\eqno (2.3.1)
$$
where $\alpha,...,\gamma$ are called the {\it principal spinors} of
$\Omega$, and the corresponding real null vectors are called the
{\it principal null directions} of $\Omega$. In the case of the Weyl
spinor, such a theorem implies that
$$
\psi_{ABCD}=\alpha_{(A} \; \beta_{B} \; \gamma_{C} \;
\delta_{D)} .
\eqno (2.3.2)
$$
The corresponding space-times can be classified as follows
(Stewart 1991).
\vskip 0.3cm
\noindent
(1) {\it Type I}. Four distinct principal null directions.
Hence the name algebraically general.
\vskip 0.3cm
\noindent
(2) {\it Type II}. Two directions coincide. Hence the name
algebraically special.
\vskip 0.3cm
\noindent
(3) {\it Type D}. Two different pairs of repeated principal
null directions exist.
\vskip 0.3cm
\noindent
(4) {\it Type III}. Three principal null directions
coincide.
\vskip 0.3cm
\noindent
(5) {\it Type N}. All four principal null directions
coincide.
\vskip 0.3cm
\noindent
Such a classification is the Petrov classification, and it
provides a relevant example of the superiority of the
two-spinor formalism in four space-time dimensions, since the
alternative ways to obtain it are far more complicated.

Within this framework (as well as in chapter three) we
need to know that $\psi_{ABCD}$ has two scalar invariants:
$$
I \equiv \psi_{ABCD} \; \psi^{ABCD} ,
\eqno (2.3.3)
$$
$$
J \equiv \psi_{AB}^{\; \; \; \; \; CD} \;
\psi_{CD}^{\; \; \; \; \; EF} \;
\psi_{EF}^{\; \; \; \; \; AB} .
\eqno (2.3.4)
$$
Type-II space-times are such that $I^{3}=6J^{2}$,
while in type-III space-times $I=J=0$. Moreover, type-D
space-times are characterized by the condition
$$
\psi_{PQR(A} \; \psi_{BC}^{\; \; \; \; \; PQ} \;
\psi_{\; \; DEF)}^{R}=0 ,
\eqno (2.3.5)
$$
while in type-N space-times
$$
\psi_{(AB}^{\; \; \; \; \; \; \; EF} \;
\psi_{CD)EF}=0 .
\eqno (2.3.6)
$$
These results, despite their simplicity, are not well known
to many physicists and mathematicians. Hence they have
been included also in this paper, to prepare the ground
for the more advanced topics of the following chapters.
\vskip 100cm
\centerline {\it CHAPTER THREE}
\vskip 1cm
\centerline {\bf CONFORMAL GRAVITY}
\vskip 1cm
\noindent
Since twistor theory enables one to reconstruct
the space-time geometry from conformally invariant geometric
objects, it is important to know the basic tools for studying
conformal gravity within the framework of general relativity.
This is achieved by defining and using the Bach and
Eastwood--Dighton tensors, here presented in two-spinor form
(relying on previous work by Kozameh, Newman and Tod).
After defining $C$-spaces and Einstein spaces, it is shown
that a space-time is conformal to an Einstein space if and only
if some equations involving the Weyl spinor, its covariant
derivatives, and the trace-free part of Ricci are satisfied.
Such a result is then extended to complex Einstein spaces.
The conformal structure of infinity of Minkowski space-time
is eventually introduced.
\vskip 100cm
\centerline {\bf 3.1 $C$-spaces}
\vskip 1cm
\noindent
Twistor theory may be viewed as the attempt to describe 
fundamental physics in terms of conformally invariant geometric
objects within a holomorphic framework. Space-time points are no
longer of primary importance, since they only appear as derived
concepts in such a scheme. To understand the following
chapters, almost entirely devoted to twistor theory and its
applications, it is therefore necessary to study the main
results of the theory of conformal gravity. They can be
understood by focusing on $C$-spaces, Einstein spaces, complex
space-times and complex Einstein spaces, as we do from now on
in this chapter.

To study $C$-spaces in a self-consistent way, we begin by recalling
some basic properties of conformal rescalings. By definition, a
{\it conformal rescaling} of the space-time metric $g$ yields the
metric $\widehat g$ as
$$
{\widehat g}_{ab} \equiv e^{2 \omega} \; g_{ab} ,
\eqno (3.1.1)
$$
where $\omega$ is a smooth scalar. Correspondingly, any tensor 
field $T$ of type $(r,s)$ is conformally weighted if
$$
{\widehat T} \equiv e^{k \omega} \; T
\eqno (3.1.2)
$$
for some integer $k$. In particular, conformal invariance of $T$
is achieved if $k=0$.

It is useful to know the transformation rules for covariant
derivatives and Riemann curvature under the rescaling (3.1.1).
For this purpose, defining
$$
F_{\; \; \; ab}^{m} \equiv 2 \delta_{\; \; a}^{m}
\; \nabla_{b} \omega
-g_{ab} \; g^{mn} \; \nabla_{n} \omega ,
\eqno (3.1.3)
$$
one finds
$$
{\widehat \nabla}_{a} \; V_{b}=\nabla_{a} \; V_{b}
-F_{\; \; \; ab}^{m} \; V_{m} ,
\eqno (3.1.4)
$$
where ${\widehat \nabla}_{a}$ denotes covariant differentiation
with respect to the metric $\widehat g$. Hence the Weyl tensor
$C_{abc}^{\; \; \; \; \; d}$, the Ricci tensor 
$R_{ab} \equiv R_{cab}^{\; \; \; \; \; c}$ and the Ricci scalar
transform as
$$
{\widehat C}_{abc}^{\; \; \; \; \; d}=C_{abc}^{\; \; \; \; \; d} ,
\eqno (3.1.5)
$$
$$
{\widehat R}_{ab}=R_{ab}+2 \nabla_{a} \omega_{b}
-2 \omega_{a} \omega_{b}
+g_{ab} \Bigr(2 \omega^{c} \omega_{c}
+\nabla^{c} \omega_{c} \Bigr) ,
\eqno (3.1.6)
$$
$$
{\widehat R}=e^{-2 \omega} \biggr[R+6 \Bigr(\nabla^{c} 
\omega_{c} + \omega^{c} \omega_{c}\Bigr)\biggr] .
\eqno (3.1.7)
$$
With our notation, $\omega_{c} \equiv \nabla_{c} \omega
=\omega_{,c}$.

We are here interested in space-times which are conformal to
$C$-spaces. The latter are a class of space-times such that
$$
{\widehat \nabla}^{f} \; {\widehat C}_{abcf}=0 .
\eqno (3.1.8)
$$
By virtue of (3.1.3) and (3.1.4) one can see that the conformal 
transform of Eq. (3.1.8) is
$$
\nabla^{f} \; C_{abcf}+\omega^{f} \; C_{abcf}=0 .
\eqno (3.1.9)
$$
This is the necessary and sufficient condition for a space-time
to be conformal to a $C$-space. Its two-spinor form is
$$
\nabla^{FA'}\psi_{FBCD}+\omega^{FA'}\psi_{FBCD}=0 .
\eqno (3.1.10)
$$
However, note that only a {\it real} solution $\omega^{FA'}$ of
Eq. (3.1.10) satisfies Eq. (3.1.9). Hence, whenever we use Eq. (3.1.10),
we are also imposing a reality 
condition (Kozameh {\it et al}. 1985).

On using the invariants defined in (2.3.3) and (2.3.4), one finds
the useful identities
$$
\psi_{ABCD} \; \psi^{ABCE}={1\over 2}I \; \delta_{D}^{\; \; E} ,
\eqno (3.1.11)
$$
$$
\psi_{ABCD} \; \psi_{\; \; \; \; PQ}^{AB} \;
\psi^{PQCE}={1\over 2} J \; \delta_{D}^{\; \; E} .
\eqno (3.1.12)
$$
The idea is now to act with $\psi^{ABCD}$ on the left-hand side
of (3.1.10) and then use (3.1.11) when $I \not = 0$. This leads to
$$
\omega^{AA'}=-{2\over I} \; \psi^{ABCD} \;
\nabla^{FA'} \; \psi_{FBCD} .
\eqno (3.1.13)
$$
By contrast, when $I=0$ but $J \not = 0$, we multiply twice 
Eq. (3.1.10) by the Weyl spinor and use (3.1.12). Hence one
finds
$$
\omega^{AA'}=-{2\over J} \; \psi_{\; \; \; \; \; EF}^{CD}
\; \psi^{EFGA} \; \nabla^{BA'} \; \psi_{BCDG} .
\eqno (3.1.14)
$$
Thus, by virtue of (3.1.13), the reality condition 
$\omega^{AA'}={\overline {\omega^{AA'}}}
={\overline \omega}^{AA'}$ implies
$$
{\overline I} \; \psi^{ABCD} \; \nabla^{FA'} \;
\psi_{FBCD} - I \; {\overline \psi}^{A'B'C'D'}
\; \nabla^{AF'} \; {\overline \psi}_{F'B'C'D'}=0 .
\eqno (3.1.15)
$$
We have thus shown that a space-time is conformally related
to a $C$-space if and only if Eq. (3.1.10) holds for some
vector $\omega^{DD'}=K^{DD'}$, and Eq. (3.1.15) holds as well.
\vskip 1cm
\centerline {\bf 3.2 Einstein spaces}
\vskip 1cm
\noindent
By definition, Einstein spaces are such that their Ricci tensor
is proportional to the metric: $R_{ab}=\lambda \; g_{ab}$. A
space-time is conformal to an Einstein space if and only if
a function $\omega$ exists (see (3.1.1)) such that (cf. (3.1.6))
$$
R_{ab}+2 \nabla_{a} \omega_{b} -2 \omega_{a} \omega_{b}
-{1\over 4} T g_{ab} =0 ,
\eqno (3.2.1)
$$
where
$$
T \equiv R + 2 \nabla^{c} \omega_{c}
-2 \omega^{c} \omega_{c} .
\eqno (3.2.2)
$$
Of course, Eq. (3.2.1) leads to restrictions on the metric. These
are obtained by deriving the corresponding integrability 
conditions. For this purpose, on taking the curl of Eq. (3.2.1)
and using the Bianchi identities, one finds
$$
\nabla^{f} \; C_{abcf}+\omega^{f} \; C_{abcf}=0 ,
$$
which coincides with Eq. (3.1.9). Moreover, acting with $\nabla^{a}$
on Eq. (3.1.9), applying the Leibniz rule, and using again (3.1.9)
to re-express $\nabla^{f} \; C_{abcf}$ as $-\omega^{f} \; C_{abcf}$,
one obtains
$$
\biggr[\nabla^{a}\nabla^{d}+\nabla^{a}\omega^{d}
-\omega^{a}\omega^{d}\biggr]C_{abcd}=0 .
\eqno (3.2.3)
$$
We now re-express $\nabla^{a}\omega^{d}$ from (3.2.1) as
$$
\nabla^{a}\omega^{d}=\omega^{a}\omega^{d}+{1\over 8}T g^{ad}
-{1\over 2}R^{ad} .
\eqno (3.2.4)
$$
Hence Eqs. (3.2.3) and (3.2.4) lead to
$$
\biggr[\nabla^{a}\nabla^{d}-{1\over 2}R^{ad}\biggr]
C_{abcd}=0 .
\eqno (3.2.5)
$$
This calculation only proves that the vanishing of the 
{\it Bach tensor}, defined as
$$
B_{bc} \equiv \nabla^{a}\nabla^{d}C_{abcd}-{1\over 2}R^{ad} \;
C_{abcd} ,
\eqno (3.2.6)
$$
is a {\it necessary} condition for a space-time to be conformal
to an Einstein space (jointly with Eq. (3.1.9)). To prove 
{\it sufficiency} of the condition, we first need the following
Lemma (Kozameh {\it et al}. 1985):
\vskip 0.3cm
\noindent
{\bf Lemma 3.2.1} Let $H^{ab}$ be a trace-free symmetric tensor.
Then, providing the scalar invariant $J$ defined in (2.3.4)
does not vanish, the only solution of the equations
$$
C_{abcd} \; H^{ad}=0 ,
\eqno (3.2.7)
$$
$$
C_{\; \; abcd}^{*} \; H^{ad}=0 ,
\eqno (3.2.8)
$$
is $H^{ad}=0$. As shown in Kozameh {\it et al}. (1985), such a Lemma
is best proved by using two-spinor methods. Hence $H_{ab}$
corresponds to the spinor field
$$
H_{AA'BB'}=\phi_{ABA'B'}={\overline \phi}_{(A'B')(AB)} ,
\eqno (3.2.9)
$$
and Eqs. (3.2.7) and (3.2.8) imply that
$$
\psi_{ABCD} \; \phi_{\; \; \; \; \; A'B'}^{CD}=0 .
\eqno (3.2.10)
$$
Note that the extra primed spinor indices $A'B'$ are irrelevant.
Hence we can focus on the simpler eigenvalue equation
$$
\psi_{ABCD} \; \varphi^{CD}= \lambda \; \varphi_{AB} .
\eqno (3.2.11)
$$
The corresponding characteristic equation for $\lambda$ is
$$
-\lambda^{3}+{1\over 2}I \lambda +{\rm det}(\psi)=0 ,
\eqno (3.2.12)
$$
by virtue of (2.3.3). Moreover, the Cayley--Hamilton theorem
enables one to re-write Eq. (3.2.12) as
$$
\psi_{AB}^{\; \; \; \; \; PQ} \;
\psi_{PQ}^{\; \; \; \; \; RS} \;
\psi_{RS}^{\; \; \; \; \; CD}
={1\over 2} I \; \psi_{AB}^{\; \; \; \; \; CD}
+{\rm det}(\psi) \delta_{(A}^{\; \; \; C} \;
\delta_{B)}^{\; \; \; D} ,
\eqno (3.2.13)
$$
and contraction of $AB$ with $CD$ yields
$$
{\rm det}(\psi)={1\over 3}J .
\eqno (3.2.14)
$$
Thus, the only solution of Eq. (3.2.10) is the trivial one unless
$J=0$ (Kozameh {\it et al}. 1985).

We are now in a position to prove sufficiency of the conditions
(cf. Eqs. (3.1.9) and (3.2.5))
$$
\nabla^{f} \; C_{abcf}+K^{f} \; C_{abcf}=0 ,
\eqno (3.2.15)
$$
$$
B_{bc}=0 .
\eqno (3.2.16)
$$
Indeed, Eq. (3.2.15) ensures that (3.1.9) is satisfied with
$\omega_{f}=\nabla_{f}\omega$ for some $\omega$. Hence Eq.
(3.2.3) holds. If one now subtracts Eq. (3.2.3) from Eq.
(3.2.16) one finds
$$
C_{abcd}\biggr(R^{ad}+2\nabla^{a}\omega^{d}-2\omega^{a}\omega^{d}
\biggr)=0 .
\eqno (3.2.17)
$$
This is indeed Eq. (3.2.7) of Lemma 3.2.1. To obtain Eq.
(3.2.8), we act with $\nabla^{a}$ on the dual of Eq. (3.1.9).
This leads to
$$
\nabla^{a}\nabla^{d}C_{\; \; abcd}^{*}
+\biggr(\nabla^{a}\omega^{d}-\omega^{a}\omega^{d}\biggr)
C_{\; \; abcd}^{*}=0 .
\eqno (3.2.18)
$$
Following Kozameh {\it et al}. (1985), the gradient of the contracted
Bianchi identity and Ricci identity is then used to derive
the additional equation
$$
\nabla^{a}\nabla^{d}C_{\; \; abcd}^{*}
-{1\over 2}R^{ad} \; C_{\; \; abcd}^{*}=0 .
\eqno (3.2.19)
$$
Subtraction of Eq. (3.2.19) from Eq. (3.2.18) now yields
$$
C_{\; \; abcd}^{*} \biggr(R^{ad}+2\nabla^{a}\omega^{d}
-2\omega^{a}\omega^{d}\biggr)=0 ,
\eqno (3.2.20)
$$
which is the desired form of Eq. (3.2.8).

We have thus completed the proof that (3.2.15) and (3.2.16) are
{\it necessary} and {\it sufficient} conditions for a space-time
to be conformal to an Einstein space. In two-spinor language,
when Einstein's equations are imposed, after a conformal rescaling
the equation for the trace-free part of Ricci becomes
(see section 2.2)
$$
\Phi_{ABA'B'}-\nabla_{BB'}\omega_{AA'}
-\nabla_{BA'}\omega_{AB'}+\omega_{AA'} \; \omega_{BB'}
+\omega_{AB'} \; \omega_{BA'}=0 .
\eqno (3.2.21)
$$
Similarly to the tensorial analysis performed so far, the
spinorial analysis shows that the integrability condition 
for Eq. (3.2.21) is
$$
\nabla^{AA'}\psi_{ABCD}+\omega^{AA'} \; \psi_{ABCD}=0 .
\eqno (3.2.22)
$$
The fundamental theorem of conformal gravity states therefore
that a space-time is conformal to an Einstein space if and
only if (Kozameh {\it et al}. 1985)
$$
\nabla^{DD'}\psi_{ABCD}+k^{DD'} \; \psi_{ABCD}=0 ,
\eqno (3.2.23)
$$
$$
{\overline I} \; \psi^{ABCD} \; \nabla^{FA'} \;
\psi_{FBCD} - I \; {\overline \psi}^{A'B'C'D'}
\; \nabla^{AF'} \; {\overline \psi}_{F'B'C'D'}=0 ,
\eqno (3.2.24)
$$
$$
B_{AFA'F'} \equiv 2 \biggr(\nabla_{\; \; A'}^{C}
\; \nabla_{\; \; F'}^{D} \; \psi_{AFCD}
+\Phi_{\; \; \; \; \; A'F'}^{CD} \;
\psi_{AFCD}\biggr)=0 .
\eqno (3.2.25)
$$
Note that reality of Eq. (3.2.25) for the Bach spinor is ensured
by the Bianchi identities.
\vskip 1cm
\centerline {\bf 3.3 Complex space-times}
\vskip 1cm
\noindent
Since this paper is devoted to complex general relativity and its
applications, it is necessary to extend the theorem expressed by
(3.2.23)--(3.2.25) to complex space-times. For this purpose, we 
find it appropriate to define and discuss such spaces in more detail
in this section. In this respect, we should say that four distinct
geometric objects are necessary to study real general relativity
and complex general relativity, here defined in four-dimensions
(Penrose and Rindler 1986, Esposito 1994).
\vskip 0.3cm
\noindent
(1) {\it Lorentzian} space-time $(M,g_{L})$. 
This is a Hausdorff four-manifold
$M$ jointly with a symmetric, non-degenerate bilinear form $g_{L}$ to
each tangent space with signature $(+,-,-,-)$ (or $(-,+,+,+)$). The
latter is then called a Lorentzian four-metric $g_{L}$.
\vskip 0.3cm
\noindent
(2) {\it Riemannian} four-space $(M,g_{R})$, where $g_{R}$ is a smooth
and {\it positive-definite} section of the bundle of symmetric
bilinear two-forms on $M$. Hence $g_{R}$ has signature $(+,+,+,+)$.
\vskip 0.3cm
\noindent
(3) {\it Complexified} space-time. This manifold originates from a
real-analytic space-time with real-analytic coordinates
$x^{a}$ and real-analytic Lorentzian metric $g_{L}$ by allowing
the coordinates to become complex, and by an holomorphic
extension of the metric coefficients into the complex domain.
In such manifolds the operation of complex conjugation, taking
any point with complexified coordinates $z^{a}$ into the point
with coordinates ${\overline {z^{a}}}$, still exists. Note that,
however, it is not possible to define reality of tensors at
{\it complex points}, since the conjugate tensor lies at the
complex conjugate point, rather than at the original point.
\vskip 0.3cm
\noindent
(4) {\it Complex} space-time. This is a {\it four-complex-dimensional}
complex-Riemannian manifold, and no four-real-dimensional subspace
has been singled out to give it a reality structure (Penrose and
Rindler 1986). In complex space-times no complex conjugation
exists, since such a map is not invariant under holomorphic
coordinate transformations.
\vskip 0.3cm
\noindent
Thus, the complex-conjugate spinors
$\lambda^{A...M}$ and ${\overline \lambda}^{A'...M'}$ of a
Lorentzian space-time are replaced by {\it independent} spinors
$\lambda^{A...M}$ and ${\widetilde \lambda}^{A'...M'}$.
This means that unprimed and primed spin-spaces become 
unrelated to one another.
Moreover, the complex scalars $\phi$ and ${\overline \phi}$ are
replaced by the pair of {\it independent} complex scalars
$\phi$ and $\widetilde \phi$. On the other hand, quantities $X$
that are originally {\it real} yield no new quantities, since the
reality condition $X={\overline X}$ becomes $X={\widetilde X}$.
For example, the covariant derivative operator $\nabla_{a}$ of
Lorentzian space-time yields no new operator
${\widetilde \nabla}_{a}$, since it is originally real. One
should instead regard $\nabla_{a}$ as a complex-holomorphic
operator. The spinors $\psi_{ABCD},\Phi_{ABC'D'}$ and the scalar
$\Lambda$ appearing in the Riemann curvature (see (2.2.26)) have as
counterparts the spinors ${\widetilde \psi}_{A'B'C'D'},
{\widetilde \Phi}_{ABC'D'}$ and the scalar $\widetilde \Lambda$.
However, by virtue of the {\it original} reality conditions in
Lorentzian space-time, one has (Penrose and Rindler 1986)
$$
{\widetilde \Phi}_{ABC'D'}=\Phi_{ABC'D'} ,
\eqno (3.3.1)
$$
$$
{\widetilde \Lambda}=\Lambda ,
\eqno (3.3.2)
$$
while the Weyl spinors $\psi_{ABCD}$ and
${\widetilde \psi}_{A'B'C'D'}$ remain independent of each other.
Hence one Weyl spinor may vanish without the other Weyl spinor
having to vanish as well. Correspondingly, a complex space-time
such that ${\widetilde \psi}_{A'B'C'D'}=0$ is called {\it right
conformally flat} or conformally anti-self-dual, whereas if
$\psi_{ABCD}=0$, one deals with a {\it left conformally flat}
or conformally self-dual complex space-time. Moreover, if the
remaining part of the Riemann curvature vanishes as well, i.e.
$\Phi_{ABC'D'}=0$ and $\Lambda=0$, the word {\it conformally}
should be omitted in the terminology described above (cf. 
chapter four). Interestingly, in a complex space-time the
principal null directions (cf. section 2.3) of the Weyl spinors
$\psi_{ABCD}$ and ${\widetilde \psi}_{A'B'C'D'}$ are independent
of each other, and one has two independent classification 
schemes at each point.
\vskip 1cm
\centerline {\bf 3.4 Complex Einstein spaces}
\vskip 1cm
\noindent
In the light of the previous discussion, the fundamental theorem
of conformal gravity in complex space-times can be stated as
follows (Baston and Mason 1987).
\vskip 0.3cm
\noindent
{\bf Theorem 3.4.1} A complex space-time is conformal to a complex
Einstein space if and only if
$$
\nabla^{DD'} \; \psi_{ABCD}+k^{DD'} \; \psi_{ABCD}=0 ,
\eqno (3.4.1)
$$
$$
{\widetilde I} \; \psi^{ABCD} \; \nabla^{FA'} \; \psi_{FBCD}
-I \; {\widetilde \psi}^{A'B'C'D'} \; \nabla^{AF'}
\; {\widetilde \psi}_{F'B'C'D'}=0 ,
\eqno (3.4.2)
$$
$$
B_{AFA'F'} \equiv 2 \biggr(\nabla_{\; \; A'}^{C} \;
\nabla_{\; \; F'}^{D} \; \psi_{AFCD}
+\Phi_{\; \; \; \; \; A'F'}^{CD} \; \psi_{AFCD}\biggr)=0 ,
\eqno (3.4.3)
$$
where $I$ is the complex scalar invariant defined in (2.3.3),
whereas $\widetilde I$ is the independent invariant defined as
$$
{\widetilde I} \equiv {\widetilde \psi}_{A'B'C'D'} \;
{\widetilde \psi}^{A'B'C'D'}.
\eqno (3.4.4)
$$
The left-hand side of Eq. (3.4.2) is called the
{\it Eastwood--Dighton spinor}, and the left-hand side of
Eq. (3.4.3) is the {\it Bach spinor}.
\vskip 10cm
\centerline {\bf 3.5 Conformal infinity}
\vskip 1cm
\noindent
To complete our introduction to conformal gravity, we find it
helpful for the reader to outline the construction of
conformal infinity for Minkowski space-time (see also an
application in section 9.5). Starting from polar local
coordinates in Minkowski, we first introduce (in $c=1$
units) the retarded coordinate $w \equiv t-r$ and the advanced
coordinate $v \equiv t+r$. To eliminate the resulting cross
term in the local form of the metric, new coordinates
$p$ and $q$ are defined implicitly as
(Esposito 1994)
$$
\tan p \equiv v , \;
\tan q \equiv w , \;
p-q \geq 0 .
\eqno (3.5.1)
$$
Hence one finds that a conformal-rescaling factor 
$\omega \equiv (\cos p)(\cos q)$ exists such that, locally,
the metric of Minkowski space-time can be
written as $\omega^{-2}{\widetilde g}$,
where
$$
{\widetilde g} \equiv -dt' \otimes dt'
+\biggr[dr' \otimes dr' +{1\over 4}
(\sin(2r'))^{2} \; \Omega_{2}\biggr],
\eqno (3.5.2)
$$
where $t' \equiv {(p+q)\over 2}, r' \equiv {(p-q)\over 2}$, and
$\Omega_{2}$ is the metric on a unit two-sphere. Although (3.5.2)
is locally identical to the metric of the Einstein static universe,
it is necessary to go beyond a local analysis. This may be
achieved by {\it analytic extension} to the whole of the Einstein
static universe. The original Minkowski space-time is then found
to be conformal to the following region of the Einstein static
universe:
$$
(t'+r') \in ]-\pi,\pi[ , \;
(t'-r') \in ]-\pi,\pi[ , \;
r' \geq 0 .
\eqno (3.5.3)
$$
By definition, the {\it boundary} of the region in (3.5.3) represents
{\it the conformal structure of infinity} of Minkowski space-time.
It consists of two null surfaces and three points, i.e.
(Esposito 1994)
\vskip 0.3cm
\noindent
(i) The null surface ${\rm SCRI}^{-} \equiv
\left \{ t'-r'=q=-{\pi \over 2} \right \}$, i.e. the 
future light cone of the point $r'=0,t'=-{\pi \over 2}$.
\vskip 0.3cm
\noindent
(ii) The null surface ${\rm SCRI}^{+} \equiv 
\left \{t'+r'=p={\pi \over 2} \right \}$, i.e. the past
light cone of the point $r'=0,t'={\pi \over 2}$.
\vskip 0.3cm
\noindent
(iii) Past timelike infinity, i.e. the point
$$
\iota^{-} \equiv \left \{r'=0,t'=-{\pi \over 2} \right \}
\Rightarrow p=q=-{\pi \over 2} .
$$
\vskip 0.3cm
\noindent
(iv) Future timelike infinity, defined as
$$
\iota^{+} \equiv \left \{r'=0,t'={\pi \over 2}
\right \} \Rightarrow p=q={\pi \over 2} .
$$
\vskip 0.3cm
\noindent
(v) Spacelike infinity, i.e. the point
$$
\iota^{0} \equiv \left \{r'={\pi \over 2},t'=0
\right \} \Rightarrow p=-q={\pi \over 2} .
$$
The extension of the SCRI formalism to curved space-times is
an open research problem, but we limit ourselves to the
previous definitions in this section.
\vskip 100cm
\centerline {\it CHAPTER FOUR}
\vskip 1cm
\centerline {\bf TWISTOR SPACES}
\vskip 1cm
\noindent
In twistor theory, $\alpha$-planes are the
building blocks of classical field theory in complexified
compactified Minkowski space-time. The $\alpha$-planes are totally
null two-surfaces $S$ in that, if $p$ is any point on $S$,
and if $v$ and $w$ are any two null tangent vectors at 
$p \in S$, the complexified Minkowski metric $\eta$ satisfies
the identity $\eta(v,w)=v_{a}w^{a}=0$. By definition, their
null tangent vectors have the two-component spinor form
$\lambda^{A}\pi^{A'}$, where $\lambda^{A}$ is varying and
$\pi^{A'}$ is fixed. Therefore, the induced metric vanishes
identically since $\eta(v,w)=\Bigr(\lambda^{A}\pi^{A'}\Bigr)
\Bigr(\mu_{A}\pi_{A'}\Bigr)=0=\eta(v,v)=
\Bigr(\lambda^{A}\pi^{A'}\Bigr)
\Bigr(\lambda_{A}\pi_{A'}\Bigr)$. One thus obtains a
conformally invariant characterization of flat space-times.
This definition can be generalized to complex or real
Riemannian space-times with non-vanishing curvature, provided
the Weyl curvature is anti-self-dual. One then finds that
the curved metric $g$ is such that $g(v,w)=0$ on $S$, and
the spinor field $\pi_{A'}$ is covariantly constant on $S$.
The corresponding holomorphic two-surfaces are called 
$\alpha$-surfaces, and they form a three-complex-dimensional
family. Twistor space is the space of all $\alpha$-surfaces,
and depends only on the conformal structure of complex
space-time.

Projective twistor space $PT$ is isomorphic to complex projective
space $CP^{3}$. The correspondence between flat space-time
and twistor space shows that complex $\alpha$-planes correspond
to points in $PT$, and real null geodesics to points in $PN$,
i.e. the space of null twistors. Moreover, a complex
space-time point corresponds to a sphere in $PT$, and a real 
space-time point to a sphere in $PN$. Remarkably, the points
$x$ and $y$ are null-separated if and only if the corresponding
spheres in $PT$ intersect. This is the twistor description of
the light-cone structure of Minkowski space-time.

A conformally invariant isomorphism exists between the complex
vector space of holomorphic solutions of $\cstok{\ }\phi=0$
on the forward tube of flat space-time, and the complex
vector space of arbitrary complex-analytic functions of three
variables, not subject to any differential equation. Moreover,
when curvature is non-vanishing, there is a one-to-one
correspondence between complex space-times with 
anti-self-dual Weyl curvature and scalar curvature
$R=24 \Lambda$, and sufficiently small deformations of flat
projective twistor space $PT$ which preserve a one-form
$\tau$ homogeneous of degree $2$ and a three-form $\rho$ 
homogeneous of degree $4$, with $\tau \wedge d\tau
=2\Lambda \rho$. Thus, to solve the anti-self-dual Einstein
equations, one has to study a geometric problem, i.e.
finding the holomorphic curves in deformed projective
twistor space.
\vskip 100cm
\centerline {\bf 4.1 $\alpha$-planes in Minkowski space-time}
\vskip 1cm
\noindent
The $\alpha$-planes provide a geometric definition of twistors
in Minkowski space-time. For this purpose, we first complexify
flat space-time, so that real coordinates 
$\Bigr(x^{0},x^{1},x^{2},x^{3}\Bigr)$ are replaced by complex
coordinates $\Bigr(z^{0},z^{1},z^{2},z^{3}\Bigr)$, and we
obtain a four-dimensional complex vector space equipped with
a non-degenerate complex-bilinear form (Ward and Wells 1990)
$$
(z,w) \equiv z^{0}w^{0}-z^{1}w^{1}-z^{2}w^{2}-z^{3}w^{3} .
\eqno (4.1.1)
$$
The resulting matrix $z^{AA'}$, which, by construction,
corresponds to
the position vector $z^{a}=\Bigr(z^{0},z^{1},z^{2},z^{3}
\Bigr)$, is no longer Hermitian as in the real case. Moreover,
we compactify such a space by identifying future null infinity
with past null infinity (Penrose 1974, Penrose and Rindler 1986,
Esposito 1994). The resulting manifold is here denoted by
$CM^{\#}$, following Penrose and Rindler (1986).

In $CM^{\#}$ with metric $\eta$, we consider two-surfaces
$S$ whose tangent vectors have the two-component spinor
form
$$
v^{a}=\lambda^{A}\pi^{A'} ,
\eqno (4.1.2)
$$
where $\lambda^{A}$ is varying and $\pi^{A'}$ is fixed.
This implies that these tangent vectors are null, since
$\eta(v,v)=v_{a}v^{a}=\Bigr(\lambda^{A}\lambda_{A}\Bigr)
\Bigr(\pi^{A'}\pi_{A'}\Bigr)=0$. Moreover, the induced 
metric on $S$ vanishes identically since any two null
tangent vectors $v^{a}=\lambda^{A}\pi^{A'}$ and
$w^{a}=\mu^{A}\pi^{A'}$ at $p \in S$ are orthogonal:
$$
\eta(v,w)=\Bigr(\lambda^{A}\mu_{A}\Bigr)
\Bigr(\pi^{A'}\pi_{A'}\Bigr)=0 ,
\eqno (4.1.3)
$$
where we have used the property $\pi^{A'}\pi_{A'}=
\varepsilon^{A'B'}\pi_{A'}\pi_{B'}=0$. By virtue of (4.1.3),
the resulting $\alpha$-plane is said to be totally null.
A twistor is then an $\alpha$-plane with constant
$\pi_{A'}$ associated to it. Note that two disjoint families
of totally null two-surfaces exist in $CM^{\#}$, since one
might choose null tangent vectors of the form
$$
u^{a}=\nu^{A}\pi^{A'} ,
\eqno (4.1.4)
$$
where $\nu^{A}$ is fixed and $\pi^{A'}$ is varying. The
resulting two-surfaces are called $\beta$-planes (Penrose
1986).

Theoretical physicists are sometimes more familiar with a
definition involving the vector space of solutions of the
differential equation
$$
{\cal D}_{A'}^{\; \; (A}\omega^{B)}=0 ,
\eqno (4.1.5)
$$
where ${\cal D}$ is the flat connection, 
and ${\cal D}_{AA'}$ the
corresponding spinor covariant derivative. The general
solution of Eq. (4.1.5) in $CM^{\#}$ takes the form
(Penrose and Rindler 1986, Esposito 1994)
$$
\omega^{A}={\Bigr(\omega^{o}\Bigr)}^{A}-i \; x^{AA'}
\pi_{A'}^{o} ,
\eqno (4.1.6)
$$
$$
\pi_{A'}=\pi_{A'}^{o} ,
\eqno (4.1.7)
$$
where $\omega_{A}^{o}$ and $\pi_{A'}^{o}$ are arbitrary
constant spinors, and $x^{AA'}$ is the spinor version of
the position vector with respect to some origin. A twistor
is then {\it represented} by the pair of spinor fields
$\Bigr(\omega^{A},\pi_{A'}\Bigr) \Leftrightarrow Z^{\alpha}$
(Penrose 1975). The twistor equation (4.1.5) is conformally
invariant. This is proved bearing in mind the spinor form 
of the flat four-metric
$$
\eta_{ab}=\varepsilon_{AB} \; \varepsilon_{A'B'} ,
\eqno (4.1.8)
$$
and making the conformal rescaling
$$
{\widehat \eta}_{ab}=\Omega^{2}\eta_{ab} ,
\eqno (4.1.9)
$$
which implies
$$
{\widehat \varepsilon}_{AB}=\Omega \varepsilon_{AB} , \;
{\widehat \varepsilon}_{A'B'}=\Omega \varepsilon_{A'B'} , \;
{\widehat \varepsilon}^{AB}=\Omega^{-1}\varepsilon^{AB} , \;
{\widehat \varepsilon}^{A'B'}=\Omega^{-1}\varepsilon^{A'B'}.
\eqno (4.1.10)
$$
Thus, defining $T_{a} \equiv 
{\cal D}_{a} \Bigr(\log \Omega \Bigr)$
and choosing ${\widehat \omega}^{B}=\omega^{B}$, one finds
(Penrose and Rindler 1986, Esposito 1994)
$$
{\widehat {\cal D}}_{AA'}{\widehat \omega}^{B}
={\cal D}_{AA'}\omega^{B}+\varepsilon_{A}^{\; \; B}
\; T_{CA'}\omega^{C} ,
\eqno (4.1.11)
$$
which implies
$$
{\widehat {\cal D}}_{A'}^{\; \; (A}{\widehat \omega}^{B)}
=\Omega^{-1}{\cal D}_{A'}^{\; \; (A}\omega^{B)} .
\eqno (4.1.12)
$$
Note that the solutions of Eq. (4.1.5) are completely determined
by the four complex components at $\rm O$ of $\omega^{A}$
and $\pi_{A'}$ in a spin-frame at $\rm O$. They are a
four-dimensional vector space over the complex numbers,
called twistor space (Penrose and Rindler 1986, Esposito
1994).

Requiring that $\nu_{A}$ be constant over the
$\beta$-planes implies that 
$\nu^{A}\pi^{A'}{\cal D}_{AA'}\nu_{B}=0$,
for each $\pi^{A'}$, i.e. $\nu^{A}{\cal D}_{AA'}\nu_{B}=0$.
Moreover, a scalar product can be defined between the 
$\omega^{A}$ field and the $\nu_{A}$-scaled $\beta$-plane:
$\omega^{A}\nu_{A}$. Its constancy over the $\beta$-plane
implies that (Penrose 1986)
$$
\nu^{A}\pi^{A'}{\cal D}_{AA'}\Bigr(\omega^{B}\nu_{B}\Bigr)=0 ,
\eqno (4.1.13)
$$
for each $\pi^{A'}$, which leads to
$$
\nu_{A}\nu_{B}\Bigr({\cal D}_{A'}^{\; \; (A}
\omega^{B)}\Bigr)=0 ,
\eqno (4.1.14)
$$
for each $\beta$-plane and hence for each $\nu_{A}$. Thus,
Eq. (4.1.14) becomes the twistor equation (4.1.5). In other
words, it is the twistor concept associated with a
$\beta$-plane which is dual to that associated with a solution
of the twistor equation (Penrose 1986).

Flat projective twistor space $PT$ can be
thought of as three-dimensional complex projective space
$CP^{3}$ (cf. example E2 in section 1.2). This means that we
take the space $C^{4}$ of complex numbers $\Bigr(z^{0},z^{1},
z^{2},z^{3}\Bigr)$ and factor out by the proportionality relation
$\Bigr(\lambda z^{0},...,\lambda z^{3}\Bigr) \sim
\Bigr(z^{0},...,z^{3}\Bigr)$, with $\lambda \in C - \{0\}$. The
homogeneous coordinates $\Bigr(z^{0},...,z^{3}\Bigr)$ are, in the
case of $PT \cong CP^{3}$, as follows:
$\Bigr(\omega^{0},\omega^{1},\pi_{0'},\pi_{1'}\Bigr) \equiv
\Bigr(\omega^{A},\pi_{A'}\Bigr)$. The $\alpha$-planes defined
in this section can be obtained from the equation (cf. (4.1.6))
$$
\omega^{A}=i \; x^{AA'}\pi_{A'} ,
\eqno (4.1.15)
$$
where $\Bigr(\omega^{A},\pi_{A'}\Bigr)$ is regarded as fixed,
with $\pi_{A'} \not =0$. This means that Eq. (4.1.15), considered
as an equation for $x^{AA'}$, has as its solution a complex
two-plane in $CM^{\#}$, whose tangent vectors take the form
in Eq. (4.1.2), i.e. we have found an $\alpha$-plane. The
$\alpha$-planes are self-dual in that, if $v$ and $u$ are any two
null tangent vectors to an $\alpha$-plane, then 
$F \equiv v \otimes u - u \otimes v$ is a self-dual bivector
since
$$
F^{AA'BB'}=\varepsilon^{AB}\phi^{(A'B')} ,
\eqno (4.1.16)
$$
where $\phi^{(A'B')}=\sigma \pi^{A'}\pi^{B'}$, with 
$\sigma \in C - \{ 0 \}$ (Ward 1981b). Note also that
$\alpha$-planes remain unchanged if we replace 
$\Bigr(\omega^{A},\pi_{A'}\Bigr)$ by
$\Bigr(\lambda \omega^{A},\lambda \pi_{A'}\Bigr)$ with
$\lambda \in C-\{0\}$, and that {\it all} $\alpha$-planes
arise as solutions of Eq. (4.1.15). If real solutions of
such equation exist, this implies that 
$x^{AA'}={\overline x}^{AA'}$. This leads to
$$
\omega^{A}{\overline \pi}_{A}+
{\overline \omega}^{A'}\pi_{A'}=i \; x^{AA'}
\Bigr(\pi_{A'}{\overline \pi}_{A}-\pi_{A'}
{\overline \pi}_{A}\Bigr)=0 ,
\eqno (4.1.17)
$$
where overbars denote complex conjugation in two-spinor
language, defined according to the rules described in
section 2.1. If (4.1.17) holds and $\pi_{A'}
\not =0$, the solution space of Eq. (4.1.15) in real Minkowski
space-time is a null geodesic, and {\it all} null geodesics
arise in this way (Ward 1981b). Moreover, if $\pi_{A'}$
vanishes, the point $\Bigr(\omega^{A},\pi_{A'}\Bigr)
=\Bigr(\omega^{A},0\Bigr)$ can be regarded as an $\alpha$-plane
at infinity in compactified Minkowski space-time. Interestingly,
Eq. (4.1.15) is the two-spinor form of the equation expressing
the incidence property of a point $(t,x,y,z)$ in Minkowski
space-time with the twistor $Z^{\alpha}$, i.e. (Penrose 1981)
$$
\pmatrix {Z^{0}\cr Z^{1}\cr}
={i\over \sqrt{2}}
\pmatrix {t+z & x+iy\cr x-iy & t-z\cr}
\pmatrix {Z^{2}\cr Z^{3}\cr} .
\eqno (4.1.18)
$$
The left-hand side of Eq. (4.1.17) may be then re-interpreted
as the twistor pseudo-norm (Penrose 1981)
$$ 
Z^{\alpha}{\overline Z}_{\alpha}=Z^{0}
{\overline {Z^{2}}}
+Z^{1}{\overline {Z^{3}}}
+Z^{2}{\overline {Z^{0}}}
+Z^{3}{\overline {Z^{1}}}
=\omega^{A}{\overline \pi}_{A}
+\pi_{A'}{\overline \omega}^{A'} ,
\eqno (4.1.19)
$$
by virtue of the property $\Bigr({\overline Z}_{0},
{\overline Z}_{1},{\overline Z}_{2},{\overline Z}_{3}
\Bigr)=\Bigr({\overline {Z^{2}}},
{\overline {Z^{3}}}, {\overline {Z^{0}}},
{\overline {Z^{1}}}\Bigr)$. Such a pseudo-norm makes it possible
to define the {\it top half} $PT^{+}$ of $PT$ by the condition
$Z^{\alpha}{\overline Z}_{\alpha}>0$, and the {\it bottom
half} $PT^{-}$ of $PT$ by the condition
$Z^{\alpha}{\overline Z}_{\alpha}<0$.

So far, we have seen that an $\alpha$-plane corresponds to a
point in $PT$, and null geodesics to points in $PN$, the
space of null twistors. However, we may also interpret (4.1.15)
as an equation where $x^{AA'}$ is fixed, and solve for
$\Bigr(\omega^{A},\pi_{A'}\Bigr)$. Within this framework,
$\pi_{A'}$ remains arbitrary, and $\omega^{A}$ is thus given
by $ix^{AA'}\pi_{A'}$. This yields a complex two-plane, and
factorization by the proportionality relation 
$\Bigr(\lambda \omega^{A},\lambda \pi_{A'}\Bigr) \sim
\Bigr(\omega^{A},\pi_{A'}\Bigr)$ leads to a complex projective
one-space $CP^{1}$, with two-sphere topology. Thus, the fixed
space-time point $x$ determines a Riemann sphere
$L_{x} \cong CP^{1}$ in $PT$. In particular, if $x$ is real,
then $L_{x}$ lies entirely within $PN$, given by those twistors
whose homogeneous coordinates satisfy Eq. (4.1.17). 
To sum up, a complex space-time point corresponds to
a sphere in $PT$, whereas a real space-time point corresponds
to a sphere in $PN$ (Penrose 1981, Ward 1981b). 

In Minkowski space-time, two points $p$ and $q$ are 
null-separated if and only if there is a null geodesic 
connecting them. In projective twistor space $PT$, this
implies that the corresponding lines $L_{p}$ and $L_{q}$
intersect, since the intersection point represents the
connecting null geodesic. To conclude this section it may
be now instructive, following Huggett and Tod (1985), to
study the relation between null twistors and null geodesics.
Indeed, given the null twistors $X^{\alpha},Y^{\alpha}$
defined by
$$
X^{\alpha} 
\equiv \Bigr(i \; x_{0}^{AC'} \; X_{C'},X_{A'}\Bigr) ,
\eqno (4.1.20)
$$
$$
Y^{\alpha} 
\equiv \Bigr(i \; x_{1}^{AC'} \; Y_{C'},Y_{A'}\Bigr) ,
\eqno (4.1.21)
$$
the corresponding null geodesics are
$$
\gamma_{X}: \; x^{AA'} \equiv x_{0}^{AA'}+\lambda \; 
{\overline X}^{A} \; X^{A'} ,
\eqno (4.1.22)
$$
$$
\gamma_{Y}: \; x^{AA'} \equiv x_{1}^{AA'}+\mu \;
{\overline Y}^{A} \; Y^{A'} .
\eqno (4.1.23)
$$
If these intersect at some point $x_{2}$, one finds
$$
x_{2}^{AA'}=x_{0}^{AA'}+\lambda \; {\overline X}^{A}
\; X^{A'}=x_{1}^{AA'}+\mu \; {\overline Y}^{A} \; Y^{A'} ,
\eqno (4.1.24)
$$
where $\lambda,\mu \in R$. Hence
$$
x_{2}^{AA'} \; {\overline Y}_{A} \; X_{A'}
=x_{0}^{AA'} \; {\overline Y}_{A} \; X_{A'}
=x_{1}^{AA'} \; {\overline Y}_{A} \; X_{A'} ,
\eqno (4.1.25)
$$
by virtue of the identities $X^{A'}X_{A'}=
{\overline Y}^{A} \; {\overline Y}_{A}=0$. Equation (4.1.25)
leads to
$$
X^{\alpha}{\overline Y}_{\alpha}=i\Bigr(x_{0}^{AA'}
\; {\overline Y}_{A} \; X_{A'}
-x_{1}^{AA'} \; {\overline Y}_{A} \; X_{A'}\Bigr)=0 .
\eqno (4.1.26)
$$
Suppose instead we are given Eq. (4.1.26). This implies 
that some real $\lambda$ and $\mu$ exist such that
$$
x_{0}^{AA'}-x_{1}^{AA'}=-\lambda \; {\overline X}^{A}
\; X^{A'}+\mu \; {\overline Y}^{A} \; Y^{A'} ,
\eqno (4.1.27)
$$
where signs on the right-hand side of (4.1.27) have been
suggested by (4.1.24). Note that (4.1.27) only holds if
$X_{A'}Y^{A'} \not =0$, i.e. if $\gamma_{X}$ and 
$\gamma_{Y}$ are not parallel. However, the whole argument can
be generalized to this case as well (our problem 4.2, Huggett and
Tod 1985), and one finds that in all cases the null geodesics
$\gamma_{X}$ and $\gamma_{Y}$ intersect if and only if
$X^{\alpha} \; {\overline Y}_{\alpha}$ vanishes.
\vskip 1cm
\centerline {\bf 4.2 $\alpha$-surfaces and twistor geometry}
\vskip 1cm
\noindent
The $\alpha$-planes defined in section 4.1 can be generalized
to a suitable class of curved complex space-times. 
By a complex space-time $(M,g)$ we mean a
four-dimensional Hausdorff manifold $M$ with holomorphic
metric $g$. Thus, with respect to a holomorphic coordinate basis
$x^{a}$, $g$ is a $4 \times 4$ matrix of holomorphic functions
of $x^{a}$, and its determinant is nowhere-vanishing (Ward
1980b, Ward and Wells 1990). Remarkably, $g$ determines a unique
holomorphic connection $\nabla$, and a holomorphic curvature
tensor $R_{\; bcd}^{a}$. Moreover, the Ricci tensor $R_{ab}$
becomes complex-valued, and the Weyl tensor 
$C_{\; bcd}^{a}$ may be split into {\it independent} holomorphic
tensors, i.e. its self-dual and anti-self-dual parts, respectively.
With our two-spinor notation, one has (see (2.1.36))
$$
C_{abcd}=\psi_{ABCD} \; \varepsilon_{A'B'} \; \varepsilon_{C'D'}
+{\widetilde \psi}_{A'B'C'D'}
\; \varepsilon_{AB} \; \varepsilon_{CD} ,
\eqno (4.2.1)
$$
where $\psi_{ABCD}=\psi_{(ABCD)}, {\widetilde \psi}_{A'B'C'D'}=
{\widetilde \psi}_{(A'B'C'D')}$. The spinors $\psi$ and
$\widetilde \psi$ are the anti-self-dual and self-dual Weyl
spinors, respectively. Following Penrose (1976a,b), Ward and 
Wells (1990), complex vacuum space-times such that
$$
{\widetilde \psi}_{A'B'C'D'}=0 , \;
R_{ab}=0 ,
\eqno (4.2.2)
$$
are called {\it right-flat} or {\it anti-self-dual}, whereas
complex vacuum space-times such that
$$
\psi_{ABCD}=0 , \;
R_{ab}=0 ,
\eqno (4.2.3)
$$
are called {\it left-flat} or {\it self-dual}. Note that this
definition only makes sense if space-time is complex or real
Riemannian, since in this case no complex conjugation relates
primed to unprimed spinors (i.e. the corresponding spin-spaces
are no longer anti-isomorphic). Hence, for example, the self-dual
Weyl spinor ${\widetilde \psi}_{A'B'C'D'}$ may vanish without
its anti-self-dual counterpart $\psi_{ABCD}$ having to vanish
as well, as in Eq. (4.2.2), or the converse may hold, as in
Eq. (4.2.3) (see section 1.1 and problem 2.3).

By definition, $\alpha$-surfaces are complex two-surfaces $S$
in a complex space-time $(M,g)$ whose tangent vectors $v$ have
the two-spinor form (4.1.2), where $\lambda^{A}$ is varying,
and $\pi^{A'}$ is a fixed primed spinor field on $S$. From this
definition, the following properties can be derived (cf.
section 4.1).

(i) tangent vectors to $\alpha$-surfaces are null;

(ii) any two null tangent vectors $v$ and $u$ to an
$\alpha$-surface are orthogonal to one another;

(iii) the holomorphic metric $g$ vanishes on $S$ in
that $g(v,u)=g(v,v)=0, \forall v,u$ (cf. (4.1.3)), so
that $\alpha$-surfaces are totally null;

(iv) $\alpha$-surfaces are self-dual, in that 
$F \equiv v \otimes u -u \otimes v$ takes the two-spinor
form (4.1.16);

(v) $\alpha$-surfaces exist in $(M,g)$ if and only if the
self-dual Weyl spinor vanishes, so that $(M,g)$ is
anti-self-dual.
\vskip 0.3cm
\noindent
Note that properties (i)--(iv), here written in a redundant
form for pedagogical reasons, are the same as in the
flat-space-time case, provided we replace the flat metric
$\eta$ with the curved metric $g$. Condition (v), however,
is a peculiarity of curved space-times. The reader may find
a detailed proof of the necessity of this condition as a 
particular case of the calculations appearing in chapter
six, where we study a holomorphic metric-compatible connection
$\nabla$ with non-vanishing torsion. To avoid repeating
ourselves, we focus instead on the sufficiency of the condition,
following Ward and Wells (1990).

We want to prove that, if $(M,g)$ is anti-self-dual, it admits
a three-complex-parameter family of self-dual $\alpha$-surfaces.
Indeed, given any point $p \in M$ and a spinor $\mu_{A'}$ at
$p$, one can find a spinor field $\pi_{A'}$ on $M$, satisfying
the equation (cf. Eq. (6.2.10))
$$
\pi^{A'}\Bigr(\nabla_{AA'}\pi_{B'}\Bigr)=\xi_{A}\pi_{B'} ,
\eqno (4.2.4)
$$
and such that
$$
\pi_{A'}(p)=\mu_{A'}(p) .
\eqno (4.2.5)
$$
Hence $\pi_{A'}$ defines a holomorphic two-dimensional
distribution, spanned by the vector
fields of the form $\lambda^{A}\pi^{A'}$, which is
integrable by virtue of (4.2.4). Thus, in particular, there
exists a self-dual $\alpha$-surface through $p$, with tangent
vectors of the form $\lambda^{A}\mu^{A'}$ at $p$. Since $p$
is arbitrary, this argument may be repeated $\forall p \in M$.
The space $\cal P$ of all self-dual $\alpha$-surfaces in
$(M,g)$ is three-complex-dimensional, and is called twistor
space of $(M,g)$.
\vskip 1cm
\centerline {\bf 4.3 Geometric theory of partial differential
equations}
\vskip 1cm
\noindent
One of the main results of twistor theory has been a deeper
understanding of the solutions of partial differential
equations of classical field theory. Remarkably, a problem
in analysis becomes a purely geometric problem (Ward 1981b,
Ward and Wells 1990). For example, in Bateman (1904) it was 
shown that the general real-analytic solution of the wave
equation $\cstok{\ }\phi=0$ in Minkowski space-time is
$$
\phi(x,y,z,t)=\int_{-\pi}^{\pi}F(x \cos \theta +y \sin \theta
+iz, y+iz \sin \theta +t \cos \theta, \theta) \; d\theta ,
\eqno (4.3.1)
$$
where $F$ is an arbitrary function of three variables, 
complex-analytic in the first two. Indeed, twistor theory
tells us that $F$ is a function on $PT$. More precisely, let
$f \Bigr(\omega^{A},\pi_{A'}\Bigr)$ be a complex-analytic
function, homogeneous of degree $-2$, i.e. such that
$$
f\Bigr(\lambda \omega^{A},\lambda \pi_{A'}\Bigr)
=\lambda^{-2} f\Bigr(\omega^{A},\pi_{A'}\Bigr) ,
\eqno (4.3.2)
$$
and possibly having singularities (Ward 1981b). We now define a
field $\phi(x^{a})$ by
$$
\phi(x^{a}) 
\equiv {1\over 2\pi i}\oint f \Bigr(i \; x^{AA'}
\pi_{A'},\pi_{B'}\Bigr)\pi_{C'} \; d\pi^{C'} ,
\eqno (4.3.3)
$$
where the integral is taken over any closed one-dimensional
contour that avoids the singularities of $f$. Such a field
satisfies the wave equation, and every solution of
$\cstok{\ }\phi=0$ can be obtained in this way. The function
$f$ has been taken to have homogeneity $-2$ since the
corresponding one-form $f \pi_{C'} \; d\pi^{C'}$ has homogeneity
zero and hence is a one-form on projective twistor space $PT$,
or on some subregion of $PT$, since it may have singularities.
The homogeneity is related to the property of $f$ of being 
a free function of three variables. Since $f$ is not defined
on the whole of $PT$, and $\phi$ does not determine $f$ uniquely,
because we can replace $f$ by $f+{\widetilde f}$, where 
$\widetilde f$ is any function such that
$$
\oint {\widetilde f} \pi_{C'} \; d\pi^{C'}=0 ,
\eqno (4.3.4)
$$
we conclude that $f$ is an element of the 
sheaf-cohomology group $H^{1}\Bigr(PT^{+},O(-2)\Bigr)$,
i.e. the complex vector space of arbitrary complex-analytic
functions of three variables, not subject to any differential
equations (Penrose 1980, Ward 1981b, Ward and Wells 1990).
Remarkably, a conformally invariant isomorphism exists between
the complex vector space of holomorphic solutions of 
$\cstok{\ }\phi=0$ on the forward tube $CM^{+}$ (i.e. the domain
of definition of positive-frequency fields), and the
sheaf-cohomology group $H^{1}\Bigr(PT^{+},O(-2)\Bigr)$.

It is now instructive to summarize some basic ideas
of sheaf-cohomology theory and its use in
twistor theory, following Penrose (1980). For this purpose,
let us begin by recalling how Cech cohomology is obtained. 
We consider a Hausdorff paracompact topological space $X$,
covered with a locally finite system of open sets $U_{i}$.
With respect to this covering, we define a {\it cochain}
with coefficients in an additive Abelian group $G$ (e.g.
$Z, R$ or $C$) in terms of elements $f_{i},f_{ij},f_{ijk}
... \in G$. These elements are assigned to the open sets
$U_{i}$ of the covering, and to their non-empty intersections,
as follows: $f_{i}$ to $U_{i}$, $f_{ij}$ to $U_{i} \cap U_{j}$,
$f_{ijk}$ to $U_{i} \cap U_{j} \cap U_{k}$ and so on. The
elements assigned to non-empty intersections are completely
antisymmetric, so that $f_{i...p}=f_{[i...p]}$. One is thus led
to define
$$
{\rm zero-cochain} \; \alpha 
\equiv \Bigr(f_{1},f_{2},f_{3},...\Bigr) ,
\eqno (4.3.5)
$$
$$
{\rm one-cochain} \; \beta 
\equiv \Bigr(f_{12},f_{23},f_{13},...
\Bigr) ,
\eqno (4.3.6)
$$
$$
{\rm two-cochain} \; \gamma 
\equiv \Bigr(f_{123},f_{124},...\Bigr) ,
\eqno (4.3.7)
$$
and the {\it coboundary operator} $\delta$:
$$
\delta \alpha \equiv \Bigr(f_{2}-f_{1},f_{3}-f_{2},f_{3}-f_{1},
... \Bigr)
\equiv \Bigr(f_{12},f_{23},f_{13},...\Bigr) ,
\eqno (4.3.8)
$$
$$ 
\delta \beta \equiv \Bigr(f_{12}-f_{13}+f_{23},f_{12}-f_{14}
+f_{24},...\Bigr) 
\equiv \Bigr(f_{123},f_{124},...\Bigr) .
\eqno (4.3.9) 
$$
By virtue of (4.3.8) and (4.3.9) one finds $\delta^{2} \alpha =
\delta^{2} \beta =...=0$. {\it Cocycles} $\gamma$ are cochains
such that $\delta \gamma=0$. {\it Coboundaries} are a particular
set of cocycles, i.e. such that $\gamma = \delta \beta$ for some
cochain $\beta$. Of course, {\it all} coboundaries are cocycles,
whereas the converse does not hold. This enables one to define 
the ${\rm p}^{\rm th}$ cohomology group as the quotient space
$$
H_{\Bigr \{U_{i}\Bigr \}}^{p}(X,G) \equiv 
{G_{CC}^{p}/ G_{CB}^{p}} ,
\eqno (4.3.10)
$$
where $G_{CC}^{p}$ is the additive group of p-cocycles,
and $G_{CB}^{p}$ is the additive group of p-coboundaries.
To avoid having a definition which depends on the covering
$\Bigr \{U_{i} \Bigr \}$, one should then take finer and finer
coverings of $X$ and settle on a {\it sufficiently fine}
covering ${\Bigr \{U_{i} \Bigr \}}^{*}$. Following Penrose
(1980), by this we mean that all the $H^{p} \Bigr(U_{i} \cap
... \cap U_{k},G \Bigr)$ vanish $\forall p >0$. One then
defines
$$
H_{{\Bigr \{U_{i}\Bigr \}}^{*}}^{p}(X,G) \equiv
H^{p}(X,G) .
\eqno (4.3.11)
$$
We always assume such a covering exists, is countable and
locally finite. Note that, rather than thinking of $f_{i}$
as an element of $G$ assigned to $U_{i}$, of $f_{ij}$ as
assigned to $U_{ij}$ and so on, we can think of $f_{i}$ as
a {\it function} defined on $U_{i}$ and taking a constant 
value $\in G$. Similarly, we can think of $f_{ij}$ as a
$G$-valued constant function defined on $U_{i} \cap U_{j}$,
and this implies it is not strictly necessary to assume that
$U_{i} \cap U_{j}$ is non-empty.

The generalization to sheaf cohomology is obtained if we do
not require the functions $f_{i},f_{ij},f_{ijk}...$ to be
constant (there are also cases when the additive group $G$
is allowed to vary from point to point in $X$). The assumption
of main interest is the holomorphic nature of the $f$'s.
A sheaf is so defined that the Cech cohomology previously
defined works as well as before (Penrose 1980). In other words,
a sheaf $S$ defines an additive group $G_{u}$ for each open set
$U \subset X$. Relevant examples are as follows.

(i) The sheaf $O$ of germs of holomorphic functions on a
complex manifold $X$ is obtained if $G_{u}$ is taken to be
the additive group of all holomorphic functions on $U$.

(ii) Twisted holomorphic functions, i.e. functions whose
values are not complex numbers, but are taken in some complex
line bundle over $X$.

(iii) A particular class of twisted functions is obtained if
$X$ is projective twistor space $PT$ (or $PT^{+}$, or
$PT^{-}$), and the functions studied are holomorphic and
homogeneous of some degree $n$ in the twistor variable, i.e.
$$
f \Bigr (\lambda \omega^{A}, \lambda \pi_{A'} \Bigr)
=\lambda^{n} f \Bigr(\omega^{A},\pi_{A'}\Bigr) .
\eqno (4.3.12)
$$
If $G_{u}$ consists of all such twisted functions on
$U \subset X$, the resulting sheaf, denoted by $O(n)$, is
the sheaf of germs of holomorphic functions twisted by
$n$ on $X$.

(iv) We can also consider vector-bundle-valued functions,
where the vector bundle $B$ is over $X$, and $G_{u}$ consists
of the cross-sections of the portion of $B$ lying above $U$.
\vskip 0.3cm
\noindent
Defining cochains and coboundary operator as before, with
$f_{i} \in G_{U_{i}}$ and so on, we obtain the ${\rm p}^{\rm th}$
cohomology group of $X$, with coefficients in the sheaf $S$, as
the quotient space
$$
H^{p}(X,S) \equiv G^{p}(S)/G_{CB}^{p}(S) ,
\eqno (4.3.13)
$$
where $G^{p}(S)$ is the group of p-cochains with coefficients
in $S$, and $G_{CB}^{p}(S)$ is the group of p-coboundaries
with coefficients in $S$. Again, we take finer and finer
coverings $\Bigr \{U_{i} \Bigr \}$ of $X$, and we settle on a
{\it sufficiently fine} covering. To understand this concept,
we recall the following definitions (Penrose 1980).
\vskip 0.3cm
\noindent
{\bf Definition 4.3.1} A {\it coherent analytic} sheaf is
locally defined by $n$ holomorphic functions factored out by
a set of $s$ holomorphic relations.
\vskip 0.3cm
\noindent
{\bf Definition 4.3.2} A Stein manifold is a holomorphically
convex open subset of $C^{n}$.
\vskip 0.3cm
\noindent
Thus, we can say that, provided $S$ is a coherent analytic
sheaf, {\it sufficiently fine} means that each of
$U_{i},U_{i}\cap U_{j},U_{i}\cap U_{j} \cap U_{k} ...$
is a Stein manifold. If $X$ is Stein and $S$ is coherent
analytic, then $H^{p}(X,S)=0, \forall p >0$.

We can now consider again the remarks following Eq. (4.3.4), i.e.
the interpretation of twistor functions as elements of
$H^{1}\Bigr(PT^{+},O(-2)\Bigr)$. Let $X$ be a part of $PT$,
e.g. the neighbourhood of a line in $PT$, or the top half
$PT^{+}$, or the closure ${\overline {PT^{+}}}$ of the top
half. We assume $X$ can be covered with two open sets 
$U_{1},U_{2}$ such that every projective line $L$ in $X$ meets
$U_{1} \cap U_{2}$ in an annular region. For us, $U_{1} \cap
U_{2}$ corresponds to the domain of definition of a twistor
function $f(Z^{\alpha})$, homogeneous of degree $n$ in the
twistor $Z^{\alpha}$ (see (4.3.12)). Then 
$f \equiv f_{12} \equiv f_{2}-f_{1}$ is a twisted function
on $U_{1} \cap U_{2}$, and defines a one-cochain $\epsilon$,
with coefficients in $O(n)$, for $X$. By construction
$\delta \epsilon =0$, hence $\epsilon$ is a cocycle. For this
covering, the one-coboundaries are functions of the form
$l_{2}-l_{1}$, where $l_{2}$ is holomorphic on $U_{2}$ and
$l_{1}$ on $U_{1}$. The equivalence between twistor functions
is just the cohomological equivalence between one-cochains
$\epsilon,\epsilon'$ that their difference should be a
coboundary: $\epsilon'-\epsilon=\delta \alpha$, with
$\alpha =\Bigr(l_{1},l_{2}\Bigr)$. This is why we view twistor
functions as defining elements of $H^{1}\Bigr(X,O(n)\Bigr)$.
Indeed, if we try to get finer coverings, we realize it is
often impossible to make $U_{1}$ and $U_{2}$ into Stein manifolds.
However, if $X={\overline {PT^{+}}}$, the covering
$\Bigr \{U_{1},U_{2} \Bigr \}$ by two sets is sufficient for
any analytic, positive-frequency field (Penrose 1980).

The most striking application of twistor theory to partial
differential equations is perhaps the geometric characterization
of anti-self-dual space-times with a cosmological constant.
For these space-times, the Weyl tensor takes the form
$$
C_{abcd}^{(A.S.D.)}=\psi_{ABCD} \; e_{A'B'} \; e_{C'D'} ,
\eqno (4.3.14)
$$
and the Ricci tensor reads
$$
R_{ab}=-2\Phi_{ab}+6\Lambda g_{ab} .
\eqno (4.3.15)
$$
With our notation, $e_{AB}$ and $e_{A'B'}$ are the curved-space
version of the $\varepsilon$-symbols (denoted again by 
$\varepsilon_{AB}$ and $\varepsilon_{A'B'}$ in Eqs. (2.1.36)
and (4.2.1)), $\Phi_{ab}$ is the trace-free part of Ricci,
$24\Lambda$ is the trace $R=R_{\; \; a}^{a}$ of Ricci
(Ward 1980b). The local structure in projective twistor space
which gives information about the metric is a pair of
differential forms: a one-form $\tau$ homogeneous of degree 2
and a three-form $\rho$ homogeneous of degree 4. Basically, 
$\tau$ contains relevant information about $e_{A'B'}$ and 
$\rho$ tells us about $e_{AB}$, hence their knowledge
determines $g_{ab}=e_{AB} \; e_{A'B'}$. The result proved in
Ward (1980b) states that a one-to-one correspondence exists 
between sufficiently local anti-self-dual solutions with
scalar curvature $R=24 \Lambda$ and sufficiently small
deformations of flat projective twistor space which preserve 
the one-form $\tau$ and the three-form $\rho$, where
$\tau \wedge d\tau=2\Lambda \rho$. We now describe how to
define the forms $\tau$ and $\rho$, whereas the explicit
construction of a class of anti-self-dual space-times is 
given in chapter five.

The geometric framework is twistor space $\cal P$ defined
at the end of section 4.2, i.e. the space of all
$\alpha$-surfaces in $(M,g)$. We take $M$ to be sufficiently
small and convex to ensure that $\cal P$ is a complex
manifold with topology $R^{4} \times S^{2}$, since every point
in an anti-self-dual space-time has such a neighbourhood
(Ward 1980b). If $Q$, represented by the pair 
$\Bigr(\alpha^{A},\beta_{A'}\Bigr)$, is any vector in $\cal P$,
then $\tau$ is defined by
$$
\tau(Q) \equiv e^{A'B'} \; \pi_{A'} \; \beta_{B'} .
\eqno (4.3.16)
$$
To make sure $\tau$ is well defined, one has to check that
the right-hand side of (4.3.16) remains covariantly constant
over $\alpha$-surfaces, i.e. is annihilated by the first-order
operator $\lambda^{A}\pi^{A'}\nabla_{AA'}$, since otherwise
$\tau$ does not correspond to a differential form on $\cal P$.
It turns out that $\tau$ is well defined provided the trace-free
part of Ricci vanishes. This is proved using spinor Ricci
identities and the equations of local twistor transport as 
follows (Ward 1980b).

Let $v$ be a vector field on the $\alpha$-surface $Z$ such
that $\epsilon v^{a}$ joins $Z$ to the neighbouring 
$\alpha$-surface $Y$. Since $\epsilon v^{a}$ acts as a connecting
vector, the Lie bracket of $v^{a}$ and $\lambda^{B}\pi^{B'}$
vanishes for all $\lambda^{B}$, i.e.
$$
\lambda^{B} \; \pi^{B'} \; \nabla_{BB'} \; v^{AA'}
-v^{BB'} \; \nabla_{BB'} \; \lambda^{A} \; \pi^{A'}=0 .
\eqno (4.3.17)
$$
Thus, after defining
$$
\beta_{A'} \equiv v^{BB'} \; \nabla_{BB'} \; \pi_{A'} ,
\eqno (4.3.18)
$$
one finds
$$
\pi_{A'} \; \lambda^{B} \; \pi^{B'} 
\; \nabla_{BB'} \; v^{AA'}
=\lambda^{A} \; \beta^{A'} \; \pi_{A'} .
\eqno (4.3.19)
$$
If one now applies the torsion-free spinor Ricci identities
(see Eqs. (6.3.17) and (6.3.18) setting ${\widetilde \chi}
={\widetilde \Sigma}=\chi=\Sigma=0$ therein), one finds that
the spinor field $\beta_{A'}(x)$ on $Z$ satisfies the
equation
$$
\lambda^{B} \; \pi^{B'} \; \nabla_{BB'} \; \beta_{A'}
=-i \; \lambda^{B} \; \pi^{B'} \; P_{ABA'B'} \; \alpha^{A} ,
\eqno (4.3.20)
$$
where $P_{ab}=\Phi_{ab}-\Lambda g_{ab}$ and $\alpha^{A}
=iv^{AC'} \; \pi_{C'}$. Moreover, Eq. (4.3.19) and the
Leibniz rule imply that
$$
\lambda^{B} \; \pi^{B'} \; \nabla_{BB'} \; \alpha^{A}
=-i \; \lambda^{A} \; \pi^{A'} \; \beta_{A'} ,
\eqno (4.3.21)
$$
since $\pi^{B'}\nabla_{BB'}\pi_{C'}=0$. Equations (4.3.20) and (4.3.21)
are indeed the equations of {\it local twistor transport}, and
Eq. (4.3.20) leads to
$$ \eqalignno{
\lambda^{C}\pi^{C'}\nabla_{CC'}\Bigr(e^{A'B'}\pi_{A'} \; 
\beta_{B'}\Bigr)&=e^{A'B'}\pi_{A'}\Bigr(\lambda^{C}\pi^{C'}
\nabla_{CC'}\beta_{B'}\Bigr)\cr
&=-i \; \lambda^{B}\pi^{B'} \pi_{C'} \; e^{C'A'}\alpha^{A}
\Bigr(\Phi_{ABA'B'}-\Lambda e_{AB} \; e_{A'B'}\Bigr)\cr
&=i \; \lambda^{B}\pi^{A'}\pi^{B'}\alpha^{A}
\Phi_{ABA'B'} ,
&(4.3.22)\cr}
$$
since $\pi^{A'}\pi^{B'}e_{A'B'}=0$. Hence, as we said before,
$\tau$ is well defined provided the trace-free part of Ricci
vanishes. Note that, strictly, $\tau$ is a twisted form rather
than a form on $\cal P$, since it is homogeneous of degree 2,
one from $\pi_{A'}$ and one from $\beta_{B'}$. By contrast,
a one-form would be independent of the scaling of $\pi_{A'}$
and $\beta_{B'}$ (Ward 1980b).
 
We are now in a position to define the three-form $\rho$, 
homogeneous of degree 4. For this purpose, let us denote by
$Q_{h}$, $h=1,2,3$ three vectors in $\cal P$, represented
by the pairs $\Bigr(\alpha_{h}^{A},\beta_{hA'}\Bigr)$. The
corresponding $\rho(Q_{1},Q_{2},Q_{3})$ is obtained by taking
$$
\rho_{123} \equiv {1\over 2} \Bigr(e^{A'B'}
\pi_{A'} \; \beta_{1B'}\Bigr)
\Bigr(e_{AB} \; \alpha_{2}^{A} \; \alpha_{3}^{B}\Bigr) ,
\eqno (4.3.23)
$$
and then anti-symmetrizing $\rho_{123}$ over $1,2,3$. This
yields
$$
\rho(Q_{1},Q_{2},Q_{3}) \equiv {1\over 6}
\Bigr(\rho_{123}-\rho_{132}+\rho_{231}-\rho_{213}
+\rho_{312}-\rho_{321}\Bigr) .
\eqno (4.3.24)
$$
The reader can check that, by virtue of Eqs. (4.3.20) and (4.3.21),
$\rho$ is well defined, since it is covariantly constant over
$\alpha$-surfaces:
$$
\lambda^{A} \; \pi^{A'} \; 
\nabla_{AA'} \; \rho(Q_{1},Q_{2},Q_{3})=0 .
\eqno (4.3.25)
$$
\vskip 100cm
\centerline {\it CHAPTER FIVE}
\vskip 1cm
\centerline {\bf PENROSE TRANSFORM FOR GRAVITATION}
\vskip 1cm
\noindent
Deformation theory of complex manifolds
is applied to construct a class of anti-self-dual
solutions of Einstein's vacuum equations, following
the work of Penrose and Ward. The hard part of the
analysis is to find the holomorphic cross-sections of
a deformed complex manifold, and the corresponding
conformal structure of an anti-self-dual
space-time. This calculation is
repeated in detail, using complex analysis and
two-component spinor techniques.

If no assumption about anti-self-duality is made,
twistor theory is by itself insufficient to characterize
geometrically a solution of the full Einstein equations.
After a brief review of alternative ideas based on the 
space of complex null geodesics of complex space-time,
and Einstein-bundle constructions, attention is focused
on the attempt by Penrose to define twistors
as charges for massless spin-${3\over 2}$ fields.
This alternative definition is considered since a vanishing
Ricci tensor provides the consistency condition for the
existence and propagation of massless spin-${3\over 2}$ fields
in curved space-time, whereas in Minkowski space-time the space
of charges for such fields is naturally identified with the
corresponding twistor space.

The two-spinor analysis of the Dirac form of such
fields in Minkowski space-time is carried out in detail by
studying their two potentials with corresponding gauge freedoms.
The Rarita--Schwinger form is also introduced, and self-dual
vacuum Maxwell fields are obtained from massless spin-${3\over 2}$
fields by spin-lowering. In curved space-time, however, 
the local expression of spin-${3\over 2}$ field strengths in terms
of the second of these
potentials is no longer possible, unless one studies
the self-dual Ricci-flat case. Thus, much more work is needed
to characterize geometrically a Ricci-flat (complex) space-time
by using this alternative concept of twistors.
\vskip 100cm
\centerline {\bf 5.1 Anti-self-dual space-times}
\vskip 1cm
\noindent
Following Ward (1978), we now use twistor-space techniques
to construct a family of anti-self-dual solutions of
Einstein's vacuum equations. Bearing in mind the
space-time twistor-space correspondence in Minkowskian
geometry described in section 4.1, we take a region
$\cal R$ of $CM^{\#}$, whose corresponding region in $PT$ is
$\widetilde {\cal R}$. Moreover, $\cal N$ is the 
non-projective version of $\widetilde {\cal R}$, which
implies ${\cal N} \subset T \subset C^{4}$. In other
words, as coordinates on $\cal N$ we may use
$\Bigr(\omega^{o},\omega^{1},\pi_{o'},\pi_{1'}\Bigr)$.
The geometrically-oriented reader may like it to know that
three important structures are associated with $\cal N$:

(i) the fibration $\Bigr(\omega^{A},\pi_{A'}\Bigr)
\rightarrow \pi_{A'}$, which implies that $\cal N$ becomes
a bundle over $C^{2}- \{0 \}$;

(ii) the two-form ${1\over 2} d\omega_{A} \wedge 
d\omega^{A}$ on each fibre;

(iii) the projective structure ${\cal N} \rightarrow
{\widetilde {\cal R}}$.
\vskip 0.3cm
\noindent
Deformations of $\cal N$ which preserve this projective
structure correspond to right-flat metrics (see section 4.2)
in $\cal R$. To obtain such deformations, cover $\cal N$ with
two patches $\cal Q$ and ${\widehat {\cal Q}}$. Coordinates on
$\cal Q$ and on ${\widehat {\cal Q}}$ are 
$\Bigr(\omega^{A},\pi_{A'}\Bigr)$ and
$\Bigr({\widehat \omega}^{A}, {\widehat \pi}_{A'}\Bigr)$
respectively. We may now {\it glue} $\cal Q$ and
${\widehat {\cal Q}}$ together according to
$$
{\widehat \omega}^{A}=\omega^{A}
+f^{A}\Bigr(\omega^{B},\pi_{B'}\Bigr) ,
\eqno (5.1.1)
$$
$$
{\widehat \pi}_{A'}=\pi_{A'} ,
\eqno (5.1.2)
$$
where $f^{A}$ is homogeneous of degree 1, holomorphic on
${\cal Q} \bigcap {\widehat {\cal Q}}$, and satisfies
$$
{\rm det} \; \biggr(\varepsilon_{A}^{\; \; B}
+{\partial f^{B} \over \partial \omega^{A}}\biggr)
=1 .
\eqno (5.1.3)
$$
Such a patching process yields a complex manifold 
${\cal N}^{D}$ which is a deformation of $\cal N$. The
corresponding right-flat space-time $\cal G$ is such that
its points correspond to the holomorphic cross-sections of
${\cal N}^{D}$. The hard part of the analysis is indeed to
find these cross-sections, but this can be done explicitly 
for a particular class of patching functions. For this
purpose, we first choose a constant spinor field
$p^{AA'B'}=p^{A(A'B')}$ and a homogeneous holomorphic
function $g(\gamma,\pi_{A'})$ of three complex variables:
$$
g\Bigr(\lambda^{3}\gamma,\lambda \pi_{A'}\Bigr)
=\lambda^{-1}g\Bigr(\gamma,\pi_{A'}\Bigr)
\; \; \forall \lambda \in C - \{0 \} .
\eqno (5.1.4)
$$
This enables one to define the spinor field
$$
p^{A} \equiv p^{AA'B'} \; \pi_{A'} \; \pi_{B'} ,
\eqno (5.1.5)
$$
and the patching function
$$
f^{A} \equiv p^{A} \; g\Bigr(p_{B}\omega^{B},\pi_{B'}\Bigr) ,
\eqno (5.1.6)
$$
and the function
$$
F(x^{a},\pi_{A'}) \equiv g \Bigr(i \; p_{A} \; x^{AC'} \; \pi_{C'},
\pi_{A'}\Bigr) .
\eqno (5.1.7)
$$
Under suitable assumptions on the singularities of $g$, $F$
may turn out to be holomorphic if $x^{a} \in {\cal R}$ and if
the ratio ${\widetilde \pi} \equiv {\pi_{o'} \over \pi_{1'}}
\in ]{1\over 2},{5\over 2}[$. It is also possible to express
$F$ as the difference of two contour integrals after
defining the differential form
$$
\Omega \equiv {\Bigr(2\pi i \rho^{A'}\pi_{A'}\Bigr)}^{-1}
\; F(x^{b},\rho_{B'}) \; \rho_{C'}d\rho^{C'} .
\eqno (5.1.8)
$$
In other words, if $\Gamma$ and $\widehat {\Gamma}$ are closed
contours on the projective $\rho_{A'}$-sphere defined by
$|{\widetilde \rho}|=1$ and
$|{\widetilde \rho}|=2$ respectively, we may define 
the function
$$
h \equiv  \oint_{\Gamma} \Omega ,
\eqno (5.1.9)
$$
holomorphic for ${\widetilde \pi}<2$, and the function
$$
{\widehat h} \equiv \oint_{\widehat \Gamma} \Omega ,
\eqno (5.1.10)
$$
holomorphic for ${\widetilde \pi}>1$. Thus, by virtue of
Cauchy's integral formula, one finds (cf. Ward 1978)
$$
F(x^{a},\pi_{A'})={\widehat h}(x^{a},\pi_{A'})
-h(x^{a},\pi_{A'}) .
\eqno (5.1.11)
$$
The basic concepts of sheaf-cohomology presented in section
4.3 are now useful to understand the deep meaning of these
formulae. For any fixed $x^{a}$, $F(x^{a},\pi_{A'})$ 
determines an element of the sheaf-cohomology group 
$H^{1}(P_{1}(C),O(-1))$, where $P_{1}(C)$ is the Riemann
sphere of projective $\pi_{A'}$ spinors and $O(-1)$ is the
sheaf of germs of holomorphic functions of $\pi_{A'}$,
homogeneous of degree $-1$. Since $H^{1}$ vanishes, $F$ is
actually a coboundary. Hence it can be split according to
(5.1.11).

In the subsequent calculations, it will be useful to write
a solution of the Weyl equation $\nabla^{AA'}\psi_{A}=0$
in the form
$$
\psi_{A} \equiv i \; \pi^{A'} \; \nabla_{AA'}h(x^{a},\pi_{C'}).
\eqno (5.1.12)
$$
Moreover, following again Ward (1978), we note that a spinor
field $\xi_{A'}^{\; \; \; B'}(x)$ can be defined by
$$
\xi_{A'}^{\; \; \; B'}\pi_{B'} \equiv
i \; p^{AB'C'} \; \pi_{B'} \; \pi_{C'}
\; \nabla_{AA'}h(x,\pi_{D'}) ,
\eqno (5.1.13)
$$
and that the following identities hold:
$$
i \; p^{AA'B'} \; \pi_{B'} \; \nabla_{AA'}h(x,\pi_{C'})
=\xi \equiv {1\over 2} \xi_{A'}^{\; \; A'} ,
\eqno (5.1.14)
$$
$$
\psi_{A} \; p^{AA'B'}=-\xi^{(A'B')} .
\eqno (5.1.15)
$$

We may now continue the analysis of 
our deformed twistor space ${\cal N}^{D}$,
written in the form (cf. (5.1.1) and (5.1.2))
$$
{\widehat \omega}^{A}=\omega^{A}+p^{A}g
\Bigr(p_{B}\omega^{B},\pi_{B'}\Bigr) ,
\eqno (5.1.16a)
$$
$$
{\widehat \pi}_{A'}=\pi_{A'} .
\eqno (5.1.16b)
$$
In the light of the split (5.1.11), holomorphic sections of
${\cal N}^{D}$ are given by
$$
\omega^{A}(x^{b},\pi_{B'})=i \; x^{AA'} \; \pi_{A'}
+p^{A} \; h(x^{b},\pi_{B'}) 
\; {\rm in} \; {\cal Q} ,
\eqno (5.1.17)
$$
$$
{\widehat \omega}^{A}(x^{b},\pi_{B'})=i \; x^{AA'} \; \pi_{A'}
+p^{A} \; {\widehat h}(x^{b},\pi_{B'})
\; {\rm in} \; {\widehat {\cal Q}} ,
\eqno (5.1.18)
$$
where $x^{b}$ are {\it complex} coordinates on $\cal G$.
The conformal structure of $\cal G$ can be computed as
follows. A vector $U=U^{BB'}\nabla_{BB'}$ at $x^{a} \in
{\cal G}$ may be represented in ${\cal N}^{D}$ by the
displacement
$$
\delta \omega^{A}
=U^{b} \; \nabla_{b} \; \omega^{A}(x^{c},\pi_{C'}) .
\eqno (5.1.19a)
$$
By virtue of (5.1.17), Eq. (5.1.19a) becomes
$$
\delta \omega^{A}=U^{BB'}\Bigr(i \; \varepsilon_{B}^{\; \; A}
\; \pi_{B'}+p^{A} \; \nabla_{BB'}h(x^{c},\pi_{C'})\Bigr) .
\eqno (5.1.19b)
$$
The vector $U$ is null, by definition, if and only if
$$
\delta \omega^{A}(x^{b},\pi_{B'})=0 ,
\eqno (5.1.20)
$$
for some spinor field $\pi_{B'}$. To prove that the solution
of Eq. (5.1.20) exists, one defines (see (5.1.14))
$$
\theta \equiv 1-\xi ,
\eqno (5.1.21)
$$
$$
\Omega_{\; \; \; \; \; AA'}^{BB'} \equiv 
\theta \; \varepsilon_{A}^{\; \; B} \;
\varepsilon_{A'}^{\; \; B'}
-\psi_{A} \; p_{A'}^{\; \; \; BB'} .
\eqno (5.1.22)
$$
We are now aiming to show that the desired solution of
Eq. (5.1.20) is given by 
$$
U^{BB'}=\Omega_{\; \; \; \; \; AA'}^{BB'}
\; \lambda^{A} \; \pi^{A'} .
\eqno (5.1.23)
$$
Indeed, by virtue of (5.1.21)--(5.1.23) one finds
$$
U^{BB'}=(1-\xi)\lambda^{B}\pi^{B'}
-\psi_{A} \; p_{A'}^{\; \; \; BB'}
\; \lambda^{A} \; \pi^{A'} .
\eqno (5.1.24)
$$
Thus, since $\pi^{B'}\pi_{B'}=0$, the calculation of
(5.1.19b) yields
$$ \eqalignno{
\delta \omega^{A}&=-\psi_{C} \; \lambda^{C} \; 
\pi^{A'} \Bigr[i \; p_{A'}^{\; \; \; AB'} \; \pi_{B'}
+p_{A'}^{\; \; \; BB'} \; p^{A} \;
\nabla_{BB'}h(x,\pi)\Bigr]\cr
&+(1-\xi)\lambda^{B} \; \pi^{B'} \; p^{A}
\; \nabla_{BB'}h(x,\pi) .
&(5.1.25)\cr}
$$
Note that (5.1.12) may be used to re-express the second line
of (5.1.25). This leads to
$$ 
\delta \omega^{A}=-\psi_{C} \; \lambda^{C} \; \Gamma^{A} ,
\eqno (5.1.26)
$$
where
$$ \eqalignno{
\Gamma^{A} & \equiv \pi^{A'}\Bigr[i \; p_{A'}^{\; \; \; AB'}
\; \pi_{B'}+p_{A'}^{\; \; \; BB'} \; p^{A} \;
\nabla_{BB'}h(x,\pi)\Bigr]
+i(1-\xi)p^{A} \cr
&=-i \; p^{AA'B'} \; \pi_{A'} \; \pi_{B'}+i \; p^{A} 
+p^{A}\Bigr[-p^{BB'A'} \; \pi_{A'} \; \nabla_{BB'}h(x,\pi)
-i\xi \Bigr]\cr
&=\Bigr[-i+i+i\xi-i\xi \Bigr]p^{A}=0 ,
&(5.1.27)\cr}
$$
in the light of (5.1.5) and (5.1.14). Hence the solution of Eq. (5.1.20)
is given by (5.1.23).

Such null vectors determine the conformal metric of $\cal G$.
For this purpose, one defines (Ward 1978)
$$
\nu_{A'}^{\; \; \; B'} \equiv \varepsilon_{A'}^{\; \; \; B'}
-\xi_{A'}^{\; \; \; B'} ,
\eqno (5.1.28)
$$
$$
\Lambda \equiv {\theta \over 2} \; \nu_{A'B'} \; \nu^{A'B'} ,
\eqno (5.1.29)
$$
$$
\Sigma_{BB'}^{\; \; \; \; \; \; CC'} \equiv
\theta^{-1} \; \varepsilon_{B}^{\; \; C} \;
\varepsilon_{B'}^{\; \; \; C'}
+\Lambda^{-1} \; \psi_{B} \; p_{A'}^{\; \; \; CC'}
\; \nu_{B'}^{\; \; \; A'} .
\eqno (5.1.30)
$$
Interestingly, $\Sigma_{b}^{\; \; c}$ is the inverse
of $\Omega_{\; \; a}^{b}$, since
$$
\Omega_{\; \; a}^{b} \; \Sigma_{b}^{\; \; c}
=\delta_{a}^{\; \; c} .
\eqno (5.1.31)
$$
Indeed, after defining
$$
H_{A'}^{\; \; \; CC'} \equiv
p_{A'}^{\; \; \; CC'}-p_{D'}^{\; \; \; CC'}
\; \xi_{A'}^{\; \; \; D'} ,
\eqno (5.1.32)
$$
$$
\Phi_{A'}^{\; \; \; CC'} \equiv
\Bigr[\theta \Lambda^{-1} \; H_{A'}^{\; \; \; CC'}
-\Lambda^{-1} \; p_{A'}^{\; \; \; BB'} \;
\psi_{B} \; H_{B'}^{\; \; \; CC'}
-\theta^{-1} \; p_{A'}^{\; \; \; CC'}\Bigr] ,
\eqno (5.1.33)
$$
a detailed calculation shows that
$$
\Omega_{\; \; \; \; \; AA'}^{BB'} \;
\Sigma_{BB'}^{\; \; \; \; \; \; CC'}
-\varepsilon_{A}^{\; \; C} \;
\varepsilon_{A'}^{\; \; \; C'}
=\psi_{A} \; \Phi_{A'}^{\; \; \; CC'} .
\eqno (5.1.34)
$$
One can now check that the right-hand side of (5.1.34)
vanishes (see problem 5.1). Hence (5.1.31) holds.
For our anti-self-dual space-time $\cal G$, the metric
$g=g_{ab}dx^{a} \otimes dx^{b}$ is such that
$$
g_{ab}=\Xi(x) \; \Sigma_{a}^{\; \; c} \;
\Sigma_{bc} .
\eqno (5.1.35)
$$
Two null vectors $U$ and $V$ at $x \in {\cal G}$ have,
by definition, the form
$$
U^{AA'} \equiv \Omega_{\; \; \; \; \; BB'}^{AA'}
\; \lambda^{B} \; \alpha^{B'} ,
\eqno (5.1.36)
$$
$$
V^{AA'} \equiv \Omega_{\; \; \; \; \; BB'}^{AA'}
\; \chi^{B} \; \beta^{B'} ,
\eqno (5.1.37)
$$
for some spinors $\lambda^{B},\chi^{B},\alpha^{B'},
\beta^{B'}$. In the deformed space ${\cal N}^{D}$, 
$U$ and $V$ correspond to two displacements
$\delta_{1}\omega^{A}$ and $\delta_{2}\omega^{A}$
respectively, as in Eq. (5.1.19{\it b}). If one defines
the corresponding skew-symmetric form
$$
{\cal S}_{\pi}(U,V) \equiv \delta_{1}\omega_{A}
\; \delta_{2}\omega^{A} ,
\eqno (5.1.38)
$$
the metric is given by
$$
g(U,V) \equiv \Bigr(\alpha^{A'} \; \beta_{A'}\Bigr)
{\Bigr(\alpha^{B'} \; \pi_{B'}\Bigr)}^{-1}
{\Bigr(\beta^{C'} \; \pi_{C'}\Bigr)}^{-1}
\; {\cal S}_{\pi}(U,V) .
\eqno (5.1.39)
$$
However, in the light of (5.1.31), (5.1.35)--(5.1.37) one finds
$$
g(U,V) \equiv
g_{ab}U^{a}V^{b}=\Xi(x)\Bigr(\lambda^{A} \; \chi_{A}\Bigr)
\Bigr(\alpha^{A'} \; \beta_{A'}\Bigr) .
\eqno (5.1.40)
$$
By comparison with (5.1.39) this leads to
$$
{\cal S}_{\pi}(U,V)=\Xi(x)\Bigr(\lambda^{A} \; \chi_{A}\Bigr)
\Bigr(\alpha^{B'} \; \pi_{B'}\Bigr)
\Bigr(\beta^{C'} \; \pi_{C'}\Bigr) .
\eqno (5.1.41)
$$
If we now evaluate (5.1.41) with $\beta^{A'}=\alpha^{A'}$,
comparison with the definition (5.1.38) and use of
(5.1.12), (5.1.13), (5.1.19b) and (5.1.36) yield
$$
\Xi=\Lambda .
\eqno (5.1.42)
$$
The anti-self-dual solution of Einstein's equations is thus
given by (5.1.30), (5.1.35) and (5.1.42).

The construction of an anti-self-dual space-time described in
this section is a particular example of the so-called non-linear
graviton (Penrose 1976a--b). In mathematical language, if $\cal M$
is a complex three-manifold, $B$ is the bundle of holomorphic 
three-forms on $\cal M$ and $H$ is the standard positive line
bundle on $P_{1}$, a non-linear graviton is the following set
of data (Hitchin 1979):

(i) $\cal M$, the total space of a holomorphic fibration
$\pi: {\cal M} \rightarrow P_{1}$;

(ii) a four-parameter family of sections, each having 
$H \oplus H$ as normal bundle (see e.g. Huggett and Tod (1985)
for the definition of normal bundle);

(iii) a non-vanishing holomorphic section $s$ of
$B \otimes \pi^{*} H^{4}$, where 
$H^{4} \equiv H\otimes H \otimes H
\otimes H$, and $\pi^{*}H^{4}$ denotes the pull-back of 
$H^{4}$ by $\pi$;

(iv) a real structure on $\cal M$ such that $\pi$ and $s$ are real.
$\cal M$ is then fibred from the real sections of the family.
\vskip 1cm
\centerline {\bf 5.2 Beyond anti-self-duality}
\vskip 1cm
\noindent
The limit of the analysis performed in section 5.1 is that it
deals with a class of solutions of (complex) Einstein equations
which is not sufficiently general. In Yasskin and Isenberg (1982)
and Yasskin (1987) the authors have examined in detail
the limits of the anti-self-dual analysis.
The two main criticisms are as follows:

(a) a right-flat space-time (cf. the analysis in Law (1985)) 
does not represent a real Lorentzian
space-time manifold. Hence it cannot be applied directly to
classical gravity (Ward 1980b);

(b) there are reasons fo expecting that the equations of a
quantum theory of gravity are much more complicated, and thus
are not solved by right-flat space-times.
\vskip 0.3cm
\noindent
However, an alternative approach due to Le Brun has become
available in the eighties (Le Brun 1985). Le Brun's approach
focuses on the space $G$ of complex null geodesics of
complex space-time $(M,g)$, 
called ambitwistor space. Thus, one deals
with a standard rank-2 holomorphic vector bundle 
$E \rightarrow G$, and in the conformal class determined by
the complex structure of $G$, a one-to-one correspondence
exists between non-vanishing holomorphic sections of $E$ and
Einstein metrics on $(M,g)$ (Le Brun 1985). The bundle $E$
is called Einstein bundle, and has also been studied in 
Eastwood (1987). The work by Eastwood adds evidence in favour
of the Einstein bundle being the correct generalization of
the non-linear-graviton construction to the non-right-flat
case (cf. Law (1985), Park (1990), Le Brun (1991),
Park (1991), our section 9.6). 
Indeed, the theorems discussed so far provide a 
characterization of the vacuum Einstein equations. However,
there is not yet an independent way of recognizing the
Einstein bundle. Thus, this is not yet a substantial progress
in solving the vacuum equations. 
Other relevant work on holomorphic ideas appears 
in Le Brun (1986), where the author proves that,
in the case of four-manifolds with self-dual Weyl curvature,
solutions of the Yang--Mills equations correspond to 
holomorphic bundles on an associated analytic space
(cf. Ward (1977), Witten (1978), Ward (1981a)).
\vskip 1cm
\centerline {\bf 5.3 Twistors as spin-${3\over 2}$ charges}
\vskip 1cm
\noindent
In this section, we describe a proposal by Penrose
to regard twistors for Ricci-flat space-times as (conserved)
{\it charges} for massless helicity-${3\over 2}$ fields
(Penrose 1990, Penrose 1991a--b--c). The new approach proposed
by Penrose is based on the following mathematical results
(Penrose 1991b):

(i) A vanishing Ricci tensor provides the consistency condition
for the existence and propagation of massless 
helicity-${3\over 2}$ fields in curved space-time
(Buchdahl 1958, Deser and Zumino 1976);

(ii) In Minkowski space-time, the space of charges for such 
fields is naturally identified with the corresponding twistor
space.
\vskip 0.3cm
\noindent
Thus, Penrose points out that if one could find the appropriate
definition of charge for massless helicity-${3\over 2}$ fields
in a Ricci-flat space-time, this should provide the concept of
twistor appropriate for vacuum Einstein equations. The corresponding
geometric program may be summarized as follows:

(1) Define a twistor for Ricci-flat space-time
$(M,g)_{RF}$;

(2) Characterize the resulting twistor space $\cal F$;

(3) Reconstruct $(M,g)_{RF}$ from $\cal F$.
\vskip 0.3cm
\noindent
We now describe, following Penrose (1990), Penrose (1991a--c),
properties and problems of this approach to twistor theory
in flat and in curved space-times.
\vskip 1cm
\centerline {\bf 5.3.1 Massless spin-${3\over 2}$ equations
in Minkowski space-time}
\vskip 1cm
\noindent
Let $(M,\eta)$ be Minkowski space-time with flat connection
$\cal D$. In $(M,\eta)$ the gauge-invariant field strength
for spin ${3\over 2}$ is represented by a totally symmetric
spinor field 
$$
\psi_{A'B'C'}=\psi_{(A'B'C')},
\eqno (5.3.1)
$$
obeying a massless free-field equation
$$
{\cal D}^{AA'} \; \psi_{A'B'C'}=0.
\eqno (5.3.2)
$$
With the conventions of Penrose, $\psi_{A'B'C'}$ describes
spin-${3\over 2}$ particles of helicity equal to
${3\over 2}$ (rather than -${3\over 2}$). The {\it Dirac form}
of this field strength is obtained by expressing {\it locally}
$\psi_{A'B'C'}$ in terms of two potentials subject to gauge
freedoms involving a primed and an unprimed spinor field.
The first potential is a spinor field symmetric in its
primed indices
$$
\gamma_{B'C'}^{A}=\gamma_{(B'C')}^{A} ,
\eqno (5.3.3)
$$
subject to the differential equation
$$
{\cal D}^{BB'} \; \gamma_{B'C'}^{A}=0,
\eqno (5.3.4)
$$
and such that
$$
\psi_{A'B'C'}={\cal D}_{AA'} \; \gamma_{B'C'}^{A}.
\eqno (5.3.5)
$$
The second potential is a spinor field symmetric in its
unprimed indices
$$
\rho_{C'}^{AB}=\rho_{C'}^{(AB)},
\eqno (5.3.6)
$$
subject to the equation
$$
{\cal D}^{CC'} \; \rho_{C'}^{AB}=0,
\eqno (5.3.7)
$$
and it yields the $\gamma_{B'C'}^{A}$ potential by
means of
$$
\gamma_{B'C'}^{A}={\cal D}_{BB'} \; \rho_{C'}^{AB}.
\eqno (5.3.8)
$$
If we introduce the spinor fields $\nu_{C'}$ and $\chi^{B}$
obeying the equations
$$
{\cal D}^{AC'} \; \nu_{C'}=0,
\eqno (5.3.9)
$$
$$
{\cal D}_{AC'} \; \chi^{A}=2i \; \nu_{C'},
\eqno (5.3.10)
$$
the gauge freedoms for the two potentials enable one
to replace them by the potentials
$$
{\widehat \gamma}_{B'C'}^{A} \equiv \gamma_{B'C'}^{A}+
{\cal D}_{B'}^{\; \; \; A} \; \nu_{C'},
\eqno (5.3.11)
$$
$$
{\widehat \rho}_{C'}^{AB} \equiv \rho_{C'}^{AB}
+\varepsilon^{AB} \; \nu_{C'}
+i \; {\cal D}_{C'}^{\; \; \; A} \; \chi^{B},
\eqno (5.3.12)
$$
without affecting the theory. Note that the right-hand side of
(5.3.12) does not contain antisymmetric parts since, despite
the explicit occurrence of the antisymmetric $\varepsilon^{AB}$,
one finds
$$
{\cal D}_{C'}^{\; \; \; [A} \; \chi^{B]}
={\varepsilon^{AB} \over 2}{\cal D}_{LC'} \; \chi^{L}
=i \; \varepsilon^{AB} \nu_{C'},
\eqno (5.3.13)
$$
by virtue of (5.3.10). Hence (5.3.13) leads to
$$
{\widehat \rho}_{C'}^{AB}=\rho_{C'}^{AB}
+i \; {\cal D}_{C'}^{\; \; \; (A} \; \chi^{B)} .
\eqno (5.3.14)
$$
The gauge freedoms are indeed given by Eqs. (5.3.11) and (5.3.12)
since in our flat space-time one finds
$$
{\cal D}^{AA'} \; {\widehat \gamma}_{A'B'}^{C}=
{\cal D}^{AA'} \; {\cal D}_{\; \; B'}^{C} \; \nu_{A'}=
{\cal D}_{\; \; B'}^{C} \; {\cal D}^{AA'} \; \nu_{A'}=0 ,
\eqno (5.3.15)
$$
by virtue of (5.3.4) and (5.3.9), and
$$ \eqalignno{
{\cal D}^{AA'} \; {\widehat \rho}_{A'}^{BC}&=
{\cal D}^{AA'} \; {\cal D}_{\; \; A'}^{C} \; \chi^{B}
={\cal D}^{CA'} \; {\cal D}_{A'}^{\; \; \; A} \; \chi^{B}\cr
&={\cal D}_{A'}^{\; \; \; A} \; {\cal D}^{CA'} \; \chi^{B}
=-{\cal D}^{AA'} \; {\cal D}_{\; \; A'}^{C} \; \chi^{B} ,
&(5.3.16a)\cr}
$$
which implies
$$
{\cal D}^{AA'} \; {\widehat \rho}_{A'}^{BC}=0 .
\eqno (5.3.16b)
$$
The result (5.3.16b) is a particular case of the application
of spinor Ricci identities to flat space-time (cf. sections
6.3 and 8.4).

We are now in a position to show that twistors can be regarded
as charges for helicity-${3\over 2}$ massless fields in Minkowski
space-time. For this purpose, following Penrose (1991a,c) let us
suppose that the field $\psi$ satisfying (5.3.1) and (5.3.2) exists
in a region ${\cal R}$ of $(M,\eta)$, surrounding a world-tube
which contains the sources for $\psi$. Moreover, we consider a
two-sphere $\cal S$ within $\cal R$ surrounding the world-tube. 
To achieve this we begin by taking a {\it dual} twistor, i.e. the
pair of spinor fields
$$
W_{\alpha} \equiv \Bigr(\lambda_{A},\mu^{A'}\Bigr) ,
\eqno (5.3.17)
$$
obeying the differential equations
$$
{\cal D}_{AA'} \; \mu^{B'}=i \; \varepsilon_{A'}^{\; \; \; B'}
\; \lambda_{A} ,
\eqno (5.3.18)
$$
$$
{\cal D}_{AA'} \; \lambda_{B}=0.
\eqno (5.3.19)
$$
Hence $\mu^{B'}$ is a solution of the complex-conjugate twistor
equation
$$
{\cal D}_{A}^{(A'} \; \mu^{B')}=0.
\eqno (5.3.20)
$$
Thus, if one defines
$$
\varphi_{A'B'} \equiv \psi_{A'B'C'} \; \mu^{C'},
\eqno (5.3.21)
$$
one finds, by virtue of (5.3.1), (5.3.2) and (5.3.20), that
$\varphi_{A'B'}$ is a solution of the self-dual vacuum
Maxwell equations
$$
{\cal D}^{AA'} \; \varphi_{A'B'}=0.
\eqno (5.3.22)
$$
Note that (5.3.21) is a particular case of the spin-lowering
procedure (Huggett and Tod 1985, Penrose and Rindler 1986). 
Moreover, $\varphi_{A'B'}$ enables one to define the self-dual
two-form
$$
F \equiv \varphi_{A'B'} \; dx_{A}^{\; \; A'} \; \wedge \;
dx^{AB'},
\eqno (5.3.23)
$$
which leads to the following {\it charge} assigned to the
world-tube:
$$
Q \equiv {i\over 4\pi} \oint F .
\eqno (5.3.24)
$$
For some twistor
$$
Z^{\alpha} \equiv \Bigr(\omega^{A},\pi_{A'}\Bigr),
\eqno (5.3.25)
$$
the charge $Q$ depends on the dual twistor $W_{\alpha}$ as
(see problem 5.3)
$$
Q =Z^{\alpha} \; W_{\alpha}=\omega^{A} \; \lambda_{A}
+\pi_{A'} \; \mu^{A'}.
\eqno (5.3.26)
$$
These equations describe the strength of the charge, for the
field $\psi$, that should be assigned to the world-tube.
Thus, a twistor $Z^{\alpha}$ arises naturally in Minkowski
space-time as the charge for a helicity $+{3\over 2}$ massless
field, whereas a dual twistor $W_{\alpha}$ is the charge for a
helicity $-{3\over 2}$ massless field (Penrose 1991c).

Interestingly, the potentials $\gamma_{A'B'}^{C}$ and 
$\rho_{A'}^{BC}$ can be used to obtain a potential for the
self-dual Maxwell field strength, since, after defining
$$
\theta_{\; \; A'}^{C} \equiv \gamma_{A'B'}^{C} \; \mu^{B'}
-i \; \rho_{A'}^{BC} \; \lambda_{B},
\eqno (5.3.27)
$$
one finds
$$ \eqalignno{
{\cal D}_{CB'} \; \theta_{\; \; A'}^{C} &=
\Bigr({\cal D}_{CB'} \; \gamma_{A'D'}^{C}\Bigr)\mu^{D'}
+\gamma_{A'D'}^{C} \Bigr({\cal D}_{CB'} \; \mu^{D'}\Bigr)
-i \Bigr({\cal D}_{CB'} \; \rho_{A'}^{BC}\Bigr)\lambda_{B} \cr
&=\psi_{A'B'D'} \; \mu^{D'}+i \; \varepsilon_{B'}^{\; \; \; D'} \;
\gamma_{A'D'}^{C} \; \lambda_{C}
-i \; \gamma_{A'B'}^{C} \; \lambda_{C} \cr
&=\psi_{A'B'D'} \; \mu^{D'}=\varphi_{A'B'},
&(5.3.28)\cr}
$$
$$ \eqalignno{
{\cal D}_{B}^{\; \; A'} \; \theta_{\; \; A'}^{C}&=
\Bigr({\cal D}_{B}^{\; \; A'} \; \gamma_{A'B'}^{C}\Bigr)
\mu^{B'}+\gamma_{A'B'}^{C}\Bigr({\cal D}_{B}^{\; \; A'}
\; \mu^{B'}\Bigr)-i\Bigr({\cal D}_{B}^{\; \; A'} \;
\rho_{A'}^{DC}\Bigr) \lambda_{D} \cr
&-i \rho_{A'}^{DC} \Bigr({\cal D}_{B}^{\; \; A'} 
\; \lambda_{D}\Bigr)=0 .
&(5.3.29)\cr}
$$
Eq. (5.3.28) has been obtained by using (5.3.5), (5.3.8),
(5.3.18) and (5.3.19), whereas (5.3.29) holds by virtue of
(5.3.3), (5.3.4), (5.3.7), (5.3.18) and (5.3.19). The one-form
corresponding to $\theta_{\; \; A'}^{C}$ is defined by
$$
A \equiv \theta_{BB'} \; dx^{BB'} ,
\eqno (5.3.30)
$$
which leads to
$$
F=2 \; dA ,
\eqno (5.3.31)
$$
by using (5.3.23) and (5.3.28).

The {\it Rarita--Schwinger form} of the field strength does not
require the symmetry (5.3.3) in $B'C'$ as we have done so far,
and the $\gamma_{B'C'}^{A}$ potential is instead subject to the
equations (Penrose 1991a--c) [cf. (8.6.3) and (8.6.4)]
$$
\varepsilon^{B'C'} \; {\cal D}_{A(A'} \; \gamma_{B')C'}^{A}
=0 ,
\eqno (5.3.32)
$$
$$
{\cal D}^{B'(B} \; \gamma_{B'C'}^{A)}=0 .
\eqno (5.3.33)
$$
Moreover, the spinor field $\nu_{C'}$ in (5.3.11) is no
longer taken to be a solution of the Weyl equation (5.3.9).

The potentials $\gamma$ and $\rho$ may or may not be global 
over $\cal S$. If $\gamma$ is global but $\rho$ is not, one
obtains a two-dimensional complex vector space parametrized 
by the spinor field $\pi_{A'}$. The corresponding subspace
where $\pi_{A'}=0$, parametrized by $\omega^{A}$, is called
$\omega$-space. Thus, following Penrose (1991c), we regard
$\pi$-space and $\omega$-space as quotient spaces defined 
as follows:
$$
\pi-{\rm space} \equiv {\rm space} 
\; {\rm of} \; {\rm global} \; 
\psi{\rm 's} / {\rm space} \; {\rm of} \; {\rm global}
\; \gamma{\rm 's},
\eqno (5.3.34)
$$
$$
\omega-{\rm space} \equiv {\rm space} 
\; {\rm of} \; {\rm global} \;
\gamma{\rm 's} / {\rm space} \; {\rm of} \; 
{\rm global} \; \rho{\rm 's}.
\eqno (5.3.35)
$$
\vskip 1cm
\centerline {\bf 5.3.2 Massless spin-${3\over 2}$ field strengths
in curved space-time}
\vskip 1cm
\noindent
The conditions for the {\it local} existence of the
$\rho_{A'}^{BC}$ potential in curved space-time are derived
by requiring that, after the gauge transformation (5.3.12)
(or, equivalently, (5.3.14)), also the 
${\widehat \rho}_{A'}^{BC}$ potential should obey the equation
$$
\nabla^{AA'} \; {\widehat \rho}_{A'}^{BC}=0,
\eqno (5.3.36)
$$
where $\nabla$ is the curved connection. 
By virtue of the spinor Ricci identity (Ward and Wells 1990)
$$
\nabla_{M'(A} \; \nabla_{\; \; \; B)}^{M'} \; \chi_{C}
=\psi_{ABDC} \; \chi^{D}-2\Lambda \; \chi_{(A} \;
\varepsilon_{B)C},
\eqno (5.3.37)
$$
the insertion of (5.3.14) into (5.3.36) yields, assuming
for simplicity that $\nu_{C'}=0$ in (5.3.10), the following
conditions (see (8.4.28)):
$$
\psi_{ABCD}=0, \;
\Lambda=0 ,
\eqno (5.3.38)
$$
which imply we deal with a vacuum self-dual (or left-flat)
space-time, since the anti-self-dual Weyl spinor has
to vanish (Penrose 1991c).

Moreover, in a complex anti-self-dual vacuum space-time 
one finds (Penrose 1991c) that spin-${3\over 2}$ field
strengths $\psi_{A'B'C'}$ can be defined according to
(cf. (5.3.5))
$$
\psi_{A'B'C'}=\nabla_{CC'} \; \gamma_{A'B'}^{C},
\eqno (5.3.39)
$$
are gauge-invariant, totally symmetric, and satisfy the massless
free-field equations (cf. (5.3.2))
$$
\nabla^{AA'} \; \psi_{A'B'C'}=0.
\eqno (5.3.40)
$$
In this case there is no obstruction to defining global 
$\psi$-fields with non-vanishing $\pi$-charge, and a global
$\pi$-space can be defined as in (5.3.34). It remains to be
seen whether the twistor space defined by $\alpha$-surfaces
may then be reconstructed (section 4.2, Penrose 1976a-b,
Ward and Wells 1990, Penrose 1991c).

Interestingly, in Penrose (1991b) it has been proposed to
interpret the potential $\gamma$ as providing a 
{\it bundle connection}. In other words, one takes the
fibre coordinates to be given by a spinor $\eta_{A'}$ and
a scalar $\mu$. For a given small $\epsilon$, one extends
the ordinary Levi--Civita connection $\nabla$ on $M$ to
bundle-valued quantities according to (Penrose 1991b)
$$
\nabla_{PP'} \pmatrix {\eta_{A'} \cr \mu \cr}
\equiv
\pmatrix {\nabla_{PP'} \; \eta_{A'} \cr
\nabla_{PP'} \; \mu}
-\epsilon \pmatrix {0 & \gamma_{PP'A'} \cr
\gamma_{PP'}^{\; \; \; \; \; \; B'} & 0 \cr}
\pmatrix {\eta_{B'} \cr \mu \cr},
\eqno (5.3.41)
$$
with gauge transformations given by
$$
\pmatrix {{\widehat \eta}_{A'} \cr {\widehat \mu} \cr}
\equiv
\pmatrix {\eta_{A'} \cr \mu \cr}
+\epsilon \pmatrix {0 & \nu_{A'} \cr
\nu^{B'} & 0 \cr}
\pmatrix {\eta_{B'} \cr \mu \cr}.
\eqno (5.3.42)
$$
Note that terms of order $\epsilon^{2}$ have been
neglected in writing (5.3.42). However, such gauge
transformations do not close under commutation, and to
obtain a theory valid to all orders in $\epsilon$ one has
to generalize to $SL(3,C)$ matrices before the commutators
close. Writing $(A)$ for the three-dimensional 
indices, so that $\eta_{(A)}$ denotes
$\pmatrix {\eta_{A'} \cr \mu \cr}$, one has a connection
defined by
$$
\nabla_{PP'} \; \eta_{(A)} \equiv
\pmatrix {\nabla_{PP'} \; \eta_{A'} \cr
\nabla_{PP'} \; \mu \cr}
-\gamma_{PP' \; (A)} 
^{\; \; \; \; \; \; \; \; \; \; \; (B)}
\; \; \eta_{(B)},
\eqno (5.3.43)
$$
with gauge transformation
$$
{\widehat \eta}_{(A)} \equiv
\eta_{(A)}
+\nu_{(A)}^{\; \; \; \; (B)}
\; \eta_{(B)} .
\eqno (5.3.44)
$$
With this notation, the 
$\nu_{(A)}^{\; \; \; \; (B)}$ are 
$SL(3,C)$-valued fields on $M$, and hence
$$
{\cal E}^{(P) \; (Q) \; (R)} \; \;
\nu_{(P)}^{\; \; \; \; (A)}
\; \; \nu_{(Q)}^{\; \; \; \; (B)}
\; \; \nu_{(R)}^{\; \; \; \; (C)}
= {\cal E}^{(A) \; (B) \; (C)},
\eqno (5.3.45)
$$
where ${\cal E}^{(P) \; (Q) \; (R)}$
are generalized Levi--Civita 
symbols. The $SL(3,C)$ definition of $\gamma$-potentials
takes the form (Penrose 1991b)
$$
\gamma_{PP' \; (A)}^{\; \; \; \; \; \; \; \; \; \; \; (B)}
\equiv
\pmatrix {\alpha_{PP'A'}^{\; \; \; \; \; \; \; \; \; \; B'}
& \beta_{PP'A'} \cr
\gamma_{PP'}^{\; \; \; \; \; \; B'} & \delta_{PP'} \cr},
\eqno (5.3.46)
$$
while the curvature is
$$
K_{pq \; (A)}^{\; \; \; \; \; \; \; \; (B)}
\equiv 2 \nabla_{[p} \; 
\gamma_{q](A)}^{\; \; \; \; \; \; \; (B)}
+2 \; \gamma_{[p \mid (A) \mid }
^{\; \; \; \; \; \; \; \; \; (C)}
\; \; \gamma_{q](C)}^{\; \; \; \; \; \; \; \; (B)}.
\eqno (5.3.47)
$$
Penrose has proposed this as a generalization of the
Rarita--Schwinger structure in Ricci-flat space-times, and he
has even speculated that a non-linear generalization of the
Rarita--Schwinger equations (5.3.32) and (5.3.33) might be
$$
{ }^{(-)}K_{PQ \; (A)}^{\; \; \; \; \; \; \; \; \; \; (B)}=0,
\eqno (5.3.48)
$$
$$
{ }^{(+)}K_{P'Q' \; (A)} 
^{\; \; \; \; \; \; \; \; \; \; \; \; (B)}
\; \; {\cal E}^{P' \; (A) \; (C)}
\; \;  
{\cal E}_{\; \; \; \; (B) \; (D)}^{Q'}=0,
\eqno (5.3.49)
$$
where ${ }^{(-)}K$ and ${ }^{(+)}K$ are the anti-self-dual
and self-dual parts of the curvature respectively, i.e.
$$
K_{pq \; (A)}^{\; \; \; \; \; \; \; \; \; (B)}
= \varepsilon_{P'Q'} \; \; 
{ }^{(-)}K_{PQ \; (A)}^{\; \; \; \; \; 
\; \; \; \; \; (B)}
+ \varepsilon_{PQ} \; \;
{ }^{(+)}K_{P'Q' \; (A)}^{\; \; \; \; \; 
\; \; \; \; \; \; \; \; \; (B)}.
\eqno (5.3.50)
$$
Following Penrose (1991b), one has
$$
{\cal E}^{P' \; (A) \; (C)}
\equiv {\cal E}^{(P) \; (A)
\; (C)}
\; e_{(P)}^{\; \; \; \; P'},
\eqno (5.3.51)
$$
$$
{\cal E}_{Q' \; (B) \; (D)}
\equiv {\cal E}_{(Q) \; (B) \; (D)}
\; \; e_{Q'}^{\; \; \; \; (Q)},
\eqno (5.3.52)
$$
the $e_{(P)}^{\; \; \; \; P'}$ and
$e_{Q'}^{\; \; \; (Q)}$ relating the bundle directions
with tangent directions in $M$. 
\vskip 100cm
\centerline {\it CHAPTER SIX}
\vskip 1cm
\centerline {\bf COMPLEX SPACE-TIMES WITH TORSION}
\vskip 1cm
\noindent
Theories of gravity with torsion are relevant since
torsion is a naturally occurring geometric property of relativistic
theories of gravitation, the gauge theory of the Poincar\'e group
leads to its presence, the constraints are second-class and the
occurrence of cosmological singularities can be less generic than
in general relativity. In a space-time manifold with non-vanishing
torsion, the Riemann tensor has 36 independent real components
at each point, rather than 20 as in general relativity. The
information of these 36 components 
is encoded in three spinor fields
and in a scalar function, having 5,9,3 and 1 complex components,
respectively. If space-time is complex, this means that, with
respect to a holomorphic coordinate basis $x^{a}$, the metric is
a $4 \times 4$ matrix of holomorphic functions of $x^{a}$, and
its determinant is nowhere-vanishing. Hence the connection and
Riemann are holomorphic as well, and the Ricci tensor becomes
complex-valued.

After a two-component spinor analysis of the curvature and of
spinor Ricci identities, the necessary condition for the 
existence of $\alpha$-surfaces in complex space-time
manifolds with non-vanishing torsion is derived. For these
manifolds, Lie brackets of vector fields and spinor Ricci
identities contain explicitly the effects of torsion.
This leads to an integrability condition for 
$\alpha$-surfaces which does not involve just the 
self-dual Weyl spinor, as in complex general
relativity, but also the torsion spinor, in a
non-linear way, and its covariant derivative. A similar
result also holds for four-dimensional, smooth real
manifolds with a positive-definite metric. Interestingly,
a particular solution of the integrability condition is
given by right conformally flat and 
right-torsion-free space-times.
\vskip 100cm
\centerline {\bf 6.1 Introduction}
\vskip 1cm
\noindent
As we know from previous chapters, after the work in Penrose (1967),
several efforts have been produced
to understand many properties of classical and quantum
field theories using twistor theory. Penrose's
original idea was that the space-time picture might be
inappropriate at the Planck length, whereas a more correct
framework for fundamental physics should be a particular complex manifold
called twistor space. In other words, concepts such as null
lines and null surfaces are more fundamental than space-time
concepts, and twistor space provides the precise mathematical
description of this idea. 

In the course of studying Minkowski space-time, twistors
can be defined either via the four-complex-dimensional vector
space of solutions to the differential equation 
(cf. Eq. (4.1.5))
$$
{\cal D}_{A'}^{\; \; (A}\omega^{B)}=0 ,
\eqno (6.1.1)
$$
or via null two-surfaces in complexified compactified 
Minkowski space $CM^{\#}$, called $\alpha$-planes.
The $\alpha$-planes (section 4.1) are such that the space-time
metric vanishes over them, and their null tangent
vectors have the two-component spinor form
$\lambda^{A}\pi^{A'}$, where $\lambda^{A}$ is
varying and $\pi^{A'}$ is fixed (i.e. fixed by Eq. (4.2.4)).
The latter definition
can be generalized to complex or real Riemannian
space-times provided that the Weyl curvature is
anti-self-dual. This leads in turn to a powerful
geometric picture, where the study of the Euclidean-time
version of the partial differential
equations of Einstein's theory is replaced by the
problem of finding the holomorphic curves 
in a complex manifold called {\it deformed (projective)
twistor space}. This finally enables one to
reconstruct the space-time metric (chapter five).
From the point of view of
gravitational physics, this is the most relevant application
of Penrose transform, which is by now
a major tool for studying the differential equations
of classical field theory (Ward and Wells 1990). 

Note that, while in differential geometry the basic
ideas of connection and curvature are local, in
complex-analytic geometry there is no local information.
Any complex manifold looks locally like $C^{n}$,
with no special features,
and any holomorphic fibre bundle is locally an
analytic product (cf. Atiyah (1988) on page 524 for a more
detailed treatment of this non-trivial point).
It is worth
bearing in mind this difference since the Penrose
transform converts problems from differential
geometry into problems of complex-analytic geometry.
We thus deal with a {\it non-local} transform, so that
local curvature information is coded into {\it global
holomorphic information}. More precisely, Penrose theory
does not hold for both anti-self-dual and self-dual
space-times, so that one only obtains a non-local
treatment of complex space-times with anti-self-dual
Weyl curvature.
However, these investigations are incomplete for at
least two reasons:

(a) anti-self-dual (or self-dual) space-times appear
a very restricted (although quite important) class
of models, and it is not clear how to
generalize twistor-space definitions to general
vacuum space-times;

(b) the fundamental theory of gravity at the Planck
length is presumably different from Einstein's 
general relativity (Hawking 1979, Esposito 1994).

In this chapter we have thus tried to extend the original
analysis appearing in the literature to a larger class
of theories of gravity, i.e. space-time models
$(M,g)$ with torsion (we are, however, not concerned with
supersymmetry). In our opinion, the main motivations for
studying these space-time models are as follows.

(1) Torsion is a peculiarity
of relativistic theories of gravitation, since the bundle
$L(M)$ of linear frames is soldered to the base $B=M$,
whereas for gauge theories other than gravitation the
bundle $L(M)$ is loosely connected to $M$.
The torsion
two-form $T$ is then defined as $T \equiv d\theta + \omega
\wedge \theta$, where $\theta$ is the soldering form and
$\omega$ is a connection one-form on $L(M)$. If $L(M)$ is
reduced to the bundle $O(M)$ of orthonormal frames,
$\omega$ is called spin-connection.

(2) The gauge theory of the Poincar\'e group naturally leads
to theories with torsion.

(3) From the point of view of constrained Hamiltonian systems,
theories with torsion are of great interest, since they are
theories of gravity with second-class constraints
(cf. Esposito (1994) and references therein).

(4) In space-time models with torsion, the occurrence of
cosmological singularities {\it can} be less generic
than in general relativity (Esposito 1992, Esposito 1994).

In the original work by Penrose and Ward, the first
(simple) problem is to characterize curved space-time
models possessing $\alpha$-surfaces. As we were saying
following Eq. (5.1.1), the necessary and sufficient condition
is that space-time be complex, or real Riemannian (i.e.
its metric is {\it positive-definite}), with anti-self-dual
Weyl curvature. This is proved by using Frobenius' theorem,
the spinor form of the Riemann curvature tensor, and spinor
Ricci identities. Our chapter is thus organized as follows.

Section 6.2 describes Frobenius' theorem and its application
to curved complex space-time models with non-vanishing
torsion. In particular, if $\alpha$-surfaces are required
to exist, one finds this is equivalent to a differential
equation involving two spinor fields $\xi_{A}$ and $w_{AB'}$,
which are completely determined by certain algebraic
relations. Section 6.3 describes the spinor form of Riemann
and spinor Ricci identities for theories with torsion.
Section 6.4 applies the formulae of section 6.3 to obtain the
integrability condition for the differential equation
derived at the end of section 6.2. The integrability condition
for $\alpha$-surfaces is then shown to
involve the self-dual Weyl spinor, the torsion spinor
and covariant derivatives of torsion.
Concluding remarks are presented in 
section 6.5.  
\vskip 1cm
\centerline {\bf 6.2 Frobenius' theorem for theories with torsion}
\vskip 1cm
\noindent
Frobenius' theorem is one of the main tools for studying
calculus on manifolds. Following 
Abraham {\it et al}. (1983), the geometric
framework and the theorem can be described as follows.
Given a manifold $M$, let $E \subset TM$ be a sub-bundle
of its tangent bundle. By definition, $E$ is {\it involutive}
if for any two $E$-valued vector fields $X$ and $Y$ defined
on $M$, their Lie bracket is $E$-valued as well.
Moreover, $E$ is {\it integrable} if $\forall m_{0} \in M$
there is a local submanifold $N \subset M$ 
through $m_{0}$, called a local
integral manifold of $E$ at $m_{0}$, whose tangent bundle
coincides with $E$ restricted to $N$. Frobenius' theorem
ensures that a sub-bundle $E$ of $TM$ is involutive if 
and only if it is integrable.

Given a complex torsion-free space-time $(M,g)$,
it is possible to pick out in $M$ a family of holomorphic
two-surfaces, called $\alpha$-surfaces, which generalize
the $\alpha$-planes of Minkowski space-time described
in section 4.1, provided that the self-dual Weyl spinor
vanishes. In the course of deriving the 
condition on the curvature enforced by the existence 
of $\alpha$-surfaces, one begins by taking a totally null
two-surface ${\hat S}$ in $M$. By definition, ${\hat S}$  
is a two-dimensional complex submanifold of $M$ such that,
$\forall p \in {\hat S}$, if $x$ and $y$ are any two tangent
vectors at $p$, then $g(x,x)=g(y,y)=g(x,y)=0$. Denoting by
$X=X^{a}e_{a}$ and $Y=Y^{a}e_{a}$ two vector fields 
tangent to ${\hat S}$, where $X^{a}$ and $Y^{a}$ have the
two-component spinor form $X^{a}=\lambda^{A}\pi^{A'}$
and $Y^{a}=\mu^{A}\pi^{A'}$, Frobenius' theorem may be
used to require that the Lie bracket of $X$ and $Y$ be a
linear combination of $X$ and $Y$, so that we
write
$$
[X,Y]=\varphi X + \rho Y ,
\eqno (6.2.1)
$$
where $\varphi$ and $\rho$ are scalar functions. 
Frobenius' theorem is indeed originally formulated for real
manifolds. If the integral submanifolds of complex
space-time are holomorphic, there are additional conditions
which are not described here.
Note also that Eq. (6.2.1) does not depend on additional structures
on $M$ (torsion, metric, etc. ...). In the torsion-free
case, it turns out that
the Lie bracket $[X,Y]$ can
also be written as $\nabla_{X}Y -\nabla_{Y}X$, and this
eventually leads to a condition which implies the vanishing
of the self-dual part of the Weyl curvature, after using
the spinorial formula for Riemann and spinor Ricci
identities.

However, for the reasons described in section 6.1,
we are here interested in models where torsion does
not vanish. Even though Frobenius' theorem (cf. (6.2.1))
does not involve torsion, the Lie bracket 
$[X,Y]$ can be also expressed using the definition 
of the torsion tensor $S$ (see comment following (6.3.3)) :
$$
[X,Y] \equiv \nabla_{X}Y-\nabla_{Y}X-2S(X,Y) .
\eqno (6.2.2)
$$
By comparison, Eqs. (6.2.1) and (6.2.2) lead to
$$
X^{a}\nabla_{a}Y^{b}-Y^{a}\nabla_{a}X^{b}=
\varphi X^{b} + \rho Y^{b}
+2S_{cd}^{\; \; \; \; b} \; X^{c} \; Y^{d}.
\eqno (6.2.3)
$$
Now, the antisymmetry $S_{ab}^{\; \; \; c}=-S_{ba}^{\; \; \; c}$
of the torsion tensor can be expressed spinorially as
$$
S_{ab}^{\; \; \; c}=\chi_{AB}^{\; \; \; \; \; CC'}
\; \varepsilon_{A'B'}
+{\widetilde \chi}_{A'B'}^{\; \; \; \; \; \; \; CC'}
\; \varepsilon_{AB} ,
\eqno (6.2.4)
$$
where the spinors $\chi$ and ${\widetilde \chi}$ are
symmetric in $AB$ and $A'B'$ respectively, and from
now on we use two-component spinor notation
(we do not write Infeld-van der Waerden symbols for
simplicity of notation). 
Thus, writing $X^{a}=\lambda^{A}\pi^{A'}$
and $Y^{a}=\mu^{A}\pi^{A'}$, one finds, using a technique
similar to the one in section 9.1 of 
Ward and Wells (1990), that Eq. (6.2.3) is equivalent to
$$
\pi^{A'} \Bigr(\nabla_{AA'}\pi_{B'}\Bigr)=
\xi_{A} \; \pi_{B'}+w_{AB'} ,
\eqno (6.2.5)
$$
for some spinor fields $\xi_{A}$ and $w_{AB'}$, if
the following conditions are imposed:
$$
-\mu^{A}\xi_{A}=\varphi ,
\eqno (6.2.6)
$$
$$
\lambda^{A}\xi_{A}=\rho ,
\eqno (6.2.7)
$$
$$
\mu_{D}\lambda^{D} \; w_{BB'}=
-2\mu_{D}\lambda^{D} \; 
{\widetilde \chi}_{C'D'BB'} \; \pi^{C'} \; \pi^{D'} .
\eqno (6.2.8)
$$
Note that, since our calculation involves two vector fields
$X$ and $Y$ tangent to ${\hat S}$, its validity is only 
local unless the surface ${\hat S}$ is parallelizable 
(i.e. the bundle $L(\hat S)$ admits a cross-section).
Moreover, since ${\hat S}$ is holomorphic by hypothesis,
also $\varphi$ and $\rho$ are holomorphic
(cf. (6.2.1)), and this affects the unprimed spinor part
of the null tangent vectors to $\alpha$-surfaces in the
light of (6.2.6) and (6.2.7). 

By virtue of Eq. (6.2.8), one finds
$$
w_{AB'}=-2 \pi^{A'} \; \pi^{C'} \; 
{\widetilde \chi}_{A'B'AC'},
\eqno (6.2.9)
$$
which implies (Esposito 1993)
$$
\pi^{A'}\Bigr(\nabla_{AA'}\pi_{B'}\Bigr)=
\xi_{A} \; \pi_{B'}-2 \pi^{A'} \; \pi^{C'} \;
{\widetilde \chi}_{A'B'AC'} .
\eqno (6.2.10)
$$
Note that, if torsion is set to zero, Eq. (6.2.10) agrees
with Eq. (9.1.2) appearing in section 9.1 of 
Ward and Wells (1990), where
complex general relativity is studied. This is the
desired necessary condition for the field $\pi_{A'}$ to
define an $\alpha$-surface in the presence of torsion
(and it may be also shown to be sufficient, as in section 4.2). 
Our next task is to derive the integrability condition for
Eq. (6.2.10). For this purpose, following 
Ward and Wells (1990), we operate with
$\pi^{B'}\pi^{C'}\nabla_{\; \; C'}^{A}$ on both sides
of Eq. (6.2.10). This leads to
$$
\pi^{B'}\pi^{C'}\nabla_{\; \; C'}^{A}
\Bigr[\pi^{A'}\Bigr(\nabla_{AA'}\pi_{B'}\Bigr)\Bigr]
=\pi^{B'}\pi^{C'}\nabla_{\; \; C'}^{A}
\Bigr[\xi_{A}\pi_{B'}
-2\pi^{A'}\pi^{D'}{\widetilde \chi}_{A'B'AD'}\Bigr].
\eqno (6.2.11)
$$
Using the Leibniz rule, (6.2.10) and the well known
property $\pi_{A'}\pi^{A'}=\xi_{A}\xi^{A}=0$, the two
terms on the right-hand side of Eq. (6.2.11) are found to be
$$
\pi^{B'} \; \pi^{C'}\Bigr[\nabla_{\; \; C'}^{A}
\Bigr(\xi_{A}\pi_{B'}\Bigr)\Bigr]=
2\xi^{A} \; \pi^{A'} \; \pi^{B'} \; \pi^{C'}
\; {\widetilde \chi}_{A'B'AC'} ,
\eqno (6.2.12)
$$
$$ \eqalignno{
\; & 
\pi^{B'} \; \pi^{C'}\Bigr[\nabla_{\; \; C'}^{A}\Bigr(
-2\pi^{A'} \; \pi^{D'} \; {\widetilde \chi}_{A'B'AD'}\Bigr)\Bigr]
=-4\xi^{A} \; \pi^{A'} \; \pi^{B'} \; \pi^{D'} \;
{\widetilde \chi}_{A'B'AD'}\cr
&+8\pi^{B'} \; \pi_{F'} \; \pi_{G'} \; {\widetilde \chi}_{A'B'AD'}
\; \pi^{(A'} \; {\widetilde \chi}^{F'D')AG'}\cr
&-2\pi^{A'} \; \pi^{B'} \; \pi^{C'} \; \pi^{D'}
\Bigr(\nabla_{\; \; C'}^{A} \; {\widetilde \chi}_{A'B'AD'}\Bigr),
&(6.2.13)\cr}
$$
where round brackets denote symmetrization over $A'$ and
$D'$ on the second line of (6.2.13).

It now remains to compute the left-hand side of Eq. (6.2.11).
This is given by 
$$ \eqalignno{
\pi^{B'} \; \pi^{C'} \; \nabla_{\; \; C'}^{A}\Bigr[\pi^{A'}
\Bigr(\nabla_{AA'}\pi_{B'}\Bigr)\Bigr]&=
\pi^{B'} \; \pi^{C'}\Bigr(\nabla_{\; \; C'}^{A}\pi^{A'}\Bigr)
\Bigr(\nabla_{AA'}\pi_{B'}\Bigr)\cr
&-\pi^{A'} \; \pi^{B'} \; \pi^{C'}
\Bigr(\cstok{\ }_{C'A'}\pi_{B'}\Bigr),
&(6.2.14)\cr}
$$
where we have defined 
$\cstok{\ }_{C'A'} \equiv \nabla_{A(C'}\nabla_{\; \; A')}^{A}$
as in section 8.4.
Using Eq. (6.2.10), the first term on the right-hand side
of (6.2.14) is easily found to be
$$ \eqalignno{
\pi^{B'} \; \pi^{C'}\Bigr(\nabla_{\; \; C'}^{A}\pi^{A'}\Bigr)
\Bigr(\nabla_{AA'}\pi_{B'}\Bigr)&=
4\pi^{B'} \; \pi^{C'} \; \pi_{F'} \; \pi_{G'} \;
{\widetilde \chi}_{A'B'AC'} \; {\widetilde \chi}^{F'A'AG'}\cr
&-2\xi^{A} \; \pi^{A'} \; \pi^{B'} \; \pi^{C'} \;
{\widetilde \chi}_{A'B'AC'}.
&(6.2.15)\cr}
$$
The second term on the
right-hand side of (6.2.14) can only be computed 
after using some fundamental
identities of spinor calculus for theories with torsion,
hereafter referred to as $U_{4}$-theories, as in
Esposito (1992), Esposito (1994).
\vskip 1cm
\centerline {\bf 6.3 Spinor Ricci identities 
for complex $U_{4}$ theory}
\vskip 1cm
\noindent
Since the results we here describe 
play a key role in obtaining the integrability
condition for $\alpha$-surfaces (cf. section 6.4), we have chosen
to summarize the main formulae in this separate section,
following Penrose (1983), Penrose and Rindler (1984).

Using abstract-index notation, the symmetric Lorentzian
metric $g$ of real Lorentzian $U_{4}$ space-times is
still expressed by (see section 2.1) 
$$
g_{ab}=\varepsilon_{AB} \; \varepsilon_{A'B'}.
\eqno (6.3.1)
$$
Moreover, the full connection still obeys the metricity
condition $\nabla g =0$, and the corresponding
spinor covariant derivative is assumed to satisfy the
additional relations
$$
\nabla_{AA'} \; \varepsilon_{BC}=0 , \; 
\nabla_{AA'} \; \varepsilon_{B'C'}=0 ,
\eqno (6.3.2)
$$
and is a linear, {\it real} operator which satisfies the
Leibniz rule. However, since torsion does not vanish,
the difference $\Bigr(\nabla_{a}\nabla_{b}-\nabla_{b}
\nabla_{a}\Bigr)$ applied to a function $f$ is equal
to $2S_{ab}^{\; \; \; c} \; \nabla_{c}f \not = 0$.
Torsion also appears explicitly in the relation defining
the Riemann tensor 
$$
\Bigr(\nabla_{a}\nabla_{b}-\nabla_{b}\nabla_{a}
-2S_{ab}^{\; \; \; c} \; \nabla_{c}\Bigr)V^{d} \equiv
R_{abc}^{\; \; \; \; \; d} \; V^{c} ,
\eqno (6.3.3)
$$
and leads to a non-symmetric Ricci tensor $R_{ab} \not =
R_{ba}$, where $R_{ab} \equiv R_{acb}^{\; \; \; \; \; c}$.
Note that in (6.3.3) the factor $2$ multiplies
$S_{ab}^{\; \; \; c}$ since we are using definition (6.2.2),
whereas in Penrose and Rindler (1984) 
a definition is used where the 
torsion tensor is $T \equiv 2S$.
The tensor $R_{abcd}$ has now $36$ independent real components
at each point, rather than $20$ as in general relativity.
The information of these $36$ components is encoded in the
spinor fields
$$
\psi_{ABCD} , \;  
\Phi_{ABC'D'} , \; 
\Sigma_{AB} ,
$$
and in the scalar function $\Lambda$, having 
$5,9,3,$ and $1$ complex components respectively, and such that
$$
\psi_{ABCD}=\psi_{(ABCD)} ,
\eqno (6.3.4)
$$
$$
\Phi_{ABC'D'}=\Phi_{(AB)(C'D')} ,
\eqno (6.3.5a)
$$
$$
\Phi_{ABC'D'}-{\overline \Phi}_{C'D'AB} \not = 0 ,
\eqno (6.3.5b)
$$
$$
\Sigma_{AB}=\Sigma_{(AB)},
\eqno (6.3.6a)
$$
$$
R_{[ab]}=\Sigma_{AB} \; \varepsilon_{A'B'}+
{\overline \Sigma}_{A'B'} \; \varepsilon_{AB},
\eqno (6.3.6b)
$$
$$
\Lambda-{\overline \Lambda} \not = 0.
\eqno (6.3.7)
$$
In (6.3.4)--(6.3.6), round (square) brackets denote, as usual, 
symmetrization (antisymmetrization), and {\it overbars}
denote complex conjugation of spinors or scalars.
The spinor $\Sigma_{AB}$ and the left-hand sides of
(6.3.5{\it b}) and (6.3.7) are determined directly by torsion
and its covariant derivative. The relations
(6.3.5{\it b}), (6.3.6{\it b}) and (6.3.7) 
express a substantial difference
with respect to general relativity, and hold in any real
Lorentzian $U_{4}$ space-time.

We are, however, interested in the case of complex
$U_{4}$ space-times (or real Riemannian, where the metric
is positive-definite), in order to compare the necessary
condition for the existence of $\alpha$-surfaces with what
holds for complex general relativity. In that case,
it is well known that the spinor covariant derivative still
obeys (6.3.2) but is now a linear, {\it complex-holomorphic} 
operator satisfying the Leibniz rule. Moreover, barred 
spinors are replaced by independent twiddled spinors (e.g.
${\widetilde \Sigma}_{A'B'}$) which are no longer complex
conjugates of unbarred (or untwiddled) spinors, since
complex conjugation is no longer available. This
also holds for real Riemannian $U_{4}$ space-times, not to
be confused with real Lorentzian $U_{4}$ space-times,
but of course, in the positive-definite
case the spinor covariant derivative is a real, rather
than complex-holomorphic operator.

For the sake of clarity, we hereafter write
$CU_{4}$, $RU_{4}$, $LU_{4}$ to denote 
complex, real Riemannian or real Lorentzian 
$U_{4}$-theory, respectively. In the light of our previous discussion, 
the spinorial form of Riemann for $CU_{4}$ and $RU_{4}$
theories is
$$ \eqalignno{
R_{abcd}&=\psi_{ABCD} \; \varepsilon_{A'B'} \; \varepsilon_{C'D'}
+{\widetilde \psi}_{A'B'C'D'} \; \varepsilon_{AB}
\; \varepsilon_{CD}\cr
&+\Phi_{ABC'D'} \; \varepsilon_{A'B'} \; \varepsilon_{CD}
+{\widetilde \Phi}_{A'B'CD} \; \varepsilon_{AB}
\; \varepsilon_{C'D'}\cr
&+\Sigma_{AB} \; \varepsilon_{A'B'}
\; \varepsilon_{CD} \; \varepsilon_{C'D'}
+{\widetilde \Sigma}_{A'B'} \; \varepsilon_{AB}
\; \varepsilon_{CD} \; \varepsilon_{C'D'}\cr
&+\Lambda \Bigr(\varepsilon_{AC} \; \varepsilon_{BD}+
\varepsilon_{AD} \; \varepsilon_{BC}\Bigr)
\varepsilon_{A'B'} \; \varepsilon_{C'D'}\cr
&+{\widetilde \Lambda} \Bigr(\varepsilon_{A'C'}
\; \varepsilon_{B'D'}+\varepsilon_{A'D'} \; \varepsilon_{B'C'}
\Bigr) \varepsilon_{AB} \; \varepsilon_{CD}.
&(6.3.8)\cr}
$$
The spinors $\psi_{ABCD}$ and ${\widetilde \psi}_{A'B'C'D'}$
appearing in (6.3.8) are called anti-self-dual and self-dual
Weyl spinors respectively as in general relativity, and
they represent the part of Riemann invariant under conformal
rescalings of the metric. This property is proved at the end
of section 4 of Penrose (1983), following Eq. (49) therein.
Note that in Penrose (1983) a class of conformal
rescalings is studied such that ${\hat g}=\Omega {\overline 
\Omega} \; g$ (where $\Omega$ is a smooth, nowhere-vanishing,
complex-valued function),
and leading to the presence of torsion. We are, however,
not interested in this method for generating torsion, and we
only study models where torsion {\it already} exists before
any conformal rescaling of the metric.

We are now in a position to compute $\cstok{\ }_{C'A'}\pi_{B'}$
appearing in (6.2.14). For this purpose, following the method in
section 4.9 of Penrose and Rindler (1984), we define the operator
$$
\cstok{\ }_{ab} \equiv 2\nabla_{[a}\nabla_{b]}
-2S_{ab}^{\; \; \; c} \; \nabla_{c},
\eqno (6.3.9)
$$
and the self-dual null bivector
$$
k^{ab} \equiv \kappa^{A} \; \kappa^{B} \; \varepsilon^{A'B'}.
\eqno (6.3.10)
$$
The Ricci identity for $U_{4}$ theories 
$$
\cstok{\ }_{ab}k^{cd}=R_{abe}^{\; \; \; \; \; c} \; k^{ed}
+R_{abe}^{\; \; \; \; \; d} \; k^{ce},
\eqno (6.3.11)
$$
then yields
$$
2 \varepsilon^{E'F'}\kappa^{(E}\cstok{\ }_{ab}\kappa^{F)}
=\Bigr(\varepsilon^{ED}\varepsilon^{E'D'}\varepsilon^{C'F'}
\kappa^{C}\kappa^{F}
+\varepsilon^{FD}\varepsilon^{F'D'}\varepsilon^{E'C'}\kappa^{E}
\kappa^{C}\Bigr)R_{abcd}.
\eqno (6.3.12)
$$
This is why, using (6.3.8) and the identity
$$
2\nabla_{[a}\nabla_{b]}=\varepsilon_{A'B'}
\cstok{\ }_{AB}+\varepsilon_{AB}\cstok{\ }_{A'B'},
\eqno (6.3.13)
$$
a lengthy calculation of the $16$ terms occurring in (6.3.12)
yields
$$ \eqalignno{
\kappa^{(C}\Bigr[\varepsilon_{A'B'}\cstok{\ }_{AB}
+\varepsilon_{AB}\cstok{\ }_{A'B'}&
-2S_{AA'BB'}^{\; \; \; \; \; \; \; \; \; \; \; \; \; HH'}
\; \nabla_{HH'}\Bigr]\kappa^{D)}\cr
&=\varepsilon_{AB} \Bigr[
{\widetilde \Phi}_{A'B'E}^{\; \; \; \; \; \; \; \; \; \; \; (C}
\kappa^{D)}\kappa^{E}+{\widetilde \Sigma}_{A'B'}
\kappa^{(C}\kappa^{D)}\Bigr]\cr
&+\varepsilon_{A'B'}\Bigr[\psi_{ABE}^{\; \; \; \; \; \; \; (C}
\kappa^{D)}\kappa^{E}\cr
&-2\Lambda \kappa^{(C}\kappa_{(B}\varepsilon_{A)}^{\; \; \; \; D)}
+\Sigma_{AB}\kappa^{(C}\kappa^{D)}\Bigr] . 
&(6.3.14)\cr}
$$
We now write explicitly the 
symmetrizations over $C$ and $D$ occurring in (6.3.14).
Thus, using (6.2.4) and comparing left- and right-hand side
of (6.3.14), one finds the equations
$$
\Bigr[\cstok{\ }_{AB} -2\chi_{AB}^{\; \; \; \; \; HH'}
\nabla_{HH'}\Bigr]\kappa^{C}=
\psi_{ABE}^{\; \; \; \; \; \; \; \; C} \; \kappa^{E}
-2\Lambda \kappa_{(A}\varepsilon_{B)}^{\; \; \; C}
+\Sigma_{AB} \; \kappa^{C} ,
\eqno (6.3.15)
$$
$$
\Bigr[\cstok{\ }_{A'B'} -2
{\widetilde \chi}_{A'B'}^{\; \; \; \; \; \; \; HH'}
\nabla_{HH'}\Bigr]\kappa^{C}=
{\widetilde \Phi}_{A'B'E}^{\; \; \; \; \; \; \; \; \; \; \; C}
\; \kappa^{E}
+{\widetilde \Sigma}_{A'B'} \; \kappa^{C} .
\eqno (6.3.16)
$$
Equations (6.3.15) and (6.3.16) are two 
of the four spinor Ricci identities
for $CU_{4}$ or $RU_{4}$ theories. The remaining spinor
Ricci identities are
$$
\Bigr[\cstok{\ }_{A'B'} -2
{\widetilde \chi}_{A'B'}^{\; \; \; \; \; \; \; HH'}
\nabla_{HH'}\Bigr]\pi^{C'}=
{\widetilde \psi}_{A'B'E'}^{\; \; \; \; \; \; \; \; \; \; \; C'}
\pi^{E'} -2{\widetilde \Lambda} \pi_{(A'}\varepsilon_{B')}^{\; \; \; \; C'}
+{\widetilde \Sigma}_{A'B'} \; \pi^{C'} ,
\eqno (6.3.17)
$$
$$
\Bigr[\cstok{\ }_{AB} -2
\chi_{AB}^{\; \; \; \; \; HH'}\nabla_{HH'}\Bigr]
\pi^{C'}=
\Phi_{ABE'}^{\; \; \; \; \; \; \; \; \; \; C'}
\; \pi^{E'}
+\Sigma_{AB} \; \pi^{C'} .
\eqno (6.3.18)
$$
\vskip 1cm
\centerline {\bf 6.4 Integrability condition for
$\alpha$-surfaces}
\vskip 1cm
\noindent
Since $\pi^{A'}\pi_{A'}=0$, insertion of (6.3.17)
into (6.2.14) and careful use of Eq. (6.2.10) yield
$$ \eqalignno{
-\pi^{A'}\pi^{B'}\pi^{C'}\Bigr(\cstok{\ }_{C'A'}
\pi_{B'}\Bigr)&=
-\pi^{A'}\pi^{B'}\pi^{C'}\pi^{D'}
{\widetilde \psi}_{A'B'C'D'}\cr 
&+4\pi^{B'}\pi^{C'}\pi_{F'} \; \pi_{G'}
\; {\widetilde \chi}_{A'B'AC'}
\; {\widetilde \chi}^{F'G'AA'}.
&(6.4.1)\cr}
$$
In the light of (6.2.11)--(6.2.15) and (6.4.1), one thus finds the
following integrability condition for Eq. (6.2.10)
in the case of $CU_{4}$ or $RU_{4}$ theories
(Esposito 1993):
$$ \eqalignno{
{\widetilde \psi}_{A'B'C'D'}&=
-4{\widetilde \chi}_{A'B'AL'} \;
{\widetilde \chi}_{C' \; \; \; \; \; \; D'}^{\; \; \; \; L'A} 
+4{\widetilde \chi}_{L'B'AC'} \;
{\widetilde \chi}_{A'D'}^{\; \; \; \; \; \; \; \; AL'}\cr
&+2\nabla_{\; \; D'}^{A}
\Bigr({\widetilde \chi}_{A'B'AC'}\Bigr) .
&(6.4.2)\cr}
$$
Note that contributions involving 
$\xi^{A}$ add up to zero.
\vskip 1cm
\centerline {\bf 6.5 Concluding remarks}
\vskip 1cm
\noindent
We have studied complex or real Riemannian space-times
with non-vanishing torsion. By analogy with complex 
general relativity, $\alpha$-surfaces have been defined
as totally null two-surfaces whose null tangent vectors have
the two-component spinor form $\lambda^{A}\pi^{A'}$,
with $\lambda^{A}$ varying and $\pi^{A'}$ fixed 
(cf. section 6.1, Ward and Wells 1990).
Using Frobenius' theorem, this leads to Eq. (6.2.10),
which differs from the equation corresponding to general
relativity by the term involving the torsion spinor.
The integrability condition for Eq. (6.2.10) is then given by
Eq. (6.4.2), which involves the self-dual Weyl spinor
(as in complex general relativity), terms 
quadratic in the torsion spinor, and the covariant
derivative of the torsion spinor. Our results (6.2.10) and (6.4.2)
are quite generic, in that they do not make use of any field
equations. We only assumed we were not studying 
supersymmetric theories of gravity.

A naturally occurring question is whether an alternative way
exists to derive our results (6.2.10) and (6.4.2). This is
indeed possible, since 
in terms of the Levi--Civita connection the necessary and
sufficient condition for the existence of 
$\alpha$-surfaces is the vanishing of the self-dual torsion-free
Weyl spinor; one has then to translate this condition into
a property of the Weyl spinor and torsion of the full
$U_{4}$-connection. 
One then finds that the integrability condition for
$\alpha$-surfaces, at first expressed using the self-dual
Weyl spinor of the Levi--Civita connection, 
coincides with Eq. (6.4.2).

We believe, however, that the more fundamental geometric
object is the full $U_{4}$-connection with torsion. This 
point of view is especially relevant when one studies the
Hamiltonian form of these theories, and is along the lines
of previous work by the author, where other properties
of $U_{4}$-theories have been studied working with the
complete $U_{4}$-connection 
(Esposito 1992, Esposito 1994). It was thus
our aim to derive Eq. (6.4.2) in a way independent of the use
of formulae relating curvature spinors of the Levi--Civita
connection to torsion and curvature spinors of the 
$U_{4}$-connection. We hope our chapter shows that this program
can be consistently developed.

Interestingly, a {\it particular} solution of Eq. (6.4.2)
is given by 
$$
{\widetilde \psi}_{A'B'C'D'}=0,
\eqno (6.5.1)
$$
$$
{\widetilde \chi}_{A'C'AB'}=0.
\eqno (6.5.2)
$$
This means that the surviving part of torsion is
$\chi_{AB}^{\; \; \; \; \; CC'} \varepsilon_{A'B'}$
(cf. (6.2.4)), which does not affect the integrability
condition for $\alpha$-surfaces, and that the $U_{4}$
Weyl curvature is anti-self-dual. Note that this is
only possible for $CU_{4}$ and $RU_{4}$ models of
gravity, since only for these theories Eqs. (6.5.1) and (6.5.2)
do not imply the vanishing of $\chi_{ACBA'}$ and
$\psi_{ABCD}$ (cf. section 6.3). By analogy with complex 
general relativity, those particular $CU_{4}$
and $RU_{4}$ space-times satisfying Eqs. (6.5.1) and (6.5.2) are
here called right conformally flat (in the light of Eq. (6.5.1)) and
{\it right-torsion-free} (in the light of Eq. (6.5.2)). Note that our
definition does not involve the Ricci tensor, and is
therefore different from Eq. (6.2.1) of Ward and
Wells (1990) (see (4.2.2)).
\vskip 100cm
\centerline {\it CHAPTER SEVEN}
\vskip 1cm
\centerline {\bf SPIN-${1\over 2}$ FIELDS IN RIEMANNIAN GEOMETRIES} 
\vskip 1cm
\noindent
Local supersymmetry leads to boundary conditions
for fermionic fields in one-loop quantum cosmology involving
the Euclidean normal $_{e}n_{A}^{\; \; A'}$ to the boundary and a pair
of independent spinor fields $\psi^{A}$ and
${\widetilde \psi}^{A'}$. This chapter studies the corresponding
classical properties, i.e. the classical boundary-value problem
and boundary terms in the variational problem. If
$\sqrt{2} \; {_{e}n_{A}^{\; \; A'}} \; \psi^{A}
\mp {\widetilde \psi}^{A'} \equiv \Phi^{A'}$ is set to zero
on a three-sphere bounding flat 
Euclidean four-space, the modes of the
massless spin-${1\over 2}$ field multiplying harmonics having
positive eigenvalues for the intrinsic 
three-dimensional Dirac operator
on $S^{3}$ should vanish on $S^{3}$. Remarkably, this coincides with
the property of the classical boundary-value problem when spectral
boundary conditions are imposed on $S^3$ in the massless case.
Moreover, the boundary term in the action functional is proportional
to the integral on the boundary of $\Phi^{A'} \; {_{e}n_{AA'}}
\; \psi^{A}$. The existence of self-adjoint extensions of the
Dirac operator subject to supersymmetric boundary conditions is
then proved. The global theory of the Dirac operator in compact
Riemannian manifolds is eventually described.
\vskip 100cm
\centerline {\bf 7.1 Dirac and Weyl equations in 
two-component spinor form}
\vskip 1cm
\noindent
Dirac's theory of massive and massless spin-${1\over 2}$
particles is still a key element of modern particle physics
and field theory. From the point of view of theoretical
physics, the description of such particles motivates indeed
the whole theory of Dirac operators. We are here concerned
with a two-component spinor analysis of the corresponding
spin-${1\over 2}$ fields in Riemannian four-geometries
$(M,g)$ with boundary. A massive spin-${1\over 2}$ Dirac
field is then described by the four independent spinor
fields $\phi^{A},\chi^{A},{\widetilde \phi}^{A'},
{\widetilde \chi}^{A'}$, and the action functional takes
the form
$$
I \equiv I_{V}+I_{B},
\eqno (7.1.1)
$$
where
$$ \eqalignno{
I_{V} &\equiv 
{i\over 2}\int_{M}\Bigr[{\widetilde \phi}^{A'}
\Bigr(\nabla_{AA'} \; \phi^{A}\Bigr)-\Bigr(\nabla_{AA'}
\; {\widetilde \phi}^{A'}\Bigr)\phi^{A}\Bigr]
\sqrt{{\rm det} \; g} \; d^{4}x \cr
&+{i\over 2}\int_{M}\Bigr[{\widetilde \chi}^{A'}
\Bigr(\nabla_{AA'} \; \chi^{A}\Bigr)-\Bigr(\nabla_{AA'}
\; {\widetilde \chi}^{A'}\Bigr)\chi^{A}\Bigr]
\sqrt{{\rm det} \; g} \; d^{4}x \cr
&+{m\over \sqrt{2}}\int_{M}\Bigr[\chi_{A}\phi^{A}
+{\widetilde \phi}^{A'}{\widetilde \chi}_{A'}\Bigr]
\sqrt{{\rm det} \; g} \; d^{4}x,
&(7.1.2)\cr}
$$
and $I_{B}$ is a suitable boundary term, necessary to
obtain a well posed variational problem. Its form is
determined once one knows which spinor fields are
fixed on the boundary (e.g. section 7.2). With our
notation, the occurrence of $i$ depends on conventions
for Infeld--van der Waerden symbols (see section 7.2).
One thus finds the field equations
$$
\nabla_{AA'} \; \phi^{A}={im\over \sqrt{2}} \;
{\widetilde \chi}_{A'},
\eqno (7.1.3)
$$
$$
\nabla_{AA'} \; \chi^{A}={im \over \sqrt{2}} \;
{\widetilde \phi}_{A'},
\eqno (7.1.4)
$$
$$
\nabla_{AA'} \; {\widetilde \phi}^{A'}=
-{im\over \sqrt{2}} \; \chi_{A},
\eqno (7.1.5)
$$
$$
\nabla_{AA'} \; {\widetilde \chi}^{A'}=
-{im\over \sqrt{2}} \; \phi_{A}.
\eqno (7.1.6)
$$
Note that this is a coupled system of first-order 
differential equations, obtained after applying
differentiation rules for anti-commuting spinor
fields. This means the spinor field acted upon by
the $\nabla_{AA'}$ operator should be always brought
to the left, hence leading to a minus sign if such a
field was not already on the left. Integration by
parts and careful use of boundary terms are also 
necessary. The equations (7.1.3)--(7.1.6) reproduce
the familiar form of the Dirac equation expressed in
terms of $\gamma$-matrices. In particular, for massless
fermionic fields one obtains the independent Weyl
equations
$$
\nabla^{AA'} \; \phi_{A}=0,
\eqno (7.1.7)
$$
$$
\nabla^{AA'} \; {\widetilde \phi}_{A'}=0,
\eqno (7.1.8)
$$
not related by any conjugation. 
\vskip 1cm
\centerline {\bf 7.2 Boundary terms for massless fermionic fields}
\vskip 1cm
\noindent
Locally supersymmetric boundary conditions have been recently
studied in quantum cosmology to understand its one-loop
properties. They involve the normal to the boundary and the field
for spin ${1\over 2}$, the normal to the boundary and the
spin-${3\over 2}$ potential for gravitinos, Dirichlet conditions
for real scalar fields, magnetic or electric field for
electromagnetism, mixed boundary conditions for the four-metric
of the gravitational field (and in particular Dirichlet conditions
on the perturbed three-metric). The aim of this section is to describe
the corresponding classical properties in the case of massless
spin-${1\over 2}$ fields.
 
For this purpose, we consider flat Euclidean four-space bounded
by a three-sphere of radius $a$. The alternative possibility 
is a more involved boundary-value problem, where flat Euclidean
four-space is bounded by two concentric three-spheres of radii
$r_{1}$ and $r_{2}$. 
The spin-${1\over 2}$ field,
represented by a pair of independent spinor fields $\psi^{A}$
and ${\widetilde \psi}^{A'}$, is expanded on a family of
three-spheres centred on the origin as 
(D'Eath and Halliwell 1987, D'Eath and Esposito 1991a, Esposito 
1994)
$$
\psi^{A}={\tau^{-{3\over 2}}\over 2\pi}
\sum_{n=0}^{\infty}\sum_{p=1}^{(n+1)(n+2)}
\sum_{q=1}^{(n+1)(n+2)} \alpha_{n}^{pq}
\Bigr[m_{np}(\tau)\rho^{nqA}+{\widetilde r}_{np}(\tau)
{\overline \sigma}^{nqA}\Bigr],
\eqno (7.2.1)
$$
$$
{\widetilde \psi}^{A'}={\tau^{-{3\over 2}}\over 2\pi}
\sum_{n=0}^{\infty}\sum_{p=1}^{(n+1)(n+2)}
\sum_{q=1}^{(n+1)(n+2)} \alpha_{n}^{pq}
\Bigr[{\widetilde m}_{np}(\tau){\overline \rho}^{nqA'}
+r_{np}(\tau)\sigma^{nqA'}\Bigr].
\eqno (7.2.2)
$$
With our notation, $\tau$ is the Euclidean-time coordinate,
the $\alpha_{n}^{pq}$ are block-diagonal matrices with
blocks $\pmatrix {1&1\cr 1&-1\cr}$, the $\rho-$ and
$\sigma$-harmonics obey the identities described in 
D'Eath and Halliwell (1987), Esposito (1994).
Last but not least, the modes $m_{np}$ and $r_{np}$ are
regular at $\tau=0$, whereas the modes ${\widetilde m}_{np}$
and ${\widetilde r}_{np}$ are singular at $\tau=0$ if the
spin-${1\over 2}$ field is massless. Bearing in mind that the
harmonics $\rho^{nqA}$ and $\sigma^{nqA'}$ have positive
eigenvalues ${1\over 2}\Bigr(n+{3\over 2}\Bigr)$ for the
three-dimensional Dirac operator on the bounding 
$S^3$ (Esposito 1994), the
decomposition (7.2.1) and (7.2.2) can be re-expressed as
$$
\psi^{A}=\psi_{(+)}^{A}+\psi_{(-)}^{A},
\eqno (7.2.3)
$$
$$
{\widetilde \psi}^{A'}={\widetilde \psi}_{(+)}^{A'}
+{\widetilde \psi}_{(-)}^{A'}.
\eqno (7.2.4)
$$
In (7.2.3) and (7.2.4), 
the $(+)$ parts correspond to the modes $m_{np}$
and $r_{np}$, whereas the $(-)$ parts correspond to the singular
modes ${\widetilde m}_{np}$ and ${\widetilde r}_{np}$,
which multiply harmonics having negative eigenvalues 
$-{1\over 2}\Bigr(n+{3\over 2}\Bigr)$ for the three-dimensional
Dirac operator on $S^3$. If one wants to find a classical
solution of the Weyl equation which is regular 
$\forall \tau \in [0,a]$, one is thus forced to set to zero
the modes ${\widetilde m}_{np}$ and ${\widetilde r}_{np}$
$\forall \tau \in [0,a]$ 
(D'Eath and Halliwell 1987). This is why, if one requires the
local boundary conditions (Esposito 1994)
$$
\sqrt{2} \; {_{e}n_{A}^{\; \; A'}} \; \psi^{A}
\mp {\widetilde \psi}^{A'}=\Phi^{A'}
\; {\rm on} \; S^{3} ,
\eqno (7.2.5)
$$
such a condition can be expressed as 
$$
\sqrt{2} \; {_{e}n_{A}^{\; \; A'}} \; \psi_{(+)}^{A}
=\Phi_{1}^{A'}
\; {\rm on} \; S^{3},
\eqno (7.2.6)
$$
$$
\mp {\widetilde \psi}_{(+)}^{A'}=\Phi_{2}^{A'}
\; {\rm on} \; S^{3},
\eqno (7.2.7)
$$
where $\Phi_{1}^{A'}$ and $\Phi_{2}^{A'}$ are the parts of the
spinor field $\Phi^{A'}$ related to the ${\overline \rho}$-
and $\sigma$-harmonics, respectively. In particular, if
$\Phi_{1}^{A'}=\Phi_{2}^{A'}=0$ 
on $S^3$, one finds
$$
\sum_{n=0}^{\infty}\sum_{p=1}^{(n+1)(n+2)}
\sum_{q=1}^{(n+1)(n+2)}\alpha_{n}^{pq}
\; m_{np}(a) \; {_{e}n_{A}^{\; \; A'}}
\; \rho_{nq}^{A}=0,
\eqno (7.2.8)
$$
$$
\sum_{n=0}^{\infty}\sum_{p=1}^{(n+1)(n+2)}
\sum_{q=1}^{(n+1)(n+2)}\alpha_{n}^{pq}
\; r_{np}(a) 
\; \sigma_{nq}^{A'}=0,
\eqno (7.2.9)
$$
where $a$ is the three-sphere radius. Since the harmonics
appearing in (7.2.8) and (7.2.9) are linearly independent, these 
relations lead to $m_{np}(a)=r_{np}(a)=0$ $\forall n,p$.
Remarkably, this simple calculation shows that the classical
boundary-value problems for regular solutions of the Weyl
equation subject to local or spectral conditions on $S^3$
share the same property provided that $\Phi^{A'}$ is set to zero
in (7.2.5): the regular modes $m_{np}$ and $r_{np}$ should vanish
on the bounding $S^3$.
 
To study the corresponding variational problem for a massless
fermionic field, we should now bear in mind that the 
spin-${1\over 2}$ action
functional in a Riemannian four-geometry takes the form 
(D'Eath and Esposito 1991a, Esposito 1994)
$$
I_{E} \equiv {i\over 2}\int_{M}\Bigr[{\widetilde \psi}^{A'}
\Bigr(\nabla_{AA'} \; \psi^{A}\Bigr)
-\Bigr(\nabla_{AA'} \; {\widetilde \psi}^{A'}\Bigr)
\psi^{A}\Bigr]\sqrt{{\rm det} \; g} \; d^{4}x+{\widehat I}_{B}.
\eqno (7.2.10)
$$
This action is {\it real}, and the factor $i$ occurs
by virtue of the convention for Infeld--van der Waerden symbols
used in D'Eath and Esposito (1991a), Esposito (1994).
In (7.2.10) ${\widehat I}_{B}$ is a suitable boundary term, to be 
added to ensure that $I_{E}$ is stationary under the boundary
conditions chosen at the various components of the boundary
(e.g. initial and final surfaces, as in 
D'Eath and Halliwell (1987)). Of course, the
variation $\delta I_{E}$ of $I_{E}$ is linear in the
variations $\delta \psi^{A}$ and $\delta {\widetilde \psi}^{A'}$.
Defining $\kappa
\equiv {2\over i}$ and $\kappa {\widehat I}_{B} \equiv I_{B}$,
variational rules for anticommuting spinor fields lead to
$$ \eqalignno{
\kappa \Bigr(\delta I_{E}\Bigr)&=
\int_{M}\Bigr[2 \delta {\widetilde \psi}^{A'}
\Bigr(\nabla_{AA'}\psi^{A}\Bigr)\Bigr]
\sqrt{{\rm det} \; g} \; d^{4}x
-\int_{M}\Bigr[\Bigr(\nabla_{AA'}{\widetilde \psi}^{A'}
\Bigr)2 \delta \psi^{A}\Bigr]
\sqrt{{\rm det} \; g} \; d^{4}x \cr
&-\int_{\partial M}\Bigr[{_{e}n_{AA'}}
\Bigr(\delta {\widetilde \psi}^{A'}\Bigr)
\psi^{A}\Bigr]\sqrt{{\rm det} \; h} \; d^{3}x
+\int_{\partial M}\Bigr[{_{e}n_{AA'}}
{\widetilde \psi}^{A'}\Bigr(\delta \psi^{A}\Bigr)
\Bigr] \sqrt{{\rm det} \; h} \; d^{3}x \cr
&+\delta I_{B},
&(7.2.11)\cr}
$$
where $I_{B}$ should be chosen in such a way that its
variation $\delta I_{B}$ combines with the sum of the two terms
on the second line of (7.2.11) so as to specify what is fixed
on the boundary (see below). Indeed, setting
$\epsilon \equiv \pm 1$ and using the boundary conditions (7.2.5)
one finds
$$
{_{e}n_{AA'}} \; {\widetilde \psi}^{A'}
={\epsilon \over \sqrt{2}}\psi_{A}
-\epsilon \; {_{e}n_{AA'}} \; \Phi^{A'}
\; {\rm on} \; S^{3}.
\eqno (7.2.12)
$$
Thus, anticommutation rules for spinor 
fields (D'Eath and Halliwell 1987) show that the
second line of Eq. (7.2.11) reads
$$ \eqalignno{
\delta I_{\partial M} & \equiv
-\int_{\partial M}\Bigr[\Bigr(\delta {\widetilde \psi}^{A'}
\Bigr){_{e}n_{AA'}}\psi^{A}\Bigr]
\sqrt{{\rm det} \; h} \; d^{3}x
+\int_{\partial M}\Bigr[{_{e}n_{AA'}}{\widetilde \psi}^{A'}
\Bigr(\delta \psi^{A}\Bigr)\Bigr]
\sqrt{{\rm det} \; h} \; d^{3}x \cr
&=\epsilon \int_{\partial M}
{_{e}n_{AA'}}\Bigr[\Bigr(\delta \Phi^{A'}\Bigr)
\psi^{A}-\Phi^{A'}\Bigr(\delta \psi^{A}\Bigr)
\Bigr] \sqrt{{\rm det} \; h} \; d^{3}x.
&(7.2.13)\cr}
$$
Now it is clear that setting 
$$
I_{B} \equiv \epsilon \, \int_{\partial M} 
\; \Phi^{A'} {_{e}n_{AA'}}
\; \psi^A \sqrt{ {\rm det} \; h} \; d^3x,
\eqno (7.2.14)
$$
enables one to specify $\Phi^{A'}$ on the boundary, since
$$
\delta \Bigr[ I_{\partial M} + I_{B} \Bigr] =
2 \epsilon \int_{\partial M} {_{e}n_{AA'}} 
\Bigr( \delta \Phi^{A'} \Bigr) \psi^A \sqrt{{\rm det }\; h}\; d^3x.
\eqno (7.2.15)
$$
Hence the action integral (7.2.10) appropriate for our boundary-value
problem is (Esposito {\it et al}. 1994)
$$ \eqalignno{
I_{E}&={i\over 2}\int_{M}\Bigr[{\widetilde \psi}^{A'}
\Bigr(\nabla_{AA'} \; \psi^{A}\Bigr)
-\Bigr(\nabla_{AA'} \; {\widetilde \psi}^{A'}\Bigr)
\psi^{A}\Bigr]\sqrt{{\rm det} \; g} \; d^{4}x \cr
&+{i\epsilon \over 2} \, \int_{\partial M} 
\; \Phi^{A'} \; {_{e}n_{AA'}}
\; \psi^A \sqrt{ {\rm det} \; h} \; d^3x .
&(7.2.16)\cr}
$$
Note that, by virtue of (7.2.5), Eq. (7.2.13) may also be
cast in the form
$$
\delta I_{\partial M}
={1\over \sqrt{2}}\int_{\partial M}
\Bigr[{\widetilde \psi}^{A'}\Bigr(\delta \Phi_{A'}\Bigr)
-\Bigr(\delta {\widetilde \psi}^{A'}\Bigr)\Phi_{A'}
\Bigr] \sqrt{{\rm det} \; h} \; d^{3}x ,
\eqno (7.2.17)
$$
which implies that an equivalent form of $I_{B}$ is
$$
I_{B} \equiv {1\over \sqrt{2}}
\int_{\partial M}{\widetilde \psi}^{A'}
\; \Phi_{A'} \sqrt{{\rm det} \; h}
\; d^{3}x .
\eqno (7.2.18)
$$

The local boundary conditions studied at the classical level
in this section, have been applied to one-loop quantum cosmology
in D'Eath and Esposito (1991a),  
Kamenshchik and Mishakov (1993), Esposito (1994). 
Interestingly, our work seems to add evidence in favour
of quantum amplitudes having to respect the properties of the
classical boundary-value problem. In other words, if fermionic
fields are massless, their one-loop properties in the presence
of boundaries coincide in the case of spectral 
(D'Eath and Halliwell 1987, D'Eath and Esposito 1991b,
Esposito 1994) or local
boundary conditions (D'Eath and Esposito 1991a, 
Kamenshchik and Mishakov 1993, Esposito 1994), 
while we find that classical modes
for a regular solution of the Weyl equation obey the same
conditions on a three-sphere boundary with spectral or local
boundary conditions, provided that the spinor field $\Phi^{A'}$
of (7.2.5) is set to zero on $S^{3}$. We also hope that the
analysis presented in Eqs. (7.2.10)--(7.2.18) may clarify the
spin-${1\over 2}$
variational problem in the case of local boundary conditions
on a three-sphere (cf. the analysis in Charap and Nelson (1983),
York (1986), Hayward (1993) for pure gravity).
\vskip 10cm
\centerline {\bf 7.3 Self-adjointness of the boundary-value problem}
\vskip 1cm
\noindent
So far we have seen that the framework for the formulation of
local boundary conditions involving normals and field strengths        
or fields is the Euclidean regime, where one deals with Riemannian metrics.
Thus, we will pay special
attention to the conjugation of $SU(2)$ spinors in Euclidean four-space.
In fact such a conjugation will play a key role in proving self-adjointness.
For this purpose, it can be useful to recall at first
some basic results about $SU(2)$ spinors on an abstract Riemannian 
three-manifold $\left(\Sigma,h\right)$. In that case, one considers
a bundle over the three-manifold, each fibre of which is isomorphic
to a two-dimensional complex vector space $W$. It is then possible to define
a nowhere vanishing antisymmetric 
$\varepsilon_{AB}$ (the usual one of section 2.1)
so as to raise and lower internal indices, and a positive-definite
Hermitian inner product on each fibre: $(\psi , \phi)={\overline \psi}^{A'}
G_{A'A}\phi^{A}$. The requirements of Hermiticity and positivity imply 
respectively that ${\overline G}_{A'A}=G_{A'A}$, ${\overline \psi}^{A'}
G_{A'A}\psi^{A}>0$,$\forall$ $\psi^{A} \not = 0$. This $G_{A'A}$ converts
primed indices to unprimed ones, and it is given by $i\sqrt{2} \; n_{AA'}$.
Given the space H of all objects
$\alpha_{\; \; B}^{A}$ such that $\alpha_{\; \; A}^{A}=0$ and
$\left({\alpha}^{\dagger}\right)_{\; \; B}^{A}=-\alpha_{\; \; B}^{A}$,
one finds there always exists a $SU(2)$ soldering form $\sigma_{\; \; A}   
^{a\; \; \; B}$ (i.e. a global isomorphism) between $H$ and the tangent
space on $\left(\Sigma,h\right)$ such that 
$h^{ab}=-\sigma_{\; \; A}^{a\; \; \; B}
\; \sigma_{\; \; B}^{b\; \; \; A}$.
Therefore one also finds $\sigma_{\; \; A}^{a\; \; \; A}=0$ and
${\left(\sigma_{\; \; A}^{a\; \; \; B}\right)}^{\dagger}=
-\sigma_{\; \; A}^{a\; \; \; B}$. One then defines $\psi^{A}$ an 
$SU(2)$ spinor on $\left(\Sigma,h\right)$. A basic remark is that     
$SU(2)$ transformations are those $SL(2,C)$ transformations which preserve
$n^{AA'}=n^{a}\sigma_{a}^{\; \; AA'}$, where $n^{a}=(1, 0, 0, 0)$
is the normal to
$\Sigma$. The Euclidean conjugation used here (not to be confused
with complex conjugation in Minkowski space-time) is such that
(see now section 2.1) 
$$
{\left(\psi_{A} +\lambda \phi_{A}\right)}^{\dagger}=
{\psi_{A}}^{\dagger}
+{\lambda}^{*} {\phi_{A}}^{\dagger} , \;             
{\left({\psi_{A}}^{\dagger}\right)}^{\dagger}=-\psi_{A},
\eqno (7.3.1)
$$
$$
{\varepsilon_{AB}}^{\dagger}=\varepsilon_{AB} , \;
{\left(\psi_{A} \; \phi_{B}\right)}^{\dagger}=
{\psi_{A}}^{\dagger} \; {\phi_{B}}^{\dagger},
\eqno (7.3.2)
$$
$$
{\left(\psi_{A}\right)}^{\dagger}\psi^{A}>0 , \; 
\forall \; \psi_{A} \not =0 .
\eqno (7.3.3)
$$
In (7.3.1) and in the following pages, the symbol $*$ denotes complex
conjugation of scalars.
How to generalize this picture to  
Euclidean four-space? For this purpose,
let us now focus our attention on states that are pairs of
spinor fields, defining 
$$
w\equiv \left(\psi^{A}, \; \widetilde \psi^{A'}\right) , \;
z\equiv \left(\phi^{A}, \; \widetilde \phi^{A'}\right),
\eqno (7.3.4)
$$
on the ball of radius $a$ in Euclidean four-space, subject always to the
boundary conditions (7.2.5). Our
$w$ and $z$ are subject also to suitable
differentiability conditions, to be specified later. Let us also define 
the operator $C$ 
$$
C\; : \left(\psi^{A}, \; \widetilde \psi^{A'}\right)\rightarrow
\left(\nabla_{\; \; B'}^{A} \; \widetilde \psi^{B'},
\nabla_{B}^{\; \; A'} \; \psi^{B}\right) ,
\eqno (7.3.5)
$$
and the {\it dagger} operation 
$$ 
{\left(\psi^{A}\right)}^{\dagger}\equiv 
\varepsilon^{AB} \; \delta_{BA'} \; 
{\overline \psi}^{A'} , \;
{\left({\widetilde \psi}^{A'}\right)}^{\dagger} \equiv
\varepsilon^{A'B'} \; \delta_{B'A} \;
{\overline {\widetilde \psi}}^{A}.
\eqno (7.3.6)
$$
The consideration of $C$ is suggested of course by the action (7.2.10).
In (7.3.6), $\delta_{BA'}$ is an identity matrix playing the same role of
$G_{AA'}$ for $SU(2)$ spinors on $\left(\Sigma,h\right)$, so that
$\delta_{BA'}$ is preserved by $SU(2)$ transformations.          
Moreover, the {\it bar} symbol
$\overline {\psi^{A}}={\overline \psi}^{A'}$ denotes the usual complex
conjugation of $SL(2,C)$ spinors. Hence one finds 
$$            
{\left({\left(\psi^{A}\right)}^{\dagger}\right)}^{\dagger}=
\varepsilon^{AC} \; \delta_{CB'} \; \overline 
{\left(\psi^{B \dagger}\right)'}   
=\varepsilon^{AC} \; \delta_{CB'} \; 
\varepsilon^{B'D'} \; \delta_{D'F} \; \psi^{F}   
=-\psi^{A},
\eqno (7.3.7)
$$
in view of the definition of $\varepsilon^{AB}$. Thus, the 
{\it dagger} operation defined in (7.3.6) is anti-involutory, because,
when applied twice to $\psi^{A}$, it yields $-\psi^{A}$.               

From now on we study commuting spinors, for simplicity
of exposition of the self-adjointness. It is easy to check that the          
{\it dagger}, also called in the literature Euclidean conjugation
(section 2.1), satisfies all properties (7.3.1)--(7.3.3). 
We can now define the scalar product 
$$
(w,z)\equiv \int_{M}\left[\psi_{A}^{\dagger}
\phi^{A}+{\widetilde \psi_{A'}}^{\dagger}
{\widetilde \phi}^{A'}\right]\sqrt{g} \; d^{4}x .
\eqno (7.3.8)
$$
This is indeed a scalar product, because it satisfies all following
properties for all vectors $u$, $v$, $w$ and $\forall \lambda \in \; C$ :
$$
(u,u)>0 , \; \forall u \not =0 ,
\eqno (7.3.9)
$$
$$
(u,v+w)=(u,v)+(u,w),
\eqno (7.3.10)
$$
$$
(u,\lambda v)=\lambda(u,v) , \;             
(\lambda u,v)={\lambda}^{*}(u,v),
\eqno (7.3.11)
$$
$$
(v,u)={(u,v)}^{*}.
\eqno (7.3.12)
$$
We are now aiming to check that $C$ or $iC$ is a symmetric operator,
i.e. that
$\left(Cz,w\right)=\left(z,Cw\right)$ or $\left(iCz,w\right)=
\left(z,iCw\right), \forall z,w$. This will be used in the course of
proving further that the symmetric operator has self-adjoint extensions.
In order to prove this result it is clear, in view
of (7.3.8), we need to know the properties of the spinor covariant
derivative acting on $SU(2)$ spinors. In the case of $SL(2,C)$ spinors
this derivative
is a linear, torsion-free map $\nabla_{AA'}$ which satisfies the Leibniz
rule, annihilates $\varepsilon_{AB}$ and is real (i.e. $\psi=\nabla_{AA'}
\theta \Rightarrow {\overline \psi}=\nabla_{AA'}{\overline \theta}$).
Moreover, we know that  
$$
\nabla^{AA'}=e_{\; \; \; \; \; \; \mu}^{AA'} \; \nabla^{\mu}=
e_{\; \; \mu}^{a} \; \sigma_{a}^{\; \; AA'} \; \nabla^{\mu}.
\eqno (7.3.13)
$$
In Euclidean four-space, we use both (7.3.13) and the relation 
$$
\sigma_{\mu AC'} \; \sigma_{\nu B}^{\; \; \; \; \; C'}
+\sigma_{\nu BC'} \; \sigma_{\mu A}^{\; \; \; \; \; C'}
= \delta_{\mu \nu}\varepsilon_{AB} ,
\eqno (7.3.14)
$$
where $\delta_{\mu\nu}$ has signature $(+,+,+,+)$. This implies that 
$\sigma_{0}=-{i\over \sqrt{2}}I$, $\sigma_{i}={\Sigma_{i}\over \sqrt{2}}$,
$\forall i=1, 2, 3$, where $\Sigma_{i}$ are the Pauli matrices. Now, in
view of (7.3.5) and (7.3.8) one finds 
$$            
(Cz,w)=
\int_{M}{\left(\nabla_{AB'} \; \phi^{A}\right)}^{\dagger}
{\widetilde \psi}^{B'}
\sqrt{g} \; d^{4}x    
+\int_{M}{\left(\nabla_{BA'}
\; {\widetilde \phi}^{A'}\right)}^{\dagger}
\psi^{B}\sqrt{g} \; d^{4}x,
\eqno (7.3.15)
$$
whereas, using the Leibniz rule to evaluate 
$$
\nabla_{\; \; B'}^{A}
\left(\phi_{A}^{\dagger}{\widetilde \psi}^{B'}\right)
$$ 
and
$$
\nabla_{B}^{\; \; A'}
\left(\left({\widetilde \phi}_{A'}\right)^{\dagger}\psi^{B}
\right),
$$
and integrating by parts, one finds 
$$ \eqalignno{
(z,Cw)&=
\int_{M}\left(\nabla_{AB'} \; \phi^{A \dagger}\right)
{\widetilde \psi}^{B'}\sqrt{g} \; d^{4}x
+\int_{M}\left(\nabla_{BA'}
\left({\widetilde \phi}^{A'}\right)^{\dagger}\right)
\psi^{B}\sqrt{g} \; d^{4}x \cr
&-\int_{\partial M}(_{e}n_{AB'})\phi^{A \dagger}
\; {\widetilde \psi}^{B'}
\sqrt{h} \; d^{3}x \cr
&-\int_{\partial M}(_{e}n_{BA'}){\left({\widetilde \phi}^{A'}\right)}
^{\dagger} \; \psi^{B}\sqrt{h} \; d^{3}x.
&(7.3.16)\cr}
$$
Now we use (7.3.6), section 2.1, the identity 
$$
{\left(_{e}n^{AA'}\phi_{A}\right)}^{\dagger}=
\varepsilon^{A'B'} \; \delta_{B'C} \;
{\overline {_{e}n^{DC'}}} \;
{\overline {\phi_{D}}}
=-\varepsilon^{A'B'} \; \delta_{B'C} \; \left(_{e}n^{CD'}\right)
{\overline \phi}_{D'},
\eqno (7.3.17)
$$
and the boundary
conditions on $S^{3}$: $\sqrt{2}\; {_{e}n^{CB'}}\psi_{C}
={\widetilde \psi}^{B'}$, $\sqrt{2}\; {_{e}n^{AA'}}\phi_{A}
={\widetilde \phi}^{A'}$. In so doing, the sum of the
boundary terms in (7.3.16) is found to vanish. This implies in turn that
equality of the volume integrands is sufficient to show that
$(Cz,w)$ and $(z,Cw)$ are equal. For                              
example, one finds in flat space, using also (7.3.6):
${\left(\nabla_{BA'}{\widetilde \phi}^{A'}\right)}^{\dagger}=\delta_{BF'}
\; {\overline \sigma}_{\; \; C}^{F'\; \; a} \; \partial_{a}
\left({\overline {\widetilde \phi}}^{C}\right)$, whereas:
$$
\left(\nabla_{BA'}\left({\widetilde \phi}^{A'}
\right)^{\dagger}\right)=
-\delta_{CF'} \; \sigma_{B}^{\; \; F'a} \; \partial_{a}
\left({\overline {\widetilde \phi}}^{C}\right).
$$ 
In other words, we are led 
to study the condition 
$$
\delta_{BF'} \; {\overline \sigma}_{\; \; \; C}^{F' \; \; \; a}=
\pm \delta_{BF'} \; \sigma_{C}^{\; \; F'a},
\eqno (7.3.18)
$$
$\forall$ $a=0, 1, 2, 3$. Now, using the relations 
$$
\sqrt{2}\; \sigma_{AA'}^{\; \; \; \; \; 0}=
\pmatrix {-i&0\cr 0&-i\cr} , \; 
\sqrt{2}\; \sigma_{AA'}^{\; \; \; \; \; 1}=
\pmatrix {0&1\cr 1&0\cr},
\eqno (7.3.19)
$$
$$
\sqrt{2}\; \sigma_{AA'}^{\; \; \; \; \; 2}=
\pmatrix {0&-i\cr i&0\cr} , \; 
\sqrt{2}\; \sigma_{AA'}^{\; \; \; \; \; 3}=
\pmatrix {1&0\cr 0&-1\cr},
\eqno (7.3.20)
$$
$$
\sigma_{\; \; A'}^{A\; \; \; a}=\varepsilon^{AB}
\; \sigma_{BA'}^{\; \; \; \; \; \; a}, \;
\sigma_{A}^{\; \; A'a}=
-\sigma_{AB'}^{\; \; \; \; \; \; a} \; \varepsilon^{B'A'},
\eqno (7.3.21)
$$
one finds that the complex conjugate of $\sigma_{\; \; A'}^{A\; \; \; a}$ is
always equal to $\sigma_{A}^{\; \; A'a}$, which is not in agreement with
the choice of the $(-)$ sign on the right-hand side of (7.3.18). This
implies in turn that the symmetric operator we are looking for is $iC$,
where $C$ has been defined in (7.3.5). The generalization to a curved       
four-dimensional Riemannian space is
obtained via the relation $e_{\; \; \; \; \; \mu}^{AA'}=e_{\; \; \mu}^{a}
\; \sigma_{a}^{\; \; AA'}$.
             
Now, it is known that every symmetric
operator has a closure, and the operator and its closure have the same closed
extensions. Moreover, a closed symmetric operator on a
Hilbert space is self-adjoint if and only if its spectrum is a subset of the
real axis. To prove
self-adjointness for our boundary-value problem,
we may recall an important result due to von
Neumann (Reed and Simon 1975). This theorem states that, given a
symmetric operator $A$ with domain $D(A)$, if a map $F : D(A)\rightarrow
D(A)$ exists such that 
$$
F(\alpha w + \beta z)=\alpha^{*}F(w)+\beta^{*}F(z),
\eqno (7.3.22)
$$
$$
(w,w)=(Fw,Fw),
\eqno (7.3.23)
$$
$$
F^{2}= \pm I,
\eqno (7.3.24)
$$
$$
FA=AF,
\eqno (7.3.25)
$$
then $A$ has self-adjoint extensions. In our case, denoting by $D$ the
operator (cf. (7.3.6))
$$
D : \left(\psi^{A}, \; {\widetilde \psi}^{A'}\right)\rightarrow
\left( {\Bigr(\psi^{A}\Bigr)}^{\dagger},
{\left({\widetilde \psi}^{A'}\right)}^{\dagger}\right),
\eqno (7.3.26)
$$
let us focus our attention on the operators 
$F \equiv iD$ and $A \equiv iC$. The 
operator $F$ maps indeed $D(A)$ into $D(A)$. 
In fact, bearing in mind the definitions
$$
G\equiv \left \{ \varphi =\Bigr(\phi^{A}, \; {\widetilde \phi}^{A'}\Bigr) :
\varphi \; {\rm is} \; {\rm at} \; {\rm least} \; 
C^{1} \right \} ,
\eqno (7.3.27)
$$
$$
D(A)\equiv \left \{ \varphi \in G : \sqrt{2} \; {_{e}n^{AA'}\phi_{A}}
=\epsilon \; {\widetilde \phi}^{A'} \; 
{\rm on} \; S^{3} \right \} ,
\eqno (7.3.28)
$$
one finds that $F$ maps $\Bigr(\phi^{A}, \; 
{\widetilde \phi}^{A'}\Bigr)$ into
$\Bigr(\beta^{A}, \; {\widetilde \beta}^{A'}\Bigr)=
\left(i {\Bigr(\phi^{A}\Bigr)}^{\dagger},
i{\Bigr({\widetilde \phi}^{A'}\Bigr)}^{\dagger}\right)$ with 
$$
\sqrt{2} \; {_{e}n^{AA'}\beta_{A}}=\gamma \; {\widetilde \beta}^{A'}
\; {\rm on}
\; S^{3} ,
\eqno (7.3.29)
$$
where $\gamma =\epsilon^{*}$. The boundary condition (7.3.29) is clearly
of the type which occurs in (7.3.28) provided that $\epsilon$ is real,
and the differentiability of
$\Bigr(\beta^{A}, \; {\widetilde \beta}^{A'}\Bigr)$ is not affected by the
action of $F$ (cf. (7.3.26)). In deriving (7.3.29), we have used the
result for ${\Bigr({_{e}n^{AA'}\phi_{A}}\Bigr)}^{\dagger}$ 
obtained in (7.3.17).
It is worth emphasizing that the requirement of self-adjointness         
enforces the choice of a real function $\epsilon$, which is a constant
in our case. Moreover, in view of
(7.3.7), one immediately sees that (7.3.22) and (7.3.24) hold when
$F=iD$, if we write (7.3.24) as $F^{2}=-I$. This is indeed a crucial
point which deserves special attention. Condition (7.3.24) is written in
Reed and Simon (1975) as $F^{2}=I$, and examples are later given (see page
144 therein) where $F$ is complex conjugation. But we are formulating
our problem in the Euclidean regime, where we have seen that the only   
possible conjugation is the {\it dagger} operation, which is
anti-involutory on spinors with an 
odd number of indices. Thus, we are here
generalizing von Neumann's theorem in the following way. If $F$ is a map
$D(A)\rightarrow D(A)$ which satisfies 
(7.3.22)--(7.3.25), then the same is clearly
true of ${\widetilde F} \equiv -iD=-F$. Hence                   
$$
-F \; D(A) \subseteq D(A) ,
\eqno (7.3.30)
$$
$$
F \; D(A) \subseteq D(A) .
\eqno (7.3.31)
$$
Acting with $F$ on both sides of (7.3.30), one finds 
$$
D(A) \subseteq F \; D(A) ,
\eqno (7.3.32)
$$
using the property $F^{2}=-I$. But 
then the relations (7.3.31) and (7.3.32) 
imply that $F \; D(A)=D(A)$, so that $F$ takes $D(A)$ into $D(A)$ also in
the case of the anti-involutory Euclidean conjugation that we called
{\it dagger}. Comparison with the proof presented at the beginning of
page 144 in Reed and Simon (1975) shows that this is all what we need so as
to generalize von Neumann's theorem to the Dirac operator acting on
$SU(2)$ spinors in Euclidean four-space (one later uses properties
(7.3.25), (7.3.22) and (7.3.23) as well to complete the proof).

It remains to verify conditions (7.3.23) and (7.3.25). First, note that 
$$ \eqalignno{
\left(Fw,Fw\right)&=
\left(iDw,iDw\right) \cr
&=\int_{M}{\left(i \; \psi_{A}^{\dagger}\right)}^{\dagger} 
i {\Bigr(\psi^{A}\Bigr)}^{\dagger} \sqrt{g} \; d^{4}x
+\int_{M}{\left(i \; {\widetilde \psi}_{A'}^{\dagger}\right)}^{\dagger}
i{\left({\widetilde \psi}^{A'}\right)^{\dagger}}
\sqrt{g} \; d^{4}x\cr
&=\left(w,w\right) ,
&(7.3.33)\cr}
$$
where we have used (7.3.7), (7.3.8) and 
the commutation property of our spinors. Second, one finds
$$             
FAw=
\left(iD\right)\left(iC\right)w
=i{\left[i\left(\nabla_{\; \; B'}^{A} \; {\widetilde \psi}^{B'},
\nabla_{B}^{\; \; A'}\psi^{B}\right)\right]}^{\dagger}    
={\left(\nabla_{\; \; B'}^{A} \; {\widetilde \psi}^{B'},
\nabla_{B}^{\; \; A'}\psi^{B}\right)}^{\dagger} ,
\eqno (7.3.34)
$$
$$             
AFw=
\left(iC\right)\left(iD\right)w
=iCi\left(\psi^{A \dagger},
{\left({\widetilde \psi}^{A'}\right)}^{\dagger}\right)    
=-\left(\nabla_{\; \; B'}^{A}
{\left({\widetilde \psi}^{B'}\right)}^{\dagger},
\nabla_{B}^{\; \; A'}\psi^{B \dagger}\right) ,
\eqno (7.3.35)
$$
which in turn implies that also (7.3.25) holds in view of
what we found just before (7.3.18) and after (7.3.21).
To sum up, we have proved that the operator
$iC$ arising in our boundary-value problem 
is symmetric and has self-adjoint extensions.
Hence the eigenvalues of $iC$ are real, and the eigenvalues $\lambda_{n}$
of $C$ are purely imaginary. 
This is what we mean by self-adjointness of our boundary-value
problem, although it remains to be seen whether there is a
unique self-adjoint extension of our first-order operator.
\vskip 1cm
\centerline {\bf 7.4 Global theory of the Dirac operator}
\vskip 1cm
\noindent
In this chapter and in other sections of our paper
there are many applications of the Dirac operator relying on
two-component spinor formalism. Hence it appears necessary to
describe some general properties of such an operator, frequently
studied in theoretical and mathematical physics.

In Riemannian four-geometries, the 
{\it total} Dirac operator may be defined
as a first-order elliptic operator mapping smooth sections of a
complex vector bundle into smooth sections of the same bundle.
Its action on the sections (i.e. the spinor fields) is given by
composition of Clifford multiplication 
(see appendix A) with covariant differentiation.
To understand these statements, we first summarize the properties of
connections on complex vector bundles, and we then introduce the 
basic properties of spin-structures which enable one to understand how
to construct the vector bundle relevant for the theory of the Dirac
operator.

A complex vector bundle (e.g. Chern (1979)) is a bundle whose fibres are
isomorphic to complex vector spaces. Denoting by $E$ the total
space, by $M$ the base space, one has the projection map
$\pi : E \rightarrow M$ and the sections $s: M \rightarrow E$
such that the composition of $\pi$ with $s$ yields the identity
on the base space: $\pi \cdot s = {\rm id}_{M}$. The sections $s$ 
represent the physical fields in our applications. Moreover, denoting
by $T$ and $T^{*}$ the tangent and cotangent bundles of $M$ respectively,
a connection $\nabla$ is a map from the space $\Gamma(E)$ of smooth
sections of $E$ into the space of smooth sections of the 
tensor-product bundle $T^{*} \otimes E$:
$$
\nabla: \Gamma(E) \rightarrow \Gamma(T^{*} \otimes E),
$$
such that the following properties hold:
$$
\nabla(s_{1}+s_{2})=\nabla s_{1}+\nabla s_{2} ,
\eqno (7.4.1)
$$
$$
\nabla(fs)=df \otimes s + f \nabla s ,
\eqno (7.4.2)
$$
where $s_{1},s_{2},s \in \Gamma(E)$ and $f$ is any $C^{\infty}$
function. The action of the connection $\nabla$ is expressed 
in terms of the connection matrix $\theta$ as
$$
\nabla s = \theta \otimes s .
\eqno (7.4.3)
$$
If one takes a section $s'$ related to $s$ by
$$
s'=h \; s ,
\eqno (7.4.4)
$$
in the light of (7.4.2)--(7.4.4) one finds by comparison that
$$
\theta'h=d \; h +h \; \theta .
\eqno (7.4.5)
$$
Moreover, the transformation law of the curvature matrix
$$
\Omega \equiv d \theta -\theta \wedge \theta ,
\eqno (7.4.6)
$$
is found to be
$$
\Omega' = h \; \Omega \; h^{-1}.
\eqno (7.4.7)
$$

We can now introduce spin-structures and the corresponding
complex vector bundle acted upon by the total Dirac operator.
Let $X$ be a compact oriented differentiable $n$-dimensional
manifold (without boundary) on which a Riemannian metric
is introduced. Let $Q$ be the principal tangential
$SO(n)$-bundle of $X$. A spin-structure of $X$ is a principal
${\rm Spin}(n)$-bundle $P$ over $X$ together with a covering map
${\widetilde \pi}: P \rightarrow Q$ such that the following
commutative structure exists. Given the Cartesian product
$P \times {\rm Spin}(n)$, one first reaches $P$ by the right 
action of ${\rm Spin}(n)$ on $P$, and one eventually arrives at
$Q$ by the projection map $\widetilde \pi$. This is equivalent
to first reaching the Cartesian product 
$Q \times SO(n)$ by the map ${\widetilde \pi} \times \rho$, and
eventually arriving at $Q$ by the right action of $SO(n)$ on $Q$.
Of course, by $\rho$ we denote the double covering 
${\rm Spin}(n) \rightarrow SO(n)$.
In other words, $P$ and $Q$ as above are principal fibre bundles
over $X$, and one has a commutative diagram with 
$P \times {\rm Spin}(n)$ and $P$ on the top,
and $Q \times SO(n)$ and $Q$ on the bottom. The projection map
from $P \times {\rm Spin}(n)$ into $Q \times SO(n)$ is
${\widetilde \pi} \times \rho$, and the projection map from
$P$ into $Q$ is $\widetilde \pi$. Horizontal arrows should be drawn
to denote the right action of ${\rm Spin}(n)$ on $P$
on the top, and of $SO(n)$ on $Q$ on the bottom. 

The group ${\rm Spin}(n)$ has a complex representation space 
$\Sigma$ of dimension $2^{n}$ called the spin-representation. 
If $G \in {\rm Spin}(n), x \in R^{n}, u \in \Sigma$, one has
therefore
$$
G(xu)=GxG^{-1} \cdot G(u)=\rho(G) x \cdot G(u),
\eqno (7.4.8)
$$
where $\rho : {\rm Spin}(n) \rightarrow SO(n)$ is the covering 
map as we said before. If $X$ is even-dimensional, i.e. $n=2l$,
the representation is the direct sum of two irreducible 
representations $\Sigma^{\pm}$ of dimension $2^{n-1}$. 
If $X$ is a ${\rm Spin}(2l)$ manifold with principal bundle $P$,
we can form the associated complex vector bundles
$$
E^{+} \equiv P \times \Sigma^{+},
\eqno (7.4.9a)
$$
$$
E^{-} \equiv P \times \Sigma^{-},
\eqno (7.4.9b)
$$
$$
E \equiv E^{+} \oplus E^{-}.
\eqno (7.4.10)
$$
Sections of these vector bundles are spinor fields on $X$.

The total Dirac operator is a first-order elliptic
differential operator $D: \Gamma(E) \rightarrow \Gamma(E)$
defined as follows. Recall first that the Riemannian metric
defines a natural $SO(2l)$ connection, and this may be used
to give a connection for $P$. One may therefore consider the
connection $\nabla$ at the beginning of this section, i.e. a
linear map from $\Gamma(E)$ into $\Gamma(T^{*} \otimes E)$.
On the other hand, the tangent and cotangent bundles of $X$ are
isomorphic, and one has the map $\Gamma (T \otimes E)
\rightarrow \Gamma(E)$ induced by {\it Clifford multiplication}
(see Ward and Wells (1990) and our appendix A 
on Clifford algebras and Clifford
multiplication). The total Dirac operator $D$ is defined to be 
the {\it composition} of these two maps. Thus, in terms of an
orthonormal base $e_{i}$ of $T$, one has {\it locally}
$$
Ds = \sum_{i} e_{i} (\nabla_{i}s) ,
\eqno (7.4.11)
$$
where $\nabla_{i}s$ is the covariant derivative of 
$s \in \Gamma(E)$ in the direction $e_{i}$, and $e_{i}( \; )$
denotes Clifford multiplication (cf. (7.3.13)). Moreover, the total
Dirac operator $D$ induces two operators
$$
D^{+}: \Gamma(E^{+}) \rightarrow \Gamma(E^{-}) ,
\eqno (7.4.12)
$$
$$
D^{-}: \Gamma(E^{-}) \rightarrow \Gamma(E^{+}),
\eqno (7.4.13)
$$
each of which is elliptic. It should be emphasized that ellipticity
of the total and partial Dirac operators only holds in Riemannian
manifolds, whereas it does not apply to the Lorentzian manifolds
of general relativity and of the original Dirac theory of
spin-${1\over 2}$ particles.
This description of the Dirac operator should be compared with the
mathematical structures presented in section 2.1.
\vskip 100cm
\centerline {\it CHAPTER EIGHT}
\vskip 1cm
\centerline {\bf SPIN-${3\over 2}$ POTENTIALS}
\vskip 1cm
\noindent
Local boundary conditions involving field strengths
and the normal to the boundary, originally studied in 
anti-de Sitter space-time, have been considered in
one-loop quantum cosmology. This chapter derives the conditions 
under which spin-lowering and spin-raising operators preserve these local
boundary conditions on a three-sphere for fields of spin
$0,{1\over 2},1,{3\over 2}$ and $2$. Moreover, the
two-component spinor analysis of the four potentials of the totally
symmetric and independent 
field strengths for spin ${3\over 2}$ is applied to
the case of a three-sphere boundary. It is shown that such
boundary conditions can only be imposed in a flat
Euclidean background,
for which the gauge freedom in the choice 
of the massless potentials remains. 

The second part of the chapter studies the two-spinor form of
the Rarita--Schwinger potentials subject to local boundary 
conditions compatible with local supersymmetry. The massless
Rarita--Schwinger field equations are studied in 
four-real-dimensional Riemannian backgrounds with boundary.
Gauge transformations on the potentials are shown to be
compatible with the field equations provided that the background
is Ricci-flat, in agreement with previous results in the
literature. However, the preservation of boundary conditions 
under such gauge transformations leads to a restriction of the
gauge freedom. The construction by Penrose of a second set of
potentials which supplement the Rarita--Schwinger potentials
is then applied. The equations for these potentials, jointly with
the boundary conditions, imply that the background four-geometry
is further restricted to be totally flat. The analysis of other
gauge transformations confirms that, in the massless case, the
only admissible class of Riemannian backgrounds with boundary is
totally flat.

In the third part of the chapter, the two-component spinor form
of massive spin-${3\over 2}$ potentials in conformally flat
Einstein four-manifolds is studied. Following earlier work
in the literature, a non-vanishing cosmological constant makes
it necessary to introduce a supercovariant derivative operator.
The analysis of supergauge transformations of potentials for
spin ${3\over 2}$ shows that the gauge freedom for massive
spin-${3\over 2}$ potentials is generated by solutions of the
supertwistor equations. The supercovariant form of a partial
connection on a non-linear bundle is then obtained, and the
basic equation obeyed by the second set of potentials
in the massive case is shown to be the
integrability condition on super $\beta$-surfaces of a 
differential operator on a vector bundle of rank three.
\vskip 100cm
\centerline {\bf 8.1 Introduction}
\vskip 1cm
\noindent
Much work in the literature has studied the quantization
of gauge theories and supersymmetric field theories in the
presence of boundaries, with application to one-loop
quantum cosmology (Moss and Poletti 1990, Poletti 1990,
D'Eath and Esposito 1991a,b, Barvinsky {\it et al}. 1992a,b,
Kamenshchik and Mishakov 1992, 1993, 1994, Esposito 1994).
In particular, in the work described in
Esposito (1994), two possible sets of local boundary conditions were 
studied. One of these, first proposed in anti-de Sitter 
space-time (Breitenlohner and Freedman 1982, Hawking 1983), 
involves the normal to the boundary
and Dirichlet or Neumann conditions for spin $0$, the normal
and the field for massless spin-${1\over 2}$ fermions, and the
normal and totally symmetric field strengths for spins
$1,{3\over 2}$ and $2$. Although more attention has been paid
to alternative local boundary conditions motivated by
supersymmetry (Poletti 1990, D'Eath and Esposito 1991a,
Kamenshchik and Mishakov 1993-94, Esposito 1994),
and studied in our sections 8.5-8.9, 
the analysis of the former boundary
conditions remains of mathematical and physical interest by
virtue of its links with twistor theory. The aim of
the first part of this chapter
is to derive the mathematical properties of
the corresponding boundary-value problems,  
since these are relevant for quantum cosmology and twistor theory.

For this purpose, sections 8.2 and 8.3 derive the conditions under which
spin-lowering and spin-raising operators preserve local boundary
conditions involving field strengths and normals. Section 8.4
applies the two-spinor form of Dirac
spin-${3\over 2}$ potentials to Riemannian four-geometries
with a three-sphere boundary. Boundary conditions on
spinor-valued one-forms describing gravitino fields are studied
in sections 8.5-8.9 for the massless Rarita--Schwinger equations.
Massive spin-${3\over 2}$ potentials are instead investigated
in sections 8.10--8.15. Concluding
remarks and open problems are presented in section 8.16.
\vskip 10cm
\centerline {\bf 8.2 Spin-lowering operators in cosmology}
\vskip 1cm
\noindent
In section 5.7 of Esposito (1994), a flat Euclidean background bounded 
by a three-sphere was studied. On the bounding $S^3$, the following
boundary conditions for a spin-$s$ field were required:
$$
2^{s} \; {_{e}}n^{AA'}... \; {_{e}}n^{LL'}
\; \phi_{A...L}= \epsilon \; {\widetilde \phi}^{A'...L'}.
\eqno (8.2.1)
$$
With our notation, ${_{e}}n^{AA'}$ is the Euclidean normal
to $S^3$ (D'Eath and Halliwell 1987, Esposito 1994), 
$\phi_{A...L}=\phi_{(A...L)}$
and ${\widetilde \phi}_{A'...L'}
={\widetilde \phi}_{(A'...L')}$ are totally symmetric
and independent (i.e. not related by any conjugation)
field strengths, which reduce to the massless 
spin-${1\over 2}$ field for $s={1\over 2}$. Moreover,
the complex scalar field $\phi$ is such that its real
part obeys Dirichlet conditions on $S^3$ and its imaginary
part obeys Neumann conditions on $S^3$, or the other way
around, according to the value of the parameter
$\epsilon \equiv \pm 1$ occurring in (8.2.1). 

In flat Euclidean four-space, we write the solutions of the
twistor equations 
$$
D_{A'}^{\; \; \;  (A} \; \omega^{B)}=0,
\eqno (8.2.2)
$$
$$
D_{A}^{\; \; (A'}\; {\widetilde \omega}^{B')}=0,
\eqno (8.2.3)
$$
as (cf. section 4.1)
$$
\omega^{A}=(\omega^{o})^{A}-i\Bigr({_{e}}x^{AA'}\Bigr)
\pi_{A'}^{o},
\eqno (8.2.4)
$$
$$
{\widetilde \omega}^{A'}=({\widetilde \omega}^{o})^{A'}
-i\Bigr({_{e}}x^{AA'}\Bigr){\widetilde \pi}_{A}^{o}.
\eqno (8.2.5)
$$
Note that, since unprimed and primed spin-spaces are no longer
anti-isomorphic in the case of 
Riemannian four-metrics, Eq. (8.2.3) is not
obtained by complex conjugation of Eq. (8.2.2). Hence the spinor
field ${\widetilde \omega}^{B'}$ is independent of 
$\omega^{B}$. This leads to distinct solutions 
(8.2.4) and (8.2.5), where the
spinor fields $\omega_{A}^{o},{\widetilde \omega}_{A'}^{o},
{\widetilde \pi}_{A}^{o},\pi_{A'}^{o}$
are covariantly constant with respect to the
flat connection $D$, whose corresponding spinor covariant 
derivative is here denoted by $D_{AB'}$. 
The following theorem can be now proved:
\vskip 0.3cm
\noindent
{\bf Theorem 8.2.1} Let $\omega^{D}$ be a solution of the twistor
equation (8.2.2) in flat Euclidean space with a three-sphere boundary,
and let ${\widetilde \omega}^{D'}$ be the solution of the
independent equation (8.2.3) in the same four-geometry with boundary.
Then a form exists of the spin-lowering operator which preserves 
the local boundary conditions on $S^3$:
$$
4 \; {_{e}n^{AA'}} \; {_{e}n^{BB'}} \; {_{e}n^{CC'}}
\; {_{e}n^{DD'}} \; \phi_{ABCD}
= \epsilon \; {\widetilde \phi}^{A'B'C'D'},
\eqno (8.2.6)
$$
$$
2^{3\over 2} \; {_{e}n^{AA'}} \; {_{e}n^{BB'}} \;
{_{e}n^{CC'}} \; \phi_{ABC} = \epsilon \; {\widetilde \phi}^{A'B'C'}.
\eqno (8.2.7)
$$
Of course, the independent field strengths appearing in
(8.2.6) and (8.2.7) are assumed to satisfy the corresponding
massless free-field equations.
\vskip 0.3cm
\noindent
{\bf Proof.} Multiplying both sides of (8.2.6) by 
${_{e}n_{FD'}}$ one gets
$$
-2 \; {_{e}n^{AA'}} \; {_{e}n^{BB'}} \;
{_{e}n^{CC'}} \; \phi_{ABCF}= \epsilon \;  
{\widetilde \phi}^{A'B'C'D'} \; {_{e}n_{FD'}}.
\eqno (8.2.8)
$$
Taking into account the total symmetry of the field strengths,
putting $F=D$ and multiplying both sides of (8.2.8) by 
$\sqrt{2} \; \omega^{D}$ one eventually gets
$$
-2^{3\over 2} \; {_{e}n^{AA'}} \; {_{e}n^{BB'}} \;
{_{e}n^{CC'}} \; \phi_{ABCD} \; \omega^{D}
=\epsilon \; \sqrt{2} \; {\widetilde \phi}^{A'B'C'D'}
\; {_{e}n_{DD'}} \; \omega^{D},
\eqno (8.2.9)
$$
$$
2^{3\over 2} \; {_{e}n^{AA'}} \; {_{e}n^{BB'}} \;
{_{e}n^{CC'}} \; \phi_{ABCD} \; \omega^{D}
= \epsilon \; {\widetilde \phi}^{A'B'C'D'} \;
{\widetilde \omega}_{D'},
\eqno (8.2.10)
$$
where (8.2.10) is obtained by inserting into (8.2.7) the definition
of the spin-lowering operator. The comparison of (8.2.9) and (8.2.10)
yields the preservation condition
$$
\sqrt{2} \; {_{e}n_{DA'}} \; \omega^{D}
=-{\widetilde \omega}_{A'}.
\eqno (8.2.11)
$$
In the light of (8.2.4) and (8.2.5), Eq. (8.2.11) is found to imply
$$
\sqrt{2} \; {_{e}n_{DA'}} \; (\omega^{o})^{D}
-i \sqrt{2} \; {_{e}n_{DA'}} \; {_{e}x^{DD'}} \; \pi_{D'}^{o} 
=-{\widetilde \omega}_{A'}^{o}
-i \; {_{e}x_{DA'}} \; 
({\widetilde \pi}^{o})^{D}.
\eqno (8.2.12)
$$
Requiring that (8.2.12) should be identically satisfied, and using
the identity ${_{e}n^{AA'}}={1\over r} \; {_{e}x^{AA'}}$ on a
three-sphere of radius $r$, one finds
$$
{\widetilde \omega}_{A'}^{o}=i \sqrt{2} \; r \; 
{_{e}n_{DA'}} \; {_{e}n^{DD'}} \; \pi_{D'}^{o}
=-{ir \over \sqrt{2}} \; \pi_{A'}^{o},
\eqno (8.2.13)
$$
$$
-\sqrt{2} \; {_{e}n_{DA'}} \; (\omega^{o})^{D}
=ir \; {_{e}n_{DA'}} \; ({\widetilde \pi}^{o})^{D}.
\eqno (8.2.14)
$$
Multiplying both sides of (8.2.14) by ${_{e}n^{BA'}}$, and then
acting with $\varepsilon_{BA}$ on both sides of the resulting
relation, one gets
$$
\omega_{A}^{o}=-{ir \over \sqrt{2}} \; 
{\widetilde \pi}_{A}^{o}.
\eqno (8.2.15)
$$
The equations (8.2.11), (8.2.13) and (8.2.15) completely solve the
problem of finding a spin-lowering operator which preserves
the boundary conditions (8.2.6) and (8.2.7) on $S^{3}$. Q.E.D.

If one requires local boundary conditions on $S^{3}$ involving
field strengths and normals also for lower spins (i.e. spin
${3\over 2}$ vs spin $1$, spin $1$ vs spin ${1\over 2}$, spin
${1\over 2}$ vs spin $0$), then by using the same technique of the
theorem just proved, one finds that the preservation condition obeyed
by the spin-lowering operator is still expressed 
by (8.2.13) and (8.2.15).
\vskip 1cm
\centerline {\bf 8.3 Spin-raising operators in cosmology}
\vskip 1cm
\noindent
To derive the corresponding preservation condition for
spin-raising operators, we begin by studying the
relation between spin-${1\over 2}$ and spin-$1$ fields.
In this case, the independent spin-$1$ field strengths
take the form (Penrose and Rindler 1986)
$$
\psi_{AB}=i \; {\widetilde \omega}^{L'}  
\Bigr(D_{BL'} \; \chi_{A}\Bigr)
-2\chi_{(A} \; {\widetilde \pi}_{B)}^{o},
\eqno (8.3.1)
$$
$$
{\widetilde \psi}_{A'B'}=-i \; \omega^{L}
\Bigr(D_{LB'} \; {\widetilde \chi}_{A'}\Bigr)
-2{\widetilde \chi}_{(A'} \; \pi_{B')}^{o},
\eqno (8.3.2)
$$
where the independent spinor fields $\Bigr(\chi_{A},
{\widetilde \chi}_{A'}\Bigr)$ represent a massless
spin-${1\over 2}$ field obeying the Weyl equations
on flat Euclidean four-space and subject to the boundary
conditions
$$
\sqrt{2} \; {_{e}}n^{AA'} \; \chi_{A}=
\epsilon \; {\widetilde \chi}^{A'}
\eqno (8.3.3)
$$
on a three-sphere of radius $r$. Thus, by requiring that (8.3.1)
and (8.3.2) should obey (8.2.1) on $S^3$ with $s=1$, and bearing 
in mind (8.3.3), one finds
$$ \eqalignno{
2\epsilon \biggr[\sqrt{2} \; {\widetilde \pi}_{A}^{o} \;
{\widetilde \chi}^{(A'} \; {_{e}}n^{AB')}
-{\widetilde \chi}^{(A'} \; \pi^{o \; B')}\biggr] 
&=i \biggr[2 \; {_{e}}n^{AA'} \; {_{e}}n^{BB'} \;
{\widetilde \omega}^{L'} \; D_{L'(B} \; \chi_{A)}\cr
&+\epsilon \; \omega^{L} \; D_{L}^{\; \; (B'} \;
{\widetilde \chi}^{A')}\biggr]
&(8.3.4)\cr}
$$
on the bounding $S^3$. It is now clear how to carry out the
calculation for higher spins. Denoting by $s$ the spin
obtained by spin-raising, and defining $n \equiv 2s$,
one finds
$$ \eqalignno{
\; & n \epsilon \biggr[\sqrt{2} \; {\widetilde \pi}_{A}^{o} \;
{_{e}}n^{A(A'} \; {\widetilde \chi}^{B'...K')}
-{\widetilde \chi}^{(A'...D'} \; \pi^{o \; K')}\biggr]\cr
&=i \biggr[2^{n\over 2} \; {_{e}}n^{AA'} ...
{_{e}}n^{KK'} \; {\widetilde \omega}^{L'} \;
D_{L'(K} \; \chi_{A...D)} 
+\epsilon \; \omega^{L} \;
D_{L}^{\; \; (K'} \; {\widetilde \chi}^{A'...D')}
\biggr]
&(8.3.5)\cr}
$$
on the three-sphere boundary. In the comparison spin-$0$ vs 
spin-${1\over 2}$, the preservation condition is not 
obviously obtained from (8.3.5). The desired result is here found
by applying the spin-raising operators to the
independent scalar fields $\phi$ and $\widetilde \phi$ (see below)
and bearing in mind (8.2.4), (8.2.5) and the boundary conditions
$$
\phi = \epsilon \; {\widetilde \phi}
\; {\rm on} \; S^{3} ,
\eqno (8.3.6)
$$
$$
{_{e}}n^{AA'}D_{AA'}\phi=-\epsilon \; {_{e}}n^{BB'}D_{BB'}
{\widetilde \phi}
\; {\rm on} \; S^{3} .
\eqno (8.3.7)
$$
This leads to the following condition on $S^3$
(cf. Eq. (5.7.23) of Esposito (1994)):
$$ \eqalignno{
0&=i\phi \biggr[{{\widetilde \pi}_{A}^{o}\over \sqrt{2}}
-\pi_{A'}^{o} \; {_{e}}n_{A}^{\; \; A'}\biggr]
-\biggr[{{\widetilde \omega}^{K'}\over \sqrt{2}}
\Bigr(D_{AK'}\phi\Bigr)
-{\omega_{A}\over 2} \; {_{e}}n_{C}^{\; \; K'}
\Bigr(D_{\; \; K'}^{C} \phi \Bigr)\biggr]\cr
&+\epsilon \; {_{e}}n_{(A}^{\; \; \; A'} \;
\omega^{B} \; D_{B)A'} \; {\widetilde \phi}.
&(8.3.8)\cr}
$$
Note that, while the preservation conditions 
(8.2.13) and (8.2.15) for
spin-lowering operators are purely algebraic, the
preservation conditions (8.3.5) and (8.3.8) for spin-raising
operators are more complicated, since they also involve 
the value at the boundary of four-dimensional covariant derivatives 
of spinor fields or scalar fields. 
Two independent scalar fields have been
introduced, since the spinor fields obtained by applying
the spin-raising operators to $\phi$ and
${\widetilde \phi}$ respectively are independent as well
in our case.
\vskip 1cm
\centerline {\bf 8.4 Dirac's spin-${3\over 2}$ potentials in cosmology}
\vskip 1cm
\noindent
In this section we focus on the totally
symmetric field strengths $\phi_{ABC}$ and
${\widetilde \phi}_{A'B'C'}$ for spin-${3\over 2}$ fields,
and we express them in terms of their potentials, rather
than using spin-raising (or spin-lowering) operators. The
corresponding theory in Minkowski space-time (and curved
space-time) is described in Penrose (1990), Penrose (1991a--c), 
and adapted here to
the case of flat Euclidean four-space with flat connection $D$.
It turns out that ${\widetilde \phi}_{A'B'C'}$ can then be
obtained locally from two potentials defined as follows. The first
potential satisfies the properties (section 5.3, Penrose 1990,
Penrose 1991a--c, Esposito and Pollifrone 1994)
$$
\gamma_{A'B'}^{C}=\gamma_{(A'B')}^{C},
\eqno (8.4.1)
$$
$$
D^{AA'} \; \gamma_{A'B'}^{C}=0 ,
\eqno (8.4.2)
$$
$$
{\widetilde \phi}_{A'B'C'}=D_{CC'} \; \gamma_{A'B'}^{C} ,
\eqno (8.4.3)
$$
with the gauge freedom of replacing it by
$$
{\widehat \gamma}_{A'B'}^{C} \equiv \gamma_{A'B'}^{C}
+D_{\; \; B'}^{C} \; {\widetilde \nu}_{A'} ,
\eqno (8.4.4)
$$
where ${\widetilde \nu}_{A'}$ satisfies the positive-helicity Weyl 
equation
$$
D^{AA'} \; {\widetilde \nu}_{A'}=0 .
\eqno (8.4.5)
$$
The second potential is defined by the conditions 
$$
\rho_{A'}^{BC}=\rho_{A'}^{(BC)} ,
\eqno (8.4.6)
$$
$$
D^{AA'} \; \rho_{A'}^{BC}=0 ,
\eqno (8.4.7)
$$
$$
\gamma_{A'B'}^{C}=D_{BB'} \; \rho_{A'}^{BC} ,
\eqno (8.4.8)
$$
with the gauge freedom of being replaced by
$$
{\widehat \rho}_{A'}^{BC} \equiv \rho_{A'}^{BC}
+D_{\; \; A'}^{C} \; \chi^{B} ,
\eqno (8.4.9)
$$
where $\chi^{B}$ satisfies the negative-helicity
Weyl equation
$$
D_{BB'} \; \chi^{B}=0 .
\eqno (8.4.10)
$$
Moreover, in flat Euclidean four-space the field strength
$\phi_{ABC}$ is expressed locally in terms of the potential
$\Gamma_{AB}^{C'}=\Gamma_{(AB)}^{C'}$, independent
of $\gamma_{A'B'}^{C}$, as
$$
\phi_{ABC}=D_{CC'} \; \Gamma_{AB}^{C'} ,
\eqno (8.4.11)
$$
with gauge freedom
$$
{\widehat \Gamma}_{AB}^{C'} \equiv \Gamma_{AB}^{C'}
+D_{\; \; B}^{C'} \; \nu_{A} .
\eqno (8.4.12)
$$
Thus, if we insert (8.4.3) and (8.4.11) into the boundary 
conditions (8.2.1) with $s={3\over 2}$, and require that
also the gauge-equivalent potentials (8.4.4) and (8.4.12)
should obey such boundary conditions on $S^3$, we
find that
$$
2^{3\over 2} \; {_{e}}n_{\; \; A'}^{A}
\; {_{e}}n_{\; \; B'}^{B}
\; {_{e}}n_{\; \; C'}^{C}
\; D_{CL'} \; D_{\; \; B}^{L'} 
\; \nu_{A}=\epsilon \;
D_{LC'} \; D_{\; \; B'}^{L}
\; {\widetilde \nu}_{A'}
\eqno (8.4.13)
$$
on the three-sphere. Note that, from now on (as already done in
(8.3.5) and (8.3.8)), covariant derivatives appearing in boundary
conditions are first taken on the background and then 
evaluated on $S^3$.
In the case of our flat background, (8.4.13) is identically
satisfied since $D_{CL'} \; D_{\; \; \; B}^{L'} \; \nu_{A}$
and $D_{LC'} \; D_{\; \; B'}^{L} \; {\widetilde \nu}_{A'}$
vanish by virtue of spinor Ricci identities. In
a curved background, however, denoting by $\nabla$ 
its curved connection, and defining
$\cstok{\ }_{AB} \equiv \nabla_{M'(A}
\nabla_{\; \; \; B)}^{M'} \; , \; \cstok{\ }_{A'B'} \equiv 
\nabla_{X(A'} \; \nabla_{\; \; B')}^{X}$,
since the spinor Ricci identities we need are 
(Ward and Wells 1990)
$$
\cstok{\ }_{AB} \; \nu_{C}=\psi_{ABDC} \; \nu^{D}
-2\Lambda \; \nu_{(A} \; \varepsilon_{B)C} ,
\eqno (8.4.14)
$$
$$
\cstok{\ }_{A'B'} \; {\widetilde \nu}_{C'}
={\widetilde \psi}_{A'B'D'C'} \; 
{\widetilde \nu}^{D'} -2 {\widetilde \Lambda}
\; {\widetilde \nu}_{(A'} \; \varepsilon_{B')C'} ,
\eqno (8.4.15)
$$
one finds that the corresponding boundary conditions
$$
2^{3\over 2} \; {_{e}}n_{\; \; A'}^{A}
\; {_{e}}n_{\; \; B'}^{B}
\; {_{e}}n_{\; \; C'}^{C}
\; \nabla_{CL'} \; \nabla_{\; \; \; B}^{L'}
\; \nu_{A}=\epsilon \; \nabla_{LC'}
\; \nabla_{\; \; B'}^{L} 
\; {\widetilde \nu}_{A'}
\eqno (8.4.16)
$$
are identically satisfied if and only if one of the 
following conditions holds: (i) $\nu_{A}=
{\widetilde \nu}_{A'}=0$; (ii) the Weyl spinors
$\psi_{ABCD},{\widetilde \psi}_{A'B'C'D'}$ and the
scalars $\Lambda,{\widetilde \Lambda}$ vanish everywhere.
However, since in a curved space-time 
with vanishing $\Lambda,{\widetilde \Lambda}$, the potentials
with the gauge freedoms (8.4.4) and (8.4.12) only exist provided
that $D$ is replaced by $\nabla$ and the trace-free part
$\Phi_{ab}$ of the Ricci tensor vanishes as well (Buchdahl 1958),
the background four-geometry is actually flat
Euclidean four-space. We require that (8.4.16) should
be identically satisfied to avoid, after a gauge
transformation, obtaining more boundary conditions than 
the ones originally imposed. The curvature of the background
should not, itself, be subject to a boundary condition.

The same result can be derived by using
the potential $\rho_{A'}^{BC}$ and its independent
counterpart $\Lambda_{A}^{B'C'}$. This spinor field
yields the $\Gamma_{AB}^{C'}$ potential by means of
$$
\Gamma_{AB}^{C'}=D_{BB'} \; \Lambda_{A}^{B'C'},
\eqno (8.4.17)
$$
and has the gauge freedom
$$
{\widehat \Lambda}_{A}^{B'C'} \equiv \Lambda_{A}^{B'C'}
+D_{\; \; A}^{C'} \; {\widetilde \chi}^{B'} ,
\eqno (8.4.18)
$$
where ${\widetilde \chi}^{B'}$ satisfies the positive-helicity
Weyl equation
$$
D_{BF'} \; {\widetilde \chi}^{F'}=0 .
\eqno (8.4.19)
$$
Thus, if also the gauge-equivalent potentials (8.4.9) and (8.4.18)
have to satisfy the boundary conditions (8.2.1) on $S^3$, one
finds
$$
2^{3\over 2} \; {_{e}}n_{\; \; A'}^{A}
\; {_{e}}n_{\; \; B'}^{B}
\; {_{e}}n_{\; \; C'}^{C}
\; D_{CL'} \; D_{BF'} \;
D_{\; \; A}^{L'} \; 
{\widetilde \chi}^{F'}
=\epsilon \; D_{LC'} \; D_{MB'} \;
D_{\; \; A'}^{L} \; \chi^{M}
\eqno (8.4.20)
$$
on the three-sphere. In our flat background, covariant derivatives
commute, hence (8.4.20) is identically satisfied by virtue of (8.4.10)
and (8.4.19). However, in the curved case the boundary conditions
(8.4.20) are replaced by
$$
2^{3\over 2} \; {_{e}}n_{\; \; A'}^{A}
\; {_{e}}n_{\; \; B'}^{B}
\; {_{e}}n_{\; \; C'}^{C}
\; \nabla_{CL'} \; \nabla_{BF'}
\; \nabla_{\; \; A}^{L'}
\; {\widetilde \chi}^{F'}
=\epsilon \; \nabla_{LC'} \;
\nabla_{MB'} \; \nabla_{\; \; A'}^{L}
\; \chi^{M}
\eqno (8.4.21)
$$
on $S^3$, if the {\it local} expressions of $\phi_{ABC}$ and
${\widetilde \phi}_{A'B'C'}$ in terms of potentials still
hold (Penrose 1990, Penrose 1991a--c). 
By virtue of (8.4.14) and (8.4.15), where $\nu_{C}$ is
replaced by $\chi_{C}$ and ${\widetilde \nu}_{C'}$ is
replaced by ${\widetilde \chi}_{C'}$, this means that
the Weyl spinors $\psi_{ABCD},{\widetilde \psi}_{A'B'C'D'}$
and the scalars $\Lambda,{\widetilde \Lambda}$ should
vanish, since one should find
$$
\nabla^{AA'} \; {\widehat \rho}_{A'}^{BC}=0, \;
\nabla^{AA'} \; {\widehat \Lambda}_{A}^{B'C'}=0 .
\eqno (8.4.22)
$$
If we assume that
$\nabla_{BF'} \; {\widetilde \chi}^{F'}=0$ and
$\nabla_{MB'} \; \chi^{M}=0$, we have to show that (8.4.21)
differs from (8.4.20) by terms involving a part of the curvature
that is vanishing everywhere. 
This is proved by using the basic rules
of two-spinor calculus and spinor Ricci identities.
Thus, bearing in mind that 
$$
\cstok{\ }^{AB} \; {\widetilde \chi}_{B'}
=\Phi_{\; \; \; \; L'B'}^{AB} \;
{\widetilde \chi}^{L'},
\eqno (8.4.23)
$$
$$
\cstok{\ }^{A'B'} \; \chi_{B}
={\widetilde \Phi}_{\; \; \; \; \; \; LB}^{A'B'}
\; \chi^{L},
\eqno (8.4.24)
$$
one finds (see (8.4.29))
$$ \eqalignno{
\nabla^{BB'} \; \nabla^{CA'} \; \chi_{B}&=
\nabla^{(BB'} \; \nabla^{C)A'} \; \chi_{B}
+\nabla^{[BB'} \; \nabla^{C]A'} \; \chi_{B} \cr
&=-{1\over 2} \nabla_{B}^{\; \; B'} \;
\nabla^{CA'} \; \chi^{B}
+{1\over 2} {\widetilde \Phi}^{A'B'LC} \; \chi_{L}.
&(8.4.25)\cr}
$$
Thus, if ${\widetilde \Phi}^{A'B'LC}$ vanishes, also the left-hand side
of (8.4.25) has to vanish since this leads to the equation
$
\nabla^{BB'} \; \nabla^{CA'} \; \chi_{B}
={1\over 2}
\nabla^{BB'} \; \nabla^{CA'} \; \chi_{B}
$. Hence (8.4.25) is identically satisfied. Similarly, the
left-hand side of (8.4.21) can be made to vanish identically 
if the additional condition $\Phi^{CDF'M'}=0$ holds. 
The conditions
$$
\Phi^{CDF'M'}=0 , \;
{\widetilde \Phi}^{A'B'CL}=0,
\eqno (8.4.26)
$$
when combined with the conditions 
$$
\psi_{ABCD}={\widetilde \psi}_{A'B'C'D'}=0, \;
\Lambda={\widetilde \Lambda}=0,
\eqno (8.4.27)
$$
arising from (8.4.22) for the local existence
of $\rho_{A'}^{BC}$ and $\Lambda_{A}^{B'C'}$ potentials,
imply that the whole Riemann curvature should vanish.
Hence, in the boundary-value problems we are interested in, 
the only admissible background four-geometry (of the Einstein
type (Besse 1987)) is flat Euclidean four-space.

Note that (8.4.25) is {\it not} an identity, since we have
already set $\Lambda$ to zero by requiring that
$$
\nabla^{AA'} {\widehat \rho}_{A'}^{BC}
=-\psi_{\; \; \; \; \; \; \; F}^{ABC} \; \chi^{F}
+\Lambda \Bigr(\chi^{A} \; \varepsilon^{CB}
+3\chi^{B} \; \varepsilon^{AC}
+\chi^{C} \; \varepsilon^{AB}\Bigr)
\eqno (8.4.28)
$$
should vanish. In general, for any solution $\chi_{B}$ of the 
Weyl equation, by virtue of the corresponding identity
$\cstok{\ }\chi_{B}=-6\Lambda \; \chi_{B}$ (see problem 2.7),
one finds
$$
\nabla^{BB'}\nabla^{CA'}\chi_{B}
={1\over 2} \nabla^{BB'}\nabla^{CA'}\chi_{B}
+{1\over 2}{\widetilde \Phi}^{A'B'LC} \; \chi_{L}
+{3\over 2}\Lambda \varepsilon^{B'A'} \; \chi^{C}.
\eqno (8.4.29)
$$
As the reader may check, the action of the 
$\cstok{\ } \equiv \nabla_{CA'}\nabla^{CA'}$ operator 
on $\chi_{B}$ is obtained by acting with the spinor
covariant derivative $\nabla_{AA'}$ on the Weyl equation
$\nabla_{B}^{\; \; A'} \chi^{B}=0$.
\vskip 1cm
\centerline {\bf 8.5 Boundary conditions in supergravity}
\vskip 1cm
\noindent
The boundary conditions studied in the previous 
sections are not appropriate if one studies 
supergravity multiplets and supersymmetry transformations at the
boundary (Esposito 1994). By contrast, it turns out
one has to impose another set of locally supersymmetric boundary
conditions, first proposed in Luckock and Moss (1989). 
These are in general mixed, and 
involve in particular Dirichlet conditions for the transverse modes of
the vector potential of electromagnetism, a mixture of Dirichlet and
Neumann conditions for scalar fields, and local boundary conditions for
the spin-${1\over 2}$ field and the spin-${3\over 2}$ potential. Using
two-component spinor notation for supergravity (D'Eath 1984), the
spin-${3\over 2}$ boundary conditions take the form
$$
\sqrt{2} \; {_{e}n_{A}^{\; \; A'}} \;
\psi_{\; \; i}^{A}=\epsilon \; 
{\widetilde \psi}_{\; \; i}^{A'} 
\; {\rm on} \; S^{3}.
\eqno (8.5.1)
$$
With our notation, $\epsilon \equiv \pm 1$,
${_{e}n_{A}^{\; \; A'}}$ is the Euclidean normal to $S^{3}$,
and $\Bigr(\psi_{\; \; i}^{A},{\widetilde \psi}_{\; \; i}^{A'}\Bigr)$
are the {\it independent} (i.e. not related by any conjugation) 
spatial components (hence $i=1,2,3$) of the spinor-valued 
one-forms appearing in the action functional of Euclidean supergravity
(D'Eath 1984, Esposito 1994).

It appears necessary to understand whether
the analysis in the previous section and in Esposito and Pollifrone
(1994) can be used to derive 
restrictions on the classical boundary-value problem corresponding
to (8.5.1). For this purpose, we study 
a Riemannian background four-geometry,
and we use the decompositions of the spinor-valued one-forms in such
a background, i.e. 
$$
\psi_{\; \; i}^{A}= {h^{- {1 \over 4}}}
\biggr[\chi^{(AB)B'}
+\varepsilon^{AB} \;
{\widetilde \phi}^{B'}\biggr]e_{BB'i} ,
\eqno (8.5.2)
$$
$$
{\widetilde \psi}_{\; \; i}^{A'}
={h^{- {1 \over 4}}}
\biggr[{\widetilde \chi}^{(A'B')B}+\varepsilon^{A'B'}\phi^{B}
\biggr]e_{BB'i} ,
\eqno (8.5.3)
$$
where $h$ is the determinant of the three-metric on $S^{3}$, and 
$e_{BB'i}$ is the spatial component of the tetrad, written
in two-spinor language. If we now reduce
the classical theory of simple supergravity to its physical
degrees of freedom by imposing the gauge conditions (Esposito 1994)
$$
e_{AA'}^{\; \; \; \; \; \; i} \; \psi_{\; \; i}^{A}=0,
\eqno (8.5.4)
$$
$$
e_{AA'}^{\; \; \; \; \; \; i} \; {\widetilde \psi}_{\; \; i}^{A'}=0,
\eqno (8.5.5)
$$
we find that the expansions of (8.5.2) and (8.5.3) 
on a family of three-spheres
centred on the origin take the forms (Esposito 1994)
$$
\psi_{\; \; i}^{A}
={h^{-{1\over 4}} \over 2\pi}
\sum_{n=0}^{\infty}\sum_{p,q=1}^{(n+1)(n+4)}
\alpha_{n}^{pq}
\biggr[m_{np}^{(\beta)}(\tau) \; \beta^{nqABB'}
+{\widetilde r}_{np}^{(\mu)}(\tau) \;
{\overline \mu}^{nqABB'}\biggr]e_{BB'i},
\eqno (8.5.6)
$$
$$
{\widetilde \psi}_{\; \; i}^{A'}
={h^{-{1\over 4}} \over 2\pi}
\sum_{n=0}^{\infty}\sum_{p,q=1}^{(n+1)(n+4)}
\alpha_{n}^{pq}
\biggr[{\widetilde m}_{np}^{(\beta)}(\tau) 
\; {\overline \beta}^{nqA'B'B}
+ r_{np}^{(\mu)}(\tau) \;
\mu^{nqA'B'B}\biggr]e_{BB'i}.
\eqno (8.5.7)
$$
With our notation, 
$\alpha_{n}^{pq}$ are block-diagonal matrices with blocks
$\pmatrix {1&1 \cr 1&-1 \cr}$, and the $\beta$- and $\mu$-harmonics
on $S^{3}$ are given by (Esposito 1994)
$$
\beta_{\; \; \; ACC'}^{nq}=\rho_{\; \; \; (ACD)}^{nq}
\; n_{\; \; C'}^{D},
\eqno (8.5.8)
$$
$$
\mu_{\; \; \; A'B'B}^{nq}=\sigma_{\; \; \; (A'B'C')}^{nq}
\; n_{B}^{\; \; C'}.
\eqno (8.5.9)
$$
In the light of (8.5.6)--(8.5.9), one gets the following 
physical-degrees-of-freedom form of the spinor-valued one-forms
of supergravity  (cf. D'Eath (1984)):
$$
\psi_{\; \; i}^{A}={h^{-{1\over 4}}}
\; \phi^{(ABC)} \; {_{e}n_{C}^{\; \; B'}} \; e_{BB'i},
\eqno (8.5.10)
$$
$$
{\widetilde \psi}_{\; \; i}^{A'}={h^{-{1\over 4}}} \;
{\widetilde \phi}^{(A'B'C')} \; {_{e}n_{\; \; C'}^{B}}
\; e_{BB'i},
\eqno (8.5.11)
$$ 
where $\phi^{(ABC)}$ and ${\widetilde \phi}^{(A'B'C')}$ 
are totally symmetric and independent spinor fields.

Within this framework, a {\it sufficient} 
condition for the validity
of the boundary conditions (8.5.1) on $S^{3}$ is
$$
\sqrt{2} \; {_{e}n_{A}^{\; \; A'}} \; {_{e}n_{C}^{\; \; B'}}
\; \phi^{(ABC)}= \epsilon \; {_{e}n_{\; \; C'}^{B}}
\; {\widetilde \phi}^{(A'B'C')}.
\eqno (8.5.12) 
$$
However, our construction does not prove that such 
$\phi^{(ABC)}$ and ${\widetilde \phi}^{(A'B'C')}$ 
can be expressed in terms
of four potentials as in Esposito and Pollifrone (1994).

It should be emphasized that our 
analysis, although motivated
by quantum cosmology, is entirely classical. Hence we have not
discussed ghost modes. 
The theory has been reduced to its physical degrees of freedom
to make a comparison with the results 
in Esposito and Pollifrone (1994), but totally
symmetric field strengths do not enable one to recover the full
physical content of simple supergravity. Hence the four-sphere
background studied in Poletti (1990) 
is not ruled out by the work in this section, and a more
careful analysis is in order (see sections 8.10--8.15). 
\vskip 1cm
\centerline {\bf 8.6 Rarita--Schwinger potentials}
\vskip 1cm
\noindent
We are here interested
in the independent spatial components $\Bigr(\psi_{\; \; i}^{A},
{\widetilde \psi}_{\; \; i}^{A'}\Bigr)$ of the gravitino field
in Riemannian backgrounds. In terms of the spatial components
$e_{AB'i}$ of the tetrad, and of spinor fields, they can be
expressed as (Aichelburg and Urbantke 1981, D'Eath 1984,
Penrose 1991)
$$
\psi_{A \; i}= \Gamma_{\; \; AB}^{C'}
\; e_{\; \; C'i}^{B} ,
\eqno (8.6.1)
$$
$$
{\widetilde \psi}_{A' \; i}= 
\gamma_{\; \; A'B'}^{C} \; 
e_{C \; \; \; i}^{\; \; B'} .
\eqno (8.6.2)
$$
A first important difference with respect to the Dirac form of
the potentials studied in Esposito and Pollifrone
(1994) is that the spinor fields 
$\Gamma_{\; \; AB}^{C'}$ and 
$\gamma_{\; \; A'B'}^{C}$ are no longer symmetric in
the second and third index. From now on, they will be 
referred to as spin-${3\over 2}$ potentials.
They obey the differential equations 
(see appendix B and cf. Rarita and Schwinger (1941),
Aichelburg and Urbantke (1981), Penrose (1991))
$$
\varepsilon^{B'C'} \; \nabla_{A(A'} \; \gamma_{\; \; B')C'}^{A}
=-3 \Lambda  \; {\widetilde \alpha}_{A'} ,
\eqno (8.6.3)
$$
$$
\nabla^{B'(B} \; \gamma_{\; \; \; B'C'}^{A)}
=\Phi_{\; \; \; \; \; \; \; \; C'}^{ABL'}
\; {\widetilde \alpha}_{L'},
\eqno (8.6.4)
$$
$$
\varepsilon^{BC} \; \nabla_{A'(A} \; \Gamma_{\; \; B)C}^{A'}
=-3\Lambda \; \alpha_{A} ,
\eqno (8.6.5)
$$
$$
\nabla^{B(B'} \; \Gamma_{\; \; \; BC}^{A')}
={\widetilde \Phi}_{\; \; \; \; \; \; \; \; \; C}^{A'B'L}
\; \alpha_{L},
\eqno (8.6.6)
$$
where $\nabla_{AB'}$ is the spinor covariant derivative corresponding
to the curved connection $\nabla$ of the background,
the spinors $\Phi_{\; \; \; \; \; C'D'}^{AB}$ and
${\widetilde \Phi}_{\; \; \; \; \; \; \; CD}^{A'B'}$ 
correspond to the trace-free part of the Ricci tensor, the
scalar $\Lambda$ corresponds to the
scalar curvature $R=24\Lambda$ of the background,
and $\alpha_{A},{\widetilde \alpha}_{A'}$ are a pair of
independent spinor fields, corresponding to the Majorana
field in the Lorentzian regime.
Moreover, the potentials are subject to the gauge transformations
(cf. section 8.9)
$$
{\widehat \gamma}_{\; \; B'C'}^{A}
\equiv \gamma_{\; \; B'C'}^{A}
+\nabla_{\; \; B'}^{A} \; \lambda_{C'},
\eqno (8.6.7)
$$
$$
{\widehat \Gamma}_{\; \; BC}^{A'}
\equiv \Gamma_{\; \; BC}^{A'}
+\nabla_{\; \; B}^{A'} \; \nu_{C}.
\eqno (8.6.8)
$$
A second important difference with respect 
to the Dirac potentials  
is that the spinor fields $\nu_{B}$ and $\lambda_{B'}$ are no
longer taken to be solutions of the Weyl equation.
They should be freely specifiable (see section 8.7).
\vskip 1cm
\centerline {\bf 8.7 Compatibility conditions}
\vskip 1cm
\noindent
Our task is now to derive compatibility conditions, by requiring
that the field equations (8.6.3)--(8.6.6) should also be satisfied by the
gauge-transformed potentials appearing on the left-hand side of
(8.6.7) and (8.6.8). For this purpose, after defining the operators
$$
\cstok{\ }_{AB} \equiv \nabla_{M'(A} 
\; \nabla_{B)}^{\; \; \; M'},
\eqno (8.7.1)
$$
$$
\cstok{\ }_{A'B'} \equiv \nabla_{F(A'} 
\; \nabla_{B')}^{\; \; \; \; F} ,
\eqno (8.7.2)
$$
we need the standard identity
$
\Omega_{[AB]} = {1\over 2} \varepsilon_{AB} \; \Omega_{C}^{\; \; C}
$
and the spinor Ricci identities 
$$
\cstok{\ }_{AB} \; \nu_{C}=\psi_{ABCD} \; \nu^{D}
-2 \Lambda \; \nu_{(A} \; \varepsilon_{B)C} ,
\eqno (8.7.3)
$$
$$
\cstok{\ }_{A'B'} \lambda_{C'}=
{\widetilde \psi}_{A'B'C'D'} \; \lambda^{D'}
-2 \Lambda \; \lambda_{(A'} \;
\varepsilon_{B')C'} ,
\eqno (8.7.4)
$$
$$
\cstok{\ }^{AB} \; \lambda_{B'}=\Phi_{\; \; \; \; M'B'}^{AB}
\; \lambda^{M'} ,
\eqno (8.7.5)
$$
$$
\cstok{\ }^{A'B'} \; \nu_{B}
={\widetilde \Phi}_{\; \; \; \; \; \; MB}^{A'B'}
\; \nu^{M} .
\eqno (8.7.6)
$$
Of course, ${\widetilde \psi}_{A'B'C'D'}$ and $\psi_{ABCD}$ are
the self-dual and anti-self-dual Weyl spinors, respectively.

Thus, on using the Eqs. (8.6.3)--(8.6.8) and (8.7.1)--(8.7.6),
the basic rules of two-spinor calculus (Penrose and Rindler 
1986, Ward and Wells 1990, Stewart 1991)
lead to the compatibility equations 
$$
3 \Lambda \; \lambda_{A'}=0,
\eqno (8.7.7)
$$
$$
\Phi_{\; \; \; \; M'}^{AB \; \; \; C'} \; \lambda^{M'}=0,
\eqno (8.7.8)
$$
$$
3\Lambda \; \nu_{A}=0,
\eqno (8.7.9)
$$
$$
{\widetilde \Phi}_{\; \; \; \; \; \; M}^{A'B' \; \; C}
\; \nu^{M}=0 .
\eqno (8.7.10)
$$
Non-trivial solutions of (8.7.7)--(8.7.10) only exist if 
the scalar curvature and the trace-free part of the Ricci
tensor vanish. Hence the gauge transformations 
(8.6.7) and (8.6.8)
lead to spinor fields $\nu_{A}$ and $\lambda_{A'}$ which are
freely specifiable {\it inside} Ricci-flat backgrounds,
while the boundary conditions (8.5.1) are preserved under the
action of (8.6.7) and (8.6.8) provided that the following 
conditions hold at the boundary:
$$
\sqrt{2} \; {_{e}n_{A}^{\; \; A'}} \;
\Bigr(\nabla^{AC'} \; \nu^{B}\Bigr)e_{BC'i}
=\pm \Bigr(\nabla^{CA'} 
\lambda^{B'}\Bigr) e_{CB'i} 
\; {\rm at} \; {\partial M}.
\eqno (8.7.11)
$$
\vskip 1cm
\centerline {\bf 8.8 Second set of
potentials in Ricci-flat backgrounds}
\vskip 1cm
\noindent
As shown by Penrose (1994), in a Ricci-flat manifold the 
Rarita--Schwinger potentials may be supplemented by a second set
of potentials. Here we use such a construction in its local form.
For this purpose, we introduce
the second set of potentials for spin ${3\over 2}$ by
requiring that locally (Penrose 1994)
$$
\gamma_{A'B'}^{\; \; \; \; \; \; \; C} \equiv  \nabla_{BB'} \; 
\rho_{A'}^{\; \; \; CB}.
\eqno (8.8.1)
$$
Of course, special attention should be payed to the index ordering
in (8.8.1), since the spin-${3\over 2}$ potentials are not
symmetric. On inserting (8.8.1) into (8.6.3), a repeated use
of symmetrizations and anti-symmetrizations leads to the equation
(hereafter $\cstok{\ } \equiv \nabla_{CF'} \nabla^{CF'}$)
$$ \eqalignno{
\; & \varepsilon_{FL} \; \nabla_{AA'} \;
\nabla^{B'(F} \; \rho_{B'}^{\; \; \; A)L}
+{1\over 2} \nabla_{\; \; A'}^{A} \;
\nabla^{B'M} \; \rho_{B'(AM)} \cr
&+ \cstok{\ }_{AM} \; \rho_{A'}^{\; \; \; (AM)}
+{3\over 8} \cstok{\ } \rho_{A'}
= 0 ,
&(8.8.2)\cr}
$$
where, following Penrose (1994), we have defined
$$
\rho_{A'} \equiv \rho_{A' C}^{\; \; \; \; \; C} ,
\eqno (8.8.3)
$$
and we bear in mind that our background has to be Ricci-flat.
Thus, if the following equation holds (Penrose 1994):
$$
\nabla^{B'(F} \; \rho_{B'}^{\; \; \; A)L}=0 ,
\eqno (8.8.4)
$$
one finds
$$
\nabla^{B'M} \; \rho_{B'(AM)}={3\over 2} \; 
\nabla_{A}^{\; \; F'} \; \rho_{F'} ,
\eqno (8.8.5)
$$
and hence Eq. (8.8.2) may be cast in the form
$$
\cstok{\ }_{AM} \; \rho_{A'}^{\; \; \; (AM)}=0.
\eqno (8.8.6)
$$
On the other hand,
a very useful identity resulting from Eq. (4.9.13)
of Penrose and Rindler (1984) enables one to show that
$$
\cstok{\ }_{AM} \; \rho_{A'}^{\; \; \; (AM)}
=-\Phi_{AMA'}^{\; \; \; \; \; \; \; \; \; \; L'} \; 
\rho_{L'}^{\; \; \; (AM)}.
\eqno (8.8.7)
$$
Hence Eq. (8.8.6) reduces to an identity by virtue of
Ricci-flatness. Moreover, we have to insert (8.8.1) into the
field equation (8.6.4) for $\gamma$-potentials. By virtue of
Eq. (8.8.4) and of the identities (cf. Penrose and Rindler (1984))
$$
\cstok{\ }^{BM} \; \rho_{B' \; \; M}^{\; \; \; A}
=-\psi^{ABLM} \; \rho_{(LM)B'}
-\Phi_{\; \; \; \; \; B'}^{BM \; \; \; D'}
\; \rho_{\; \; MD'}^{A}
+4\Lambda \; \rho_{\; \; \; \; \; \; \; B'}^{(AB)} ,
\eqno (8.8.8)
$$
$$
\cstok{\ }^{B'F'} \; \rho_{B'}^{\; \; \; (AB)}
=3 \Lambda \; \rho^{(AB)F'}
+{\widetilde \Phi}_{\; \; \; \; \; \; \; L}^{B'F' \; \; \; A}
\; \rho_{\; \; \; \; \; \; \; B'}^{(LB)}
+{\widetilde \Phi}_{\; \; \; \; \; \; \; \; \; \; L}^{B'F'B}
\; \rho_{\; \; \; \; \; \; \; B'}^{(AL)} ,
\eqno (8.8.9)
$$
this leads to the equation
$$ 
\psi^{ABLM} \; \rho_{(LM)C'}=0 ,
\eqno (8.8.10)
$$
where we have again used the Ricci-flatness condition.

Of course, potentials supplementing the $\Gamma$-potentials
may also be constructed locally. On defining (cf. (8.8.1))
$$
\Gamma_{AB}^{\; \; \; \; \; C'} \equiv
\nabla_{B'B} \; \theta_{A}^{\; \; C'B'} ,
\eqno (8.8.11)
$$
$$
\theta_{A} \equiv \theta_{AC'}^{\; \; \; \; \; C'},
\eqno (8.8.12)
$$
and requiring that (Penrose 1994, Esposito 1995)
$$
\nabla^{B(F'} \; \theta_{B}^{\; \; A')L'}=0,
\eqno (8.8.13)
$$
one finds
$$
\nabla^{BM'} \; \theta_{B(A'M')}={3\over 2}
\nabla_{A'}^{\; \; \; F} \; \theta_{F},
\eqno (8.8.14)
$$
and a similar calculation yields an identity and the equation
$$ 
{\widetilde \psi}^{A'B'L'M'}
\; \theta_{(L'M')C}=0.
\eqno (8.8.15)
$$
Note that Eqs. (8.8.10) and (8.8.15) relate explicitly the
second set of potentials to 
the curvature of the background. This inconsistency
is avoided if one of the following conditions holds
(Esposito, Gionti {\it et al}. 1995):
\vskip 0.3cm
\noindent
(i) The whole conformal curvature of the background vanishes.
\vskip 0.3cm
\noindent
(ii) $\psi^{ABLM}$ and $\theta_{(L'M')C}$, or
${\widetilde \psi}^{A'B'L'M'}$ and 
$\rho_{(LM)C'}$, vanish.
\vskip 0.3cm
\noindent
(iii) The symmetric parts of the 
$\rho$- and $\theta$-potentials vanish.
\vskip 0.3cm
\noindent
In the first case one finds that the only admissible 
background is again flat Euclidean four-space with boundary,
as in Esposito and Pollifrone (1994). By contrast, in the other cases, 
left-flat, right-flat or Ricci-flat backgrounds are still
admissible, provided that the $\rho$- and
$\theta$-potentials take the form
$$
\rho_{A'}^{\; \; \; CB}=\varepsilon^{CB} \; 
{\widetilde \alpha}_{A'} ,
\eqno (8.8.16)
$$
$$
\theta_{A}^{\; \; C'B'}=\varepsilon^{C'B'} \;
\alpha_{A},
\eqno (8.8.17)
$$
where $\alpha_{A}$ and ${\widetilde \alpha}_{A'}$ solve
the Weyl equations
$$
\nabla^{AA'} \; \alpha_{A}=0,
\eqno (8.8.18)
$$
$$
\nabla^{AA'} \; {\widetilde \alpha}_{A'}=0.
\eqno (8.8.19)
$$
Eqs. (8.8.16)--(8.8.19) ensure also the validity of Eqs.
(8.8.4) and (8.8.13).

However, if one requires the preservation of Eqs. (8.8.4)
and (8.8.13) under the following gauge transformations for 
$\rho$- and $\theta$-potentials (the order of the indices $AL$, $A'L'$
is of crucial importance):
$$
{\widehat \rho}_{B'}^{\; \; \; AL} \equiv
\rho_{B'}^{\; \; \; AL}+\nabla_{B'}^{\; \; \; A}
\; \mu^{L} ,
\eqno (8.8.20)
$$
$$
{\widehat \theta}_{B}^{\; \; A'L'} \equiv
\theta_{B}^{\; \; A'L'}+\nabla_{B}^{\; \; A'}
\; \sigma^{L'},
\eqno (8.8.21)
$$
one finds compatibility conditions in Ricci-flat backgrounds
of the form
$$
\psi_{AFLD} \; \mu^{D}=0,
\eqno (8.8.22)
$$
$$
{\widetilde \psi}_{A'F'L'D'} \; \sigma^{D'}=0.
\eqno (8.8.23)
$$
Thus, to ensure {\it unrestricted} gauge freedom (except at
the boundary) for the second set of potentials, one is forced
to work with flat Euclidean backgrounds. The boundary
conditions (8.5.1) play a role in this respect, since they make
it necessary to consider both $\psi_{i}^{A}$ and
${\widetilde \psi}_{i}^{A'}$, and hence both
$\rho_{B'}^{\; \; \; AL}$ and $\theta_{B}^{\; \; A'L'}$.
Otherwise, one might use Eq. (8.8.22) to set to zero the 
anti-self-dual Weyl spinor only, {\it or} Eq. (8.8.23) to set to
zero the self-dual Weyl spinor only, so that self-dual
(left-flat) or anti-self-dual (right-flat) Riemannian 
backgrounds with boundary would survive.
\vskip 1cm
\centerline {\bf 8.9 Other gauge transformations}
\vskip 1cm
\noindent
In the massless case, flat Euclidean backgrounds with
boundary are really the only possible choice for
spin-${3\over 2}$ potentials with a gauge freedom. 
To prove this,
we have also investigated an alternative set of gauge
transformations for spin-${3\over 2}$ potentials, written in
the form (cf. (8.6.7) and (8.6.8))
$$
{\widehat \gamma}_{\; \; B'C'}^{A} \equiv
\gamma_{\; \; B'C'}^{A}+\nabla_{\; \; C'}^{A} 
\; \lambda_{B'},
\eqno (8.9.1)
$$
$$
{\widehat \Gamma}_{\; \; \; BC}^{A'} \equiv
\Gamma_{\; \; \; BC}^{A'}+\nabla_{\; \; \; C}^{A'}
\; \nu_{B}.
\eqno (8.9.2)
$$
These gauge transformations {\it do not} correspond to the
usual formulation of the Rarita--Schwinger system, but we
will see that they can be interpreted in terms of
familiar physical concepts.

On imposing that the field equations (8.6.3)--(8.6.6) should be
preserved under the action of (8.9.1) and (8.9.2), and setting to
zero the trace-free part of the Ricci spinor (since it is
inconsistent to have gauge fields $\lambda_{B'}$ and
$\nu_{B}$ which depend explicitly 
on the curvature of the background) one finds compatibility
conditions in the form of differential equations, i.e. 
(cf. Esposito (1995))
$$
\cstok{\ }\lambda_{B'}=-2\Lambda \; \lambda_{B'},
\eqno (8.9.3)
$$
$$
\nabla^{(A(B'} \; \nabla^{C')B)} \lambda_{B'}=0,
\eqno (8.9.4)
$$
$$
\cstok{\ }\nu_{B}=-2\Lambda \; \nu_{B},
\eqno (8.9.5)
$$
$$
\nabla^{(A'(B} \; \nabla^{C)B')} \; \nu_{B}=0.
\eqno (8.9.6)
$$
In a flat Riemannian four-manifold with flat connection $D$,
covariant derivatives commute and $\Lambda=0$. Hence it is
possible to express $\lambda_{B'}$ and $\nu_{B}$ as
solutions of the Weyl equations
$$
D^{AB'} \; \lambda_{B'}=0 ,
\eqno (8.9.7)
$$
$$
D^{BA'} \; \nu_{B}=0,
\eqno (8.9.8)
$$
which agree with the flat-space version of (8.9.3)--(8.9.6).
The boundary conditions (8.5.1) are then preserved under the
action of (8.9.1) and (8.9.2) if $\nu_{B}$ and $\lambda_{B'}$
obey the boundary conditions (cf. (8.7.11))
$$
\sqrt{2} \; {_{e}n_{A}^{\; \; A'}}
\Bigr(D^{BC'} \; \nu^{A}\Bigr)e_{BC'i}
=\pm \Bigr(D^{CB'} \; \lambda^{A'}\Bigr)
e_{CB'i} \; {\rm at} \; 
{\partial M} .
\eqno (8.9.9)
$$

In the curved case, on defining
$$
\phi^{A} \equiv \nabla^{AA'} \; \lambda_{A'} ,
\eqno (8.9.10)
$$
$$
{\widetilde \phi}^{A'} \equiv \nabla^{AA'} \; \nu_{A} ,
\eqno (8.9.11)
$$
equations (8.9.4) and (8.9.6) imply that 
these spinor fields solve the equations (cf. Esposito (1995))
$$
\nabla_{C'}^{\; \; \; (A} \; \phi^{B)}=0,
\eqno (8.9.12)
$$
$$
\nabla_{C}^{\; \; (A'} \; {\widetilde \phi}^{B')}=0.
\eqno (8.9.13)
$$
Moreover, Eqs. (8.9.3), (8.9.5) and the spinor Ricci identities
imply that
$$
\nabla_{AB'} \; \phi^{A}=2\Lambda \; \lambda_{B'},
\eqno (8.9.14)
$$
$$
\nabla_{BA'} \; {\widetilde \phi}^{A'}=2\Lambda \; \nu_{B}.
\eqno (8.9.15)
$$
Equations (8.9.12) and (8.9.13) are the twistor
equations (Penrose and Rindler 1986) in Riemannian four-geometries. 
The consistency conditions for the existence of non-trivial
solutions of such equations in curved Riemannian four-manifolds 
are given by (Penrose and Rindler 1986)
$$
\psi_{ABCD}=0,
\eqno (8.9.16)
$$
and
$$
{\widetilde \psi}_{A'B'C'D'}=0,
\eqno (8.9.17)
$$
respectively. 

Further consistency conditions for our problem are derived
by acting with covariant differentiation on the twistor
equation, i.e.
$$
\nabla_{A'}^{\; \; \; C} \; \nabla^{AA'} \; \phi^{B}
+\nabla_{A'}^{\; \; \; C} \; \nabla^{BA'} \; \phi^{A}=0.
\eqno (8.9.18)
$$
While the complete symmetrization in $ABC$ yields Eq. (8.9.16),
the use of Eq. (8.9.18), jointly with the spinor Ricci identities
of section 8.7, yields
$$
\cstok{\ }\phi^{B}=2\Lambda \; \phi^{B},
\eqno (8.9.19)
$$
and an analogous equation is found for ${\widetilde \phi}^{B'}$.
Thus, since Eq. (8.9.12) implies 
$$
\nabla_{C'}^{\; \; \; A} \; \phi^{B}=\varepsilon^{AB}
\; \pi_{C'},
\eqno (8.9.20)
$$
we may obtain from (8.9.20) the equation 
$$
\nabla^{BA'} \; \pi_{A'}=2\Lambda \; \phi^{B},
\eqno (8.9.21)
$$
by virtue of the spinor Ricci identities and of Eq. (8.9.19).
On the other hand, in the light of (8.9.20), Eq. (8.9.14)
leads to
$$
\nabla_{AB'} \; \phi^{A}=2\pi_{B'}
=2\Lambda \; \lambda_{B'}.
\eqno (8.9.22)
$$
Hence $\pi_{A'}=\Lambda \; \lambda_{A'}$, and the 
definition (8.9.10) yields
$$
\nabla^{BA'} \; \pi_{A'}=\Lambda \; \phi^{B}.
\eqno (8.9.23)
$$
By comparison of Eqs. (8.9.21) and (8.9.23), one gets the
equation $\Lambda \; \phi^{B}=0$. If $\Lambda \not = 0$,
this implies that $\phi^{B}$, $\pi_{B'}$ and
$\lambda_{B'}$ have to vanish, and there is no gauge
freedom fou our model. This inconsistency is avoided
if and only if $\Lambda=0$, and the corresponding 
background is forced to be totally flat, since we have
already set to zero the trace-free part of the Ricci 
spinor and the whole conformal curvature. The same argument
applies to ${\widetilde \phi}^{B'}$ and to the gauge field
$\nu_{B}$. The present analysis corrects the statements
made in section 8.8 of Esposito (1995), where it was not realized 
that, in our {\it massless} model, a non-vanishing cosmological
constant is incompatible with a gauge freedom for the
spin-${3\over 2}$ potential. More precisely, if one sets
$\Lambda=0$ from the beginning in Eqs. (8.9.3) and (8.9.5), the
system (8.9.3)--(8.9.6) admits solutions of the Weyl equation in
Ricci-flat manifolds. These backgrounds are further
restricted to be totally flat on considering the Eqs.
(8.8.10) and (8.8.15) for an arbitrary 
form of the $\rho$- and $\theta$-potentials.
As already pointed out at the end of section 8.8,
the boundary conditions (8.5.1) play a role, since otherwise
one might focus on right-flat or left-flat Riemannian
backgrounds with boundary.

Yet other gauge transformations can be studied (e.g. the
ones involving gauge fields $\lambda_{B'}$ and $\nu_{B}$
which solve the twistor equations), but they are all
incompatible with a non-vanishing cosmological 
constant in the massless case.
\vskip 10cm
\centerline {\bf 8.10 The superconnection}
\vskip 1cm
\noindent
In the massless case, the two-spinor form of the 
Rarita--Schwinger equations is the one given in Eqs.
(8.6.3)--(8.6.6) with vanishing right-hand sides,
where $\nabla_{AA'}$ is the spinor covariant derivative
corresponding to the connection $\nabla$ of the background.
In the massive case, however, the appropriate connection,
hereafter denoted by $S$, has an additional term which couples
to the cosmological constant $\lambda=6\Lambda$.
In the language of
$\gamma$-matrices, the new covariant derivative $S_{\mu}$ to be
inserted {\it in the field equations} 
(Townsend 1977) takes the form 
$$
S_{\mu} \equiv \nabla_{\mu}+f(\Lambda)\gamma_{\mu},
\eqno (8.10.1)
$$
where $f(\Lambda)$ vanishes at $\Lambda=0$, and $\gamma_{\mu}$
are the curved-space $\gamma$-matrices. 
Since, following Esposito and Pollifrone (1996), 
we are interested in the two-spinor formulation of the problem,
we have to bear in mind the action of $\gamma$-matrices on any
spinor $\varphi \equiv \Bigr(\beta^{C},
{\widetilde \beta}_{C'}\Bigr)$. Note that unprimed and primed
spin-spaces are no longer (anti-)isomorphic in the case of
positive-definite four-metrics, since there is no complex
conjugation which turns primed spinors into unprimed spinors,
or the other way around (Penrose and Rindler 1986). 
Hence $\beta^{C}$ and
${\widetilde \beta}_{C'}$ are totally unrelated. With this
understanding, we write the supergauge transformations for
massive spin-${3\over 2}$ potentials in the form
(cf. (8.6.7) and (8.6.8))
$$
{\widehat \gamma}_{\; \; B'C'}^{A} \equiv
\gamma_{\; \; B'C'}^{A}+S_{\; \; B'}^{A} 
\; \lambda_{C'},
\eqno (8.10.2)
$$
$$
{\widehat \Gamma}_{\; \; \; BC}^{A'} \equiv
\Gamma_{\; \; \; BC}^{A'}+S_{\; \; \; B}^{A'} \; \nu_{C},
\eqno (8.10.3)
$$
where the action of $S_{AA'}$ on the gauge fields  
$\Bigr(\nu^{B},\lambda_{B'}\Bigr)$ is defined by (cf. (8.10.1))
$$
S_{AA'} \; \nu_{B} \equiv \nabla_{AA'} \; \nu_{B}
+f_{1}(\Lambda)\varepsilon_{AB} \; \lambda_{A'},
\eqno (8.10.4)
$$
$$
S_{AA'} \; \lambda_{B'} \equiv \nabla_{AA'} \; \lambda_{B'}
+f_{2}(\Lambda) \varepsilon_{A'B'} \; \nu_{A}.
\eqno (8.10.5)
$$
With our notation, $R=24 \Lambda$ is the scalar curvature,
$f_{1}$ and $f_{2}$ are two functions which vanish at
$\Lambda=0$, whose form will be determined later by a
geometric analysis. 

The action of $S_{AA'}$ on a many-index
spinor $T_{B'...F'}^{A...L}$ can be obtained by expanding
such a $T$ as a sum of products of spin-vectors, 
i.e. (Penrose and Rindler 1984)
$$
T_{B'...F'}^{A...L}=\sum_{i} \alpha_{(i)}^{A} ...
\beta_{(i)}^{L} \; \gamma_{B'}^{(i)} 
... \delta_{F'}^{(i)},
\eqno (8.10.6)
$$
and then applying the Leibniz rule and the definitions 
(8.10.4) and (8.10.5), where $\alpha_{(i)}^{A}$ has an independent
partner ${\widetilde \alpha}_{(i)}^{A'}$, ... ,
$\gamma_{B'}^{(i)}$ has an independent partner
${\widetilde \gamma}_{B}^{(i)}$, ... , and so on. 
Thus, one has for example
$$ \eqalignno{
\Bigr(S_{AA'}-\nabla_{AA'}\Bigr) 
\; T_{BCE'}&=\sum_{i}\Bigr[f_{1} \varepsilon_{AB}
\; {\widetilde \alpha}_{A'}^{(i)} \; \beta_{C}^{(i)}
\; \gamma_{E'}^{(i)}
+ f_{1} \varepsilon_{AC} \; \alpha_{B}^{(i)} \; 
{\widetilde \beta}_{A'}^{(i)} \;
\gamma_{E'}^{(i)} \cr
&+ f_{2} \varepsilon_{A'E'} \; \alpha_{B}^{(i)} \; \beta_{C}^{(i)}
\; {\widetilde \gamma}_{A}^{(i)}\Bigr].
&(8.10.7)\cr}
$$

A further requirement is that $S_{AA'}$ should annihilate
the curved $\varepsilon$-spinors. Hence in our analysis we
always assume that
$$
S_{AA'} \; \varepsilon_{BC}=0,
\eqno (8.10.8)
$$
$$
S_{AA'} \; \varepsilon_{B'C'}=0.
\eqno (8.10.9)
$$
In the light of the definitions and assumptions presented
so far, one can write the Rarita--Schwinger equations
with non-vanishing cosmological constant
$\lambda=6\Lambda$, i.e. 
$$
\varepsilon^{B'C'} \; S_{A(A'} \;
\gamma_{\; \; B')C'}^{A}=\Lambda 
\; {\widetilde F}_{A'},
\eqno (8.10.10)
$$
$$
S^{B'(B} \; \gamma_{\; \; \; B'C'}^{A)}=0,
\eqno (8.10.11)
$$
$$
\varepsilon^{BC} \; S_{A'(A} \; \Gamma_{\; \; \; B)C}^{A'}
=\Lambda \; F_{A},
\eqno (8.10.12)
$$
$$
S^{B(B'} \; \Gamma_{\; \; \; \; BC}^{A')}=0 .
\eqno (8.10.13)
$$
With our notation, $F_{A}$ and ${\widetilde F}_{A'}$ are
spinor fields proportional to the traces of the second set
of potentials for spin ${3\over 2}$. These will be studied in
section 8.13.
\vskip 1cm
\centerline {\bf 8.11 Gauge freedom of the second kind}
\vskip 1cm
\noindent
The gauge freedom of the second kind is the one which does
not affect the potentials after a gauge 
transformation. This 
requirement corresponds to the case analyzed in Siklos (1985), 
where it is pointed out that, while the Lagrangian of
$N=1$ supergravity is invariant under gauge transformations 
with arbitrary spinor fields $\Bigr(\nu^{A},\lambda_{A'}\Bigr)$,
the actual {\it solutions} are only invariant if the 
supercovariant derivatives (8.10.4) and (8.10.5) vanish.

On setting to zero $S_{AA'} \; \nu_{B}$ and 
$S_{AA'} \; \lambda_{B'}$, one gets a coupled set of
equations which are the Euclidean version of the
Killing-spinor equation (Siklos 1985), i.e.
$$
\nabla_{\; \; \; B}^{A'} \; \nu_{C}=-f_{1}(\Lambda)
\lambda^{A'} \; \varepsilon_{BC},
\eqno (8.11.1)
$$
$$
\nabla_{\; \; B'}^{A} \; \lambda_{C'}=-f_{2}(\Lambda)
\nu^{A} \; \varepsilon_{B'C'}.
\eqno (8.11.2)
$$
What is peculiar of Eqs. (8.11.1) and (8.11.2) is that their
right-hand sides involve spinor fields which are,
themselves, solutions of the twistor equation. Hence one
deals with a special type of twistors, which do not exist
in a generic curved manifold. Equation (8.11.1) can
be solved for $\lambda^{A'}$ as
$$
\lambda_{C'}={1\over 2f_{1}(\Lambda)}\nabla_{C'}^{\; \; \; B}
\; \nu_{B}.
\eqno (8.11.3)
$$
The insertion of (8.11.3) into Eq. (8.11.2) and the use of spinor
Ricci identities (see (8.7.3)--(8.7.6)) 
yields the second-order equation
$$
\cstok{\ }\nu_{A}
+(6\Lambda+8f_{1}f_{2})\nu_{A}=0 .
\eqno (8.11.4)
$$
On the other hand, Eq. (8.11.1) implies the twistor equation
$$
\nabla_{\; \; \; (B}^{A'} \; \nu_{C)}=0.
\eqno (8.11.5)
$$
Covariant differentiation of Eq. (8.11.5), jointly with spinor Ricci
identities, leads to (see Eq. (8.9.19))
$$
\cstok{\ }\nu_{A}-2\Lambda \nu_{A}=0.
\eqno (8.11.6)
$$
By comparison of Eqs. (8.11.4) and (8.11.6) one finds the condition
$f_{1}f_{2}=-\Lambda$. The integrability condition of Eq. (8.11.5)
is given by (Penrose and Rindler 1986)
$$
\psi_{ABCD} \; \nu^{D}=0,
\eqno (8.11.7)
$$
which implies that our manifold is conformally left-flat. 

The condition $f_{1}f_{2}=-\Lambda$ is also obtained by 
comparison of first-order equations, since for example
$$
\nabla^{AA'} \; \nu_{A}=2f_{1}\lambda^{A'}
=-2{\Lambda \over f_{2}} \lambda^{A'} .
\eqno (8.11.8)
$$
The first equality in (8.11.8) results from Eq. (8.11.1), while the
second one is obtained since the twistor equations also
imply that (see Eq. (8.11.2))
$$
\nabla^{AA'} \Bigr(-f_{2}\nu_{A}\Bigr)
=2\Lambda \; \lambda^{A'}.
\eqno (8.11.9)
$$
Analogous results are obtained on considering the
twistor equation resulting from Eq. (8.11.2), i.e.
$$
\nabla_{\; \; (B'}^{A} \; \lambda_{C')}=0.
\eqno (8.11.10)
$$
The integrability condition of Eq. (8.11.10) is
$$
{\widetilde \psi}_{A'B'C'D'} \; \lambda^{D'}=0 .
\eqno (8.11.11)
$$
Since our gauge fields cannot be four-fold principal
spinors of the Weyl spinor (cf. Lewandowski (1991)), Eqs. (8.11.7) and 
(8.11.11) imply that our background geometry is conformally flat.
\vskip 10cm
\centerline {\bf 8.12 Compatibility conditions}
\vskip 1cm
\noindent
We now require that the field equations (8.10.10)--(8.10.13) should
be preserved under the action of the supergauge transformations
(8.10.2) and (8.10.3). This is the procedure one follows in the massless
case, and is a milder requirement with respect to the analysis
of section 8.11.

If $\nu^{B}$ and $\lambda_{B'}$ are twistors, but not necessarily
Killing spinors, they obey the Eqs. (8.11.5) and (8.11.10), which
imply that, for some independent spinor fields $\pi^{A}$ and
${\widetilde \pi}^{A'}$, one has
$$
\nabla_{\; \; \; B}^{A'} \; \nu_{C}
=\varepsilon_{BC} \; {\widetilde \pi}^{A'},
\eqno (8.12.1)
$$
$$
\nabla_{\; \; B'}^{A} \; \lambda_{C'}
=\varepsilon_{B'C'} \; \pi^{A}.
\eqno (8.12.2)
$$
In the compatibility equations, whenever one has terms of the
kind $S_{AA'} \; \nabla_{\; \; B'}^{A} \; \lambda_{C'}$, it is
therefore more convenient to symmetrize and anti-symmetrize over
$B'$ and $C'$. A repeated use of this algorithm leads to a
considerable simplification of the lengthy calculations. For
example, the preservation condition of Eq. (8.10.10) has the 
general form
$$ 
3f_{2}\Bigr(\nabla_{AA'} \; \nu^{A}+2f_{1}\lambda_{A'}\Bigr)
+\varepsilon^{B'C'}\biggr[S_{AA'}\Bigr(\nabla_{\; \; B'}^{A}
\; \lambda_{C'}\Bigr)+S_{AB'}\Bigr(\nabla_{\; \; A'}^{A}
\; \lambda_{C'}\Bigr)\biggr]=0 .
\eqno (8.12.3)
$$
By virtue of Eq. (8.12.2), Eq. (8.12.3) becomes
$$
f_{2}\Bigr(\nabla_{AA'} \; \nu^{A}+2f_{1}\lambda_{A'}\Bigr)
+S_{AA'} \; \pi^{A}=0.
\eqno (8.12.4)
$$
Following (8.10.4) and (8.10.5), the action of the
supercovariant derivative on $\pi_{A},{\widetilde \pi}_{A'}$
yields
$$
S_{AA'} \; \pi_{B} \equiv \nabla_{AA'} \; \pi_{B}
+f_{1}(\Lambda)\varepsilon_{AB} 
\; {\widetilde \pi}_{A'},
\eqno (8.12.5)
$$
$$
S_{AA'} \; {\widetilde \pi}_{B'} \equiv 
\nabla_{AA'} \; {\widetilde \pi}_{B'}
+f_{2}(\Lambda)\varepsilon_{A'B'} \; \pi_{A}.
\eqno (8.12.6)
$$
Equations (8.12.4) and (8.12.5), jointly with the equations
$$
\cstok{\ }\lambda_{A'}-2\Lambda 
\; \lambda_{A'}=0 ,
\eqno (8.12.7)
$$
$$
\nabla^{AA'} \; \pi_{A}=2\Lambda \; \lambda^{A'},
\eqno (8.12.8)
$$
which result from Eq. (8.12.2), lead to
$$
(f_{1}+f_{2}){\widetilde \pi}_{A'}
+(f_{1}f_{2}-\Lambda)\lambda_{A'}=0.
\eqno (8.12.9)
$$
Moreover, the preservation of Eq. (8.10.11) under (8.10.2) leads to
the equation
$$
S^{B'(A} \; \pi^{B)} + f_{2} \nabla^{B'(A} \;
\nu^{B)}=0,
\eqno (8.12.10)
$$
which reduces to
$$
\nabla^{B'(A} \; \pi^{B)}=0,
\eqno (8.12.11)
$$
by virtue of (8.12.1) and (8.12.5). Note that a supertwistor is
also a twistor, since
$$
S^{B'(A} \; \pi^{B)}=\nabla^{B'(A} \; \pi^{B)},
\eqno (8.12.12)
$$
by virtue of the definition (8.12.5). It is now clear that,
for a gauge freedom generated by twistors, the preservation
of Eqs. (8.10.12) and (8.10.13) under (8.10.3) leads to the compatibility 
equations
$$
(f_{1}+f_{2})\pi_{A}+(f_{1}f_{2}-\Lambda)\nu_{A}=0,
\eqno (8.12.13)
$$
$$
\nabla^{B(A'} \; {\widetilde \pi}^{B')}=0,
\eqno (8.12.14)
$$
where we have also used the equation 
(see Eqs. (8.11.6) and (8.12.1))
$$
\nabla^{AA'} \; {\widetilde \pi}_{A'}=2\Lambda \; \nu^{A}.
\eqno (8.12.15)
$$
Note that, if $f_{1}+f_{2} \not = 0$,
one can solve Eqs. (8.12.9) and (8.12.13) as
$$
{\widetilde \pi}_{A'}={(\Lambda-f_{1}f_{2})\over (f_{1}+f_{2})}
\lambda_{A'},
\eqno (8.12.16)
$$
$$
\pi_{A}={(\Lambda-f_{1}f_{2})\over (f_{1}+f_{2})}\nu_{A},
\eqno (8.12.17)
$$
and hence one deals again with Euclidean Killing spinors as
in section 8.11. However, if
$$
f_{1}+f_{2}=0,
\eqno (8.12.18)
$$
$$
f_{1}f_{2}-\Lambda=0,
\eqno (8.12.19)
$$
the spinor fields ${\widetilde \pi}_{A'}$ and $\lambda_{A'}$
become {\it unrelated}, as well as $\pi_{A}$ and $\nu_{A}$. 
This is a crucial point. Hence one may have 
$f_{1}=\pm \sqrt{-\Lambda}$, $f_{2}=\mp \sqrt{-\Lambda}$, and
one finds a more general structure (Esposito and Pollifrone 1996).

In the generic case, we do not assume that $\nu^{B}$ and
$\lambda_{B'}$ obey any equation. This means that, on the
second line of Eq. (8.12.3), it is more convenient to
express the term in square brackets as 
$2S_{A(A'} \; \nabla_{\; \; B')}^{A} \; \lambda_{C'}$. The
rule (8.10.7) for the action of $S_{AA'}$ on spinors with
many indices leads therefore to the compatibility 
conditions 
$$ 
3f_{2} \Bigr(\nabla_{AA'} \; \nu^{A}+2f_{1}\lambda_{A'}
\Bigr)-6\Lambda \; \lambda_{A'} 
+4f_{1}{\widetilde P}_{(A'B')}^{\; \; \; \; \; \; \; \; B'}
+3f_{2}{\widetilde Q}_{A'}=0 ,
\eqno (8.12.20)
$$
$$ 
3f_{1} \Bigr(\nabla_{AA'} \; \lambda^{A'}
+2f_{2} \nu_{A}\Bigr)-6\Lambda \; \nu_{A} 
+4f_{2}P_{(AB)}^{\; \; \; \; \; \; B}
+3f_{1}Q_{A}=0 ,
\eqno (8.12.21)
$$
$$
\Phi_{\; \; \; \; C'D'}^{AB} \; \lambda^{D'}
+f_{2}U_{\; \; \; \; \; \; C'}^{(AB)}
-f_{2}\nabla_{C'}^{\; \; \; (A} \; \nu^{B)}=0,
\eqno (8.12.22)
$$
$$
{\widetilde \Phi}_{\; \; \; \; \; \; CD}^{A'B'} \; \nu^{D}
+f_{1}{\widetilde U}_{\; \; \; \; \; \; \; \; C}^{(A'B')}
-f_{1}\nabla_{C}^{\; \; (A'} \; \lambda^{B')}=0 ,
\eqno (8.12.23)
$$
where the detailed form of 
$P,{\widetilde P},Q,{\widetilde Q},U,{\widetilde U}$ 
is not strictly necessary,
but we can say that they do not depend explicitly on the
trace-free part of the Ricci spinor, or on the Weyl spinors.
Note that, in the massless limit $f_{1}=f_{2}=0$, the Eqs.
(8.12.20)--(8.12.23) reduce to the familiar form of compatibility
equations which admit non-trivial solutions only in Ricci-flat
backgrounds.

Our consistency analysis still makes it necessary to set to
zero $\Phi_{\; \; \; \; C'D'}^{AB}$ (and hence 
${\widetilde \Phi}_{\; \; \; \; \; \; CD}^{A'B'}$ 
by reality (Penrose and Rindler 1984)).
The remaining contributions to (8.12.20)--(8.12.23) should then
become algebraic relations by virtue of the twistor equation.
This is confirmed by the analysis of gauge freedom for
the second set of potentials in section 8.13.
\vskip 1cm
\centerline {\bf 8.13 Second set of potentials}
\vskip 1cm
\noindent
According to the prescription of section 8.10, which replaces
$\nabla_{AA'}$ by $S_{AA'}$ in the field
equations (Townsend 1977), we now {\it assume} that the super 
Rarita--Schwinger equations corresponding to (8.8.4) 
and (8.8.13) are (see section 8.15)
$$
S^{B'(F} \; \rho_{B'}^{\; \; \; A)L}=0,
\eqno (8.13.1)
$$
$$
S^{B(F'} \; \theta_{B}^{A')L'}=0,
\eqno (8.13.2)
$$
where the second set of potentials are subject locally to the
supergauge transformations
$$
{\widehat \rho}_{B'}^{\; \; \; AL} \equiv
\rho_{B'}^{\; \; \; AL}+S_{B'}^{\; \; \; A} \; \mu^{L},
\eqno (8.13.3)
$$
$$
{\widehat \theta}_{B}^{\; \; A'L'} \equiv 
\theta_{B}^{\; \; A'L'}+S_{B}^{\; \; A'} \; \zeta^{L'}.
\eqno (8.13.4)
$$
The analysis of the gauge freedom of the second kind is 
analogous to the one in section 8.11, since equations
like (8.10.4) and (8.10.5) now apply to $\mu_{L}$ and $\zeta_{L'}$.
Hence we do not repeat this investigation.

A more general gauge freedom of the twistor type relies on
the supertwistor equations (see Eq. (8.12.12))
$$
S_{B'}^{\; \; \; (A} \; \mu^{L)}
=\nabla_{B'}^{\; \; \; (A} \; \mu^{L)}=0,
\eqno (8.13.5)
$$
$$
S_{B}^{\; \; (A'} \; \zeta^{L')}=\nabla_{B}^{\; \; (A'} \;
\zeta^{L')}=0.
\eqno (8.13.6)
$$
Thus, if one requires preservation of the super Rarita--Schwinger
equations (8.13.1) and (8.13.2) under the supergauge transformations
(8.13.3) and (8.13.4), one finds the preservation conditions
$$
S^{B'(F} \; S_{B'}^{\; \; \; A)} \; \mu^{L}=0,
\eqno (8.13.7)
$$
$$
S^{B(F'} \; S_{B}^{\; \; A')} \; \zeta^{L'}=0,
\eqno (8.13.8)
$$
which lead to
$$
(f_{1}+f_{2})\pi_{F}+(f_{1}f_{2}-\Lambda)\mu_{F}=0,
\eqno (8.13.9)
$$
$$
(f_{1}+f_{2}){\widetilde \pi}_{F'}
+(f_{1}f_{2}-\Lambda)\zeta_{F'}=0.
\eqno (8.13.10)
$$
Hence we can repeat the remarks following Eqs. 
(8.12.16)--(8.12.19). Again, it is essential that $\pi_{F},\mu_{F}$
and ${\widetilde \pi}_{F'},\zeta_{F'}$ may be unrelated if
(8.12.18) and (8.12.19) hold. In the massless case this is impossible, 
and hence there is no gauge freedom compatible with a
non-vanishing cosmological constant.

If one does not assume the validity of Eqs. (8.13.5) and (8.13.6),
the general preservation equations (8.13.7) and (8.13.8) lead instead
to the compatibility conditions 
$$ 
\eqalignno{
\; & \psi_{\; \; \; \; \; \; D}^{AFL} \; \mu^{D}
-2\Lambda \; \mu^{(A} \; \varepsilon^{F)L}
+2f_{2} \omega^{(AF)L} 
+f_{1} \varepsilon^{L(A} \; T^{F)} \cr
&+f_{1} \varepsilon^{L(A} \; S^{F)B'} \; \zeta_{B'}=0,
&(8.13.11)\cr}
$$
$$
\eqalignno{
\; & {\widetilde \psi}_{\; \; \; \; \; \; \; \; \; D'}^{A'F'L'}
\; \zeta^{D'} -2\Lambda \; \zeta^{(A'} \; \varepsilon^{F')L'}
+2f_{1}{\widetilde \omega}^{(A'F')L'} 
+f_{2} \varepsilon^{L'(A'} \; {\widetilde T}^{F')} \cr
&+f_{2} \varepsilon^{L'(A'} \; 
{\widetilde S}^{F')B} \; \mu_{B}=0.
&(8.13.12)\cr}
$$
If we now combine the compatibility equations (8.12.20)--(8.12.23)
with (8.13.11) and (8.13.12), and require that the gauge fields
$\nu_{A},\lambda_{A'},\mu_{A},\zeta_{A'}$ should not depend
explicitly on the curvature of the background, we find
that the trace-free part of the Ricci spinor has to
vanish, and the Riemannian four-geometry is forced to be
conformally flat, since under our assumptions the equations
$$
\psi_{AFLD} \; \mu^{D}=0 ,
\eqno (8.13.13)
$$
$$
{\widetilde \psi}_{A'F'L'D'} \; \zeta^{D'}=0,
\eqno (8.13.14)
$$
force the anti-self-dual and self-dual Weyl spinors to vanish.
Equations (8.13.13) and (8.13.14) are just the integrability
conditions for the existence of non-trivial solutions of the
supertwistor equations (8.13.5) and (8.13.6). Hence the spinor fields
$\omega,S,T,{\widetilde \omega},{\widetilde S}$ 
and $\widetilde T$ in
(8.13.11) and (8.13.12) are such that these equations reduce to 
(8.13.9) and (8.13.10). In other words, for massive spin-${3\over 2}$
potentials, the gauge freedom is indeed
generated by solutions of the twistor equations in conformally
flat Einstein four-manifolds.

Last, on inserting the local equations (8.8.1) and (8.8.11) into 
the second half of the Rarita--Schwinger equations,
and then replacing
$\nabla_{AA'}$ by $S_{AA'}$, one finds equations whose
preservation under the supergauge transformations (8.13.3) and (8.13.4)
is again guaranteed if the supertwistor equations 
(8.13.5) and (8.13.6) hold.
\vskip 1cm
\centerline {\bf 8.14 Non-linear superconnection}
\vskip 1cm
\noindent
As a first step in the proof
that Eqs. (8.13.1) and (8.13.2) arise naturally as 
integrability conditions of a suitable connection, 
we introduce a partial superconnection 
$W_{A'}$ (cf. Penrose (1994)) acting on unprimed spinor  
fields $\eta_{D}$ defined on the Riemannian background.
 
With our notation
$$
W_{A'} \; \eta_{D} \equiv \eta^{A} \; S_{AA'}
\; \eta_{D} - \eta_{B} \; \eta_{C} \; 
\rho_{A'}^{\; \; \; BC}
\; \eta_{D}.
\eqno (8.14.1)
$$
Writing
$$ 
W_{A'}=\eta^{A} \; \Omega_{AA'},
\eqno (8.14.2)
$$
where the operator $\Omega_{AA'}$ acts on spinor fields 
$\eta_{D}$, we obtain 
$$
\eta^{A} \; \Omega_{AA'}
=\eta^{A} \; S_{AA'}-\eta_{B} \; \eta_{C} \;
\rho_{A'}^{\; \; \; BC}.
\eqno (8.14.3)
$$
Following Penrose (1994), we require that $\Omega_{AA'}$ 
should provide a genuine 
superconnection on the spin-bundle, 
so that it acts in any direction.
Thus, from (8.14.3) one can take (cf. Penrose (1994))
$$
\Omega_{AA'} \equiv S_{AA'}-\eta^{C} \; \rho_{A'AC}=
S_{AA'}-\eta^{C} \; \rho_{A'(AC)}
+{1\over 2}\eta_{A} \; \rho_{A'}.
\eqno (8.14.4)
$$
Note that (8.14.4) makes it necessary to know the trace $\rho_{A'}$, 
while in (8.14.1) only the symmetric part of 
$\rho_{A'}^{\; \; \; BC}$  survives. 
Thus we can see that, independently of the analysis in the 
previous sections, the definition of $\Omega_{AA'}$ picks out 
a potential of the Rarita--Schwinger type (Penrose 1994).
\vskip 1cm
\centerline {\bf 8.15 Integrability condition}
\vskip 1cm
\noindent
In section 8.14 we have introduced a  
superconnection $\Omega_{AA'}$ which acts on a bundle
with non-linear fibres, where the term $-\eta^{C} \;
\rho_{A'AC}$ is responsible for the non-linear nature
of $\Omega_{AA'}$ (see (8.14.4)). Following Penrose (1994),
we now pass to a description in terms of a vector bundle 
of rank three. On introducing the local coordinates
$(u_{A},\xi)$, where
$$
u_{A}=\xi \; \eta_{A},
\eqno (8.15.1)
$$
the action of the new operator
${\widetilde \Omega}_{AA'}$ reads (cf. Penrose (1994))
$$
{\widetilde\Omega}_{AA'}(u_{B},\xi) 
\equiv  \Bigr(S_{AA'} \; u_{B}, 
S_{AA'} \; \xi-u^{C} \; \rho_{A'AC}\Bigr).
\eqno (8.15.2)
$$
Now we are able to prove that Eqs. (8.13.1) and (8.13.2) are 
integrability conditions.

The super $\beta$-surfaces are totally null two-surfaces 
whose tangent vector has the form $u^{A} \; \pi^{A'}$,
where $\pi^{A'}$ is varying and $u^{A}$ obeys the equation
$$
u^{A} \; S_{AA'} \; u_{B}=0,
\eqno (8.15.3)
$$ 
which means that $u^{A}$ is supercovariantly constant
over the surface. On defining
$$
\tau_{A'} \equiv u_{B} \; u_{C} \; \rho_{A'}^{\; \; \; BC},
\eqno (8.15.4)
$$
the condition for ${\widetilde \Omega}_{AA'}$ to be
integrable on super $\beta$-surfaces is (cf. Penrose (1994))
$$
u^{A} \; {\widetilde \Omega}_{AA'} \; \tau^{A'}=
u_{A} \; u_{B} \; u_{C} \; 
S^{A'(A} \; \rho_{A'}^{\; \; \; B)C}= 0,
\eqno (8.15.5)
$$
by virtue of the Leibniz rule and of (8.15.2)--(8.15.4).
Equation (8.15.5) implies
$$
S^{A'(A} \; \rho_{A'}^{\; \; \; B)C}=0 ,
\eqno (8.15.6)
$$
which is indeed Eq. (8.13.1). Similarly, on studying
super $\alpha$-surfaces defined by the equation
$$
{\widetilde u}^{A'} \; S_{AA'} \; {\widetilde u}_{B'}=0,
\eqno (8.15.7)
$$
one obtains Eq. (8.13.2). Thus, although Eqs. (8.13.1) and (8.13.2) 
are naturally suggested by the local theory of 
spin-${3\over 2}$ potentials, they have a deeper geometric
origin, as shown.
\vskip 1cm
\centerline {\bf 8.16 Results and open problems}
\vskip 1cm
\noindent
The consideration of boundary conditions is essential if one wants
to obtain a well-defined formulation of physical theories
in quantum cosmology (Hartle and Hawking 1983, 
Hawking 1984). In particular, one-loop
quantum cosmology (Esposito 1994a, Esposito {\it et al}. 1997)
makes it necessary to study 
spin-${3\over 2}$ potentials about four-dimensional
Riemannian backgrounds with boundary. Following Esposito (1994),
Esposito and Pollifrone (1994), we have first derived the conditions
(8.2.13), (8.2.15), (8.3.5) and (8.3.8) under which spin-lowering
and spin-raising operators preserve the local boundary conditions
studied in Breitenlohner and Freedman (1982), Hawking (1983),
Esposito (1994). Note that, for spin 0, we have introduced a pair of
independent scalar fields on the real Riemannian section of a
complex space-time, following Hawking (1979), rather than a single 
scalar field, as done in Esposito (1994). Setting $\phi \equiv
\phi_{1}+i \phi_{2}, {\widetilde \phi} \equiv \phi_{3}+i \phi_{4}$,
this choice leads to the boundary conditions
$$
\phi_{1}=\epsilon \; \phi_{3} \; {\rm on} \; S^{3} ,
\eqno (8.16.1)
$$
$$
\phi_{2}=\epsilon \; \phi_{4} \; {\rm on} \; S^{3} ,
\eqno (8.16.2)
$$
$$
{_{e}}n^{AA'}D_{AA'}\phi_{1}=-\epsilon \; {_{e}n}^{AA'}
D_{AA'}\phi_{3} \; {\rm on} \; S^{3} ,
\eqno (8.16.3)
$$
$$
{_{e}}n^{AA'}D_{AA'}\phi_{2}=-\epsilon \; {_{e}n}^{AA'}
D_{AA'}\phi_{4} \; {\rm on} \; S^{3} ,
\eqno (8.16.4)
$$
and it deserves further study.

We have then focused on the Dirac potentials for spin-${3\over 2}$
field strengths in flat or curved Riemannian four-space bounded
by a three-sphere. Remarkably, it turns out that local boundary
conditions involving field strengths and normals can only be imposed
in a flat Euclidean background, for which the gauge freedom in the
choice of the potentials remains. In Penrose (1991c) it was found that 
$\rho$ potentials exist {\it locally} only in the self-dual 
Ricci-flat case, whereas $\gamma$ potentials may be introduced in the
anti-self-dual case. Our result may be interpreted as a further
restriction provided by (quantum) cosmology. What happens is that the
boundary conditions (8.2.1) fix at the boundary a spinor field
involving {\it both} the field strength $\phi_{ABC}$ and the field
strength ${\widetilde \phi}_{A'B'C'}$. The local existence of potentials
for the field strength $\phi_{ABC}$, jointly with the occurrence of
a boundary, forces half of the Riemann curvature of the background 
to vanish. Similarly, the remaining half of such Riemann curvature 
has to vanish on considering the field strength 
${\widetilde \phi}_{A'B'C'}$. Hence the background four-geometry can
only be flat Euclidean space. This is different from the analysis in
Penrose (1990), Penrose (1991a,b), 
since in that case one is not dealing 
with boundary conditions forcing us to consider both $\phi_{ABC}$
and ${\widetilde \phi}_{A'B'C'}$.

A naturally occurring question is whether the Dirac potentials can
be used to perform one-loop calculations for spin-${3\over 2}$ 
field strengths subject to (8.2.1) on $S^{3}$. This problem may
provide another example of the fertile interplay between twistor
theory and quantum cosmology (Esposito 1994), and its solution might
shed new light on one-loop quantum cosmology and on the quantization
program for gauge theories in the presence of boundaries. For this
purpose, it is necessary to study Riemannian background 
four-geometries bounded by two three-surfaces (cf. Kamenshchik and
Mishakov (1994)). Moreover, the consideration of non-physical degrees
of freedom of gauge fields, set to zero in our classical analysis,
is necessary to achieve a covariant quantization scheme. 

Sections 8.6--8.9 have studied Rarita--Schwinger potentials in
four-dimensional Riemannian backgrounds with boundary, to complement
the analysis of Dirac's potentials appearing in section 8.4.
Our results are as follows.
First, the gauge transformations (8.6.7)
and (8.6.8) are compatible with the
massless Rarita--Schwinger equations provided that the 
background four-geometry is Ricci-flat
(Deser and Zumino 1976). However, the presence
of a boundary restricts the gauge freedom, since the boundary
conditions (8.5.1) are preserved under 
the action of (8.6.7) and (8.6.8)
only if the boundary conditions (8.7.11) hold.

Second, the Penrose construction of 
a second set of potentials in Ricci-flat
four-manifolds shows that the admissible backgrounds may be
further restricted to be totally flat, or left-flat, or 
right-flat, unless these potentials take the special 
form (8.8.16) and (8.8.17). Hence the potentials
supplementing the Rarita--Schwinger potentials have a very clear 
physical meaning in Ricci-flat four-geometries with boundary:
they are related to the spinor fields $\Bigr(\alpha_{A},
{\widetilde \alpha}_{A'}\Bigr)$ corresponding to the Majorana
field in the Lorentzian version of Eqs. (8.6.3)--(8.6.6). [One 
should bear in mind that, in real Riemannian 
four-manifolds, the only admissible
spinor conjugation is Euclidean conjugation, which is anti-involutory
on spinor fields with an odd number 
of indices (Woodhouse 1985). Hence no Majorana
field can be defined in real Riemannian four-geometries.]

Third, to ensure unrestricted gauge freedom for the $\rho$- and
$\theta$-potentials, one is forced to work with flat Euclidean
backgrounds, when the boundary conditions (8.5.1) are imposed.
Thus, the very restrictive results obtained 
in Esposito and Pollifrone (1994)
for massless Dirac potentials with the boundary conditions
(8.2.7) are indeed confirmed also for massless Rarita--Schwinger 
potentials subject to the supersymmetric boundary conditions
(8.5.1). Interestingly, a formalism originally
motivated by twistor theory 
has been applied to classical
boundary-value problems relevant for one-loop quantum cosmology.

Fourth, the gauge transformations 
(8.9.1) and (8.9.2) with non-trivial
gauge fields are compatible with the field equations 
(8.6.3)--(8.6.6) if and only if the background is totally flat.
The corresponding gauge fields solve the Weyl equations
(8.9.7) and (8.9.8), subject to the boundary conditions (8.9.9).
Indeed, it is well known that the Rarita--Schwinger 
description of a massless spin-${3\over 2}$ field is
equivalent to the Dirac description in a special choice of
gauge (Penrose 1994). In such a gauge, the spinor fields 
$\lambda_{B'}$ and $\nu_{B}$ solve the Weyl equations,
and this is exactly what we find in section 8.9 on choosing
the gauge transformations (8.9.1) and (8.9.2).

Moreover, some interesting problems are found to arise:
\vskip 0.3cm
\noindent
(i) Can one relate Eqs. (8.8.4) and (8.8.13) to the theory of
integrability conditions relevant for massless fields in
curved backgrounds (see Penrose (1994))?
What happens when such equations do not hold?
\vskip 0.3cm
\noindent
(ii) Is there an underlying global theory of Rarita--Schwinger
potentials? In the affirmative case, what are the key features
of the global theory? 
\vskip 0.3cm
\noindent
(iii) Can one reconstruct the Riemannian four-geometry from the
twistor space in Ricci-flat or conformally flat backgrounds with
boundary, or from whatever is going to replace twistor
space?
\vskip 0.3cm
\noindent
Thus, the results and problems presented in our chapter seem to add
evidence in favour of a deep link existing between twistor
geometry, quantum cosmology and modern field theory.

In the sections 8.10--8.15, we
have given an entirely two-spinor description of massive
spin-${3\over 2}$ potentials in Einstein four-geometries.
Although the supercovariant derivative
(8.10.1) was well known in the literature, following the work
in Townsend (1977), and its Lorentzian version was already applied in
Perry (1984) and Siklos (1985), 
the systematic analysis of spin-${3\over 2}$
potentials with the local form of their supergauge 
transformations was not yet available in the literature, to
the best of our knowledge, before the work in
Esposito and Pollifrone (1996).

Our first result is the two-spinor
proof that, for massive spin-${3\over 2}$ potentials, the
gauge freedom is generated by solutions of
the supertwistor equations in conformally flat Einstein
four-manifolds. Moreover, we have shown that the first-order
equations (8.13.1) and (8.13.2),
whose consideration is suggested by the local
theory of massive spin-${3\over 2}$ potentials, admit a
deeper geometric interpretation as integrability
conditions on super $\beta$- and super $\alpha$-surfaces 
of a connection on a rank-three vector bundle.
One now has to find explicit solutions
of Eqs. (8.10.10)--(8.10.13), and the supercovariant form
of $\beta$-surfaces studied in our chapter deserves 
a more careful consideration. 
Hence we hope that our work can lead to a better understanding
of twistor geometry and consistent supergravity theories
in four dimensions. For other work on spin-${3\over 2}$ potentials
and supercovariant derivatives, the reader is referred to
Tod (1983), Torres del Castillo (1989), Torres del Castillo (1990),
Torres del Castillo (1992), Frauendiener (1995), Izquierdo and
Townsend (1995), Tod (1995), Frauendiener 
{\it et al}. (1996), Tod (1996).
\vskip 100cm
\centerline {\it CHAPTER NINE}
\vskip 1cm
\centerline {\bf UNDERLYING MATHEMATICAL STRUCTURES}
\vskip 1cm
\noindent
This chapter begins with a review of four
definitions of twistors in curved space-time proposed
by Penrose in the seventies, i.e. local twistors,
global null twistors, hypersurface 
twistors and asymptotic twistors.
The Penrose transform for gravitation is then re-analyzed,
with emphasis on the double-fibration picture. Double
fibrations are also used to introduce the ambitwistor
correspondence, and the Radon transform in complex analysis
is mentioned. Attention is then focused on the Ward picture of
massless fields as bundles, which has motivated the 
analysis by Penrose of a second set of potentials which 
supplement the Rarita--Schwinger potentials in curved
space-time (chapter eight). The boundary conditions 
studied in chapters seven and eight have been recently
applied in the quantization program of field theories.
Hence the chapter ends with a review of progress made in
studying bosonic fields subject to boundary conditions
respecting BRST invariance and local supersymmetry.
Interestingly, it remains to be seen whether the 
methods of spectral geometry
can be applied to obtain an explicit proof of gauge
independence of quantum amplitudes.
\vskip 100cm
\centerline {\bf 9.1 Introduction}
\vskip 1cm
\noindent
This review chapter is written for those readers who are
more interested in the mathematical foundations of
twistor theory (see appendices C and D). 
In Minkowski space-time, twistors are defined
as the elements of the vector space of solutions of the
differential equation (4.1.5), or as $\alpha$-planes. The
latter concept, more geometric, has been extended to
curved space-time through the totally null surfaces called
$\alpha$-surfaces, whose integrability condition (in the
absence of torsion) is the vanishing of the self-dual Weyl
spinor. To avoid having to set to zero half of the conformal
curvature of complex space-time, yet another definition of
twistors, i.e. charges for massless spin-${3\over 2}$ fields 
in Ricci-flat space-times, has been proposed by Penrose.

The first part of this chapter supplements these efforts by
describing various definitions of twistors in curved space-time.
Each of these ideas has its merits and its drawbacks. To compare
local twistors at different points of space-time one is led to
introduce local twistor transport (cf. section 4.3) along a curve,
which moves the point with respect to which the twistor is defined,
but not the twistor itself. 

On studying the space of null twistors, 
a closed two-form and a one-form are naturally obtained, but their
definition cannot be extended to non-null twistors unless one
studies Minkowski space-time. In other words, one deals with
a symplectic structure which remains invariant, since a
non-rotating congruence of null geodesics remains non-rotating
in the presence of curvature. However, the attempt to obtain
an invariant complex structure fails, since a shear-free
congruence of null geodesics acquires shear in the presence
of conformal curvature. 

If an analytic space-time with analytic
hypersurface $\cal S$ in it are given, one can, however, 
construct an hypersurface twistor space relative to $\cal S$.
The differential equations describing the geometry of
hypersurface twistors encode, by construction, the information
on the complex structure, which here retains a key role. The
differential forms introduced in the theory of global null
twistors can also be expressed in the language of hypersurface
twistors. However, the whole construction relies on the choice
of some analytic (spacelike) hypersurface in curved space-time.

To overcome this difficulty, asymptotic twistors are introduced
in asymptotically flat space-times. One is thus led to combine
the geometry of future and past null infinity, which are null
hypersurfaces, with the differential equations of hypersurface
twistors and with the local twistor description. Unfortunately,
it is unclear how to achieve such a synthesis in a generic
space-time.

In the second part, attention is focused on the
geometry of conformally invariant operators, and on the
description of the Penrose transform in a more abstract
mathematical language, i.e. in terms of a double fibration
of the projective primed spin-bundle over twistor space and
space-time, respectively. The ambitwistor correspondence
of Le Brun is then introduced, in terms of a holomorphic
double fibration, and a mention is made of the Radon transform,
i.e. an integral transform which associates to a real-valued
function on $R^{2}$ its integral along a straight line in 
$R^{2}$. Such a mathematical construction is very important for
modern twistor theory, by virtue of its links with the
abstract theory of the Penrose transform.

Ward's construction of twisted photons and massless fields
as bundles is described in section 9.9, since it enables
one to understand the geometric structures underlying
the theory of spin-${3\over 2}$ potentials used in
section 8.8. In particular, Eq. (8.8.4) is related to
a class of integrability conditions arising from the
generalization of Ward's construction, as is shown
in Penrose (1994). Remarkably, this sheds new light on the
differential equations describing the local theory of
spin-${3\over 2}$ potentials (cf. section 8.15).

Since the boundary conditions of chapters seven and eight
are relevant for the elliptic boundary-value problems 
occurring in modern attempts to obtain a mathematically
consistent formulation of quantum field theories in the
presence of boundaries, recent progress on these problems
is summarized in section 9.10. While the conformal anomalies
for gauge fields in Riemannian manifolds with boundary
have been correctly evaluated after many years of dedicated
work by several authors, it remains to be seen whether the
{\it explicit} (i.e. not formal) proof of gauge independence
of quantum amplitudes can be obtained. It appears exciting that 
gauge independence of quantum amplitudes might be related
to the invariance under homotopy of the residue of a
meromorphic function, obtained from the eigenvalues of the
elliptic operators of the problem.
\vskip 1cm
\centerline {\bf 9.2 Local twistors}
\vskip 1cm
\noindent
A {\it local} twistor $Z^{\alpha}$ at $P \in \cal M$ is
represented by a pair of spinors $\omega^{A},\pi_{A'}$
at $P$:
$$
Z^{\alpha} \longleftrightarrow 
\Bigr(\omega^{A},\pi_{A'}\Bigr),
\eqno (9.2.1)
$$
with respect to the metric $g$ on $\cal M$. After a
conformal rescaling ${\widehat g} \equiv \Omega^{2} g$
of the metric, the representation of $Z^{\alpha}$ 
changes according to the rule
$$
\Bigr({\widehat \omega}^{A},{\widehat \pi}_{A'}\Bigr)
=\Bigr(\omega^{A},\pi_{A'}
+i \; T_{AA'} \; \omega^{A}\Bigr),
\eqno (9.2.2)
$$
where $T_{AA'} \equiv \nabla_{AA'}\log(\Omega)$. The
comparison of local twistors at {\it different points}
of $\cal M$ makes it necessary to introduce the
{\it local twistor transport} along a curve $\tau$ in
$\cal M$ with tangent vector $t$. This does not lead
to a displacement of the twistor along $\tau$, but
moves the {\it point} with respect to which the twistor
is defined. On defining the spinor
$$
P_{AA'BB'} \equiv {1\over 12}R \; g_{AA'BB'}
-{1\over 2}R_{AA'BB'},
\eqno (9.2.3)
$$
the equations of local twistor transport are
(cf. Eqs. (4.3.20) and (4.3.21))
$$
t^{BB'} \; \nabla_{BB'} \; \omega^{A}
=-i \; t^{AB'} \; \pi_{B'},
\eqno (9.2.4)
$$
$$
t^{BB'} \; \nabla_{BB'} \; \pi_{A'}
=-i \; P_{BB'AA'} \; t^{BB'} \; \omega^{A}.
\eqno (9.2.5)
$$

A more general concept is the one of covariant derivative
in the $t$-direction of a {\it local twistor field} on
$\cal M$ according to the rule
$$ \eqalignno{
t^{BB'} \; \nabla_{BB'} \; Z^{\alpha} & \longleftrightarrow
\Bigr(t^{BB'} \; \nabla_{BB'} \; \omega^{A}
+i \; t^{AB'} \; \pi_{B'},\cr
& t^{BB'} \; \nabla_{BB'} \; \pi_{A'}
+i \; P_{BB'AA'} \; t^{BB'}
\; \omega^{A}\Bigr) .
&(9.2.6)\cr}
$$
After a conformal rescaling of the metric, both $Z^{\alpha}$
and its covariant derivative change according to (9.2.2).
In particular, this implies that local twistor transport
is conformally invariant.

The presence of conformal curvature is responsible for a
local twistor not returning to its original state after
being carried around a small loop by local twistor transport.
In fact, as shown in Penrose (1975), denoting by 
$[t,u]$ the Lie bracket of $t$ and $u$, one finds
$$ 
\Bigr[t^{p}\nabla_{p},u^{q}\nabla_{q}\Bigr]Z^{\beta}
-[t,u]^{p}\nabla_{p}Z^{\beta}
\longleftrightarrow
t^{PP'} \; u^{QQ'} 
\left \{S_{PP'QQ'}^{B},V_{PP'QQ'B'} \right \} ,
\eqno (9.2.7)
$$
where
$$
S_{PP'QQ'}^{B} \equiv
\varepsilon_{P'Q'} \; \psi_{PQA}^{\; \; \; \; \; \; \; B}
\; \omega^{A} ,
\eqno (9.2.8)
$$
$$ \eqalignno{
V_{PP'QQ'B'} & \equiv
-i \Bigr(\varepsilon_{PQ} \; \nabla_{AA'}
\; {\widetilde \psi}_{B'P'Q'}^{\; \; \; \; \; \; \; \; \; \; \; A'}
+ \varepsilon_{P'Q'} \; \nabla_{BB'} \;
\psi_{APQ}^{\; \; \; \; \; \; \; B}\Bigr)\omega^{A}\cr
&-\varepsilon_{PQ} \; 
{\widetilde \psi}_{P'Q'B'}^{\; \; \; \; \; \; \; \; \; \; \; A'}
\; \pi_{A'} .
&(9.2.9)\cr}
$$
Equation (9.2.7) implies that, for these twistors to be defined
globally on space-time, our $({\cal M},g)$ should be
conformally flat.

In a Lorentzian space-time $({\cal M},g)_{L}$, one can
define local twistor transport of dual twistors 
$W_{\alpha}$ by complex conjugation of Eqs. (9.2.4) and (9.2.5).
On re-interpreting the complex conjugate of $\omega^{A}$
(resp. $\pi_{A'}$) as some spinor $\pi^{A'}$ (resp. 
$\omega_{A}$), this leads to
$$
t^{BB'} \; \nabla_{BB'} \; \pi^{A'}
=i \; t^{BA'} \; \omega_{B},
\eqno (9.2.10)
$$
$$
t^{BB'} \; \nabla_{BB'} \; \omega_{A}
=i \; P_{BB'AA'} \; t^{BB'} \; \pi^{A'}.
\eqno (9.2.11)
$$
Moreover, in $({\cal M},g)_{L}$ the covariant derivative in
the $t$-direction of a {\it local dual twistor field} is
also obtained by complex conjugation of (9.2.6), and leads to
$$ \eqalignno{
t^{BB'} \; \nabla_{BB'} \; W_{\alpha}& \longleftrightarrow
\Bigr(t^{BB'} \; \nabla_{BB'} \; \omega_{A}
-i \; P_{BB'AA'} \; t^{BB'} \; \pi^{A'}, \cr
&t^{BB'} \; \nabla_{BB'} \; \pi^{A'}
-i \; t^{BA'} \; \omega_{B} \Bigr).
&(9.2.12)\cr}
$$
One thus finds
$$
t^{b}\nabla_{b} \Bigr(Z^{\alpha}W_{\alpha}\Bigr)
=Z^{\alpha} \; t^{b}\nabla_{b} \; W_{\alpha}
+W_{\alpha} \; t^{b}\nabla_{b} \; Z^{\alpha},
\eqno (9.2.13)
$$
where the left-hand side denotes the ordinary derivative 
of the scalar $Z^{\alpha}W_{\alpha}$ along $\tau$. This
implies that, if local twistor transport of $Z^{\alpha}$
and $W_{\alpha}$ is preserved along $\tau$, their scalar
product is covariantly constant along $\tau$.
\vskip 1cm
\centerline {\bf 9.3 Global null twistors}
\vskip 1cm
\noindent
To define global null twistors one is led to consider null
geodesics $Z$ in curved space-time, and the $\pi_{A'}$ spinor
parallelly propagated along $Z$. The corresponding momentum
vector $p_{AA'}={\overline \pi}_{A} \; \pi_{A'}$ is then
tangent to $Z$. Of course, we want the resulting space 
$\cal N$ of null twistors to be physically meaningful.
Following Penrose (1975), the space-time $({\cal M},g)$ is
taken to be globally hyperbolic to ensure that $\cal N$ is a
Hausdorff manifold (see section 1.2). Since the space of
unscaled null geodesics is five-dimensional, and the freedom
for $\pi_{A'}$ is just a complex multiplying factor, the
space of null twistors turns out to be seven-dimensional.
Global hyperbolicity of $\cal M$ is indeed the strongest
causality assumption, and it ensures that Cauchy surfaces
exist in $\cal M$ (Hawking and Ellis 1973, Esposito 1994,
and references therein).

On $\cal N$ a closed two-form $\omega$ exists, i.e.
$$
\omega \equiv dp_{a}\wedge dx^{a}.
\eqno (9.3.1)
$$
Although $\omega$ is initially defined on the cotangent
bundle $T^{*}{\cal M}$, it actually yields a two-form on $\cal N$
if it is taken to be constant under the rescaling
$$
\pi_{A'} \rightarrow e^{i\theta} \; \pi_{A'},
\eqno (9.3.2)
$$
with real parameter $\theta$. Such a two-form may be
viewed as the rotation of a congruence, since it can
be written as
$$
\omega = \nabla_{[b} \; p_{c]} \;
dx^{b} \wedge dx^{c},
\eqno (9.3.3)
$$
where $\nabla_{[b} \; p_{c]}$ yields the rotation
of the field $p$ on $\cal M$, for a congruence of
geodesics. Our two-form $\omega$ may be obtained by
exterior differentiation of the one-form
$$
\phi \equiv p_{a} \; dx^{a},
\eqno (9.3.4)
$$
i.e.
$$
\omega = d\phi .
\eqno (9.3.5)
$$
Note that $\phi$ is defined on the space of null twistors 
and is constant under the rescaling (9.3.2). Penrose has
proposed an interpretation of $\phi$ as measuring the
time-delay in a family of scaled null geodesics
(Penrose 1975).

The main problem is how to extend these definitions to
non-null twistors. Indeed, this is possible in Minkowski
space-time, where
$$
\omega = i \; dZ^{\alpha} \wedge d{\overline Z}_{\alpha},
\eqno (9.3.6)
$$
$$
\phi = i \; Z^{\alpha} \; d{\overline Z}_{\alpha}.
\eqno (9.3.7)
$$
It is clear that Eqs. (9.3.6) and (9.3.7), if viewed as
definitions, do not depend on the twistor $Z^{\alpha}$
being null (in Minkowski). Alternative choices for
$\phi$ are
$$
\phi_{1} \equiv -i \; {\overline Z}_{\alpha}
\; dZ^{\alpha},
\eqno (9.3.8)
$$
$$
\phi_{2} \equiv {i\over 2} 
\Bigr(Z^{\alpha} \; d{\overline Z}_{\alpha}
-{\overline Z}_{\alpha} \; dZ^{\alpha}
\Bigr).
\eqno (9.3.9)
$$
The {\it invariant structure} of (flat) twistor space is
then given by the one-form $\phi$, the two-form $\omega$,
and the scalar $s \equiv {1\over 2} \; Z^{\alpha}  \; 
{\overline Z}_{\alpha}$. Although one might be tempted to
consider only $\phi$ and $s$ as basic structures, since
exterior differentiation yields $\omega$ as in (9.3.5),
the two-form $\omega$ is very important since it provides 
a symplectic structure for flat twistor space
(cf. Tod (1977)). However, if 
one restricts $\omega$ to the space of null twistors,
one first has to factor out the phase circles
$$
Z^{\beta} \rightarrow e^{i\theta} \; Z^{\beta},
\eqno (9.3.10)
$$
$\theta$ being real, to obtain again a symplectic
structure. On restriction to $\cal N$, the triple
$(\omega,\phi,s)$ has an invariant meaning also in
curved space-time, hence its name. 

Suppose now that there are two regions $M_{1}$ and
$M_{2}$ of Minkowski space-time separated by a region
of curved space-time (Penrose 1975). In each flat region,
one can define $\omega$ and $\phi$ on twistor space
according to (9.3.6) and (9.3.7), and then re-express them
as in (9.3.1), (9.3.4) on the space $\cal N$ of null
twistors in curved space-time. If there are regions of
$\cal N$ where {\it both} definitions are valid, the
flat-twistor-space definitions should agree with the 
curved ones in these regions of $\cal N$. However, it is
unclear how to carry a {\it non-null} twistor from
$M_{1}$ to $M_{2}$, if in between them there is a
region of curved space-time.

It should be emphasized that, although one has a good 
definition of {\it invariant structure} on the space
$\cal N$ of null twistors in curved space-time, with the
corresponding symplectic structure, such a construction
of global null twistors does not enable one to introduce
a complex structure. The underlying reason is that a
{\it non-rotating} congruence of null geodesics remains
non-rotating on passing through a region of curved
space-time. By contrast, a {\it shear-free} congruence of 
null geodesics acquires shear on passing through a region
of conformal curvature. This is why the symplectic structure
is invariant, while {\it the complex structure is not invariant 
and is actually affected by the conformal curvature}.

Since twistor theory relies instead on holomorphic ideas
and complex structures in a conformally invariant framework,
it is necessary to introduce yet another definition of 
twistors in curved space-time, where the complex structure
retains its key role. This problem is studied in the following
section.
\vskip 1cm
\centerline {\bf 9.4 Hypersurface twistors}
\vskip 1cm
\noindent
Given some hypersurface $\cal S$ in space-time, we are going
to construct a twistor space $T(\cal S)$, relative to
$\cal S$, with an associated complex structure. On going from
$\cal S$ to a different hypersurface ${\cal S}'$, the
corresponding twistor space $T({\cal S}')$ turns out to be
a complex manifold different from $T(\cal S)$. For any 
$T(\cal S)$, its elements are the hypersurface twistors. To
construct these mathematical structures, we follow again Penrose
(1975) and we focus on an analytic space-time $\cal M$, with analytic
hypersurface $\cal S$ in $\cal M$. These assumptions enable one
to consider the corresponding complexifications $C{\cal M}$
and $C{\cal S}$. We know from chapter four that any twistor
$Z^{\alpha}$ in $\cal M$ defines a totally null plane $CZ$
and a spinor $\pi_{A'}$ such that the tangent vector to 
$CZ$ takes the form $\xi^{A} \; \pi^{A'}$. Since $\pi_{A'}$
is constant on $CZ$, it is also constant along the complex
curve $\gamma$ giving the intersection $CZ \cap C{\cal S}$.
The geometric objects we are interested in are the normal
$n$ to $C{\cal S}$ and the tangent $t$ to $\gamma$. Since,
by construction, $t$ has to be orthogonal to $n$:
$$
n_{AA'} \; t^{AA'}=0,
\eqno (9.4.1)
$$
it can be written in the form
$$
t^{AA'}=n^{AB'} \; \pi_{B'} \; \pi^{A'},
\eqno (9.4.2)
$$
which clearly satisfies (9.4.1) by virtue of the identity
$\pi_{B'} \; \pi^{B'}=0$. Thus, for $\pi_{A'}$ to be
constant along $\gamma$, the following equation should
hold:
$$
t^{AA'} \; \nabla_{AA'} \; \pi_{C'}
=n^{AB'} \; \pi_{B'} \; \pi^{A'} \; \nabla_{AA'} \; \pi_{C'}
=0.
\eqno (9.4.3)
$$
Note that Eq. (9.4.3) also provides a differential equation for
$\gamma$ (i.e., for a given normal, the direction of $\gamma$
is fixed by (9.4.2)), and the solutions of (9.4.3) on
$C{\cal S}$ are the elements of the hypersurface twistor 
space $T(\cal S)$. Since no complex conjugation
is involved in deriving Eq. (9.4.3), the resulting $T(\cal S)$
is a complex manifold (see section 3.3).
 
It is now helpful to introduce some notation. We write
$Z^{(h)}$ for any element of $T(\cal S)$, and we remark that
if $Z^{(h)} \in T(\cal S)$ corresponds to $\pi_{A'}$ along
$\gamma$ satisfying (9.4.3), then $\rho Z^{(h)} \in T(\cal S)$
corresponds to $\rho \pi_{A'}$ along the same curve $\gamma$,
$\forall \rho \in C$ (Penrose 1975). This means one may
consider the space $PT(\cal S)$ of {\it equivalence classes}
of proportional hypersurface twistors, and regard it as the
space of curves $\gamma$ defined above. The zero-element
$0^{(h)} \in T(\cal S)$, however, does not correspond to any
element of $PT(\cal S)$. For each $Z^{(h)} \in T(\cal S)$,
$0Z^{(h)}$ is defined as $0^{(h)} \in T(\cal S)$. If the curve
$\gamma$ contains a real point of $\cal S$, the corresponding
hypersurface twistor $Z^{(h)} \in T(\cal S)$ is said to be
{\it null}. Of course, one may well ask how many real points
of $\cal S$ can be found on $\gamma$. It turns out that,
if the complexification $C{\cal S}$ 
of ${\cal S}$ is suitably chosen, only
one real point of $\cal S$ can lie on each of the curves
$\gamma$. The set $P{\cal N}(\cal S)$ of such curves is
five-real-dimensional, and the corresponding set 
${\cal N}(\cal S)$, i.e. the $\gamma$-curves with $\pi_{A'}$
spinor, is seven-real-dimensional. Moreover, the hypersurface
twistor space is four-complex-dimensional, and the space
$PT(\cal S)$ of equivalence classes defined above is
three-complex-dimensional.

The space ${\cal N}(\cal S)$ of null hypersurface twistors
has two remarkable properties:
\vskip 0.3cm
\noindent
(i) ${\cal N}(\cal S)$ may be identified with the space 
$\cal N$ of global null twistors defined in section 9.3. To
prove this one points out that the spinor $\pi_{A'}$ at the
real point of $\gamma$ (for $Z^{(h)} \in {\cal N}(\cal S)$)
defines a null geodesic in $\cal M$. Such a null geodesic
passes through that point in the real null direction given
by $v^{AA'} \equiv {\overline \pi}^{A} \; \pi^{A'}$. Parallel
propagation of $\pi_{A'}$ along this null geodesic yields a
unique element of $\cal N$. On the other hand, each global 
null twistor in $\cal N$ defines a null geodesic and a
$\pi_{A'}$. Such a null geodesic intersects $\cal S$ at a unique
point. A unique $\gamma$-curve in $C{\cal S}$ exists, passing
through this point $x$ and defined uniquely by $\pi_{A'}$
at $x$.
\vskip 0.3cm
\noindent
(ii) The hypersurface $\cal S$ enables one to supplement
the elements of ${\cal N}(\cal S)$ by some non-null
twistors, giving rise to the four-complex-dimensional
manifold $T(\cal S)$. Unfortunately, the whole construction
depends on the particular choice of (spacelike Cauchy)
hypersurface in $({\cal M},g)$.
 
The holomorphic operation
$$
Z^{(h)} \rightarrow \rho \; Z^{(h)}, \;
Z^{(h)} \in T(\cal S),
$$
enables one to introduce homogeneous holomorphic functions
on $T(\cal S)$. Setting to zero these functions gives rise 
to regions of $CT(\cal S)$ corresponding to congruences
of $\gamma$-curves on $\cal S$. A congruence of null geodesics
in $\cal M$ is defined by $\gamma$-curves on $\cal S$ having
real points. Consider now $\pi_{A'}$ as a spinor field on
$C(\cal S)$, subject to the scaling 
$\pi_{A'} \rightarrow \rho \; \pi_{A'}$. On making this
scaling, the new field $\beta_{A'} \equiv \rho \; \pi_{A'}$
no longer solves Eq. (9.4.3), since the following term 
survives on the left-hand side:
$$
E_{C'} \equiv n^{AB'} \; \pi_{B'} \;
\pi_{C'} \; \pi^{A'} \; \nabla_{AA'} \rho.
\eqno (9.4.4)
$$
This suggests to consider the weaker condition
$$
n^{AB'} \; \pi_{B'} \biggr(\pi^{A'} \; \pi^{C'} \;
\nabla_{AA'} \; \pi_{C'} \biggr)=0
\; {\rm on} \; {\cal S},
\eqno (9.4.5)
$$
since $\pi^{C'}$ has a vanishing contraction with $E_{C'}$.
Equation (9.4.5) should be regarded as an equation for the spinor
field $\pi_{A'}$ restricted to $\cal S$. Following
Penrose (1975), round brackets have been used to emphasize the
role of the spinor field 
$$
B_{A} \equiv \pi^{A'} \; \pi^{C'} \;
\nabla_{AA'} \; \pi_{C'},
$$
whose vanishing leads to a shear-free congruence of null
geodesics with tangent vector $v^{AA'} \equiv
{\overline \pi}^{A} \; \pi^{A'}$.

A careful consideration of extensions and restrictions of
spinor fields enables one to write an equivalent form
of Eq. (9.4.5). In other words, if we extend $\pi_{A'}$
to a spinor field on the whole of $\cal M$, Eq. (9.4.5) holds
if we replace $n^{AB'} \; \pi_{B'}$ by
${\overline \pi}^{A}$. This implies that the same
equation holds on $\cal S$ if we omit $n^{AB'} \; \pi_{B'}$.
Hence one eventually deals with the equation
$$
\pi^{A'} \; \pi^{C'} \; \nabla_{AA'} \; \pi_{C'}=0.
\eqno (9.4.6)
$$
Since it is well known in general relativity that conformal
curvature is responsible for a shear-free congruence of null
geodesics to acquire shear, the previous analysis proves that
the complex structure of hypersurface twistor space is affected
by the particular choice of $\cal S$ unless the space-time is
conformally flat.

The {\it dual hypersurface twistor space} $T^{*}(\cal S)$ may
be defined by interchanging primed and unprimed indices in
Eq. (9.4.3), i.e.
$$
n^{BA'} \; {\widetilde \pi}_{B} \; {\widetilde \pi}^{A}
\; \nabla_{AA'} \; {\widetilde \pi}_{C}=0.
\eqno (9.4.7)
$$
In agreement with the notation used in our paper and
proposed by Penrose, the {\it tilde} symbol denotes spinor
fields not obtained by complex conjugation of the spinor 
fields living in the complementary spin-space, since, in a
complex manifold, complex conjugation is not invariant under
holomorphic coordinate transformations. Hence the complex
nature of $T(\cal S)$ and $T^{*}(\cal S)$ is responsible for 
the spinor fields in (9.4.3) and (9.4.7) being totally
independent. Equation (9.4.7) defines a unique complex curve
$\widetilde \gamma$ in $C{\cal S}$ through each point
of $C{\cal S}$. The geometric interpretation of
$n^{BA'} \; {\widetilde \pi}_{B} \; {\widetilde \pi}^{A}$
is in terms of the tangent direction to the curve
$\widetilde \gamma$ for any choice of ${\widetilde \pi}_{A}$.
The curve $\widetilde \gamma$ and the spinor field 
${\widetilde \pi}_{A}$ solving Eq. (9.4.7) define a dual
hypersurface twistor ${\widetilde Z}_{(h)} \in T^{*}(\cal S)$.
Indeed, the {\it complex conjugate} ${\overline Z}_{(h)}$ of
the hypersurface twistor $Z^{(h)} \in T(\cal S)$ may also be
defined if the following conditions hold:
$$
{\widetilde \pi}_{A}={\overline \pi}_{A} , \;
{\widetilde \gamma}=\gamma .
\eqno (9.4.8)
$$
The {\it incidence} between $Z^{(h)} \in T(\cal S)$ and
${\widetilde Z}_{(h)} \in T^{*}(S)$ is instead defined by
the condition
$$
Z^{(h)} \; {\widetilde Z}_{(h)}=0,
\eqno (9.4.9)
$$
where $(h)$ is not an index, but a label to denote
{\it hypersurface} twistors (instead of the dot used in
Penrose (1975)). Thus, $\gamma$ and
$\widetilde \gamma$ have a point of $C{\cal S}$ in common.
Null hypersurface twistors are then defined by the
condition
$$
Z^{(h)} \; {\overline Z}_{(h)}=0.
\eqno (9.4.10)
$$
However, it is hard to make sense of the (scalar) product
$Z^{(h)} \; {\widetilde Z}_{(h)}$ for arbitrary elements of
$T(\cal S)$ and $T^{*}(\cal S)$, respectively.

We are now interested in holomorphic maps
$$
F: T^{*}({\cal S}) \times T({\cal S}) \rightarrow C.
\eqno (9.4.11)
$$
Since $T(\cal S)$ and $T^{*}(\cal S)$ are both 
four-complex-dimensional, the space 
$T^{*}({\cal S}) \times T({\cal S})$ is eight-complex-dimensional.
A seven-complex dimensional subspace ${\widetilde N}(\cal S)$
can be singled out in $T^{*}({\cal S}) 
\times T({\cal S})$, on considering 
those pairs $\Bigr({\widetilde Z}_{(h)},Z^{(h)}\Bigr)$ such that
Eq. (9.4.9) holds. One may want to study these holomorphic 
maps in the course of writing contour-integral formulae for
solutions of the massless free-field equations, where the
integrand involves a homogeneous function $F$ acting on twistors
and dual twistors. Omitting the details (Penrose 1975),
we only say that, when the space-time point $y$ under consideration
does not lie on $C{\cal S}$, one has to reinterpret $F$ as
a function of $U_{(h)} \in T^{*}({\cal S}')$,
$X^{(h)} \in T({\cal S}')$, where the hypersurface
${\cal S}'$, or $C{\cal S}'$, is chosen to pass through the
point $y$. 

A naturally occurring question is how to deal with the
one-form $\phi$ and the two-form $\omega$ introduced in
section 9.3. Indeed, if the space-time is analytic, such
forms $\phi$ and $\omega$ can be complexified. On making a
complexification, two one-forms $\phi$ and $\widetilde \phi$
are obtained, which take the same values on $C{\cal N}$, but
whose functional forms are different. For $Z^{(h)} \in
T({\cal S})$, $W_{(h)} \in T^{*}({\cal S})$,
$X^{(h)} \in T({\cal S}')$, $U_{(h)} \in T^{*}({\cal S}')$,
$\cal S$ and ${\cal S}'$ being two different hypersurfaces in
$\cal M$, one has (Penrose 1975)
$$
\omega = i \; dZ^{(h)} \wedge dW_{(h)}
=i \; dX^{(h)} \wedge dU_{(h)},
\eqno (9.4.12)
$$
$$
\phi = i \; Z^{(h)} \; dW_{(h)} = i \; X^{(h)}
\; dU_{(h)},
\eqno (9.4.13)
$$
$$
{\widetilde \phi}=-i \; W_{(h)} \; dZ^{(h)}
=-i \; U_{(h)} \; dX^{(h)}.
\eqno (9.4.14)
$$
Hence one is led to ask wether the passage from a
$\Bigr(W_{(h)},Z^{(h)}\Bigr)$ description on $\cal S$ to a
$\Bigr(U_{(h)},X^{(h)}\Bigr)$ description on ${\cal S}'$
can be regarded as a canonical transformation. This is
achieved on introducing the equivalence relations
(Penrose 1975)
$$
\Bigr(W_{(h)},Z^{(h)}\Bigr) \equiv 
\Bigr(\rho^{-1} \; W_{(h)},\rho \; Z^{(h)} \Bigr),
\eqno (9.4.15)
$$
$$
\Bigr(U_{(h)},X^{(h)}\Bigr) \equiv 
\Bigr(\sigma^{-1} \; U_{(h)},\sigma \; X^{(h)} \Bigr),
\eqno (9.4.16)
$$
which yield a six-complex-dimensional space $S_{6}$
(see problem 9.2).
\vskip 1cm
\centerline {\bf 9.5 Asymptotic twistors}
\vskip 1cm
\noindent
Although in the theory of hypersurface twistors the complex
structure plays a key role, their definition depends on an
arbitrary hypersurface $\cal S$, and the attempt to define
the scalar product $Z^{(h)} \; W_{(h)}$ faces great
difficulties. The concept of asymptotic twistor tries to
overcome these limitations by focusing on asymptotically
flat space-times. Hence the emphasis is on null hypersurfaces,
i.e. ${\rm SCRI}^{+}$ and ${\rm SCRI}^{-}$ (cf. section 3.5),
rather than on spacelike hypersurfaces. Since the construction
of hypersurface twistors is independent of conformal rescalings
of the metric, while future and past null infinity have 
well known properties (Hawking and Ellis 1973), the theory of
asymptotic twistors appears well defined. Its key features
are as follows.

First, one complexifies future null infinity
${\cal I}^{+}$ to get $C{\cal I}^{+}$. Hence its complexified
metric is described by complexified coordinates
$\eta,{\widetilde \eta},u$, where $\eta$ and
${\widetilde \eta}$ are totally independent (cf. section
3.5). The corresponding planes $\eta={\rm constant}$,
${\widetilde \eta}={\rm constant}$, are totally null planes
(in that the complexified metric of $C{\cal I}^{+}$ vanishes
over them) with a topological twist (Penrose 1975).

Second, note that for any null hypersurface, its normal has
the spinor form
$$
n^{AA'}=\iota^{A} \; {\widetilde \iota}^{A'}.
\eqno (9.5.1)
$$
Thus, if ${\widetilde \iota}^{B'} \; \pi_{B'}
\not = 0$, the insertion of (9.5.1) into
Eq. (9.4.3) yields
$$
\iota^{A} \; \pi^{A'} \; \nabla_{AA'} \; \pi_{C'}=0.
\eqno (9.5.2)
$$
Similarly, if $\iota^{B} \; {\widetilde \pi}_{B}
\not = 0$, the insertion of (9.5.1) into the Eq. (9.4.7)
for dual hypersurface twistors leads to
$$
{\widetilde \pi}^{A} \; {\widetilde \iota}^{A'}
\; \nabla_{AA'} \; {\widetilde \pi}_{C}=0.
\eqno (9.5.3)
$$
These equations tell us that the $\gamma$-curves are null
geodesics on $C{\cal I}^{+}$, lying entirely in the
${\widetilde \eta}={\rm constant}$ planes, while the 
${\widetilde \gamma}$ curves are null geodesics lying 
in the $\eta={\rm constant}$ planes.

By definition, an {\it asymptotic twistor} is an element
$Z^{(a)} \in T({\cal I}^{+})$, and corresponds to a
null geodesic $\gamma$ in $C{\cal I}^{+}$ with tangent
vector $\iota^{A} \; \pi^{A'}$, where $\pi_{A'}$ undergoes
parallel propagation along $\gamma$. By contrast, a 
{\it dual asymptotic twistor} is an element 
${\widetilde Z}_{(a)} \in T^{*}({\cal I}^{+})$,
and corresponds to a null geodesic 
$\widetilde \gamma$ in $C{\cal I}^{+}$ with tangent
vector ${\widetilde \pi}^{A} \; {\widetilde \iota}^{A'}$,
where ${\widetilde \pi}_{A}$ undergoes parallel 
propagation along $\widetilde \gamma$.

It now remains to be seen how to define the scalar product
$Z^{(a)} \; {\widetilde Z}_{(a)}$. For this purpose, denoting
by $\lambda$ the intersection of the 
${\widetilde \eta}={\rm constant}$ plane containing $\gamma$
with the $\eta={\rm constant}$ plane containing
$\widetilde \gamma$, we assume for simplicity that $\lambda$
intersects $C{\cal I}^{+}$ in such a way that a continuous
path $\beta$ exists in $\gamma \cup \lambda \cup 
{\widetilde \gamma}$, unique up to homotopy, connecting 
$Q \in \gamma$ to ${\widetilde Q} \in {\widetilde \gamma}$.
One then gives a local twistor description of $Z^{(a)}$
as $\Bigr(0,\pi_{A'}\Bigr)$ at $Q$, and one carries this
along $\beta$ by local twistor transport (section 9.2) to
$\widetilde Q$. At the point $\widetilde Q$, the local twistor
obtained in this way has the usual scalar product with the
local twistor description $\Bigr({\widetilde \pi}_{A},0\Bigr)$
at $\widetilde Q$ of ${\widetilde Z}_{(a)}$. By virtue of
Eqs. (9.2.4), (9.2.5) and (9.2.13), such a definition of
scalar product is independent of the choice made to locate
$Q$ and $\widetilde Q$, and it also applies on going from
$\widetilde Q$ to $Q$. Thus, the theory of asymptotic twistors
combines in an essential way the asymptotic structure of 
space-time with the properties of local twistors and hypersurface
twistors. Note also that $Z^{(a)} \; {\widetilde Z}_{(a)}$
has been defined as a holomorphic function on some open subset
of $T({\cal I}^{+}) \times T^{*}({\cal I}^{+})$ containing
$C{\cal N}({\cal I}^{+})$. Hence one can take derivatives with
respect to $Z^{(a)}$ and ${\widetilde Z}_{(a)}$ so as to obtain
the differential forms in (9.4.12)--(9.4.14). If 
$W_{(a)} \in T^{*}({\cal I}^{+})$,
$Z^{(a)} \in T({\cal I}^{+})$,
$U_{(a)} \in T^{*}({\cal I}^{-})$,
$X^{(a)} \in T({\cal I}^{-})$, one can write
$$
\omega = i \; dZ^{(a)} \wedge dW_{(a)}
=i \; dX^{(a)} \wedge dU_{(a)},
\eqno (9.5.4)
$$
$$
\phi = i \; Z^{(a)} \; dW_{(a)}
=i \; X^{(a)} \; dU_{(a)} ,
\eqno (9.5.5)
$$
$$
{\widetilde \phi}=-i \; W_{(a)} \;
dZ^{(a)}=-i \; U_{(a)} \; dX^{(a)}.
\eqno (9.5.6)
$$
The asymptotic twistor space at future null infinity is
also very useful in that its global complex structure
enables one to study the outgoing radiation field arising
from gravitation (Penrose 1975).
\vskip 1cm
\centerline {\bf 9.6 Penrose transform}
\vskip 1cm
\noindent
As we know from chapter four, on studying the massless
free-field equations in Minkowski space-time, the Penrose
transform provides the homomorphism (Eastwood 1990)
$$
{\cal P}:H^{1}(V,{\cal O}(-n-2)) \rightarrow
\Gamma(U,Z_{n}).
\eqno (9.6.1)
$$
With the notation in (9.6.1), $U$ is an open subset of
compactified complexified Minkowski space-time, $V$ is the 
corresponding open subset of projective twistor space,
$O(-n-2)$ is the sheaf of germs (appendix D) of holomorphic
functions homogeneous of degree $-n-2$, $Z_{n}$ is the sheaf
of germs of holomorphic solutions of the massless free-field
equations of helicity ${n\over 2}$. Although the Penrose
transform may be viewed as a geometric way of studying the
partial differential equations of mathematical physics, the
main problem is to go beyond flat space-time and reconstruct
a generic curved space-time from its twistor space or from
some more general structures. Here, following Eastwood (1990),
we study a four-complex-dimensional conformal manifold $M$,
which is assumed to be geodesically convex. For a given choice
of spin-structure on $M$, let $F$ be the projective primed
spin-bundle over $M$ with local coordinates $x^{a},\pi_{A'}$.
After choosing a metric in the conformal class, the corresponding
metric connection is lifted horizontally to a differential
operator $\nabla_{AL'}$ on spinor fields on $F$.

Denoting by $\phi_{B}$ a spinor field on $M$ of conformal weight
$w$, a conformal rescaling ${\widehat g}=\Omega^{2}g$ of the
metric leads to a change of the operator according to the rule
$$
{\widehat \nabla}_{AL'} \; \phi_{B}
=\nabla_{AL'} \; \phi_{B}
-Y_{BL'} \; \phi_{A}+w \; Y_{AL'} \; \phi_{B}
+\pi_{L'} \; Y_{AB'} \; {\partial \phi_{B}\over \partial
\pi_{B'}} ,
\eqno (9.6.2)
$$
where $Y_{AL'} \equiv \Omega^{-1}\nabla_{AL'}\Omega$.
In particular, on {\it functions} of weight $w$ one finds
$$
{\widehat \nabla}_{AL'}\phi=\nabla_{AL'}\phi
+w \; Y_{AL'}\phi+\pi_{L'} \; Y_{AB'} \;
{\partial \phi \over \partial \pi_{B'}} .
\eqno (9.6.3)
$$
Thus, if the conformal weight vanishes, acting with 
$\pi^{A'}$ on both sides of (9.6.3) and defining
$$
\nabla_{A} \equiv \pi^{A'} \; \nabla_{AA'} ,
\eqno (9.6.4)
$$
one obtains
$$
{\widehat \nabla}_{A}\phi=\nabla_{A}\phi .
\eqno (9.6.5)
$$
This means that $\nabla_{A}$ is a conformally invariant
operator on ordinary functions and hence may be regarded as
an invariant distribution on the projective spin-bundle
$F$ (Eastwood 1990). From chapters four and six we know
that such a distribution is integrable if and only if the
self-dual Weyl spinor ${\widetilde \psi}_{A'B'C'D'}$
vanishes. One can then integrate the distribution on $F$
to give a new space $P$ as the space of leaves. This leads
to the double fibration familiar to the mathematicians 
working on twistor theory:
$$
P {\mathrel{\mathop{\longleftarrow}^{\mu}}} F 
{\mathrel{\mathop{\longrightarrow}^{\nu}}} M.
\eqno (9.6.6)
$$
In (9.6.6) $P$ is the twistor space of $M$, and the submanifolds
$\nu(\mu^{-1}(z))$ of $M$, for $z \in P$, are the $\alpha$-surfaces
in $M$ (cf. chapter four). Each point $x \in M$ is known to give
rise to a line $L_{x} \equiv \mu(\nu^{-1}(x))$ in $P$, whose
points {\it correspond} to the $\alpha$-surfaces through $x$ as
described in chapter four. The conformally anti-self-dual complex
space-time $M$ with its conformal structure is then recovered
from its twistor space $P$, and an explicit construction has been
given in section 5.1.

To get a deeper understanding of this non-linear-graviton
construction, we now introduce the Einstein bundle $E$. For
this purpose, let us consider a function $\phi$ of conformal
weight 1. Equation (9.6.3) implies that, under a conformal rescaling
of the metric, $\nabla_{A}\phi$ rescales as
$$
{\widehat \nabla}_{A}\phi=\nabla_{A}\phi+Y_{A}\phi.
\eqno (9.6.7)
$$
Thus, the transformation rule for $\nabla_{A}\nabla_{B}\phi$ is
$$
{\widehat \nabla}_{A}{\widehat \nabla}_{B}\phi
=\nabla_{A}{\widehat \nabla}_{B}\phi-Y_{B}{\widehat \nabla}_{A}\phi
=\nabla_{A}\nabla_{B}\phi
+\Bigr[(\nabla_{A}Y_{B})-Y_{B}Y_{A}\Bigr]\phi .
\eqno (9.6.8)
$$
Although $\nabla_{A}\nabla_{B}\phi$ is not conformally
invariant, Eq. (9.6.8) suggests how to modify our operator
to make it into a conformally invariant operator. For this
purpose, denoting by $\Phi_{ABA'B'}$ the trace-free 
part of the Ricci spinor, and defining
$$
\Phi_{AB} \equiv \pi^{A'} \; \pi^{B'} \;
\Phi_{ABA'B'} .
\eqno (9.6.9)
$$
we point out that, under a conformal rescaling, $\Phi_{AB}$
transforms as
$$
{\widehat \Phi}_{AB}=\Phi_{AB}-\nabla_{A}Y_{B}+Y_{A}Y_{B}.
\eqno (9.6.10)
$$
Equations (9.6.8) and (9.6.10) imply that the conformally invariant
operator we are looking for is (Eastwood 1990)
$$
D_{AB} \equiv \nabla_{A}\nabla_{B}+\Phi_{AB} ,
\eqno (9.6.11)
$$
acting on functions of weight 1. In geometric language,
$\nabla_{A}$ and $D_{AB}$ act along the fibres of $\mu$. A
vector bundle $E$ over $P$ is then obtained by considering
the vector space of functions defined on $\mu^{-1}Z$ such
that $D_{AB}\phi=0$ and having conformal weight 1. Such a
space is indeed three-dimensional, since $\alpha$-surfaces
inherit from the conformal structure on $M$ a flat projective
structure, and $D_{AB}$ in (9.6.11) is a projectively invariant
differential operator (Eastwood 1990, and earlier analysis 
by Bailey cited therein).

Remarkably, the Penrose transform establishes an isomorphism
between the space of smooth sections $\Gamma(P,E)$
($E$ being our Einstein bundle on $P$) and the space of
functions $\phi$ of conformal weight 1 on $M$ such that
$$
\nabla_{(A}^{(A'} \; \nabla_{B)}^{B')}\phi
+\Phi_{AB}^{A'B'}\phi=0.
\eqno (9.6.12)
$$
The proof is obtained by first pointing out that, in the light
of the definition of $E$, $\Gamma(P,E)$ is isomorphic to the
space of functions $\phi$ of conformal weight 1 on the
spin-bundle $F$ such that
$$
\nabla_{A}\nabla_{B}\phi+\Phi_{AB}\phi=0.
\eqno (9.6.13)
$$
The next step is the remark that the fibres of 
$\nu: F \rightarrow M$ are Riemann spheres and hence
are compact, which implies that $\phi(x^{a},\pi_{A'})$ is
a function of $x^{a}$ only. The resulting equation on the
spin-bundle $F$ is
$$
\pi^{A'}\; \pi^{B'} \; \nabla_{AA'} \; \nabla_{BB'}\phi
+\pi^{A'} \; \pi^{B'} \; \Phi_{ABA'B'}\phi=0.
\eqno (9.6.14)
$$
At this stage, the contribution of $\pi^{A'} \; \pi^{B'}$
has been factorized, which implies we are left with Eq.
(9.6.12). Conformal invariance of the equation on $M$ is
guaranteed by the use of the conformally invariant 
operator $D_{AB}$.

From the point of view of gravitational physics, what is
important is the resulting isomorphism between nowhere
vanishing sections of $E$ over $P$ and Einstein metrics
in the conformal class on $M$. Of course, the Einstein 
condition means that the Ricci tensor is proportional to
the metric, and hence the trace-free part of Ricci vanishes:
$\Phi_{ab}=0$. To prove this basic property one points out
that, since $\phi$ may be chosen to be nowhere vanishing,
${\widehat \phi}$ can be set to 1, so that Eq. (9.6.13) implies
${\widehat \Phi}_{ab}=0$, which is indeed the Einstein 
condition. The converse also holds (Eastwood 1990).

Moreover, a pairing between solutions of differential 
equations can be established. To achieve this, note first
that the tangent bundle of $P$ corresponds to solutions
of the differential equation (Eastwood 1990)
$$
\nabla_{(A} \; \omega_{B)}=0,
\eqno (9.6.15)
$$
where $\omega_{B}$ is homogeneous of degree 1 in $\pi_{A'}$
and has conformal weight 1. Now the desired pairing is between
solutions of Eq. (9.6.15) where $\omega_{B} \in {\cal O}_{B}(-1)[1]$
as above, and solutions of
$$
\nabla_{A}\nabla_{B}\phi+\Phi_{AB}\phi=0
\; \phi \in {\cal O}[1].
\eqno (9.6.16)
$$
Following again Eastwood (1990), we now consider a function $f$
which is conformally invariant, and constant along the fibres
of $\mu: F \rightarrow P$. Since $f$ is defined as
$$
f \equiv 2 \omega^{A}\nabla_{A}\phi-\phi\nabla_{A}\omega^{A},
\eqno (9.6.17)
$$
its conformal invariance is proved by inserting (9.6.7)
into the transformation rule
$$
{\widehat f}=2\omega^{A}{\widehat \nabla}_{A}\phi
-\phi {\widehat \nabla}_{A}\omega^{A}.
\eqno (9.6.18)
$$
The constancy of $f$ along the fibres of $\mu$ is proved
in two steps. First, the Leibniz rule, Eq. (8.7.3) and
Eq. (9.6.15) imply that
$$
\nabla_{B}f=2\omega^{A}\nabla_{B}\nabla_{A}\phi
-\phi \nabla_{B}\nabla_{A}\omega^{A}.
\eqno (9.6.19)
$$
Second, using an identity for $\nabla_{B}\nabla_{A}\omega^{A}$
and then applying again Eq. (9.6.15) one finds (Eastwood 1990)
$$
\nabla_{B}\nabla_{A}\omega^{A}=\varepsilon_{BA} \; \Phi_{C}^{A}
\; \omega^{C}+{1\over 2}\nabla_{A} \; \delta_{B}^{A}
\; \nabla_{C}\omega^{C},
\eqno (9.6.20)
$$
which implies
$$
\nabla_{B}\nabla_{A}\omega^{A}=-2\Phi_{AB} \; \omega^{A}.
\eqno (9.6.21)
$$
Thus, Eqs. (9.6.16), (9.6.19) and (9.6.21) lead to
$$
\nabla_{B}f=2\omega^{A}\Bigr(\nabla_{A}\nabla_{B}
+\Phi_{AB}\Bigr)\phi=0.
\eqno (9.6.22)
$$
\rightline {Q.E.D.}

The results presented so far may be combined to show that an
Einstein metric in the given conformal class on $M$ corresponds
to a nowhere vanishing one-form $\tau$ on twistor space $P$,
homogeneous of degree two (cf. section 4.3). One then considers
$\tau \wedge d\tau$, which can be written as $2\Lambda \rho$ for
some function $\Lambda$. This $\Lambda$ is indeed the cosmological
constant, since the holomorphic functions in $P$ are necessarily
constant.
\vskip 1cm
\centerline {\bf 9.7 Ambitwistor correspondence}
\vskip 1cm
\noindent
In this section we consider again
a complex space-time $({\cal M},g)$, where $\cal M$ is
a four-complex-dimensional complex manifold, and $g$ is a
holomorphic non-degenerate symmetric two-tensor on $\cal M$
(i.e. a complex-Riemannian metric). A family of null geodesics
can be associated to $({\cal M},g)$ by considering those inextendible,
connected, one-dimensional complex submanifolds
$\gamma \subset {\cal M}$ such that any tangent vector field 
$v \in \Gamma(\gamma,O(T\gamma))$ satisfies 
(Le Brun 1990)
$$
\nabla_{v}v= \sigma v,
\eqno (9.7.1)
$$
$$
g(v,v)=0,
\eqno (9.7.2)
$$
where $\sigma$ is a proportionality parameter and 
$\nabla$ is the Levi-Civita connection of $g$. These curves
determine completely the conformal class of the complex metric $g$,
since a vector is null if and only if it is tangent to some null
geodesic $\gamma$. Conversely, the conformal class determines the
set of null geodesics (Le Brun 1990). We now denote by $\cal N$ 
the set of null geodesics of $({\cal M},g)$, and by $\cal Q$ the
hypersurface of null covectors defined by
$$
{\cal Q} \equiv \left \{[\phi] \in PT^{*}{\cal M}:
g^{-1}(\phi,\phi)=0 \right \}.
\eqno (9.7.3)
$$
A {\it quotient} map $q:{\cal Q} \rightarrow {\cal N}$ can
be given as the map assigning, to each point of $\cal Q$, the
leaf through it. If $\cal N$ is equipped with the quotient 
topology, and if $({\cal M},g)$ is geodesically convex, 
$\cal N$ is then Hausdorff and has a {\it unique} complex
structure making $q$ into a holomorphic map of maximal rank. The
corresponding complex manifold $\cal N$ is, by definition,
the {\it ambitwistor space} of $({\cal M},g)$. 

Denoting by $p:{\cal Q} \rightarrow {\cal M}$ the restriction
to $\cal Q$ of the canonical projection 
$\pi: PT^{*}{\cal M} \rightarrow {\cal M}$, one has a 
holomorphic double fibration
$$
{\cal N} {\mathrel{\mathop{\longleftarrow}^{q}}} 
{\cal Q}
{\mathrel{\mathop{\longrightarrow}^{p}}}
{\cal M},
\eqno (9.7.4)
$$
the {\it ambitwistor correspondence}, which relates complex
space-time to its space of null geodesics. For example, in the
case of the four-quadric $Q_{4} \subset P_{5}$, obtained by
conformal compactification of
$$
\biggr(C^{4},\sum_{j=1}^{4}(dz^{j})^{\otimes 2}\biggr),
$$
the corresponding ambitwistor space is (Le Brun 1990)
$$
{\cal A} \equiv \left \{\Bigr([Z^{\alpha}],[W_{\alpha}]\Bigr)
\in P_{3} \times P_{3}:
\sum_{\alpha=1}^{4}Z^{\alpha}W_{\alpha}=0 \right \}.
\eqno (9.7.5)
$$
Ambitwistor space has been used as an attempt to go beyond the
space of $\alpha$-surfaces, i.e. twistor space (chapter four).
However, we prefer to limit ourselves to a description of the
main ideas, to avoid becoming too technical. Hence the reader
is referred to Le Brun's original papers appearing in the
bibliography for a thorough analysis of ambitwistor geometry.
\vskip 1cm
\centerline {\bf 9.8 Radon transform}
\vskip 1cm
\noindent
In the mathematical literature, the analysis of the Penrose
transform is frequently supplemented by the study of the Radon
transform, and the former is sometimes referred to as the
Radon--Penrose transform. Indeed, the transform introduced in
Radon (1917) associates to a real-valued function $f$ on $R^{2}$
the following integral:
$$
(Rf)(L) \equiv \int_{L}f ,
\eqno (9.8.1)
$$
where $L$ is a straight line in $R^{2}$. On inverting the Radon
and Penrose transforms, however, one appreciates there is a
substantial difference between them (Bailey {\it et al}. 1994). 
In other words, (9.8.1) is invertible in that the value of the
original function at a particular point may be recovered from its
integrals along all cycles passing near that point. By contrast,
in the Penrose transform, the original data in a neighbourhood
of a particular cycle can be recovered from the transform
restricted to that neighbourhood. Hence the Radon transform is
globally invertible, while {\it the Penrose transform may be inverted
locally}. [I am grateful to Mike Eastwood for making it possible 
for me to study the work appearing in Bailey et al.
(1994). No original result obtained in Bailey {\it et al}. (1994) has 
been even mentioned in this section]
\vskip 1cm
\centerline {\bf 9.9 Massless fields as bundles}
\vskip 1cm
\noindent
In the last part of chapter eight, motivated by our early work
on one-loop quantum cosmology, we have 
studied a second set of potentials
for gravitino fields in curved Riemannian backgrounds with
non-vanishing cosmological constant. Our analysis is a direct
generalization of the work in Penrose (1994), where the author
studies the Ricci-flat case and relies on the analysis of
twisted photons appearing in Ward (1979). Thus, we here review the
mathematical foundations of these potentials in the simpler
case of Maxwell theory.

With the notation in Ward (1979), $B$ is the primed spin-bundle
over space-time, and $\Bigr(x^{a},\pi_{A'}\Bigr)$ are
coordinates on $B$. Of course, $x^{a}$ are space-time coordinates
and $\pi_{A'}$ are coordinates on primed spin-space. Moreover,
we introduce the Euler vector field on $B$:
$$
T \equiv \pi^{A'} \; {\partial \over \partial \pi_{A'}}.
\eqno (9.9.1)
$$
A function $f$ on $B$ such that $Tf=0$ is homogeneous of degree
zero in $\pi_{A'}$ and hence is defined on the projective
spin-bundle. We are now interested in the two-dimensional 
distribution spanned by the two vector fields 
$\pi^{A'}\nabla_{AA'}$. The integral surfaces of such a 
distribution are the elements of non-projective twistor
space $T$. To deform $T$ without changing $PT$, Ward replaced
$\pi^{A'}\nabla_{AA'}$ by $\pi^{A'}\nabla_{AA'}-\psi_{A}T$,
$\psi_{0}$ and $\psi_{1}$ being two functions on $B$. By virtue
of Frobenius' theorem (cf. section 6.2), the necessary and
sufficient condition for the integrability of the new distribution
is the validity of the equation
$$
\pi^{A'}\; \nabla_{AA'} \; \psi^{A}
-\psi_{A} \; T \; \psi^{A}=0,
\eqno (9.9.2)
$$
for {\it all} values of $\pi^{A'}$. In geometric language,
if Eq. (9.9.2) holds $\forall \pi^{A'}$, a four-dimensional space
$T'$ of integral surfaces exists, and $T'$ is a holomorphic
bundle over projective twistor space $PT$. One can also say
that $T'$ is a {\it deformation} of flat twistor space $T$.
If $\psi_{A}$ takes the form
$$
\psi_{A}(x,\pi_{A'})=i \; \Phi_{A}^{\; \; A'...L'}(x)
\; \pi_{A'} ... \pi_{L'},
\eqno (9.9.3)
$$
then Eq. (9.9.2) becomes
$$
\nabla^{A(A'} \; \Phi_{A}^{\; \; B'...M')}=0.
\eqno (9.9.4)
$$
Thus, the spinor field $\Phi_{A}^{\; \; A'...L'}$ is a potential
for a massless free field
$$
\phi_{AB...M} \equiv \nabla_{(B}^{B'} ... \nabla_{M}^{\; \; M'}
\; \Phi_{A)B'...M'},
\eqno (9.9.5)
$$
since the massless free-field equations
$$
\nabla^{AA'} \; \phi_{AB...M}=0
\eqno (9.9.6)
$$
result from Eq. (9.9.4).

In the particular case of Maxwell theory, suppose that
$$
\psi_{A}=i \; \Phi_{A}^{\; \; A'}(x) \; \pi_{A'},
\eqno (9.9.7)
$$
with (cf. Eq. (8.8.4))
$$
\nabla^{A(A'} \; \Phi_{A}^{\; \; B')}=0.
\eqno (9.9.8)
$$
Note that here the space $T'$ of integral surfaces is a 
principal fibre bundle over $PT$ with group the 
non-vanishing complex numbers. Following Ward (1979), here $PT$
is just the {\it neighbourhood} of a line in $CP^{3}$, but
not the whole of $CP^{3}$.

For the mathematically-oriented reader, we should say that, 
in the language of sheaf cohomology, one has the exact sequence
$$
... \rightarrow
H^{1}(PT,Z)\rightarrow H^{1}(PT,{\cal O})
\rightarrow H^{1}(PT,{\cal O}^{*})
\rightarrow H^{2}(PT,Z) \rightarrow ... \; .
\eqno (9.9.9)
$$
If $PT$ has $R^{4} \times S^{2}$ topology (see section 4.3),
then $H^{1}(PT,Z)=0$. 
Moreover, $H^{1}(PT,{\cal O})$ is isomorphic to the space
of left-handed Maxwell fields $\phi_{AB}$ satisfying the
massless free-field equations
$$
\nabla^{AA'}\phi_{AB}=0.
\eqno (9.9.10)
$$
$H^{1}(PT,{\cal O}^{*})$ is the space of line bundles over
$PT$, and $H^{2}(PT,Z) \cong Z$ is the space of possible 
Chern classes of such bundles. Thus, the space of left-handed
Maxwell fields is isomorphic to the space of deformed line
bundles $T'$. To realize this correspondence, we should bear
in mind that a twistor determines an $\alpha$-surface, jointly
with a primed spinor field $\pi_{A'}$ propagated over the
$\alpha$-surface (chapter four). The usual propagation is
parallel transport:
$$
\pi^{A'} \; \nabla_{AA'} \; \pi_{B'}=0.
\eqno (9.9.11)
$$
However, in the deformed case, the propagation equation
is taken to be
$$
\pi^{A'}\Bigr(\nabla_{AA'}+i \; \Phi_{AA'}\Bigr)\pi_{B'}=0.
\eqno (9.9.12)
$$
Remarkably, the integrability condition for Eq. (9.9.12)
is Eq. (9.9.8) (Ward 1979 and our problem 9.4). 
This property suggests that
also Eq. (8.8.4) may be viewed as an integrability condition.
In Penrose (1994), this geometric interpretation
has been investigated for spin-${3\over 2}$ fields. It appears
striking that the equations of the local theory of spin-${3\over 2}$
potentials lead naturally to equations which
can be related to integrability conditions.
Conversely, from some suitable integrability conditions,
one may hope of constructing a local theory of 
potentials for gauge fields. The interplay between
these two points of view deserves further consideration.
\vskip 1cm
\centerline {\bf 9.10 Quantization of field theories}
\vskip 1cm
\noindent
The boundary conditions studied in chapters seven and eight 
are a part of the general set which should be imposed
on bosonic and fermionic fields to respect
BRST invariance and local supersymmetry. In this chapter devoted
to mathematical foundations we describe some recent progress
on these issues, but we do not repeat our early analysis
appearing in Esposito (1994).

The way in which quantum fields respond to the presence of
boundaries is responsible for many interesting physical
effects such as, for example, the Casimir effect, and the
quantization program of spinor fields, gauge fields and 
gravitation in the presence of boundaries is currently leading
to a better understanding of modern quantum field theories
(Esposito {\it et al}. 1997).
The motivations for this investigation come from at least three
areas of physics and mathematics, i.e. 
\vskip 0.3cm
\noindent
(i) {\it Cosmology}. One wants to understand what is the quantum
state of the universe, and how to formulate boundary conditions
for the universe (Esposito 1994 and references therein).
\vskip 0.3cm
\noindent
(ii) {\it Field Theory}. It appears necessary to get a deeper
understanding of different quantization techniques in field
theory, i.e. the reduction to physical degrees of freedom 
before quantization, or the Faddeev--Popov Lagrangian method,
or the Batalin--Fradkin--Vilkovisky extended phase space.
Moreover, perturbative properties of supergravity theories 
and conformal anomalies in field theory deserve further investigation,
especially within the framework of semiclassical evaluation
of path integrals in field theory via zeta-function
regularization.
\vskip 0.3cm
\noindent
(iii) {\it Mathematics}. A (pure) mathematician may regard
quantum cosmology as a problem in cobordism theory
(i.e. when a compact manifold may be regarded as the boundary 
of another compact manifold), and
one-loop quantum cosmology as a relevant application of
the theory of eigenvalues in Riemannian geometry, of 
self-adjointness theory, and of the analysis of asymptotic
heat kernels for manifolds with boundary.

On using zeta-function regularization (Esposito 1994), the $\zeta(0)$
value yields the scaling of quantum amplitudes and the one-loop
divergences of physical theories. The choices to be made concern
the quantization technique, the background four-geometry, the 
boundary three-geometry, the boundary conditions respecting 
Becchi--Rouet--Stora--Tyutin invariance and local supersymmetry,
the gauge condition, the regularization algorithm. 
We are here interested in the mode-by-mode analysis of 
BRST-covariant Faddeev--Popov amplitudes for 
Euclidean Maxwell theory, which relies
on the expansion of the electromagnetic potential in
harmonics on the boundary three-geometry. In the case of
three-sphere boundaries, one has (Esposito 1994)
$$
A_{0}(x,\tau)=\sum_{n=1}^{\infty}R_{n}(\tau)Q^{(n)}(x),
\eqno (9.10.1)
$$
$$
A_{k}(x,\tau)=\sum_{n=2}^{\infty}
\biggr[f_{n}(\tau)S_{k}^{(n)}(x)
+g_{n}(\tau)P_{k}^{(n)}(x)\biggr],
\eqno (9.10.2)
$$
where $Q^{(n)}(x),S_{k}^{(n)}(x)$ and $P_{k}^{(n)}(x)$ 
are scalar, transverse and longitudinal vector harmonics
on $S^{3}$, respectively.

Magnetic conditions set to zero at the boundary the
gauge-averaging functional, the tangential components of the
potential, and the ghost field, i.e.
$$
[\Phi(A)]_{\partial M}=0 , \;
[A_{k}]_{\partial M}=0, \;
[\epsilon]_{\partial M}=0.
\eqno (9.10.3)
$$
Alternatively, electric conditions set to zero at the boundary the
normal component of the potential, the normal derivative of
tangential components of the potential, and the normal
derivative of the ghost field, i.e. 
$$
[A_{0}]_{\partial M}=0, \;
\left[{\partial A_{k}\over \partial \tau}\right]_{\partial M}=0, \;
\left[{\partial \epsilon \over \partial \tau}\right]_{\partial M}=0.
\eqno (9.10.4)
$$
One may check that these boundary conditions are compatible
with BRST transformations, and do not give rise to additional
boundary conditions after a gauge transformation
(Esposito {\it et al}. 1997).

By using zeta-function regularization and flat Euclidean
backgrounds, the effects of relativistic gauges are as
follows (Esposito and Kamenshchik 1994, 
Esposito {\it et al}. 1997, and references therein).
\vskip 0.3cm
\noindent
(i) In the Lorenz gauge, the mode-by-mode analysis of one-loop
amplitudes agrees with the results of the Schwinger--DeWitt 
technique, both in the one-boundary case (i.e. the disk) and 
in the two-boundary case (i.e. the ring).
\vskip 0.3cm
\noindent
(ii) In the presence of boundaries, the effects of gauge modes
and ghost modes {\it do not} cancel each other.
\vskip 0.3cm
\noindent
(iii) When combined with the contribution of physical degrees
of freedom, i.e. the transverse part of the potential, this lack
of cancellation is exactly what one needs to achieve agreement
with the results of the Schwinger--DeWitt technique.
\vskip 0.3cm
\noindent
(iv) Thus, physical degrees of freedom are, by themselves,
insufficient to recover the full information about one-loop
amplitudes.
\vskip 0.3cm
\noindent
(v) Moreover, even on taking into account physical, non-physical
and ghost modes, the analysis of relativistic gauges different 
from the Lorenz gauge yields gauge-independent amplitudes 
only in the two-boundary case.
\vskip 0.3cm
\noindent
(vi) Gauge modes obey a coupled set of second-order eigenvalue
equations. For some particular choices of gauge
conditions it is possible to decouple such a set of differential
equations, by means of two functional matrices which diagonalize
the original operator matrix.
\vskip 0.3cm
\noindent
(vii) For arbitrary choices of relativistic gauges, gauge modes
remain coupled. The explicit proof of gauge independence of 
quantum amplitudes becomes a problem in homotopy theory. 
Hence there seems to be a deep relation between the 
Atiyah--Patodi--Singer theory of Riemannian four-manifolds with
boundary (Atiyah {\it et al}. 1976), the
zeta-function, and the BKKM function (Barvinsky {\it et al}. 1992b):
$$
I(M^{2},s) \equiv \sum_{n=n_{0}}^{\infty}
d(n) \; n^{-2s} \; \log \Bigr[f_{n}(M^{2})\Bigr].
\eqno (9.10.5)
$$
In (9.10.5), $d(n)$ is the degeneracy of the eigenvalues 
parametrized by the integer $n$, and $f_{n}(M^{2})$ is the
function occurring in the equation obeyed by the eigenvalues
by virtue of the boundary conditions, after taking out false roots.
The analytic continuation of (9.10.5) to the whole complex-$s$ plane 
is given by 
$$
``I(M^{2},s)"={I_{\rm pole}(M^{2})\over s} 
+I^{R}(M^{2})+O(s),
\eqno (9.10.6)
$$
and enables one to evaluate $\zeta(0)$ as 
$$
\zeta(0)=I_{\rm log}+I_{\rm pole}(\infty)-I_{\rm pole}(0),
\eqno (9.10.7)
$$
$I_{\rm log}$ being the coefficient of $\log(M)$ appearing in
$I^{R}$ as $M \rightarrow \infty$.

A detailed mode-by-mode study of perturbative quantum gravity
about a flat Euclidean background bounded by two concentric
three-spheres, including non-physical degrees of freedom and ghost
modes, leads to one-loop amplitudes in agreement with the
covariant Schwinger--DeWitt method
(Esposito, Kamenshchik {\it et al}. 1994). 
This calculation provides
the generalization of the previous analysis of fermionic fields
and electromagnetic fields (Esposito 1994). The basic idea is to expand
the metric perturbations $h_{00},h_{0i}$ and $h_{ij}$ on a 
family of three-spheres centred on the origin, and then use the
de Donder gauge-averaging functional in the Faddeev--Popov
Euclidean action. The resulting eigenvalue equation for metric
perturbations about a flat Euclidean background:
$$
\cstok{\ } h_{\mu \nu}^{(\lambda)}
+\lambda \; h_{\mu \nu}^{(\lambda)}=0,
\eqno (9.10.8)
$$
gives rise to seven coupled eigenvalue equations for metric
perturbations. On considering also the ghost one-form 
$\varphi_{\mu}$, and imposing the mixed boundary conditions
of Luckock, Moss and Poletti,
$$
[h_{ij}]_{\partial M} =0,
\eqno (9.10.9a)
$$
$$
[h_{i0}]_{\partial M}=0,
\eqno (9.10.9b)
$$
$$
[\varphi_{0}]_{\partial M}=0,
\eqno (9.10.9c)
$$
$$
\left[{\partial h_{00}\over \partial \tau}
+{6\over \tau}h_{00}-{\partial \over \partial \tau}
\Bigr(g^{ij}h_{ij}\Bigr)\right]_{\partial M}=0,
\eqno (9.10.10)
$$
$$
\left[{\partial \varphi_{i}\over \partial \tau}
-{2\over \tau} \varphi_{i} \right]_{\partial M}=0,
\eqno (9.10.11)
$$
the analysis in Esposito, Kamenshchik {\it et al}. 1994 
has shown that the full $\zeta(0)$
vanishes in the two-boundary problem, while the contributions
of ghost modes and gauge modes {\it do not} cancel each
other, as it already happens for Euclidean Maxwell theory.

The main open problem seems to be the explicit proof of gauge
independence of one-loop amplitudes for relativistic gauges, in
the case of flat Euclidean space bounded by two concentric
three-spheres. For this purpose, one may have to show that, for 
coupled gauge modes, $I_{\rm log}$ and the difference
$I_{\rm pole}(\infty)-I_{\rm pole}(0)$ are not affected by a
change in the gauge parameters. Three steps are in order:
\vskip 0.3cm
\noindent
(i) To relate the regularization at large $x$ 
used in Esposito (1994) to the BKKM regularization relying
on the function (9.10.5).
\vskip 0.3cm
\noindent
(ii) To evaluate $I_{\rm log}$ from an asymptotic analysis
of coupled eigenvalue equations. 
\vskip 0.3cm
\noindent
(iii) To evaluate $I_{\rm pole}(\infty)-I_{\rm pole}(0)$ by
relating the analytic continuation to the whole complex-$s$ 
plane of the difference $I(\infty,s)-I(0,s)$,
to the analytic continuation of the zeta-function.
\vskip 0.3cm
The last step might involve a non-local, integral transform 
relating the BKKM function to the zeta-function, and a
non-trivial application of the Atiyah--Patodi--Singer spectral
analysis of Riemannian four-manifolds with boundary 
(Atiyah {\it et al}. 1976). In other words,
one might have to prove that, {\it in the two-boundary problem
only}, $I_{\rm pole}(\infty)-I_{\rm pole}(0)$ resulting
from coupled gauge modes is the residue of a meromorphic 
function, invariant under a smooth variation in the gauge
parameters of the matrix of elliptic self-adjoint operators
appearing in the system 
$$
{\widehat {\cal A}}_{n}g_{n}+{\widehat {\cal B}}_{n}R_{n}=0, \;
\forall n \geq 2 ,
\eqno (9.10.12)
$$
$$
{\widehat {\cal C}}_{n}g_{n}+{\widehat {\cal D}}_{n}R_{n}=0, \;
\forall n \geq 2,
\eqno (9.10.13)
$$
where one has
$$
{\widehat {\cal A}}_{n} \equiv
{d^{2}\over d\tau^{2}}+{1\over \tau}{d\over d\tau}
-{\gamma_{3}^{2}\over \alpha}{(n^{2}-1)\over \tau^{2}}
+\lambda_{n},
\eqno (9.10.14)
$$
$$
{\widehat {\cal B}}_{n} \equiv
- \Bigr(1+{\gamma_{1}\gamma_{3}\over \alpha}\Bigr)  
(n^{2}-1){d\over d\tau}
- \Bigr(1+{\gamma_{2}\gamma_{3}\over \alpha}\Bigr) 
{(n^{2}-1)\over \tau},
\eqno (9.10.15)
$$
$$
{\widehat {\cal C}}_{n} \equiv
\Bigr(1+{\gamma_{1}\gamma_{3}\over \alpha}\Bigr)
{1 \over \tau^{2}}{d\over d\tau}
+{\gamma_{3}\over \alpha}(\gamma_{1}-\gamma_{2})
{1\over \tau^{3}},
\eqno (9.10.16)
$$
$$
{\widehat {\cal D}}_{n} \equiv
{\gamma_{1}^{2}\over \alpha}{d^{2}\over d\tau^{2}}
+{3\gamma_{1}^{2}\over \alpha}{1\over \tau}
{d\over d\tau}
+\left[{\gamma_{2}\over \alpha}(2\gamma_{1}-\gamma_{2})
-(n^{2}-1)\right]{1\over \tau^{2}}+\lambda_{n}.
\eqno (9.10.17)
$$
With our notation, $\gamma_{1},\gamma_{2}$ and $\gamma_{3}$
are dimensionless parameters which enable one to study the
most general gauge-averaging functional. This may be written in
the form (the boundary being given by three-spheres)
$$
\Phi(A) \equiv \gamma_{1} { }^{(4)}\nabla^{0}A_{0}
+{\gamma_{2}\over 3}A_{0} \; {\rm Tr}(K)
-\gamma_{3} { }^{(3)}\nabla^{i}A_{i},
\eqno (9.10.18)
$$
where $K$ is the extrinsic-curvature tensor of the boundary.

Other relevant research problems are the mode-by-mode analysis
of one-loop amplitudes for gravitinos, including gauge modes
and ghost modes studied within the Faddeev--Popov formalism.
Last, but not least, the mode-by-mode analysis of linearized gravity
in the unitary gauge in the one-boundary case, 
and the mode-by-mode analysis of
one-loop amplitudes in the case of curved backgrounds, appear 
to be necessary to complete the picture outlined so far.
The recent progress on problems with boundaries, 
however, seems to strengthen the evidence in favour
of new perspectives being in sight in quantum field theory
(Avramidi and Esposito 1998a,b, 1999).
\vskip 100cm
\centerline {\it CHAPTER TEN}
\vskip 1cm
\centerline {\bf OLD AND NEW IDEAS IN COMPLEX GENERAL RELATIVITY}
\vskip 1cm
\noindent
The analysis of (conformally) right-flat 
space-times of the previous chapters has its counterpart in the
theory of heaven spaces developed by Plebanski. This chapter
begins with a review of weak heaven spaces, strong heaven spaces,
heavenly tetrads and heavenly equations. An outline is also
presented of the work by McIntosh, Hickman and other authors on
complex relativity and real solutions. 
The last section is instead 
devoted to modern developments in complex general relativity.
In particular, the analysis of
real general relativity based on multisymplectic techniques
has shown that boundary terms may occur in the constraint
equations, unless some boundary conditions are imposed.
The corresponding form of such boundary terms in complex
general relativity is here studied. A complex Ricci-flat
space-time is recovered provided that some boundary conditions
are imposed on two-complex-dimensional surfaces.
One then finds that the holomorphic multimomenta should
vanish on an arbitrary three-complex-dimensional surface,
to avoid having
restrictions at this surface on the spinor fields
expressing the invariance of the theory under holomorphic
coordinate transformations. The Hamiltonian constraint 
of real general relativity is then replaced by a geometric
structure linear in the holomorphic multimomenta, and a link
with twistor theory is found. Moreover, a deep relation
emerges between complex space-times which are not
anti-self-dual and two-complex-dimensional surfaces 
which are not totally null. 
\vskip 100cm
\centerline {\bf 10.1 Introduction}
\vskip 1cm
\noindent
One of the most recurring themes of this paper is the
analysis of complex or real Riemannian manifolds where half
of the conformal curvature vanishes and the vacuum Einstein
equations hold. Chapter five has provided an explicit construction
of such anti-self-dual space-times, and the underlying 
Penrose-transform theory has been presented in chapters four
and nine. However, alternative ways exist to construct these
solutions of the Einstein equations, and hence this chapter
supplements the previous chapters by describing the work
in Plebanski (1975). By using the tetrad formalism and some
basic results in the theory of partial differential equations,
the so-called {\it heaven spaces} and {\it heavenly tetrads}
are defined and constructed in detail. A brief review is then
presented of the work by Hickman, McIntosh {\it et al}. on complex
relativity and real solutions.

The last section of this chapter is instead devoted to new
ideas in complex general relativity. First, the multisymplectic
form of such a theory is outlined. Hence one deals with
jet bundles described, locally, by a holomorphic coordinate
system with holomorphic tetrad, holomorphic connection one-form,
multivelocities corresponding to the tetrad and multivelocities
corresponding to the connection, both of holomorphic nature
(Esposito and Stornaiolo 1995).
Remarkably, the equations of complex general relativity are all
linear in the holomorphic multimomenta, and the anti-self-dual
space-times relevant for twistor theory turn out to be a
particular case of this more general structure. Moreover, the
analysis of two-complex-dimensional surfaces in the generic
case is shown to maintain a key role in complex general relativity.
\vskip 10cm
\centerline {\bf 10.2 Heaven spaces}
\vskip 1cm
\noindent
In his theory of heaven spaces, Plebanski studies a 
four-dimensional {\it analytic} manifold $M_{4}$ with metric
given in terms of tetrad vectors as (Plebanski 1975)
$$
g=2e^{1}e^{2}+2e^{3}e^{4}=g_{ab} \; e^{a} e^{b}
\; \in \Lambda^{1} \otimes \Lambda^{1}.
\eqno (10.2.1)
$$
The definition of the $2 \times 2$ matrices
$$
\tau^{AB'} \equiv \sqrt{2} 
\pmatrix {e^{4}&e^{2}\cr e^{1}&-e^{3} \cr}
\eqno (10.2.2)
$$
enables one to re-express the metric as
$$
g=-{\rm det} \; \tau^{AB'}
={1\over 2} \; \varepsilon_{AB} \; \varepsilon_{C'D'} \;
\tau^{AC'} \; \tau^{BD'}.
\eqno (10.2.3)
$$
Moreover, since the manifold is analytic, there exist two
{\it independent} sets of $2 \times 2$ complex matrices
with unit determinant: $L_{\; \; \; A}^{A'} \in \; SL(2,C)$
and ${\widetilde L}_{\; \; \; B}^{B'} \in \;
{\widetilde {SL}}(2,C)$. On defining a new set of tetrad vectors
such that
$$
\sqrt{2} \pmatrix {e^{4'}&e^{2'}\cr e^{1'}&-e^{3'}\cr}
=L_{\; \; \; A}^{A'} \;
{\widetilde L}_{\; \; \; B'}^{B}
\; \tau^{AB'},
\eqno (10.2.4)
$$
the metric is still obtained as $2e^{1'}e^{2'}+2e^{3'}e^{4'}$.
Hence the tetrad gauge group may be viewed as
$$
{\cal G} \equiv SL(2,C) \times {\widetilde {SL}}(2,C).
\eqno (10.2.5)
$$

A key role in the following analysis is played by a pair of
differential forms whose spinorial version is obtained from
the wedge product of the matrices in (10.2.2), i.e.
$$
\tau^{AB'}\wedge \tau^{CD'}=S^{AC} \; \varepsilon^{B'D'}
+\varepsilon^{AC} \; {\widetilde S}^{B'D'},
\eqno (10.2.6)
$$
where
$$
S^{AB} \equiv {1\over 2} \; \varepsilon_{R'S'} \;
\tau^{AR'} \wedge \tau^{BS'}
={1\over 2} \; e^{a}\wedge e^{b} \; S_{ab}^{\; \; \; AB},
\eqno (10.2.7)
$$
$$
{\widetilde S}^{A'B'} \equiv {1\over 2} \;
\varepsilon_{RS} \; \tau^{RA'} \wedge \tau^{SB'}
={1\over 2} \; e^{a} \wedge e^{b} \;
{\widetilde S}_{ab}^{\; \; \; A'B'}.
\eqno (10.2.8)
$$
The forms $S^{AB}$ and ${\widetilde S}^{A'B'}$ are self-dual
and anti-self-dual respectively, in that the action of the 
Hodge-star operator on them leads to (Plebanski 1975)
$$
{ }^{*}S^{AB}=S^{AB},
\eqno (10.2.9)
$$
$$
{ }^{*}{\widetilde S}^{A'B'}=-{\widetilde S}^{A'B'}.
\eqno (10.2.10)
$$
To obtain the desired spinor description of the curvature, we
introduce the antisymmetric connection forms
$\Gamma_{ab}=\Gamma_{[ab]}$ through the first structure
equations
$$
de^{a}=e^{b} \wedge \Gamma_{\; \; b}^{a}.
\eqno (10.2.11)
$$
The spinorial counterpart of $\Gamma_{ab}$ is given by
$$
\Gamma_{AB} \equiv -{1\over 4} \Gamma_{ab} \; 
S_{\; \; \; AB}^{ab},
\eqno (10.2.12)
$$
$$
{\widetilde \Gamma}_{A'B'} \equiv -{1\over 4}
\Gamma_{ab} \; {\widetilde S}_{\; \; \; A'B'}^{ab},
\eqno (10.2.13)
$$
which implies
$$
\Gamma_{ab}=-{1\over 2}S_{ab}^{\; \; \; AB} 
\; \Gamma_{AB} -{1\over 2} {\widetilde S}_{ab}^{\; \; \; A'B'}
\; {\widetilde \Gamma}_{A'B'}.
\eqno (10.2.14)
$$
To appreciate that $\Gamma_{AB}$ and ${\widetilde \Gamma}_{A'B'}$
are actually independent, the reader may find it useful to check
that (Plebanski 1975)
$$
\Gamma_{AB}=-{1\over 2} \pmatrix 
{2\Gamma_{42}& {\Gamma_{12}+\Gamma_{34}}\cr
{\Gamma_{12}+\Gamma_{34}} & 2\Gamma_{31}\cr},
\eqno (10.2.15)
$$
$$
{\widetilde \Gamma}_{A'B'}=-{1\over 2} \pmatrix
{2\Gamma_{41} & {-\Gamma_{12}+\Gamma_{34}}\cr
{-\Gamma_{12}+\Gamma_{34}}& 2\Gamma_{32}\cr}.
\eqno (10.2.16)
$$
The action of exterior differentiation on
$\tau^{AB'},S^{AB},{\widetilde S}^{A'B'}$ shows that
$$
d\tau^{AB'}=\tau^{AL'} 
\wedge {\widetilde \Gamma}_{\; \; \; L'}^{B'}
+\tau^{LB'} \wedge \Gamma_{\; \; L}^{A},
\eqno (10.2.17)
$$
$$
dS^{AB}=-3S^{(AB} \; \Gamma_{\; \; \; C}^{C)},
\eqno (10.2.18)
$$
$$
d{\widetilde S}^{A'B'}=-3{\widetilde S}^{(A'B'} \; 
{\widetilde \Gamma}_{\; \; \; \; C'}^{C')},
\eqno (10.2.19)
$$
and two {\it independent} curvature forms are obtained as
$$ \eqalignno{
R_{\; \; B}^{A}& \equiv
d\Gamma_{\; \; B}^{A}+\Gamma_{\; \; L}^{A} \wedge 
\Gamma_{\; \; B}^{L} \cr
&=-{1\over 2} \; \psi_{\; \; BCD}^{A} \; S^{CD}
+{R\over 24} \; S_{\; \; B}^{A}
+{1\over 2} \; \Phi_{\; \; BC'D'}^{A}
\; {\widetilde S}^{C'D'},
&(10.2.20)\cr}
$$
$$ \eqalignno{
{\widetilde R}_{\; \; \; B'}^{A'} & \equiv
d{\widetilde \Gamma}_{\; \; \; B'}^{A'}
+{\widetilde \Gamma}_{\; \; \; L'}^{A'}
\wedge {\widetilde \Gamma}_{\; \; \; B'}^{L'} \cr
&=-{1\over 2} \; {\widetilde \psi}_{\; \; \; B'C'D'}^{A'}
\; {\widetilde S}^{C'D'}+{R\over 24} \; 
{\widetilde S}_{\; \; \; B'}^{A'}
+{1\over 2} \; \Phi_{CD \; \; \; \; B'}^{\; \; \; \; \; \; A'}
\; S^{CD} .
&(10.2.21)\cr}
$$
The spinors and scalars in (10.2.20) and (10.2.21) have the same
meaning as in the previous chapters. With the conventions in
Plebanski (1975), the Weyl spinors are obtained as
$$
\psi_{ABCD}={1\over 16} \; S_{\; \; \; AB}^{ab}
\; C_{abcd} \; S_{\; \; \; CD}^{cd}
=\psi_{(ABCD)} ,
\eqno (10.2.22)
$$
$$
{\widetilde \psi}_{A'B'C'D'}={1\over 16} \;
{\widetilde S}_{\; \; \; A'B'}^{ab}
\; C_{abcd} \;
{\widetilde S}_{\; \; \; C'D'}^{cd}
={\widetilde \psi}_{(A'B'C'D')} ,
\eqno (10.2.23)
$$
and conversely the Weyl tensor is
$$
C_{abcd}={1\over 4} \; S_{ab}^{\; \; \; AB}
\; \psi_{ABCD} \; S_{cd}^{\; \; \; CD}
+{1\over 4} \; {\widetilde S}_{ab}^{\; \; \; A'B'}
\; {\widetilde \psi}_{A'B'C'D'}
\; {\widetilde S}_{cd}^{\; \; \; C'D'}.
\eqno (10.2.24)
$$
The spinor version of the Petrov classification 
(section 2.3) is hence obtained
by stating that $k^{A}$ and $\omega^{A'}$ are the two types of
P-spinors if and only if the {\it independent} conditions hold:
$$
\psi_{ABCD} \; k^{A} \; k^{B} \; k^{C} \; k^{D}=0,
\eqno (10.2.25)
$$
$$
{\widetilde \psi}_{A'B'C'D'} \; \omega^{A'} \;
\omega^{B'} \; \omega^{C'} \; \omega^{D'}=0.
\eqno (10.2.26)
$$
For our purposes, we can omit the details about the principal null
directions, and focus instead on the classification of spinor fields
and analytic manifolds under consideration. Indeed, Plebanski
proposed to call all objects which are ${\widetilde {SL}}(2,C)$
scalars and are geometric objects with respect to $SL(2,C)$,
the {\it heavenly objects} (e.g. $S^{AB},\Gamma_{AB},\psi_{ABCD}$).
Similarly, objects which are $SL(2,C)$ scalars and behave like
geometric objects with respect to ${\widetilde {SL}}(2,C)$ belong
to the complementary world, i.e. the set of {\it hellish objects}
(e.g. ${\widetilde S}^{A'B'},{\widetilde \Gamma}_{A'B'},
{\widetilde \psi}_{A'B'C'D'}$). Last, spinor fields with (abstract)
indices belonging to both primed and unprimed spin-spaces are the
{\it earthly objects}.

With the terminology of Plebanski, a {\it weak heaven} space is
defined by the condition 
$$
{\widetilde \psi}_{A'B'C'D'}=0,
\eqno (10.2.27)
$$
and corresponds to the {\it conformally right-flat} space
of chapter three. Moreover, a {\it strong heaven} space is
a four-dimensional analytic manifold where a choice of null
tetrad exists such that
$$
{\widetilde \Gamma}_{A'B'}=0.
\eqno (10.2.28)
$$
One then has {\it a forteriori}, by virtue of (10.2.21),
the conditions (Plebanski 1975)
$$
{\widetilde \psi}_{A'B'C'D'}=0, \;
\Phi_{ABC'D'}=0, \;
R=0 .
\eqno (10.2.29)
$$
The vacuum Einstein equations are then automatically fulfilled 
in a strong heaven space, which turns out to be a right-flat
space-time in modern language. Of course, strong heaven spaces
are non-trivial if and only if the anti-self-dual Weyl spinor
$\psi_{ABCD}$ does not vanish, otherwise they reduce to flat
four-dimensional space-time.
\vskip 1cm
\centerline {\bf 10.3 First heavenly equation}
\vskip 1cm
\noindent
A space which is a strong heaven according to (10.2.28) is
characterized by a key function $\Omega$ which obeys the
so-called first heavenly equation. The basic ideas are as
follows. In the light of (10.2.19) and (10.2.28),
$d{\widetilde S}^{A'B'}$ vanishes, and hence, {\it in a simply
connected region}, an element $U^{A'B'}$ of the bundle
$\Lambda^{1}$ exists such that locally
$$
{\widetilde S}^{A'B'}=dU^{A'B'}.
\eqno (10.3.1)
$$
Thus, since
$$
{\widetilde S}^{1'1'}=2 e^{4} \wedge e^{1},
\eqno (10.3.2)
$$
$$
{\widetilde S}^{2'2'}=2 e^{3} \wedge e^{2},
\eqno (10.3.3)
$$
$$
{\widetilde S}^{1'2'}=-e^{1} \wedge e^{2}
+e^{3} \wedge e^{4},
\eqno (10.3.4)
$$
Eq. (10.3.1) leads to
$$
2 e^{4} \wedge e^{1}=dU^{1'1'},
\eqno (10.3.5)
$$
$$
2 e^{3} \wedge e^{2}=dU^{2'2'}.
\eqno (10.3.6)
$$
Now the Darboux theorem holds in our complex manifold, and
hence scalar functions $p,q,r,s$ exist such that
$$
2 e^{4} \wedge e^{1} = 2 dp \wedge dq = 2 d(p \; dq + d\tau),
\eqno (10.3.7)
$$
$$
2 e^{3} \wedge e^{2} = 2 dr \wedge ds = 2 d(r \; ds + d\sigma),
\eqno (10.3.8)
$$
$$
e^{1}\wedge e^{2} \wedge e^{3} \wedge e^{4}=
dp \wedge dq \wedge dr \wedge ds.
\eqno (10.3.9)
$$
The form of the {\it heavenly tetrad} in these coordinates is
$$
e^{1}=A \; dp +B \; dq,
\eqno (10.3.10)
$$
$$
e^{2}=G \; dr + H \; ds ,
\eqno (10.3.11)
$$
$$
e^{3}=E \; dr + F \; ds ,
\eqno (10.3.12)
$$
$$
e^{4}=-C \; dp -D \; dq .
\eqno (10.3.13)
$$
If one now inserts (10.3.10)--(10.3.13) into (10.3.7)--(10.3.9),
one finds that 
$$
AD-BC=EH-FG=1 ,
\eqno (10.3.14)
$$
which is supplemented by a set of equations resulting from the
condition $d{\widetilde S}^{1'2'}=0$. These
equations imply the existence of a function, the 
{\it first key function}, such that (Plebanski 1975)
$$
AG-CE=\Omega_{pr},
\eqno (10.3.15)
$$
$$
BG-DE=\Omega_{qr},
\eqno (10.3.16)
$$
$$
AH-CF=\Omega_{ps},
\eqno (10.3.17)
$$
$$
BH-DF=\Omega_{qs}.
\eqno (10.3.18)
$$
Thus, $E,F,G,H$ are given by
$$
E=B \; \Omega_{pr}-A \; \Omega_{qr},
\eqno (10.3.19)
$$
$$
F=B \; \Omega_{ps}-A \; \Omega_{qs},
\eqno (10.3.20)
$$
$$
G=D \; \Omega_{pr}-C \; \Omega_{qr},
\eqno (10.3.21)
$$
$$
H=D \; \Omega_{ps}-C \; \Omega_{qs}.
\eqno (10.3.22)
$$
The request of compatibility of (10.3.19)--(10.3.22) with (10.3.14)
leads to the {\it first heavenly equation}
$$
{\rm det} \; \pmatrix
{\Omega_{pr} & \Omega_{ps} \cr
\Omega_{qr} & \Omega_{qs} \cr}=1.
\eqno (10.3.23)
$$
\vskip 1cm
\centerline {\bf 10.4 Second heavenly equation}
\vskip 1cm
\noindent
A more convenient description of the heavenly tetrad is obtained
by introducing the coordinates
$$
x \equiv \Omega_{p} , \;
y \equiv \Omega_{q},
\eqno (10.4.1)
$$
and then defining
$$
A \equiv -\Omega_{pp} , \;
B \equiv -\Omega_{pq}, \;
C \equiv -\Omega_{qq}.
\eqno (10.4.2)
$$
The corresponding heavenly tetrad reads (Plebanski 1975)
$$
e^{1}=dp ,
\eqno (10.4.3)
$$
$$
e^{2}=dx+A \; dp + B \; dq ,
\eqno (10.4.4)
$$
$$
e^{3}=-dy-B \; dp -C \; dq,
\eqno (10.4.5)
$$
$$
e^{4}=-dq.
\eqno (10.4.6)
$$
Now the closure condition for ${\widetilde S}^{2'2'}:
d{\widetilde S}^{2'2'}=0$, leads to the equations
$$
A_{x}+B_{y}=0,
\eqno (10.4.7)
$$
$$
B_{x}+C_{y}=0,
\eqno (10.4.8)
$$
$$
\Bigr(AC-B^{2}\Bigr)_{x}+B_{q}-C_{p}=0,
\eqno (10.4.9)
$$
$$
\Bigr(AC-B^{2}\Bigr)_{y}-A_{q}+B_{p}=0.
\eqno (10.4.10)
$$
By virtue of (10.4.7) and (10.4.8), a function $\theta$ exists
such that
$$
A=-\theta_{yy} , \;
B=\theta_{xy} , \;
C=-\theta_{xx}.
\eqno (10.4.11)
$$
On inserting (10.4.11) into (10.4.9) and (10.4.10) one finds
$$
\partial_{w} \Bigr(\theta_{xx} \; \theta_{yy}
-\theta_{xy}^{2}+\theta_{xp}+\theta_{yq}\Bigr)=0 ,
\eqno (10.4.12)
$$
where $w=x,y$. Thus, one can write that
$$
\theta_{xx}\theta_{yy}-\theta_{xy}^{2}+\theta_{xp}
+\theta_{yq}=f_{p}(p,q),
\eqno (10.4.13)
$$
where $f$ is an arbitrary function of $p$ and $q$. This 
suggests defining the function
$$
\Theta \equiv \theta-xf ,
\eqno (10.4.14)
$$
which implies
$$
f_{p}=\Theta_{xx}\Theta_{yy}-\Theta_{xy}^{2}
+\Theta_{xp}+\Theta_{yq}+f_{p},
$$
and hence
$$
\Theta_{xx} \; \Theta_{yy} -\Theta_{xy}^{2}
+\Theta_{xp}+\Theta_{yq}=0.
\eqno (10.4.15)
$$
Equation (10.4.15) ensures that all forms 
${\widetilde S}^{A'B'}$ are closed, and is called the
{\it second heavenly equation}. Plebanski was able to find
heavenly metrics of all possible algebraically degenerate
types. An example is given by the function
$$
\Theta \equiv {\beta \over 2\alpha (\alpha-1)}
x^{\alpha} \; y^{1-\alpha}.
\eqno (10.4.16)
$$
The reader may check that such a
solution is of the type $[2-2] \otimes [-]$ if 
$\alpha=-1,2$, and is of the type $[2-1-1] \otimes [-]$ 
whenever $\alpha \not = -1,2$ (Plebanski 1975). More work on
related topics and on yet other ideas in complex general relativity
can be found in Plebanski and Hacyan (1975), Finley and Plebanski
(1976), Newman (1976), Plebanski 
and Schild (1976), Ko {\it et al}. (1977),
Boyer {\it et al}. (1978), Hansen {\it et al}. (1978), 
Tod (1980), Tod and Winicour (1980),
Finley and Plebanski (1981), Ko {\it et al}. (1981),
Sparling and Tod (1981), Bergmann and Smith (1991), Plebanski and
Przanowski (1994), Plebanski and Garcia--Compean (1995a,b).
\vskip 1cm
\centerline {\bf 10.5 Complex relativity and real solutions}
\vskip 1cm
\noindent
Another research line has dealt with real solutions of Einstein's
field equations as seen from the viewpoint of complex relativity
(Hall {\it et al}. 1985, McIntosh and Hickman 1985, Hickman and McIntosh
1986a,b, McIntosh {\it et al}. 1988). In particular, Hickman and 
McIntosh (1986a) integrated Einstein's vacuum equations in complex
relativity in a number of cases when the Weyl tensor is of type
$N \otimes N$, i.e. the left and right Weyl spinors are each of
type $N$. Three of the five metrics obtained were found to be 
complexified versions of Robinson--Trautman and two families of
plane-fronted wave real-type $N$ vacuum metrics, whereas the other
two metrics were shown to have no real slices. Moreover, in
Hickman and McIntosh (1986b) the authors integrated the vacuum Einstein
equations for integrable double Kerr--Schild (hereafter, IDKS)
spaces, and were able to show that the vacuum equations can be 
reduced to a single hyperheavenly equation (cf. section 10.4) in
terms of two potentials.

This section is devoted to a review of the fifth paper in the series,
by McIntosh {\it et al}. (1988). To begin, recall that the metric of 
IDKS spaces can be written as
$$
g=g_{0}+P\theta^{2} \otimes \theta^{2}+2R \theta^{2} \otimes 
\theta^{4}+Q \theta^{4} \otimes \theta^{4} ,
\eqno (10.5.1)
$$
where $P,Q,R$ are complex parameters, 
$g_{0}$ is a Minkowski metric,
$\theta^{2}$ and $\theta^{4}$ span an integrable codistribution and
are null with respect to both $g$ and $g_{0}$. When the condition
$$
PQ-R^{2}=0
\eqno (10.5.2)
$$
is fulfilled, the IDKS metric (10.5.1) reduces to an integrable 
single Kerr--Schild (hereafter, ISKS) metric with a null vector
${\bf l}$, and the tetrad can be aligned so that $g$ can be written 
in the form
$$
g=g_{0}+P \theta^{2} \otimes \theta^{2} ,
\eqno (10.5.3)
$$
where $P$ is complex and ${\bf l} \cdot \theta^{2}=0$.

Interestingly, a metric which is of 
the form (10.5.1) and hence is IDKS 
with respect to $g_{0}$, may be ISKS with respect to some other 
flat-space background metric, and hence may be expressed in the
form (10.5.3) {\it for some other} $g_{0}$. An intriguing problem
is the freedom of transformations which keep a particular metric
in the form (10.5.1) or (10.5.3). There is indeed a combined 
problem of coordinate freedom and tetrad freedom in choosing 
$\theta^{2}$ and $\theta^{4}$, or $\theta^{2}$.

A generalized form of the IDKS metric can be written, in local
coordinates $(u,v,x,y)$, with the help of the following tetrad:
$$
\theta^{1} \equiv dx+(G_{y}+y^{-1}{\cal G}_{y})du
+(F_{y}+y^{-1}{\cal F}_{y})dv ,
\eqno (10.5.4)
$$
$$
\theta^{2} \equiv y du ,
\eqno (10.5.5)
$$
$$
\theta^{3} \equiv dy-(G_{x}+y^{-1}{\cal G}_{x})du
-(F_{x}+y^{-1}{\cal F}_{x})dv ,
\eqno (10.5.6)
$$
$$
\theta^{4} \equiv dv+xdu ,
\eqno (10.5.7)
$$
where, denoting by $H$ and $\Omega$ two functions of the variables
$(u,v,x,y)$, one has
$$
F \equiv H_{x} ,
\eqno (10.5.8)
$$
$$
G \equiv xH_{x}+yH_{y}-3H ,
\eqno (10.5.9)
$$
$$
{\cal F} \equiv \Omega_{x} ,
\eqno (10.5.10)
$$
$$
{\cal G} \equiv x\Omega_{x}+y\Omega_{y}
-\Omega ,
\eqno (10.5.11)
$$
with the understanding that subscripts denote partial derivatives
of the function with respect to the variable occurring in the
subscript, e.g. $\Omega_{x} \equiv {\partial \Omega 
\over \partial x}$. The basis dual to (10.5.4)--(10.5.7) is
$$
D \equiv \partial_{x} ,
\eqno (10.5.12)
$$
$$
\delta \equiv \partial_{y} ,
\eqno (10.5.13)
$$
$$ \eqalignno{
\bigtriangleup & \equiv y^{-1} \left \{ \partial_{u}
-x \partial_{v}-\Bigr[G_{y}-xF_{y}+y^{-1}({\cal G}_{y}
-x {\cal F}_{y})\Bigr] \partial_{x} \right . \cr
& \left . + \Bigr[G_{x}-xF_{x}+y^{-1}({\cal G}_{x}
-x{\cal F}_{x})\Bigr]\partial_{y} \right \} ,
&(10.5.14)\cr}
$$
$$
{\widetilde \delta} \equiv \partial_{v}
-(F_{y}+y^{-1}{\cal F}_{y})\partial_{x}
+(F_{x}+x y^{-1}{\cal F}_{x})\partial_{y} .
\eqno (10.5.15)
$$
The non-vacuum IDKS metric can then be written as
$$ \eqalignno{
g&= g_{0}+2 \Bigr[xG_{x}+yG_{y}+y^{-1}(x{\cal G}_{x}
+y{\cal G}_{y})\Bigr]du \otimes du \cr 
&+4(G_{x}+F_{x}+y^{-1}{\cal G}_{x})du \otimes dv
+2(F_{x}+y^{-1}{\cal F}_{x}) dv \otimes dv ,
&(10.5.16)\cr}
$$
where
$$
g_{0}=2 \Bigr[ydx \otimes du-dy \otimes (dv+xdu)\Bigr].
\eqno (10.5.17)
$$
On evaluating the left connection one-forms for the tetrad
(10.5.4)--(10.5.7), one finds that the non-vanishing tetrad 
components of the left Weyl tensor are 
$$
\Psi_{2}=2y^{-3}{\cal F}_{x} ,
\eqno (10.5.18)
$$
$$
\Psi_{3}=({\widetilde \delta}+y^{-1}F_{x})\gamma
-(\bigtriangleup +y^{-1}F_{y})\alpha
-\lambda y^{-1} ,
\eqno (10.5.19)
$$
$$
\Psi_{4}=({\widetilde \delta}+4y^{-2}{\cal F}_{x}
+y^{-1}F_{x})\nu -[\bigtriangleup +y^{-1}F_{y}
+2y^{-2}{\cal F}_{y}]\lambda ,
\eqno (10.5.20)
$$
where
$$
\gamma \equiv y^{-2}{\cal F}_{y} , \;
\alpha \equiv y^{-2}{\cal F}_{x},
\eqno (10.5.21)
$$
$$
\lambda \equiv y^{-1}\Bigr[\Sigma_{x}+y^{-2}
({\cal F}_{x}G_{x}-F_{x}{\cal G}_{x})\Bigr] ,
\eqno (10.5.22)
$$
$$ \eqalignno{
\nu & \equiv y^{-1} \left \{ \Sigma_{y}
+y^{-2}\Bigr[{\cal F}_{x}(G_{y}+y^{-1}{\cal G}_{y}) \right . \cr
& \left . -{\cal G}_{x}(F_{y}+y^{-1}{\cal F}_{y})
+({\cal G}_{v}-{\cal F}_{u})\Bigr] \right \} ,
&(10.5.23)\cr}
$$
having denoted by $\Sigma$ the function
$$ \eqalignno{
\Sigma & \equiv (F_{x}+y^{-1}{\cal F}_{x})(G_{y}+{\cal G}_{y})
-(F_{y}+y^{-1}{\cal F}_{y})(G_{x}+y^{-1}{\cal G}_{x}) \cr
&+(G+y^{-1}{\cal G})_{v}-(F+y^{-1}{\cal F})_{u} .
&(10.5.24)\cr}
$$

Moreover, from the evaluation of the right connection one-forms,
one finds that the right Weyl tensor components are given by
$$
{\widetilde \Psi}_{0}=H_{xxxx}+y^{-1}\Omega_{xxxx} ,
\eqno (10.5.25)
$$
$$
{\widetilde \Psi}_{1}=H_{xxxy}+y^{-1}\Omega_{xxxy} ,
\eqno (10.5.26)
$$
$$
{\widetilde \Psi}_{2}=H_{xxyy}+y^{-1}\Omega_{xxyy} ,
\eqno (10.5.27)
$$
$$
{\widetilde \Psi}_{3}=H_{xyyy}+y^{-1}\Omega_{xyyy} ,
\eqno (10.5.28)
$$
$$
{\widetilde \Psi}_{4}=H_{yyyy}+y^{-1}\Omega_{yyyy} .
\eqno (10.5.29)
$$
The vacuum field equations are obtained for the following form
of $\Omega, {\cal F}$ and ${\cal G}$:
$$
\Omega=-{1\over 2}Lx^{2} ,
\eqno (10.5.30)
$$
$$
{\cal F}=-Lx ,
\eqno (10.5.31)
$$
$$
{\cal G}=-{1\over 2}Lx^{2} ,
\eqno (10.5.32)
$$
where $L$ is an arbitrary function of $u$ and $v$. The field
equations reduce then to the {\it Plebanski--Robinson equation}
$$
\Sigma=S-L H_{yy}=\lambda^{0}(u,v)x+\nu^{0}(u,v)y ,
\eqno (10.5.33)
$$
with the function $S$ given by
$$
S \equiv G_{y}F_{x}-G_{x}F_{y}+G_{v}-F_{u} ,
\eqno (10.5.34)
$$
whereas $\lambda^{0}$ and $\nu^{0}$ are arbitrary functions 
of $u$ and $v$.

Following McIntosh {\it et al}. (1988) one 
should stress that, for a given
metric and for a particular coordinate and tetrad frame, $H$ is
not unique. Both the metric described by (10.5.16) and (10.5.17),
and the Plebanski--Robinson equation (10.5.33), are invariant under
the transformation
$$
H \rightarrow H+f(u,v)y^{3}+g(u,v) .
\eqno (10.5.35)
$$
Moreover, the metric (10.5.16) is linear in $H$ and $\Omega$. This
implies that, for some known vacuum metrics (e.g. Schwarzschild)
$H$ can be written in the form
$$
H=H_{m}+H_{0} ,
\eqno (10.5.36)
$$
where $H_{0}$ is the $H$ function for a form of the flat-space
metric and is proportional to the curvature constant, whereas
$H_{m}$ is proportional to the mass constant.

Interestingly, different coordinate versions of flat-space metrics
are obtained when dealing with various forms of both complex and
complexified metrics. In McIntosh {\it et al}. (1988), three forms of 
$H$ are derived which generate flat space and are hence denoted 
by $H_{0}$. They are as follows.
\vskip 0.3cm
\noindent
(i) {\bf First form of $H_{0}$}.
$$
H_{0}={k\over 4}(x^{2}-2y^{2}) ,
\eqno (10.5.37)
$$
where $k$ is a real parameter. The resulting metric can be 
written as
$$
g=g_{0}+k(2y^{2}-x^{2})du \otimes du
+k dv \otimes dv ,
\eqno (10.5.38)
$$
where the metric $g_{0}$ reads
$$
g_{0}=2 \Bigr[ydx \otimes du -x du \otimes dy
-dy \otimes dv \Bigr] .
\eqno (10.5.39)
$$
The metric $g$ is an IDKS metric with respect to $g_{0}$, and
$du$ and $dv$ span an integrable codistribution.
\vskip 0.3cm
\noindent
(ii) {\bf Second form of $H_{0}$} 
$$
H_{0}=0 .
\eqno (10.5.40)
$$
The corresponding metric can be written in the form
$$
g=g_{0}=2 \Bigr[d\xi \otimes d\eta -d\zeta \otimes 
d{\widetilde \zeta}\Bigr] ,
\eqno (10.5.41)
$$
with coordinate transformation
$$
\xi \sqrt{2k}=-\left({x\over \sqrt{2}}-y+kv \right),
\eqno (10.5.42)
$$
$$
\eta \sqrt{2k}=\left({x\over \sqrt{2}}+y-kv \right),
\eqno (10.5.43)
$$
$$
\zeta \sqrt{2k}=\left({x\over \sqrt{2}}+y \right)
e^{ku\sqrt{2}} ,
\eqno (10.5.44)
$$
$$
{\widetilde \zeta}\sqrt{2k}=-\left({x\over \sqrt{2}}
-y \right)e^{-ku \sqrt{2}} .
\eqno (10.5.45)
$$
\vskip 0.3cm
\noindent
(iii) {\bf Third form of $H_{0}$}
$$
H_{0}={k\over 2}{(UX+V)^{2}\over U^{4}} ,
\eqno (10.5.46)
$$
where the coordinates $(X,Y,U,V)$ replace $(x,y,u,v)$. One
then finds that
$$
g=g_{0}+2k \Bigr[d(V/U) \otimes d(V/U)
-2H_{0} dU \otimes dU \Bigr] ,
\eqno (10.5.47)
$$
with the metric $g_{0}$ having the form
$$
g_{0}=2 \Bigr[YdX \otimes dU -X dU \otimes dY
-dY \otimes dV \Bigr] .
\eqno (10.5.48)
$$
The coordinate transformation which 
relates $X,Y,U,V$ and $\xi, \eta,
\zeta$ and $\widetilde \zeta$ used in (10.5.42)--(10.5.45)
can be shown to be
$$
\xi=X ,
\eqno (10.5.49)
$$
$$
\eta=UY-k {(2V+UX)\over U} ,
\eqno (10.5.50)
$$
$$
\zeta=Y-k {(V+UX)\over U^{2}} ,
\eqno (10.5.51)
$$
$$
{\widetilde \zeta}=V+UX .
\eqno (10.5.52)
$$
\vskip 1cm
\centerline {\bf 10.6 Multimomenta in 
complex general relativity}
\vskip 1cm
\noindent
Among the various approaches to the quantization of the
gravitational field, much insight has been gained by the
use of twistor theory and Hamiltonian techniques. 
For example, it is by now well known
how to reconstruct an anti-self-dual space-time from
deformations of flat projective twistor space
(chapter five), and the various
definitions of twistors in curved space-time enable one to
obtain relevant information about complex space-time geometry
within a holomorphic, conformally invariant framework
(chapter nine).
Moreover, the recent approaches to canonical gravity described
in Ashtekar (1991) have led to many exact solutions of the quantum 
constraint equations of general relativity, although their
physical relevance for the quantization program remains
unclear. A basic difference between the Penrose formalism
and the Ashtekar formalism is as follows.
The twistor program refers to a four-complex-dimensional
complex-Riemannian manifold with holomorphic metric, 
holomorphic connection and holomorphic curvature tensor,
where the complex Einstein equations are imposed. By contrast,
in the recent approaches to canonical gravity, one studies
complex tetrads on a four-real-dimensional Lorentzian
manifold, and real general relativity may be recovered
provided that one is able to impose suitable reality conditions.
The aim of this section is to describe a new property of
complex general relativity within the holomorphic framework
relevant for twistor theory, whose derivation results from
recent attempts to obtain a manifestly covariant formulation
of Ashtekar's program (Esposito {\it et al}. 1995, Esposito and
Stornaiolo 1995).

Indeed, it has been recently shown in 
Esposito {\it et al}. (1995) that the constraint
analysis of general relativity may be performed by using
multisymplectic techniques, without relying on a 3+1 split
of the space-time four-geometry. The constraint equations 
have been derived while paying attention to boundary terms, and
the Hamiltonian constraint turns out to be linear in the
{\it multimomenta} (see below). While the latter property is more
relevant for the (as yet unknown) quantum theory of gravitation,
the former result on boundary terms deserves further thinking 
already at the classical level, and is the object of our
investigation. 

We here write the Lorentzian space-time four-metric as
$$
g_{ab}=e_{a}^{\; \; {\hat c}} \; e_{b}^{\; \; {\hat d}}
\; \eta_{{\hat c}{\hat d}},
\eqno (10.6.1)
$$
where $e_{a}^{\; \; {\hat c}}$ is the tetrad and $\eta$ is
the Minkowski metric. In first-order formalism, the tetrad 
$e_{a}^{\; \; {\hat c}}$ and the connection one-form
$\omega_{a}^{\; \; {\hat b}{\hat c}}$ are regarded as
independent variables. In Esposito {\it et al}. (1995)
it has been shown that, on
using jet-bundle formalism and covariant multimomentum 
maps, the constraint equations of real general relativity
hold on an {\it arbitrary} three-real-dimensional 
hypersurface $\Sigma$ provided that one of the following
three conditions holds:
\vskip 0.3cm
\noindent
(i) $\Sigma$ has no boundary;
\vskip 0.3cm
\noindent
(ii) the multimomenta
$$
{\tilde p}_{\; \; \; {\hat c}{\hat d}}^{ab}
\equiv e \Bigr(e_{\; \; {\hat c}}^{a} \;
e_{\; \; {\hat d}}^{b}
-e_{\; \; {\hat c}}^{b} \; e_{\; \; {\hat d}}^{a}
\Bigr)
$$ 
vanish at $\partial \Sigma$, $e$ being the
determinant of the tetrad;
\vskip 0.3cm
\noindent
(iii) an element of the algebra $o(3,1)$
corresponding to the gauge group, represented by
the antisymmetric $\lambda^{{\hat a}{\hat b}}$,
vanishes at $\partial \Sigma$, and the connection one-form 
$\omega_{a}^{\; \; {\hat b}{\hat c}}$ or $\xi^{b}$ vanishes
at $\partial \Sigma$, $\xi$ being a vector field describing
diffeomorphisms on the base-space.
\vskip 0.3cm
\noindent
In other words, boundary terms may occur in the constraint
equations of real general relativity, and they result from 
the total divergences of 
$$
\sigma^{ab} \equiv {\tilde p}_{\; \; \; {\hat c}{\hat d}}^{ab}
\; \lambda^{{\hat c}{\hat d}},
\eqno (10.6.2)
$$
$$
\rho^{ab} \equiv {\tilde p}_{\; \; \; {\hat c}{\hat d}}^{ab}
\; \omega_{f}^{\; \; {\hat c}{\hat d}}
\; \xi^{f},
\eqno (10.6.3)
$$
integrated over $\Sigma$.

In two-component spinor language, denoting by 
$\tau_{\; \; BB'}^{{\hat a}}$ the Infeld--van der Waerden
symbols, the two-spinor version of the tetrad reads
$$
e_{\; \; BB'}^{a} \equiv e_{\; \; {\hat a}}^{a} \;
\tau_{\; \; BB'}^{{\hat a}},
\eqno (10.6.4)
$$
which implies that $\sigma^{ab}$ defined in (10.6.2) takes the form
$$
\sigma^{ab}=e \Bigr(e_{\; \; CC'}^{a} \;
e_{\; \; DD'}^{b}-e_{\; \; DD'}^{a} \;
e_{\; \; CC'}^{b}\Bigr)
\tau_{{\hat a}}^{\; \; CC'} \;
\tau_{{\hat b}}^{\; \; DD'}
\; \lambda^{{\hat a}{\hat b}}.
\eqno (10.6.5)
$$
Thus, on defining the spinor field
$$
\lambda^{CC'DD'} \equiv \tau_{{\hat a}}^{\; \; CC'}
\; \tau_{{\hat b}}^{DD'} \; \lambda^{{\hat a}{\hat b}}
\equiv \Lambda_{1}^{(CD)} \; \varepsilon^{C'D'}
+\Lambda_{2}^{(C'D')} \; \varepsilon^{CD},
\eqno (10.6.6)
$$
the first of the boundary conditions in (iii) is satisfied
provided that 
$$
\Lambda_{1}^{(CD)}=0
$$ 
at $\partial \Sigma$ in real
general relativity, since then $\Lambda_{2}^{(C'D')}$ is obtained
by complex conjugation of $\Lambda_{1}^{(CD)}$, and hence the
condition $\Lambda_{2}^{(C'D')}=0$ at $\partial \Sigma$ leads to
no further information.

In the {\it holomorphic} framework, however, no complex conjugation
relating primed to unprimed spin-space can be defined, since
such a map is not invariant under holomorphic coordinate 
transformations (chapter three). Hence spinor fields belonging to
unprimed or primed spin-space are {\it totally independent},
and the first of the boundary conditions in (iii) reads
$$
\Lambda^{(CD)}=0 
\; {\rm at} \; \partial \Sigma_{c} ,
\eqno (10.6.7)
$$
$$
{\widetilde \Lambda}^{(C'D')}=0
\; {\rm at} \; \partial \Sigma_{c} ,
\eqno (10.6.8)
$$
where $\partial \Sigma_{c}$ is a two-complex-dimensional
complex surface, bounding the three-complex-dimensional
surface $\Sigma_{c}$, and the {\it tilde} is used to
denote {\it independent} spinor fields, not related
by any conjugation.

Similarly, $\rho^{ab}$ defined in (10.6.3) takes the form
$$
\rho^{ab}=e \Bigr(e_{\; \; CC'}^{a} \;
e_{\; \; DD'}^{b}
-e_{\; \; DD'}^{a} \; e_{\; \; CC'}^{b} \Bigr)
\Bigr(\Omega_{f}^{(CD)} \; \varepsilon^{C'D'}
+{\widetilde \Omega}_{f}^{(C'D')} \;
\varepsilon^{CD}\Bigr)\xi^{f} ,
\eqno (10.6.9)
$$
and hence the second of the boundary conditions in (iii)
leads to the {\it independent} boundary conditions
$$
\Omega_{f}^{(CD)}=0 
\; {\rm at} \; \partial \Sigma_{c},
\eqno (10.6.10)
$$
$$
{\widetilde \Omega}_{f}^{(C'D')}=0
\; {\rm at} \; \partial \Sigma_{c},
\eqno (10.6.11)
$$
in complex general relativity.
 
The resulting picture of complex general relativity is
highly non-trivial. One starts from a one-jet bundle $J^{1}$
which, in local coordinates, is described by a holomorphic
coordinate system, with holomorphic tetrad, holomorphic
connection one-form $\omega_{a}^{\; \; {\hat b}{\hat c}}$,
multivelocities corresponding to the tetrad and 
multivelocities corresponding to 
$\omega_{a}^{\; \; {\hat b}{\hat c}}$, both of holomorphic
nature. The intrinsic form of the field equations, which is a
generalization of a mathematical structure already existing in
classical mechanics, leads to the complex vacuum Einstein
equations $R_{ab}=0$, and to a condition on the covariant
divergence of the multimomenta. Moreover, the covariant
multimomentum map, 
evaluated on a section of $J^{1}$ and integrated on an
arbitrary three-complex-dimensional surface $\Sigma_{c}$,
reflects the
invariance of complex general relativity under all holomorphic
coordinate transformations. Since space-time is now a complex
manifold, one deals with holomorphic coordinates which are
all on the same footing, and hence no time coordinate can be
defined. Thus, the counterpart of the constraint equations
results from the holomorphic
version of the covariant multimomentum map, but cannot be related
to a Cauchy problem as in the Lorentzian theory. 
In particular, the Hamiltonian constraint
of Lorentzian general relativity is replaced by a geometric
structure which 
is linear in the holomorphic multimomenta, provided that 
two boundary terms can be set to
zero (of course, our multimomenta are holomorphic by
construction, since in complex general relativity the
tetrad is holomorphic). 
For this purpose, one of the following three conditions
should hold:
\vskip 0.3cm
\noindent
(i) $\Sigma_{c}$ has no boundary;
\vskip 0.3cm
\noindent
(ii) the holomorphic multimomenta vanish at 
$\partial \Sigma_{c}$;
\vskip 0.3cm
\noindent
(iii) the equations (10.6.7) and (10.6.8) hold at $\partial \Sigma_{c}$,
as well as the equations (10.6.10) and (10.6.11).
The latter equations may be replaced by the condition
$u^{AA'}=0$ at $\partial \Sigma_{c}$, where $u$ is a
holomorphic vector field describing holomorphic coordinate
transformations on the base-space, i.e. on complex space-time.
\vskip 0.3cm
Note that it is not {\it a priori} obvious that the
three-complex-dimensional surface $\Sigma_{c}$ has no 
boundary. Hence one really has to consider the boundary
conditions (ii) or (iii) in the holomorphic framework.
They imply that the holomorphic multimomenta have to
vanish everywhere on $\Sigma_{c}$ (by virtue of a 
well known result in complex analysis), or the elements
of $o(4,C)$ have to vanish everywhere on $\Sigma_{c}$,
jointly with the self-dual and anti-self-dual parts of the
connection one-form. The latter of these conditions may be
replaced by the vanishing of the holomorphic vector field
$u$ on $\Sigma_{c}$. In other words, if $\Sigma_{c}$ has a
boundary, unless the holomorphic multimomenta vanish on
the whole of $\Sigma_{c}$, there are restrictions 
at $\Sigma_{c}$ on the
spinor fields expressing the holomorphic nature of the 
theory and its invariance under all holomorphic 
coordinate transformations. Indeed, already in real Lorentzian
four-manifolds one faces a choice between boundary conditions on
the multimomenta and restrictions on the invariance group
resulting from boundary effects. We choose the former,
following Esposito and Stornaiolo (1995), and emphasize their role in
complex general relativity. Of course, the spinor fields
involved in the boundary conditions are instead non-vanishing
on the four-complex-dimensional space-time.

Remarkably, to ensure that the holomorphic multimomenta
${\tilde p}_{\; \; \;{\hat c}{\hat d}}^{ab}$ vanish at
$\partial \Sigma_{c}$, and hence on $\Sigma_{c}$ as well,
the determinant $e$ of the tetrad should vanish at
$\partial \Sigma_{c}$, or
$e^{-1} \; {\tilde p}_{\; \; \; {\hat c}{\hat d}}^{ab}$
should vanish at $\partial \Sigma_{c}$. The former case
admits as a subset the totally null two-complex-dimensional
surfaces known as $\alpha$-surfaces and $\beta$-surfaces
(chapter four). Since the integrability condition for 
$\alpha$-surfaces is expressed by the vanishing of the
self-dual Weyl spinor, our formalism enables one to recover
the anti-self-dual (also called right-flat) space-time
relevant for twistor theory, where both the Ricci spinor and
the self-dual Weyl spinor vanish. However, if 
$\partial \Sigma_{c}$ is not totally null, the resulting theory
does not correspond to twistor theory. 
The latter case implies that the tetrad vectors are turned
into holomorphic vectors $u_{1},u_{2},u_{3},u_{4}$ such that
one of the following conditions holds
at $\partial \Sigma_{c}$, and hence on $\Sigma_{c}$ as well:
(i) $u_{1}=u_{2}=u_{3}=u_{4}=0$;
(ii) $u_{1}=u_{2}=u_{3}=0, u_{4} \not = 0$;
(iii) $u_{1}=u_{2}=0, u_{3}=\gamma u_{4}, \gamma \in {\cal C}$;
(iv) $u_{1}=0, \gamma_{2}u_{2}=\gamma_{3}u_{3}
=\gamma_{4}u_{4}, \gamma_{i} \in {\cal C}, i=2,3,4$;
(v) $\gamma_{1}u_{1}=\gamma_{2}u_{2}=\gamma_{3}u_{3}
=\gamma_{4}u_{4}, \gamma_{i} \in {\cal C}, i=1,2,3,4$.
 
It now appears important to
understand the relation between complex general relativity
derived from jet-bundle theory and complex general
relativity as in the Penrose twistor program. 
For this purpose, one has to study the topology and the
geometry of the space of two-complex-dimensional surfaces
$\partial \Sigma_{c}$ in the generic case. This leads to
a deep link between complex space-times which are not
anti-self-dual and two-complex-dimensional surfaces 
which are not totally null. In other words, on going beyond
twistor theory, one finds that the analysis of 
two-complex-dimensional surfaces still plays a key role.
Last, but not least, one has to solve equations which are now
linear in the {\it holomorphic multimomenta}, both in
classical and in quantum gravity (these equations correspond
to the constraint equations of the Lorentzian theory). 
Hence this analysis seems to add evidence in favour of new
perspectives being in sight in relativistic theories of gravitation.

For other recent developments in complex, spinor and twistor geometry, 
we refer the reader to the work in Lewandowski {\it et al}. (1990, 1991),
Dunajski and Mason (1997), Nurowski (1997), 
Tod and Dunajski (1997), Penrose (1997), Dunajski (1999), 
Frauendiener and Sparling (1999).
\vskip 100cm
\centerline {\bf APPENDIX A: Clifford algebras}
\vskip 1cm
\noindent
In section 7.4 we have defined the total Dirac operator in
Riemannian geometries as the first-order elliptic operator
whose action on the sections is given by composition of
Clifford multiplication with covariant differentiation.
Following Ward and Wells (1990), this appendix presents a
self-contained description of Clifford algebras and Clifford
multiplication.

Let $V$ be a real vector space equipped with an inner product
$\langle \; , \; \rangle$, defined 
by a non-degenerate quadratic form $Q$
of signature $(p,q)$. Let $T(V)$ be the tensor algebra of $V$
and consider the ideal $\cal I$ in $T(V)$ generated by
$x \otimes x +Q(x)$. By definition, $\cal I$ consists of sums
of terms of the kind 
$a \otimes \Bigr \{x \otimes x + Q(x) \Bigr \} 
\otimes b$, $x \in V, a,b \in T(V)$. The quotient space
$$
Cl(V) \equiv Cl(V,Q) \equiv T(V)/{\cal I}
\eqno ({\rm A}.1)
$$
is the Clifford algebra of the vector space $V$ equipped with
the quadratic form $Q$. The product induced by the tensor product
in $T(V)$ is known as Clifford multiplication or the Clifford
product and is denoted by $x \cdot y$, for $x,y \in Cl(V)$.
The dimension of $Cl(V)$ is $2^{n}$ if dim$(V)=n$. A basis for
$Cl(V)$ is given by the scalar 1 and the products
$$
e_{i_{1}} \cdot e_{i_{2}} \cdot e_{i_{n}} \; \; \; \; \; \;
i_{1}<...<i_{n} ,
$$
where $\Bigr \{e_{1},...,e_{n}\Bigr \}$ is an orthonormal
basis for $V$. Moreover, the products satisfy
$$
e_{i} \cdot e_{j}+e_{j} \cdot e_{i}=0
\; i \not = j ,
\eqno ({\rm A}.2)
$$
$$
e_{i} \cdot e_{i}=-2 \langle e_{i},e_{i} \rangle
\; i=1,...,n .
\eqno ({\rm A}.3)
$$
As a vector space, $Cl(V)$ is isomorphic to $\Lambda^{*}(V)$,
the Grassmann algebra, with
$$
e_{i_{1}} ... e_{i_{n}} \longrightarrow
e_{i_{1}} \wedge ... \wedge e_{i_{n}} .
$$
There are two natural {\it involutions} on $Cl(V)$. The first,
denoted by $\alpha:Cl(V) \rightarrow Cl(V)$, is induced by the
involution $x \rightarrow -x$ defined on $V$, which extends to
an automorphism of $Cl(V)$. The eigenspace of $\alpha$ with
eigenvalue $+1$ consists of the even elements of $Cl(V)$, and the
eigenspace of $\alpha$ of eigenvalue $-1$ consists of the odd
elements of $Cl(V)$.

The second involution is a mapping $x \rightarrow x^{t}$, induced
on generators by
$$
\Bigr(e_{i_{1}} ... e_{i_{p}}\Bigr)^{t}
=e_{i_{p}} ... e_{i_{1}} ,
$$
where $e_{i}$ are basis elements of $V$. Moreover, we define
$x \rightarrow {\overline x}$, a third involution of
$Cl(V)$, by ${\overline x} \equiv \alpha(x^{t})$. 

One then defines $Cl^{*}(V)$ to be the group of invertible 
elements of $Cl(V)$, and the Clifford group $\Gamma(V)$ is
the subgroup of $Cl^{*}(V)$ defined by
$$
\Gamma(V) \equiv \biggr \{x \in Cl^{*}(V):
y \in V \Rightarrow \alpha(x)yx^{-1} \in V \biggr \} .
\eqno ({\rm A}.4)
$$
One can show that the map $\rho : V \rightarrow V$
given by $\rho(x)y=\alpha(x)yx^{-1}$ is an isometry of $V$
with respect to the quadratic form $Q$. The map 
$x \rightarrow \| x \| \equiv 
x{\overline x}$ is the square-norm map,
and enables one to define a remarkable subgroup of
the Clifford group, i.e.
$$
{\rm Pin}(V) \equiv \biggr \{x \in \Gamma(V):
\| x \| =1 \biggr \} .
\eqno ({\rm A}.5)
$$
\vskip 100cm
\centerline {\bf APPENDIX B: Rarita--Schwinger equations}
\vskip 1cm
\noindent
Following Aichelburg and Urbantke (1981), 
one can express the $\Gamma$-potentials of (8.6.1) as
$$
\Gamma_{\; \; BB'}^{A}=\nabla_{BB'} \; \alpha^{A} .
\eqno ({\rm B}.1)
$$
Thus, acting with $\nabla_{CC'}$ on both sides of (B.1), 
symmetrizing over $C'B'$
and using the spinor Ricci identity (8.7.6), one finds
$$
\nabla_{C(C'} \; \Gamma_{\; \; \; \; \; B')}^{AC}
={\widetilde \Phi}_{B'C'L}^{\; \; \; \; \; \; \; \; \; \; A} \;
\alpha^{L} .
\eqno ({\rm B}.2)
$$
Moreover, acting with $\nabla_{C}^{\; \; C'}$ on both sides of (B.1),
putting $B'=C'$ (with contraction over this index), and using the
spinor Ricci identity (8.7.4) leads to
$$
\varepsilon^{AB} \; \nabla_{(C}^{\; \; \; C'} \; 
\Gamma_{\mid A \mid B)C'}=-3\Lambda \; \alpha_{C} .
\eqno ({\rm B}.3)
$$
Equations (B.1)--(B.3) rely on the conventions 
in Aichelburg and Urbantke (1981). However,
to achieve agreement with the conventions in Penrose (1994)
and in our paper, the equations (8.6.3)--(8.6.6) are obtained
by defining (cf. (B.1))
$$
\Gamma_{B \; \; \; \; B'}^{\; \; \; A}
\equiv \nabla_{BB'} \; \alpha^{A} ,
\eqno ({\rm B}.4)     
$$
and similarly for the $\gamma$-potentials of (8.6.2) (for the effect of
torsion terms, see comments following equation (21) in 
Aichelburg and Urbantke (1981)).
\vskip 100cm
\centerline {\bf APPENDIX C: Fibre bundles}
\vskip 1cm
\noindent
The basic idea in fibre-bundle theory is to deal with
topological spaces which are locally, but not necessarily
globally, a product of two spaces. This appendix begins
with the definition of fibre bundles and the reconstruction
theorem for bundles, jointly with a number of examples,
following Nash and Sen (1983). A more formal presentation
of some related topics is then given, for completeness.

A fibre bundle may be defined as the collection of the
following five mathematical objects:
\vskip 0.3cm
\noindent
(1) A topological space $E$ called the total space.
\vskip 0.3cm
\noindent
(2) A topological space $X$, i.e. the base space, and a
projection $\pi: E \rightarrow X$ of $E$ onto $X$.
\vskip 0.3cm
\noindent
(3) A third topological space $F$, i.e. the fibre.
\vskip 0.3cm
\noindent
(4) A group $G$ of homeomorphisms of $F$, called the
structure group.
\vskip 0.3cm
\noindent
(5) A set $\left \{ U_{\alpha} \right \}$ of open
coordinate neighbourhoods which cover $X$. These
reflect the {\it local} product structure of $E$.
Thus, a homeomorphism $\phi_{\alpha}$ is given
$$
\phi_{\alpha}: \pi^{-1}(U_{\alpha})
\rightarrow U_{\alpha} \times F ,
\eqno ({\rm C}.1)
$$
such that the composition of the projection map
$\pi$ with the inverse of $\phi_{\alpha}$ yields
points of $U_{\alpha}$, i.e.
$$
\pi \; \phi_{\alpha}^{-1}(x,f)=x
\; x \in U_{\alpha} , \;
f \in F .
\eqno ({\rm C}.2)
$$

To see how this abstract definition works, let us focus on
the M\"{o}bius strip, which can be obtained by twisting
ends of a rectangular strip before joining them. In this
case, the base space $X$ is the circle $S^1$, while the
fibre $F$ is a line segment. For any $x \in X$, the action
of $\pi^{-1}$ on $x$ yields the fibre over $x$. The
structure group $G$ appears on going from local coordinates
$\Bigr(U_{\alpha},\phi_{\alpha}\Bigr)$ to local coordinates
$\Bigr(U_{\beta},\phi_{\beta}\Bigr)$. If $U_{\alpha}$ and
$U_{\beta}$ have a non-empty intersection, then
$\phi_{\alpha} \circ \phi_{\beta}^{-1}$ is a continuous
invertible map
$$
\phi_{\alpha} \circ \phi_{\beta}^{-1}:
\Bigr(U_{\alpha} \cap U_{\beta}\Bigr) \times F
\rightarrow 
\Bigr(U_{\alpha} \cap U_{\beta}\Bigr) \times F .
\eqno ({\rm C}.3)
$$
For fixed $x \in U_{\alpha} \cap U_{\beta}$, such a map
becomes a map $h_{\alpha \beta}$ from $F$ to $F$. This is,
by definition, the transition function, and yields a
homeomorphism of the fibre $F$. The structure group
$G$ of $E$ is then defined as the set of all these maps
$h_{\alpha \beta}$ for {\it all} choices of local
coordinates $\Bigr(U_{\alpha},\phi_{\alpha}\Bigr)$.
Here, it consists of just two elements 
$\left \{e,h \right \}$. This is best seen on considering
the covering $\left \{ U_{\alpha} \right \}$ which is
given by two open arcs of $S^{1}$ denoted by $U_{1}$
and $U_{2}$. Their intersection consists of two disjoint
open arcs $A$ and $B$, and hence the transition functions
$h_{\alpha \beta}$ are found to be
$$
h_{12}(x)=e \; {\rm if}
\; x \in A , \;
h \; {\rm if} \;  
x \in B ,
\eqno ({\rm C}.4)
$$
$$
h_{12}(x)=h_{21}^{-1}(x) ,
\eqno ({\rm C}.5)
$$
$$
h_{11}(x)=h_{22}(x)=e .
\eqno ({\rm C}.6)
$$
To detect the group $G=\left \{e,h \right \}$ it is
enough to move the fibre once round the M\"{o}bius
strip. By virtue of this operation, $F$ is reflected
in its midpoint, which implies that the group element
$h$ is responsible for such a reflection. Moreover,
on squaring up the reflection one obtains the identity
$e$, and hence $G$ has indeed just two elements.

So far, our definition of a bundle involves the total
space, the base space, the fibre, the structure group
and the set of open coordinate neighbourhoods covering
the base space. However, the essential information about
a fibre bundle can be obtained from a smaller set of
mathematical objects, i.e. the base space, the fibre,
the structure group and the transition functions 
$h_{\alpha \beta}$. Following again Nash and Sen (1983)
we now prove the reconstruction theorem for bundles,
which tells us how to obtain the total space $E$, the
projection map $\pi$ and the homeomorphisms $\phi_{\alpha}$
from $\biggr(X,F,G,\left \{ h_{\alpha \beta} \right \}
\biggr)$. 

First, $E$ is obtained from an equivalence relation,
as follows. One considers the set $\widetilde E$ defined
as the union of all products of the form
$U_{\alpha} \times F$, i.e.
$$
{\widetilde E} \equiv \bigcup_{\alpha} U_{\alpha} \times F .
\eqno ({\rm C}.7)
$$
One here writes $(x,f)$ for an element of $\widetilde E$,
where $x \in U_{\alpha}$. An equivalence relation $\sim$
is then introduced by requiring that, given 
$(x,f) \in U_{\alpha} \times F$ and 
$(x',f') \in U_{\beta} \times F$, these elements are
equivalent,
$$
(x,f) \sim (x',f') ,
\eqno ({\rm C}.8)
$$
if
$$
x=x' \; \; \; \; {\rm and} \; \; \; \;
h_{\alpha \beta}(x)f=f' .
\eqno ({\rm C}.9)
$$
This means that the transition functions enable one to
pass from $f$ to $f'$, while the points $x$ and $x'$
coincide. The desired total space $E$ is hence given as
$$
E \equiv {\widetilde E}/ \sim ,
\eqno ({\rm C}.10)
$$
i.e. $E$ is the set of all equivalence classes under
$\sim$.

Second, denoting by $[(x,f)]$ the equivalence class 
containing the element $(x,f)$ of $U_{\alpha} \times F$,
the projection $\pi: E \rightarrow X$ is defined as the
map
$$
\pi : [(x,f)] \rightarrow x .
\eqno ({\rm C}.11)
$$
In other words, $\pi$ maps the equivalence class $[(x,f)]$ 
into $x \in U_{\alpha}$.

Third, the function $\phi_{\alpha}$ is defined (indirectly) by
giving its inverse
$$
\phi_{\alpha}^{-1}: U_{\alpha} \times F \rightarrow
\pi^{-1}(U_{\alpha}) .
\eqno ({\rm C}.12)
$$
Note that, by construction, $\phi_{\alpha}^{-1}$ 
satisfies the condition
$$
\pi \; \phi_{\alpha}^{-1}(x,f)=x \in U_{\alpha} ,
\eqno ({\rm C}.13)
$$
and this is what we actually need, despite one might be
tempted to think in terms of $\phi_{\alpha}$ rather than
its inverse.

The readers who are not familiar with fibre-bundle theory
may find it helpful to see an application of this reconstruction
theorem. For this purpose, we focus again on the M\"{o}bius
strip. Thus, our data are the base $X=S^{1}$, a line segment
representing the fibre, the structure group
$\left \{e,h \right \}$, where $h$ is responsible for $F$ being
reflected in its midpoint, and the transition functions
$h_{\alpha \beta}$ in (C.4)--(C.6). Following the definition
(C.8) and (C.9) of equivalence relation, and bearing in mind that
$h_{12}=h$, one finds
$$
f=f' \; {\rm if} \;
x \in A ,
\eqno ({\rm C}.14)
$$
$$
hf=f' \; {\rm if} \; 
x \in B ,
\eqno ({\rm C}.15)
$$
where $A$ and $B$ are the two open arcs whose disjoint
union gives the intersection of the covering arcs 
$U_{1}$ and $U_{2}$. In the light of (C.14) and (C.15), if
$x \in A$ then the equivalence class $[(x,f)]$ consists of
$(x,f)$ only, whereas, if $x \in B$, $[(x,f)]$ consists of
two elements, i.e. $(x,f)$ and $(x,hf)$. Hence it should
be clear how to construct the total space $E$ by using 
equivalence classes, according to (C.10). What happens 
can be divided into three steps (Nash and Sen 1983):
\vskip 0.3cm
\noindent
(i) The base space splits into two, and one has the 
covering arcs $U_{1},U_{2}$ and the intersection regions
$A$ and $B$.
\vskip 0.3cm
\noindent
(ii) The space $\widetilde E$ defined in (C.7) splits
into two. The regions $A \cap F$ are glued together 
without a twist, since the equivalence class 
$[(x,f)]$ has only the element $(x,f)$ if $x \in A$.
By contrast, a twist is necessary to glue together the
regions $B \cap F$, since $[(x,f)]$ consists of two 
elements if $x \in B$. The identification of $(x,f)$
and $(x,hf)$ under the action of $\sim$, makes it
necessary to glue with twist the regions $B \cap F$.
\vskip 0.3cm
\noindent
(iii) The bundle $E \equiv {\widetilde E}/ \sim$ has
been obtained. Shaded regions may be drawn, which are
isomorphic to $A \cap F$ and $B \cap F$, respectively.
\vskip 0.3cm
If we now come back to the general theory of fibre
bundles, we should mention some important properties of 
the transition functions $h_{\alpha \beta}$. They obey
a set of compatibility conditions, where repeated indices
are not summed over, i.e.
$$
h_{\alpha \alpha}(x)=e , \;
x \in U_{\alpha} ,
\eqno ({\rm C}.16)
$$
$$
h_{\alpha \beta}(x)=(h_{\beta \alpha}(x))^{-1} , \;
x \in U_{\alpha} \cap U_{\beta} ,
\eqno ({\rm C}.17)
$$
$$
h_{\alpha \beta}(x) \; h_{\beta \gamma}(x)
=h_{\alpha \gamma}(x) , \;
x \in U_{\alpha} \cap U_{\beta} \cap U_{\gamma} .
\eqno ({\rm C}.18)
$$
A simple calculation can be now made which shows that
any bundle can be actually seen as an equivalence
class of bundles. The underlying argument is as follows.
Suppose two bundles $E$ and $E'$ are given, with the same
base space, fibre, and group. Moreover, let 
$\left \{\phi_{\alpha},U_{\alpha} \right \}$ and
$\left \{\psi_{\alpha},U_{\alpha} \right \}$ be the
sets of coordinates and coverings for $E$ and $E'$,
respectively. The map
$$
\lambda_{\alpha} \equiv \phi_{\alpha} \circ
\psi_{\alpha}^{-1}: U_{\alpha} \times F
\rightarrow U_{\alpha} \times F
$$ 
is now required to be a homeomorphism of $F$ belonging
to the structure group $G$. Thus, if one combines the
definitions
$$
\lambda_{\alpha}(x) \equiv \phi_{\alpha} \circ 
\psi_{\alpha}^{-1}(x) ,
\eqno ({\rm C}.19)
$$
$$
h_{\alpha \beta}(x) \equiv 
\phi_{\alpha} \circ \phi_{\beta}^{-1}(x) ,
\eqno ({\rm C}.20)
$$
$$
h_{\alpha \beta}'(x) \equiv 
\psi_{\alpha} \circ \psi_{\beta}^{-1}(x) ,
\eqno ({\rm C}.21)
$$
one finds
$$
\lambda_{\alpha}^{-1}(x)h_{\alpha \beta}(x)\lambda_{\beta}(x)
=\psi_{\alpha} \circ \phi_{\alpha}^{-1} \circ
\phi_{\alpha} \circ \phi_{\beta}^{-1} \circ
\phi_{\beta} \circ \psi_{\beta}^{-1}(x)
=h_{\alpha \beta}'(x) .
\eqno ({\rm C}.22)
$$
Thus, since $\lambda_{\alpha}$ belongs to the structure
group $G$ by hypothesis, as the transition function
$h_{\alpha \beta}$ varies, both 
$\lambda_{\alpha}^{-1} \; h_{\alpha \beta} \;
\lambda_{\beta}$ and $h_{\alpha \beta}'$ generate all 
elements of $G$. The only difference between the bundles
$E$ and $E'$ lies in the assignment of coordinates, 
and the equivalence of such bundles is expressed by
(C.22). The careful reader may have noticed that in our
argument the coverings of the base space for $E$ and $E'$
have been taken to coincide. However, this restriction
is unnecessary. One may instead consider coordinates and
coverings given by $\left \{\phi_{\alpha},U_{\alpha}
\right \}$ for $E$, and by $\left \{\psi_{\alpha},V_{\alpha}
\right \}$ for $E'$. The equivalence of $E$ and $E'$ is then
defined by requiring that the homeomorphism 
$\phi_{\alpha} \circ \psi_{\beta}^{-1}(x)$ should
coincide with an element of the structure group $G$ for
$x \in U_{\alpha} \cap V_{\beta}$ (Nash and Sen 1983).

Besides the M\"{o}bius strip, the naturally occurring examples
of bundles are the tangent and cotangent bundles and the
frame bundle. The tangent bundle $T(M)$ is defined as the
collection of all tangent spaces $T_{p}(M)$, for all points
$p$ in the manifold $M$, i.e.
$$
T(M) \equiv \bigcup_{p \in M} (p,T_{p}(M)) .
\eqno ({\rm C}.23)
$$
By construction, the base space is $M$ itself, and the fibre
at $p \in M$ is the tangent space $T_{p}(M)$. Moreover, the
projection map $\pi: T(M) \rightarrow M$ associates to any
tangent vector $\in T_{p}(M)$ the point $p \in M$. Note
that, if $M$ is $n$-dimensional, the fibre at $p$ is an
$n$-dimensional vector space isomorphic to $R^{n}$. The
{\it local} product structure of $T(M)$ becomes evident if one
can construct a homeomorphism $\phi_{\alpha}:\pi^{-1}(U_{\alpha})
\rightarrow U_{\alpha} \times R^{n}$. Thus, we are expressing
$T(M)$ in terms of points of $M$ and tangent vectors at such points.
This is indeed the case since, for a tangent vector $V$ at $p$,
its expression in local coordinates is
$$
V = \left . b^{i}(p) 
{\partial \over \partial x^{i}} \right |_{p} .
\eqno ({\rm C}.24)
$$
Hence the desired $\phi_{\alpha}$ has to map $V$ into the pair
$\Bigr(p,b^{i}(p)\Bigr)$. Moreover, the structure group is
the general linear group $GL(n,R)$, whose action on elements
of the fibre should be viewed as the action of a matrix on a
vector.

The {\it frame} bundle of $M$ requires taking a total space
$B(M)$ as the set of all frames at all points in $M$. Such
(linear) frames $b$ at $x \in M$ are, of course, an ordered
set $\Bigr(b_{1},b_{2},...,b_{n}\Bigr)$ of basis vectors for
the tangent space $T_{x}(M)$. The projection 
$\pi: B(M) \rightarrow M$ acts by mapping a base $b$ into
the point of $M$ to which $b$ is attached. Denoting by $u$
an element of $GL(n,R)$, the $GL(n,R)$ action on $B(M)$ is
defined by
$$
\Bigr(b_{1},...,b_{n}\Bigr)u \equiv 
\Bigr(b_{j}u_{j1},...,b_{j}u_{jn}\Bigr) .
\eqno ({\rm C}.25)
$$
The coordinates for a differentiable structure on $B(M)$ 
are $\Bigr(x^{1},...,x^{n};u_{i}^{j}\Bigr)$,
where $x^{1},...,x^{n}$ are coordinate functions in a
coordinate neighbourhood $V \subset M$, while $u_{i}^{j}$
appear in the representation of the map
$$
\gamma: V \times GL(n,R) \rightarrow \pi^{-1}(V) ,
\eqno ({\rm C}.26)
$$
by means of the rule (Isham 1989)
$$
(x,u) \rightarrow \biggr(u_{1}^{j}(\partial_{j})_{x},
...,u_{n}^{j}(\partial_{j})_{x}\biggr) .
$$

To complete our introduction to fibre bundles, we now define
cross-sections, sub-bundles, vector bundles, and connections
on principal bundles, following Isham (1989).
\vskip 0.3cm
\noindent
(i) Cross-sections are very important from the point of
view of physical applications, since in classical field
theory the physical fields may be viewed as sections of
a suitable class of bundles. The idea is to deal with
functions defined on the base space and taking values
in the fibre of the bundle. Thus, given a bundle 
$(E,\pi,M)$, a {\it cross-section} is a map 
$s: M \rightarrow E$ such that the image of each point
$x \in M$ lies in the fibre $\pi^{-1}(x)$ over $x$:
$$
\pi \circ s = {\rm id}_{M} .
\eqno ({\rm C}.27)
$$
In other words, one has the projection map from $E$
to $M$, and the cross-section from $M$ to $E$, and
their composition yields the identity on the base 
space. In the particular case of a product bundle, a
cross-section defines a unique function
${\widehat s}: M \rightarrow F$ given by
$$
s(x)=(x,{\widehat s}(x)) , \;
\forall x \in M .
\eqno ({\rm C}.28)
$$
\vskip 0.3cm
\noindent
(ii) The advantage of introducing the sub-bundle $E'$
of a given bundle $E$ lies in the possibility to refer to
a mathematical structure less complicated than the original.
Let $(E,\pi,M)$ be a fibre bundle with fibre $F$.
A sub-bundle of $(E,\pi,M)$ is a sub-space of $E$ with
the extra property that it always contains complete fibres
of $E$, and hence is itself a fibre bundle. The formal
definition demands that the following conditions on
$(E',\pi',M')$ should hold:
$$
E' \subset E ,
\eqno ({\rm C}.29)
$$
$$
M' \subset M ,
\eqno ({\rm C}.30)
$$
$$
\pi' = \pi \mid_{E} .
\eqno ({\rm C}.31)
$$
In particular, if $T \equiv (E,\pi,M)$ is a sub-bundle of
the product bundle $(M \times F,{\rm pr}_{1},M)$, then
cross-sections of $T$ have the form $s(x)=(x,{\widehat s}(x))$,
where ${\widehat s}: M \rightarrow F$ is a function such that,
$\forall x \in M$, $(x,{\widehat s}(x)) \in E$. For example,
the tangent bundle $TS^{n}$ of the $n$-sphere $S^{n}$ may be
viewed as the sub-bundle of $S^{n} \times R^{n+1}$
(Isham 1989)
$$
E(TS^{n}) \approx \left \{(x,y) \in S^{n} \times R^{n+1}:
x \cdot y =0 \right \} .
\eqno ({\rm C}.32)
$$
Cross-sections of $TS^{n}$ are vector fields on the
$n$-sphere. It is also instructive to introduce the normal
bundle $\nu(S^{n})$ of $S^{n}$, i.e. the set of all vectors
in $R^{n+1}$ which are normal to points on $S^{n}$ 
(Isham 1989):
$$
E(\nu(S^{n})) \equiv
\left \{ (x,y) \in S^{n} \times R^{n+1}:
\exists k \in R: y=kx \right \} .
\eqno ({\rm C}.33)
$$
\vskip 0.3cm
\noindent
(iii) In the case of vector bundles, the fibres are 
isomorphic to a vector space, and the space of cross-sections
has the structure of a vector space. Vector bundles are
relevant for theoretical physics, since gauge theory may be
formulated in terms of vector bundles (Ward and Wells 1990),
and the space of cross-sections can replace the space of
functions on a manifold (although, in this respect, the 
opposite point of view may be taken). By definition, 
a $n$-dimensional real (resp. complex) vector bundle
$(E,\pi,M)$ is a fibre bundle in which each fibre is
isomorphic to a $n$-dimensional real (resp. complex)
vector space. Moreover, $\forall x \in M$, a neighbourhood
$U \subset M$ of $x$ exists, jointly with a {\it local}
trivialization $\rho: U \times R^{n} \rightarrow
\pi^{-1}(U)$ such that, $\forall y \in U$,
$\rho : \{ y \} \times R^{n} \rightarrow \pi^{-1}(y)$ is
a {\it linear map}. 

The simplest examples are the product space $M \times R^{n}$,
and the tangent and cotangent bundles of a manifold $M$.
A less trivial example is given by the normal bundle 
(cf. (C.33)). If $M$ is a $m$-dimensional sub-manifold of
$R^{n}$, its {\it normal bundle} is a $(n-m)$-dimensional
vector bundle $\nu(M)$ over $M$, with total space (Isham 1989)
$$
E(\nu(M)) \equiv
\left \{ (x,v) \in M \times R^{n}:
v \cdot w =0, \forall w \in T_{x}(M) \right \} ,
\eqno ({\rm C}.34)
$$
and projection map $\pi : E(\nu(M)) \rightarrow M$
defined by $\pi(x,v) \equiv x$. Last, but not least,
we mention the {\it canonical real line bundle}
$\gamma_{n}$ over the real projective space $RP^{n}$, with
total space
$$
E(\gamma_{n}) \equiv
\left \{ ([x],v) \in RP^{n} \times R^{n+1}:
v=\lambda \; x, \lambda \in R \right \} ,
\eqno ({\rm C}.35)
$$
where $[x]$ denotes the line passing through 
$x \in R^{n+1}$. The projection map
$\pi: E(\gamma_{n}) \rightarrow RP^{n}$ is defined by
the condition
$$
\pi ([x],v) \equiv [x] .
\eqno ({\rm C}.36)
$$
Its inverse is therefore the line in $R^{n+1}$ passing
through $x$. Note that $\gamma_{n}$ is a one-dimensional
vector bundle.
\vskip 0.3cm
\noindent
(iv) In Nash and Sen (1983), principal bundles are defined
by requiring that the fibre $F$ should be (isomorphic to)
the structure group. However, a more precise definition, such
as the one given in Isham (1989), relies on the theory of Lie
groups. Since it is impossible to describe such a theory
in a short appendix, we refer the reader to Isham (1989) and
references therein for the theory of Lie groups, and we limit
ourselves to the following definitions.
\vskip 0.3cm
A bundle $(E,\pi,M)$ is a $G$-bundle if $E$ is a right
$G$-space and if $(E,\pi,M)$ is isomorphic to the
bundle $(E,\sigma,E/G)$, where $E/G$ is the orbit space
of the $G$-action on $E$, and $\sigma$ is the usual
projection map. Moreover, if $G$ acts freely on $E$,
then $(E,\pi,M)$ is said to be a {\it principal}
$G$-bundle, and $G$ is the structure group of the 
bundle. Since $G$ acts freely on $E$ by hypothesis, each
orbit is homeomorphic to $G$, and hence one has a
fibre bundle with fibre $G$ (see earlier remarks).

To define connections in a principal bundle, with the
associated covariant differentiation, one has to look
for vector fields on the bundle space $P$ that point 
from one fibre to another. The first basic remark is
that the tangent space $T_{p}(P)$ at a point 
$p \in P$ admits a natural direct-sum decomposition 
into two sub-spaces $V_{p}(P)$ and $H_{p}(P)$, and the
connection enables one to obtain such a split of
$T_{p}(P)$. Hence the elements of $T_{p}(P)$ are uniquely
decomposed into a sum of components lying in
$V_{p}(P)$ and $H_{p}(P)$ by virtue of the connection.
The first sub-space, $V_{p}(P)$, is defined as
$$
V_{p}(P) \equiv \left \{t \in T_{p}(P):
\pi_{*}t=0 \right \} ,
\eqno ({\rm C}.37)
$$
where $\pi: P \rightarrow M$ is the projection map from
the total space to the base space. The elements of
$V_{p}(P)$ are, by construction, {\it vertical} vectors
in that they point along the fibre. The desired vectors,
which point away from the fibres, lie instead in the
horizontal sub-space $H_{p}(P)$. By definition, a {\it connection}
in a principal bundle $P \rightarrow M$ with group $G$ is a
smooth assignment, to each $p \in P$, of a {\it horizontal}
sub-space $H_{p}(P)$ of $T_{p}(P)$ such that
$$
T_{p}(P) \approx V_{p}(P) \oplus H_{p}(P) .
\eqno ({\rm C}.38)
$$
By virtue of (C.38), a connection is also called, within
this framework, a {\it distribution}. Moreover, the
decomposition (C.38) is required to be compatible with the
right action of $G$ on $P$.

The constructions outlined in this appendix are the first
step towards a geometric and intrinsic formulation of
gauge theories, and they are frequently applied also in
twistor theory (sections 5.1--5.3, 9.6 and 9.7). 
\vskip 100cm
\centerline {\bf APPENDIX D: Sheaf theory}
\vskip 1cm
\noindent
In chapter four we have given an elementary introduction
to sheaf cohomology. However, to understand the language of
section 9.6, it may be helpful to supplement our early
treatment by some more precise definitions. This is here
achieved by relying on Chern (1979).

The definition of a sheaf of Abelian groups involves two
topological spaces $\cal S$ and $M$, jointly with a map
$\pi: {\cal S} \rightarrow M$. The sheaf of Abelian groups
is then the pair $({\cal S},\pi)$ such that:
\vskip 0.3cm
\noindent
(i) $\pi$ is a local homeomorphism;
\vskip 0.3cm
\noindent
(ii) $\forall x \in M$, the set $\pi^{-1}(x)$, i.e. the 
{\it stalk} over $x$, is an Abelian group;
\vskip 0.3cm
\noindent
(iii) the group operations are continuous in the topology
of $\cal S$.
\vskip 0.3cm
Denoting by $U$ an open set of $M$, a {\it section} of the
sheaf $\cal S$ over $U$ is a continuous map 
$f: U \rightarrow S$ such that its composition with $\pi$
yields the identity (cf. appendix C). The set $\Gamma(U,{\cal S})$
of all (smooth) sections over $U$ is an Abelian group, since if
$f,g \in \Gamma(U,{\cal S})$, one can define $f-g$ by the
condition $(f-g)(x) \equiv f(x)-g(x), x \in U$. The {\it zero}
of $\Gamma(U,{\cal S})$ is the zero section assigning the zero
of the stalk $\pi^{-1}(x)$ to every $x \in U$.

The next step is the definition of {\it presheaf} of
Abelian groups over $M$. This is obtained on considering 
the homomorphism between sections over $U$ and sections
over $V$, for $V$ an open subset of $U$. More precisely,
by a presheaf of Abelian groups over $M$ we mean
(Chern 1979):
\vskip 0.3cm
\noindent
(i) a basis for the open sets of $M$;
\vskip 0.3cm
\noindent
(ii) an Abelian group $S_{U}$ assigned to each open set $U$
of the basis;
\vskip 0.3cm
\noindent
(iii) a homomorphism $\rho_{VU}:S_{U} \rightarrow S_{V}$ 
associated to each inclusion $V \subset U$, such that
$$
\rho_{WV} \; \rho_{VU}=\rho_{WU}
\; {\rm whenever} 
\; W \subset V \subset U .
$$
The sheaf is then obtained from the presheaf by a limiting
procedure (cf. chapter four).
For a given complex manifold $M$, the following sheaves play
a very important role (cf. section 9.6):
\vskip 0.3cm
\noindent
(i) The sheaf ${\cal A}_{pq}$ of germs of complex-valued
$C^{\infty}$ forms of type $(p,q)$. In particular, the sheaf
of germs of complex-valued $C^{\infty}$ functions is denoted
by ${\cal A}_{00}$.
\vskip 0.3cm
\noindent
(ii) The sheaf $C_{pq}$ of germs of complex-valued 
$C^{\infty}$ forms of type $(p,q)$, closed under the
operator ${\overline {\partial}}$. The sheaf of germs of
holomorphic functions (i.e. zero-forms) is denoted by
${\cal O}=C_{00}$. This is the most important sheaf in
twistor theory (as well as in the theory of complex manifolds,
cf. Chern (1979)).
\vskip 0.3cm
\noindent
(iii) The sheaf ${\cal O}^{*}$ of germs of nowhere-vanishing
holomorphic functions. The group operation is the 
multiplication of germs of holomorphic functions.
\vskip 0.3cm
Following again Chern (1979), we complete this brief review
by introducing {\it fine sheaves}. They are fine in that they
admit a partition of unity subordinate to {\it any} locally finite
open covering, and play a fundamental role in cohomology, since
the corresponding cohomology groups $H^{q}(M,{\cal S})$
vanish $\forall q \geq 1$. Partitions of unity of a sheaf of
Abelian groups, subordinate to the locally finite open
covering $\cal U$ of $M$, are a collection of sheaf
homomorphisms $\eta_{i}: {\cal S} \rightarrow {\cal S}$
such that:
\vskip 0.3cm
\noindent
(i) $\eta_{i}$ is the zero map in an open neighbourhood
of $M-U_{i}$;
\vskip 0.3cm
\noindent
(ii) $\sum_{i}\eta_{i}$ equals the identity map of the 
sheaf $({\cal S},\pi)$.
\vskip 0.3cm
The sheaf of germs of complex-valued $C^{\infty}$ forms
is indeed fine, while ${\cal C}_{pq}$ and the constant
sheaf are not fine.
\vskip 100cm
\parindent=0pt
\everypar{\hangindent=20pt \hangafter=1}
\centerline {\bf BIBLIOGRAPHY}
\vskip 1cm
[1] Abraham R. , Marsden J. E. and Ratiu T. (1983)
{\it Manifolds, Tensor Analysis, and Applications}
(London: Addison--Wesley).

[2] Aichelburg P. C. and Urbantke H. K. (1981) 
Necessary and Sufficient Conditions for Trivial Solutions
in Supergravity, {\it Gen. Rel. Grav.}
{\bf 13}, 817--828.

[3] Ashtekar A. (1988) {\it New Perspectives in Canonical
Gravity} (Naples: Bibliopolis).

[4] Ashtekar A. (1991) {\it Lectures on Non-Perturbative 
Canonical Gravity} (Singapore: World Scientific).

[5] Atiyah M. F. , Patodi V. K. and Singer I. M. (1976)
Spectral Asymmetry and Riemannian Geometry III,
{\it Math. Proc. Camb. Phil. Soc.} {\bf 79}, 71--99.

[6] Atiyah M. F. (1988) {\it Collected Works, Vol. 5: 
Gauge Theories} (Oxford: Clarendon Press).

[7] Avramidi I. G. and Esposito G. (1998a) New Invariants in
the One-Loop Divergences on Manifolds with Boundary,
{\it Class. Quantum Grav.} {\bf 15}, 281--297.

[8] Avramidi I. G. and Esposito G. (1998b) Lack of Strong
Ellipticity in Euclidean Quantum Gravity, {\it Class.
Quantum Grav.} {\bf 15}, 1141--1152. 

[9] Avramidi I. G. and Esposito G. (1999) Gauge Theories 
on Manifolds with Boundary, {\it Commun. Math. Phys.}
{\bf 200}, 495--543. 

[10] Bailey T. N. and Baston R. J. (1990) {\it Twistors in
Mathematics and Physics} (Cambridge: Cambridge 
University Press).

[11] Bailey T. N. , Eastwood M. G. , Gover A. R. and Mason L. J.
(1994) {\it Complex Analysis and the Funk Transform}
(preprint).

[12] Barvinsky A. O. , Kamenshchik A. Yu. , Karmazin I. P. and
Mishakov I. V. (1992a) One-Loop Quantum Cosmology: The 
Contributions of Matter Fields to the Wavefunction of
the Universe, {\it Class. Quantum Grav.}
{\bf 9}, L27--L32.

[13] Barvinsky A. O. , Kamenshchik A. Yu. and Karmazin I. P. (1992b)
One-Loop Quantum Cosmology: $\zeta$-Function Technique for the
Hartle--Hawking Wave Function of the Universe,
{\it Ann. Phys.} {\bf 219}, 201--242.

[14] Baston R. J. and Mason L. J. (1987) 
Conformal Gravity, the Einstein Equations and Spaces
of Complex Null Geodesics, {\it Class. Quantum
Grav.} {\bf 4}, 815--826.

[15] Bateman H. (1904) The Solution of Partial Differential
Equations by Means of Definite Integrals, 
{\it Proc. Lond. Math. Soc.} {\bf 1(2)}, 451--458.

[16] Bergmann P. G. and Smith G. J. (1991) Complex Phase Spaces
and Complex Gauge Groups in General Relativity, {\it Phys. Rev.}
{\bf D 43}, 1157--1161.

[17] Besse A. L. (1987) {\it Einstein Manifolds} (Berlin:
Springer-Verlag).

[18] Boyer C. P. , Finley J. D. and Plebanski J. F. (1978)
Complex General Relativity, $\cal H$ and ${\cal {HH}}$ Spaces:
a Survey, Communicaciones Tecnicas, Serie Naranja: Investigaciones,
Vol. {\bf 9}, No. 174.

[19] Breitenlohner P. and Freedman D. Z. (1982) 
Stability in Gauged Extended Supergravity,
{\it Ann. Phys.} {\bf 144}, 249--281.

[20] Buchdahl H. A. (1958) On the Compatibility of Relativistic
Wave Equations for Particles of Higher Spin in the Presence
of a Gravitational Field,
{\it Nuovo Cimento} {\bf 10}, 96--103.

[21] Charap J. M. and Nelson J. E. (1983) 
Surface Integrals and the Gravitational Action,
{\it J. Phys.} {\bf A 16}, 1661--1668.

[22] Chern S. S. (1979) {\it Complex Manifolds without
Potential Theory} (Berlin: Springer-Verlag).

[23] D'Eath P. D. (1984) Canonical Quantization of
Supergravity, {\it Phys. Rev.} {\bf D 29}, 2199--2219.

[24] D'Eath P. D. and Halliwell J. J. (1987) 
Fermions in Quantum Cosmology,
{\it Phys. Rev.} {\bf D 35}, 1100--1123.

[25] D'Eath P. D. and Esposito G. (1991a) 
Local Boundary Conditions for the Dirac Operator and One-Loop
Quantum Cosmology, {\it Phys. Rev.} {\bf D 43}, 3234--3248.

[26] D'Eath P. D. and Esposito G. (1991b) 
Spectral Boundary Conditions in One-Loop Quantum Cosmology,
{\it Phys. Rev.} {\bf D 44}, 1713--1721.

[27] Deser S. and Zumino B. (1976) Consistent Supergravity, 
{\it Phys. Lett.} {\bf B 62}, 335--337.

[28] Dunajski M. and Mason L. J. (1997) A Recursion Operator
for ASD Vacuums and Zero-Rest-Mass Fields on ASD Backgrounds,
{\it Twistor Newsletter} n. {\bf 33}, 24--29.

[29] Dunajski M. (1999) The Twisted Photon Associated to 
Hyper-Hermitian Four-Manifolds, {\it J. Geom. Phys.} {\bf 30},
266--281.

[30] Eastwood M. (1987) The Einstein Bundle of a Non-Linear
Graviton, {\it Twistor Newsletter} n. {\bf 24}, 3--4.

[31] Eastwood M. (1990) The Penrose Transform, 
in {\it Twistors in Mathematics and Physics},
eds. T. N. Bailey and R. J. Baston (Cambridge: Cambridge 
University Press) 87--103.

[32] Esposito G. (1992) Mathematical Structures of Space-Time,
{\it Fortschr. Phys.} {\bf 40}, 1--30.

[33] Esposito G. (1993) $\alpha$-Surfaces for Complex 
Space-Times with Torsion, 
{\it Nuovo Cimento} {\bf B 108}, 123--125.

[34] Esposito G. (1994) {\it Quantum Gravity, Quantum Cosmology
and Lorentzian Geometries}, Lecture Notes in Physics, New
Series m: Monographs, Vol. m12, second corrected and
enlarged edition (Berlin: Springer--Verlag).

[35] Esposito G. and Kamenshchik A. Yu. (1994) 
Coulomb Gauge in One-Loop Quantum Cosmology,
{\it Phys. Lett.} {\bf B 336}, 324--329.

[36] Esposito G. , Kamenshchik A. Yu. , Mishakov I. V.
and Pollifrone G. (1994) Gravitons in One-Loop Quantum
Cosmology: Correspondence Between Covariant and Non-Covariant
Formalisms, {\it Phys. Rev.} {\bf D 50}, 6329--6337.

[37] Esposito G. , Morales-T\'ecotl H. and Pollifrone G. (1994)
Boundary Terms for Massless Fermionic Fields,
{\it Found. Phys. Lett.} {\bf 7}, 303--308.

[38] Esposito G. and Pollifrone G. (1994) 
Spin-Raising Operators and Spin-${3\over 2}$ Potentials
in Quantum Cosmology, {\it Class. Quantum
Grav.} {\bf 11}, 897--903; Twistors ans Spin-${3\over 2}$
Potentials in Quantum Gravity, in {\it Twistor Theory}, ed. S.
Huggett (New York: Marcel Dekker) 35--54.

[39] Esposito G. (1995) {\it Complex General Relativity},
Fundamental Theories of Physics, Vol. 69, First Edition
(Dordrecht: Kluwer).

[40] Esposito G. , Gionti G. , Kamenshchik A. Yu. , 
Mishakov I. V. and Pollifrone G. (1995) Spin-3/2 Potentials
in Backgrounds with Boundary, {\it Int. J. Mod. Phys.}
{\bf D 4}, 735--747.

[41] Esposito G. , Gionti G. and Stornaiolo C. (1995) 
Space-Time Covariant Form of Ashtekar's Constraints, 
{\it Nuovo Cimento} {\bf B 110}, 1137--1152.

[42] Esposito G. and Stornaiolo C. (1995) Boundary Terms
in Complex General Relativity, {\it Class. Quantum Grav.}
{\bf 12}, 1733--1738.

[43] Esposito G. and Pollifrone G. (1996) Twistors in
Conformally Flat Einstein Four-Manifolds, {\it Int. J.
Mod. Phys.} {\bf D 5}, 481--493.

[44] Esposito G. , Kamenshchik A. Yu. and Pollifrone G. 
(1997) {\it Euclidean Quantum Gravity on Manifolds with
Boundary}, Fundamental Theories of Physics, Vol. 85
(Dordrecht: Kluwer).

[45] Finley J. D. and Plebanski J. F. (1976) Further Heavenly
Metrics and Their Symmetries, {\it J. Math. Phys.} 
{\bf 17}, 585--596.

[46] Finley J. D. and Plebanski J. F. (1981) All Algebraically
Degenerate $\cal H$ Spaces, via ${\cal {HH}}$ Spaces,
{\it J. Math. Phys.} {\bf 22}, 667--674.

[47] Frauendiener J. (1994) Another View at the
Spin-3/2 Equation, {\it Twistor Newsletter} 
n. {\bf 37}, 7--9.

[48] Frauendiener J. (1995) On Spin-3/2 Systems in Ricci-Flat
Space-Times, {\it J. Math. Phys.} {\bf 36}, 3012--3022.

[49] Frauendiener J. , Ghosh J. and Newman E. T. (1996)
Twistors and the Asymptotic Behaviour of Massless Spin-3/2
Fields, {\it Class. Quantum Grav.} {\bf 13}, 461--480.

[50] Frauendiener J. and Sparling G. A. J. (1999) On a Class
of Consistent Linear Higher Spin Equations on Curved Manifolds,
{\it J. Geom. Phys.} {\bf 30}, 54--101.

[51] Gibbons G. W. and Hawking S. W. (1993) {\it Euclidean
Quantum Gravity} (Singapore: World Scientific).

[52] Hall G. S. , Hickman M. S. and McIntosh C. B. G. (1985)
Complex Relativity and Real Solutions II: Classification of
Complex Bivectors and Metric Classes, {\it Gen. Rel. Grav.}
{\bf 17}, 475--491.

[53] Hansen R. O. , Newman, E. T. , Penrose R. and Tod K. P.
(1978) The Metric and Curvature Properties of $\cal H$ Space,
{\it Proc. Roy. Soc. Lond.}, {\bf A 363}, 445--468.

[54] Hartle J. B. and Hawking S. W. (1983) Wave Function 
of the Universe, {\it Phys. Rev.} {\bf D 28}, 2960--2975.

[55] Hawking S. W. and Ellis G. F. R. (1973) {\it The Large-Scale
Structure of Space-Time} (Cambridge: Cambridge University
Press).

[56] Hawking S. W. (1979) The Path-Integral Approach to
Quantum Gravity, in {\it General Relativity, an Einstein
Centenary Survey}, eds. S. W. Hawking and W. Israel
(Cambridge: Cambridge University Press) 746--789.

[57] Hawking S. W. (1983) The Boundary Conditions for Gauged
Supergravity, {\it Phys. Lett.} {\bf B 126}, 175--177.

[58] Hawking S. W. (1984) The Quantum State of the Universe,
{\it Nucl. Phys.} {\bf B 239}, 257--276.

[59] Hayward G. (1993) Gravitational Action for Space-Times
with Non-Smooth Boundaries, 
{\it Phys. Rev.} {\bf D 47}, 3275--3280.

[60] Hickman M. S. and McIntosh C. B. G. (1986a) Complex
Relativity and Real Solutions. III. Real Type-$N$ 
Solutions from Complex $N \otimes N$ Ones, {\it Gen. Rel. Grav.}
{\bf 18}, 107--136.

[61] Hickman M. S. and McIntosh C. B. G. (1986b) Complex
Relativity and Real Solutions. IV. Perturbations of Vacuum
Kerr--Schild Spaces, {\it Gen. Rel. Grav.} {\bf 18},
1275--1290.

[62] Hitchin N. J. (1979) Polygons and Gravitons, 
{\it Proc. Camb. Phil. Soc.} {\bf 85}, 465--476.

[63] Huggett S. A. (1985) The Development of Ideas in Twistor
Theory, {\it Int. J. Theor. Phys.} {\bf 24}, 391--400.

[64] Huggett S. A. and Tod K. P. (1985) {\it An Introduction to
Twistor Theory}, second enlarged edition published in 1994 
(Cambridge: Cambridge University Press).

[65] Isham C. J. (1989) {\it Modern Differential Geometry
for Physicists} (Singapore: World Scientific).

[66] Izquierdo J. M. and Townsend P. K. (1995) Supersymmetric
Spacetimes in (2+1) adS-Supergravity Models, 
{\it Class. Quantum Grav.} {\bf 12}, 895--924.

[67] Kamenshchik A. Yu. and Mishakov I. V. (1992) 
$\zeta$-Function Technique for Quantum Cosmology:
The Contributions of Matter Fields to the Hartle--Hawking
Wave Function of the Universe,
{\it Int. J. Mod. Phys.} {\bf A 7}, 3713--3746.

[68] Kamenshchik A. Yu. and Mishakov I. V. (1993) 
Fermions in One-Loop Quantum Cosmology,
{\it Phys. Rev.} {\bf D 47}, 1380--1390.

[69] Kamenshchik A. Yu. and Mishakov I. V. (1994) 
Fermions in One-Loop Quantum Cosmology II. The Problem of
Correspondence Between Covariant and Non-Covariant
Formalisms, {\it Phys. Rev.} {\bf D 49}, 816--824.

[70] Ko M. , Newman E. T. and Penrose R. (1977) The K\"{a}hler
Structure of Asymptotic Twistor Space, {\it J. Math. Phys.}
{\bf 18}, 58--64.

[71] Ko M. , Ludvigsen M. , Newman E. T. and Tod K. P. (1981)
The Theory of $\cal H$-Space, {\it Phys. Rep.} {\bf 71}, 51--139.

[72] Kozameh C. N. , Newman E. T. and Tod K. P. (1985)
Conformal Einstein Spaces,
{\it Gen. Rel. Grav.} {\bf 17}, 343--352.

[73] Law P. R. (1985) Twistor Theory and the Einstein Equations,
{\it Proc. Roy. Soc. London} {\bf A 399}, 111--134.

[74] Le Brun C. R. (1985) Ambi-Twistors and Einstein's
Equations, {\it Class. Quantum Grav.}
{\bf 2}, 555--563.

[75] Le Brun C. R. (1986) Thickenings and Gauge Fields, 
{\it Class. Quantum Grav.} {\bf 3}, 1039--1059.

[76] Le Brun C. R. (1990) Twistors, Ambitwistors and
Conformal Gravity, in {\it Twistors in Mathematics
and Physics}, eds. T. N. Bailey and R. J. Baston
(Cambridge: Cambridge University Press) 71--86.

[77] Le Brun C. R. (1991) Thickenings and Conformal Gravity, 
{\it Commun. Math. Phys.} {\bf 139}, 1--43.

[78] Lewandowski J. , Nurowski P. and Tafel J. (1990) Einstein's
Equations and Realizability of CR Manifolds, 
{\it Class. Quantum Grav.} {\bf 7}, L241--L246.

[79] Lewandowski J. , Nurowski P. and Tafel J. (1991) Algebraically
Special Solutions of the Einstein Equations with Pure Radiation
Fields, {\it Class. Quantum Grav.} {\bf 8}, 493--501.

[80] Lewandowski J. (1991) Twistor Equation in a Curved
Space-Time, {\it Class. Quantum Grav.} {\bf 8}, L11--L17.

[81] Luckock H. C. and Moss I. G. (1989) The Quantum
Geometry of Random Surfaces and Spinning Membranes,
{\it Class. Quantum Grav.} {\bf 6}, 1993--2027.

[82] Manin Y. I. (1988) {\it Gauge Field Theory and Complex
Geometry} (Berlin: Springer-Verlag).

[83] Mason L. J. and Hughston L. P. (1990) {\it Further Advances
in Twistor Theory, Vol. I: The Penrose Transform and Its
Applications} (Harlow: Longman Scientific and Technical).

[84] Mason L. J. and Penrose R. (1994) 
Spin-${3\over 2}$ Fields and Local Twistors, 
{\it Twistor Newsletter} n. {\bf 37}, 1--6.

[85] Mason L. J. and Woodhouse N. M. J. (1996) {\it Integrability,
Self-Duality, and Twistor Theory}, London Mathematical Society
Monographs, New Series, Vol. 15 (Oxford: Clarendon Press).

[86] McIntosh C. B. G. and Hickman M. S. (1985) Complex
Relativity and Real Solutions. I: Introduction, 
{\it Gen. Rel. Grav.} {\bf 17}, 111--132.

[87] McIntosh C. B. G. , Hickman M. S. and Lun A. W.--C. (1988)
Complex Relativity and Real Solutions. V. The Flat Space
Background, {\it Gen. Rel. Grav.} {\bf 20}, 647--657.

[88] Moss I. G. and Poletti S. (1990) Boundary Conditions
for Quantum Cosmology, {\it Nucl. Phys.} {\bf B 341}, 155--166.

[89] Nash C. and Sen S. (1983) {\it Topology and Geometry
for Physicists} (London: Academic Press).

[90] Newman E. T. (1976) Heaven and Its Properties, 
{\it Gen. Rel. Grav.} {\bf 7}, 107--111.

[91] Nurowski P. (1997) Twistor Bundles, Einstein Equations and
Real Structures, {\it Class. Quantum Grav.} {\bf 14},
A261--A290.

[92] Park Q. H. (1990) Self-Dual Gravity as a Large-N Limit
of the 2D Non-Linear Sigma Model, 
{\it Phys. Lett.} {\bf B 238}, 287--290.

[93] Park Q. H. (1991) Self-Dual Yang--Mills (+Gravity)
as a 2D Sigma Model,
{\it Phys. Lett.} {\bf B 257}, 105--110.

[94] Penrose R. (1960) A Spinor Approach to General Relativity,
{\it Ann. Phys.} {\bf 10}, 171--201.

[95] Penrose R. (1967) Twistor Algebra, 
{\it J. Math. Phys.} {\bf 8}, 345--366.

[96] Penrose R. (1968) Twistor Quantization and Curved
Space-Time, {\it Int. J. Theor. Phys.} {\bf 1}, 61--99.

[97] Penrose R. and MacCallum M. A. H. (1973) 
Twistor Theory: an Approach to the Quantization of Fields
and Space-Time, {\it Phys. Rep.} {\bf 6 C}, 241--316.

[98] Penrose R. (1974) Relativistic Symmetry Groups, 
in {\it Group Theory in Nonlinear Problems},
ed. A. O. Barut (Dordrecht: Reidel Publishing Company) 1--58.

[99] Penrose R. (1975) Twistor Theory, Its Aims and
Achievements, in {\it Quantum Gravity, an Oxford Symposium},
eds. C. J. Isham, R. Penrose and D. W. Sciama (Oxford:
Clarendon Press) 268--407.

[100] Penrose R. (1976a) Non-Linear Gravitons and Curved
Twistor Theory, {\it Gen. Rel. Grav.} {\bf 7}, 31--52.

[101] Penrose R. (1976b) The Non-Linear Graviton, 
{\it Gen. Rel. Grav.} {\bf 7}, 171--176.

[102] Penrose R. (1977) The Twistor Programme,
{\it Rep. Math. Phys.} {\bf 12}, 65--76.

[103] Penrose R. (1980) On the Twi\-stor De\-scrip\-ti\-ons of
Mass\-less Fiel\-ds, in {\it Com\-plex-Ma\-ni\-fold Tech\-ni\-ques in
Theoretical Physics}, eds. D. Lerner and P. Sommers
(London: Pitman) 55--91.

[104] Penrose R. and Ward R. S. (1980) Twistors for Flat
and Curved Space-Time, in {\it General Relativity
and Gravitation, Vol. II}, ed. A. Held (New York: Plenum
Press) 283--328.

[105] Penrose R. (1981) Some Remarks on Twistors and 
Curved-Space Quantization, in {\it Quantum Gravity 2, A Second 
Oxford Symposium}, eds. C. J. Isham, R. Penrose and D. W. Sciama
(Oxford: Clarendon Press) 578--592.

[106] Penrose R. (1983) Spinors and Torsion in General
Relativity, {\it Found. Phys.} {\bf 13}, 325--339.

[107] Penrose R. and Rindler W. (1984) {\it Spinors and Space-Time,
Vol. I: Two-Spinor Calculus and Relativistic Fields} 
(Cambridge: Cambridge University Press).

[108] Penrose R. and Rindler W. (1986) {\it Spinors and Space-Time,
Vol. II: Spinor and Twistor Methods in Space-Time Geometry}
(Cambridge: Cambridge University Press).

[109] Penrose R. (1986) Twistors in General Relativity, 
in {\it General Relativity and Gravitation},
ed. M. A. H. MacCallum (Cambridge: Cambridge University Press)
158--176.

[110] Penrose R. (1987) On the Origins of Twistor Theory, 
in {\it Gravitation and Geometry}, eds. 
W. Rindler and A. Trautman (Naples: Bibliopolis) 341--361.

[111] Penrose R. (1990) Twistor Theory for Vacuum Space-Times:
a New Approach, {\it Twistor Newsletter} n. {\bf 31}, 6--8.

[112] Penrose R. (1991a) Twistors as Charges for Spin
${3\over 2}$ in Vacuum, {\it Twistor Newsletter} 
n. {\bf 32}, 1--5.

[113] Penrose R. (1991b) Twistors as Spin-${3\over 2}$
Charges Continued: $SL(3,C)$ Bundles, {\it Twistor Newsletter} 
n. {\bf 33}, 1--6.

[114] Penrose R. (1991c) Twistors as Spin-${3\over 2}$ Charges, 
in {\it Gravitation and Modern Cosmology},
eds. A. Zichichi, V. de Sabbata and N. S\'anchez (New York:
Plenum Press) 129--137.

[115] Penrose R. (1994) Twistors and the Einstein Equations, 
in {\it Twistor Theory}, ed. S. Huggett
(New York: Marcel Dekker) 145--157.
 
[116] Penrose R. (1997) Googly Maps as Forms, 
{\it Twistor Newsletter} n. {\bf 43}, 1--4. 

[117] Perry M. J. (1984) The Positive Mass Theorem and Black
Holes, in {\it Asymptotic Behaviour of Mass and Space-Time
Geometry}, ed. F. J. Flaherty (Berlin: Springer--Verlag) 31--40.

[118] Plebanski J. F. (1975) Some Solutions of Complex Einstein
Equations, {\it J. Math. Phys.} {\bf 16}, 2395--2402.

[119] Plebanski J. F. and Hacyan S. (1975) Null Geodesic 
Surfaces and Goldberg--Sachs Theorem in Complex Riemannian 
Spaces, {\it J. Math. Phys.} {\bf 16}, 2403--2407.

[120] Plebanski J. F. and Schild A. (1976) Complex Relativity
and Double KS Metrics, {\it Nuovo Cimento} {\bf B 35}, 35--53.

[121] Plebanski J. F. and Przanowski M. (1994) Evolution
Hyperheavenly Equations, {\it J. Math. Phys.} 
{\bf 35}, 5990--6000.

[122] Plebanski J. F. and Garcia-Compean H. (1995a) A Note
on Weak $\cal H$ Spaces, Heavenly Equations and Their 
Evolution Weak Heavenly Equations, {\it Acta Phys. Polon.}
{\bf B 26}, 3--18.

[123] Plebanski J. F. and Garcia-Compean H. (1995b) A
$Q$-Deformed Version of the Heavenly Equations, {\it Int. J.
Mod. Phys.} {\bf A 10}, 3371--3379.

[124] Poletti S. (1990) Does the Boundary Contribution to the
One-Loop $\beta$-Function Vanish for Gauged Supergravity? 
{\it Phys. Lett.} {\bf B 249}, 249--254.

[125] Radon J. (1917) \"{U}ber die Bestimmung von 
Funktionen durch ihre Integralwerte l\"{a}ngs gewisser
Mannigfaltigkeiten, {\it S\"{a}chs. Akad. Wiss. Leipzig,
Math.-Nat. Kl.} {\bf 69}, 262--277.

[126] Reed M. and Simon B. (1975) {\it Methods of Modern
Mathematical Physics, Volume Two: Fourier Analysis, 
Self-Adjointness} (New York: Academic Press).

[127] Ruse H. S. (1937) On the Geometry of Dirac's Equations
and Their Expression in Tensor Form, {\it Proc. Roy. Soc.
Edinburgh} {\bf 57}, 97--127.

[128] Siklos S. T. C. (1985) Lobatchevski Plane Gravitational
Waves, in {\it Galaxies, Axisymmetric Systems and Relativity},
ed. M. A. H. MacCallum (Cambridge: Cambridge University Press)
247--274.

[129] Sparling G. A. J. and Tod K. P. (1981) An Example of
an $\cal H$ Space, {\it J. Math. Phys.} {\bf 22}, 331--332.

[130] Stewart J. M. (1991) {\it Advanced General Relativity}
(Cambridge: Cambridge University Press).

[131] Tod K. P. (1977) Some Symplectic Forms Arising in
Twistor Theory, {\it Rep. Math. Phys.} {\bf 11}, 339--346.

[132] Tod K. P. (1980) Curved Twistor Spaces and $\cal H$-Space,
{\it Surveys in High Energy Physics} {\bf 1}, 299--312.

[133] Tod K. P. and Winicour J. (1980) Singularities at Real 
Points of $\cal H$ Space, {\it Gen. Rel. Grav.} {\bf 12},
99--108.

[134] Tod K. P. (1983) All Metrics Admitting Supercovariantly
Constant Spinors, {\it Phys. Lett.} {\bf B 121}, 241--244.

[135] Tod K. P. (1995) More on Supercovariantly Constant 
Spinors, {\it Class. Quantum Grav.} {\bf 12}, 1801--1820.

[136] Tod K. P. (1996) The Rarita--Schwinger Equation for
Einstein--Maxwell Space-Times, {\it Twistor Newsletter}
{\bf 40}, 20--21.

[137] Tod K. P. and Dunajski M. (1997) `Special' Einstein--Weyl
Spaces from the Heavenly Equation, {\it Twistor Newsletter}
n. {\bf 33}, 13.

[138] Torres del Castillo G. F. (1989) Debye Potentials for
Rarita--Schwinger Fields in Curved Space-Times, 
{\it J. Math. Phys.} {\bf 30}, 1323--1328.

[139] Torres del Castillo G. F. and Silva--Ortigoza G. (1990)
Rarita--Schwinger Fields in the Kerr Geometry, 
{\it Phys. Rev.} {\bf D 42}, 4082--4086.

[140] Torres del Castillo G. F. and Silva--Ortigoza G. (1992)
Spin-3/2 Perturbations of the Kerr--Newman Solution, 
{\it Phys. Rev.} {\bf D 46}, 5395--5398.

[141] Townsend P. K. (1977) Cosmological Constant in
Supergravity, {\it Phys. Rev.} {\bf D 15}, 2802--2805.

[142] Veblen O. (1933) Geometry of Two-Component Spinors,
{\it Proc. Nat. Acad. Sci.} {\bf 19}, 462--474.

[143] Ward R. S. (1977) On Self-Dual Gauge Fields,
{\it Phys. Lett.} {\bf 61A}, 81--82.

[144] Ward R. S. (1978) A Class of Self-Dual Solutions 
of Einstein's Equations, 
{\it Proc. Roy. Soc. London} {\bf A 363}, 289--295.

[145] Ward R. S. (1979) The Twisted Photon: Massless
Fields as Bundles, in {\it Advances in Twistor Theory},
eds. L. P. Hughston and R. S. Ward 
(London: Pitman) 132--135.

[146] Ward R. S. (1980a) The Self-Dual Yang--Mills and
Einstein Equations, in {\it Complex-Manifold Techniques in 
Theoretical Physics}, eds. D. Lerner and P. Sommers
(London: Pitman) 12--34.

[147] Ward R. S. (1980b) Self-Dual Space-Times with
Cosmological Constant, {\it Commun. Math. Phys.} {\bf 78}, 1--17.

[148] Ward R. S. (1981a) Ans\"{a}tze for Self-Dual Yang--Mills
Fields, {\it Commun. Math. Phys.} {\bf 80}, 563--574.

[149] Ward R. S. (1981b) The Twistor Approach to Differential
Equations, in {\it Quantum Gravity 2, A Second 
Oxford Symposium}, eds. C. J. Isham, R. Penrose and D. W.
Sciama (Oxford: Clarendon Press) 593--610.

[150] Ward R. S. and Wells R. O. (1990) {\it Twistor Geometry and 
Field Theory} (Cambridge: Cambridge University Press).

[151] Witten E. (1978) An Interpretation of Classical Yang--Mills
Theory, {\it Phys. Lett.} {\bf B 77}, 394--398.

[152] Woodhouse N. M. J. (1985) Real Methods in Twistor
Theory, {\it Class. Quantum Grav.}
{\bf 2}, 257--291.

[153] Yasskin P. B. and Isenberg J. A. (1982) 
Non-Self-Dual Non-Linear Gravitons,
{\it Gen. Rel. Grav.} {\bf 14}, 621--627.

[154] Yasskin P. B. (1987) An Ambitwistor Approach for
Gravity, in {\it Gravitation and Geometry}, eds. W. 
Rindler and A. Trautman (Naples: Bibliopolis) 477--495.

[155] York J. W. (1986) Boundary Terms in the Action Principles
of General Relativity, 
{\it Found. Phys.} {\bf 16}, 249--257.
\vskip 100cm
\noindent
{\bf DEDICATION.} The present paper is dedicated to Michela Foa.
\bye